%% file: main.tex
\documentclass[11pt,english]{article}
\pdfoutput=1
\usepackage[T1]{fontenc}
\usepackage{amssymb}
\usepackage{amsmath}
\usepackage[unicode=true,
bookmarks=true,bookmarksnumbered=false,bookmarksopen=false,
 breaklinks=false,pdfborder={0 0 1},backref=false,colorlinks=true]
 {hyperref}
\usepackage[hang,flushmargin]{footmisc}
\usepackage{relsize}
\usepackage{color}

\usepackage{amsfonts}
\usepackage{graphicx}
\usepackage{longtable}

\usepackage{lineno}
\usepackage{tcolorbox}
\tcbuselibrary{breakable}
\usepackage{placeins}

\DeclareFontShape{OT1}{cmr}{mx}{n}%
{<->cmr10}{}

\usepackage{tocloft}
\usepackage{etoolbox}

\hypersetup{
    colorlinks=true,
    linkcolor=blue,  
    urlcolor=blue,
    citecolor=blue
}

\makeatletter
\pretocmd{\l@section}{\hypersetup{linkcolor=black}}{}{}
\apptocmd{\l@section}{\hypersetup{linkcolor=blue}}{}{}
\pretocmd{\l@subsection}{\hypersetup{linkcolor=black}}{}{}
\apptocmd{\l@subsection}{\hypersetup{linkcolor=blue}}{}{}
\makeatother

\usepackage{amsmath,amssymb,amsthm,mathtools}
\numberwithin{equation}{section}
\usepackage{tikz-cd}
\let\sl\relax

\usepackage[normalem]{ulem}
\definecolor{JBDComment}{rgb}{0.50,0.20,0.65}

\definecolor{CCComment}{rgb}{0.00,0.50,0.30}

\definecolor{JPMComment}{rgb}{0.85,0.33,0.10}

\definecolor{RTComment}{rgb}{0.86,0.10,0.10}

\newcommand{\AuthorEdits}[1]{%
  \expandafter\newcommand\csname#1replace\endcsname[2]{\textcolor{#1Comment}{\ifstrempty{##1}{##2}{\sout{##1}\ifstrempty{##2}{}{ ##2}}}}%
}
\forcsvlist{\AuthorEdits}{JBD,CC,JPM,RT}

\newcommand{\Hom}[0]{\text{Hom}}
\newcommand{\C}[0]{\mathcal C}
\newcommand{\id}[0]{\textup{id}}

\newcommand{\zee}[1]{\mathbb Z_{#1}}

\newcommand{\sl}[1]{\mathrm{SL}(2,#1)}
\newcommand{\mods}[0]{\mathfrak s}
\newcommand{\modt}[0]{\mathfrak t}
\newcommand{\irr}[0]{\text{Irr}}
\newcommand{\Uone}{\text{U}(1)}
\newcommand{\sVec}{\text{sVec}}
\newcommand{\SU}{\text{SU}}
\newcommand{\Spin}{\text{Spin}}
\newcommand{\Sp}{\text{Sp}}

\newcommand{\E}[1]{e^{2\pi i #1}}

\usepackage[
  backend=biber,
  style=phys,
  biblabel=brackets,
  sorting=none,
  eprint=true,
  doi=true,
  url=false,
  isbn=false,
  giveninits=true,
  maxbibnames=10,
]{biblatex}
\usepackage{subcaption}
\usepackage{cleveref}
\crefname{figure}{Fig.}{Figs.}
\Crefname{figure}{Fig.}{Figs.}
\crefname{equation}{Eq.}{Eqs.}
\Crefname{equation}{Eq.}{Eqs.}
\crefname{section}{Section}{Sections}
\Crefname{section}{Section}{Sections}
\crefname{subsection}{Section}{Sections}
\Crefname{subsection}{Section}{Sections}
\crefname{appendix}{Appendix}{Appendices}
\Crefname{appendix}{Appendix}{Appendices}
\AddToHook{cmd/appendix/after}{%
  \crefalias{subsection}{appendix}%
  \crefalias{subsubsection}{appendix}%
}
\AtBeginDocument{\let\LocalAliasLabel\label}

\input{figures/tikz-diagrams/rigidity-diagrams.tex}
\input{figures/tikz-diagrams/quantum-dimension-diagrams.tex}
\input{figures/tikz-diagrams/f-move-diagrams.tex}
\input{figures/tikz-diagrams/pentagon-diagrams.tex}
\input{figures/tikz-diagrams/r-move-diagrams.tex}
\input{figures/tikz-diagrams/monodromy-diagrams.tex}
\input{figures/tikz-diagrams/hexagon-diagrams.tex}
\input{figures/tikz-diagrams/ribbon-twist-diagrams.tex}

\input{figures/tikz-diagrams/balancing-isotopy-diagrams.tex}

\input{figures/tikz-diagrams/frobenius-schur-diagrams.tex}
\input{figures/tikz-diagrams/s-matrix-diagrams.tex}
\input{figures/tikz-diagrams/fermion-grading-diagrams.tex}

\begin{document}
\hypersetup{citecolor=blue,linkcolor=blue,urlcolor=blue}
\interfootnotelinepenalty=10000
\baselineskip=18pt

\vspace*{2cm}
\thispagestyle{empty}

\begin{center}
{\fontsize{25pt}{0pt}  \bf
The Anyon Atlas
}


\bigskip\vspace{1cm}{
{\large Jonah Berean-Dutcher$^{\,1,\,2}$, Clay C\'ordova$^{\,3}$, Julio Parra-Martinez$^{\,2}$, and Richard Tuddenham$^{\,3}$}
} \\[7mm]
{$^{1}$\textit{University of British Columbia, Vancouver, V6T 1Z1, Canada} \\
$^{2}$\textit{Institut des Hautes \'Etudes Scientifiques, 91440 Bures-sur-Yvette, France} \\
$^{3}$\textit{Leinweber Institute for Theoretical Physics  \& Enrico Fermi Institute, University of Chicago.}}
\end{center}

\bigskip
\centerline{\large\bf Abstract}
\begin{quote} \small
We classify possible topological quantum field theories in 2+1 dimensions with up to 22 anyons in the bosonic case and up to 26 in the fermionic case, through the modular data of modular and supermodular tensor categories. We construct such data via congruence representations of $\sl{\zee{}}$ using a numerical optimization algorithm that scales to high rank. We chart the relations between the resulting theories, including anyon condensations and Chern--Simons realizations. We find examples of modular data that do not appear to arise from any Chern--Simons theory by any sequence of condensations of anyons or gaugings of ordinary global symmetries. These examples give qualitatively new topological quantum field theories, which go beyond any notion of gauge theory. For examples in this class with 11 and 12 anyon types, we solve the pentagon and hexagon equations explicitly, showing that these modular data are realized by full unitary topological quantum field theories. We also introduce \href{http://www.AnyonAtlas.org}{\texttt{AnyonAtlas.org}}, an online resource for exploring the data.
\end{quote}
\setcounter{footnote}{0}

\vfill
{\small \textit{E-mail:} \href{mailto:jbd@phas.ubc.ca}{\texttt{jbd@phas.ubc.ca}}, \href{mailto:clayc@uchicago.edu}{\texttt{clayc@uchicago.edu}}, \href{mailto:julio@ihes.fr}{\texttt{julio@ihes.fr}}, \href{mailto:rtuddenh@uchicago.edu}{\texttt{rtuddenh@uchicago.edu}}}\par\medskip
\today
\newpage
\tableofcontents

\flushbottom

\section{Introduction}
\label{sec:intro}


Anyons are quasiparticles in two spatial dimensions with exchange statistics that are neither bosonic nor fermionic~\cite{Leinaas:1977fm, Wilczek:1982wy}. They are the excitations of topologically ordered phases of matter, whose long-distance description is believed to be captured by a topological quantum field theory (TQFT) in 2+1 dimensions. These TQFTs play a central role in many areas of physics and mathematics, from condensed matter and quantum information to high-energy physics and low-dimensional topology. In condensed matter physics, fractional quantum Hall states are the best-established examples of topological order~\cite{Laughlin:1983fy, Halperin:1984fn, Arovas:1984qr, Moore:1991ks}, and experiments have measured anyon statistics and realized topological order in moiré materials and on quantum processors~\cite{Bartolomei:2020qfr, Nakamura:2020mok, Cai:2023signatures, Park:2023observation, Lu:2023zow, Andersen:2022xmz, Iqbal:2023wvm}. In quantum information, braiding non-abelian anyons is the basis of fault-tolerant topological quantum computation~\cite{Kitaev:1997wr, Freedman:2001eqc, Nayak:2008zza}. In high-energy physics, Chern--Simons theories are related to rational conformal field theories~\cite{wittenQuantumFieldTheory1989, Moore:1988qv}, 2+1d TQFTs describe infrared phases of gauge theories and enter their dualities~\cite{Komargodski:2017keh, seibergDualityWeb2+12016}, and, as symmetry TFTs, they encode finite and non-invertible symmetries of 1+1d quantum field theories~\cite{Gaiotto:2020iye,Apruzzi:2021nmk,Freed:2022qnc,Kaidi:2022cpf}. In mathematics, 2+1d TQFTs produce invariants of knots and 3-manifolds~\cite{wittenQuantumFieldTheory1989, reshetikhinInvariants3manifoldsLink1991}.

In each of these settings, the anyons are characterized by a few physical properties: the number of anyon types and their topological spins, the ways in which anyons fuse, the associativity of fusion and the phases acquired when anyons are exchanged or braided around one another. Together with the chiral central charge $c_-$, which can be measured through the thermal Hall conductance~\cite{Banerjee:2017ebi}, these properties specify the corresponding 2+1d TQFT.   Mathematically, the 2+1d TQFTs describing a given set of anyons correspond to a unitary modular tensor category (MTC)~\cite{kitaevAnyonsExactlySolved2006}, or a supermodular category when some local excitations are fermions~\cite{lanTheory2+1DFermionic2016,bruillardFermionicModularCategories2017}. This is a rigid mathematical structure which we review in \cref{app:modular-tensor-categories}.  It is characterized by a rank equal to the number of anyon types, the twists $\theta_a$ encoding the topological spins, the fusion multiplicities $N_{ab}^c$, and a set of $R$- and $F$-symbols encoding respectively anyon exchange and the associativity of fusion. These data must furthermore satisfy the pentagon and hexagon equations, which express the consistency of fusion and braiding. At fixed rank there are only finitely many modular and supermodular categories~\cite{bruillardRankfinitenessModularCategories2015,jonesRankfinitenessGcrossedBraided2019}. This enables a classification of modular categories of a given rank, and hence of 2+1d TQFTs with a fixed number of simple anyons. 

Given the important role of 2+1d TQFTs/MTCs across physics and mathematics, in this paper we revisit the problem of classifying them by the number of anyons and of charting the relations between them. We extend the classification to bosonic theories with up to 22 anyons and to fermionic theories with up to 26 anyons, and determine the candidate anyon condensations between them. The resulting map of the space of these theories is a first step towards building an atlas of 2+1d TQFTs, which  we call the {\it Anyon Atlas}.

\subsection{Classification of Modular Data}

A complete classification of modular categories presents a substantial challenge. For instance, it is broader than the classification of finite groups, since each finite group $G$ has at least one associated MTC given by a $G$-gauge theory (or Dijkgraaf--Witten theory)~\cite{dijkgraafTopologicalGaugeTheories1990}. A practical route to such a classification is through the modular data $(S,T)$. The diagonal matrix $T$ collects the topological spins $\theta_a$, and the matrix $S$ encodes the mutual braiding of pairs of anyons:
\begin{equation}
    T_{ab}=\delta_{ab}\theta_a,
    \qquad
    S_{ab}=\vcenter{\hbox{\HopfLinkAmplitude{a}{b}}}\,.
\end{equation}
Here $\theta_a$ is the phase acquired by an anyon of type $a$ under a $2\pi$ rotation:
\begin{equation}
    \StandardTwistEquation\,.
    \label{eq:topological-twist-intro}
\end{equation}
Together $S$ and $T$ describe how the degenerate ground states on a torus transform under modular transformations. The modular data is the primary gauge-independent data of the theory, unlike, say, the $F$-symbols, which depend on a choice of basis for the vector spaces associated to anyon junctions. In fact, the modular data determines the $R$-symbols in a suitable gauge~\cite{ngRecoveringRsymbolsModular2024}, so a theory is fixed by its modular data up to a choice of $F$-symbols satisfying the pentagon and hexagon equations with these $R$-symbols. Two inequivalent theories may nonetheless share the same modular data (such theories are called modular isotopes~\cite{mignardModularCategoriesAre2021}), and a set of modular data may satisfy all known conditions yet admit no categorification. Crucially, rank-finiteness also holds for modular data satisfying the conditions we test, since every property used in~\cite{bruillardRankfinitenessModularCategories2015} can be checked on the modular data. Thus, at each rank one is guaranteed to find a finite list of admissible modular data, where some a priori unknown fraction might fail to admit a categorification.  This last issue can be mitigated for many examples by recognizing a given $(S,T)$ as related to a familiar example such as a Chern--Simons theory where alternative constructions ensure that a full unitary MTC exists.

\begin{figure}[tb!]
    \centering
    \begin{subfigure}[t]{\textwidth}
        \centering
        \includegraphics[width=\textwidth]{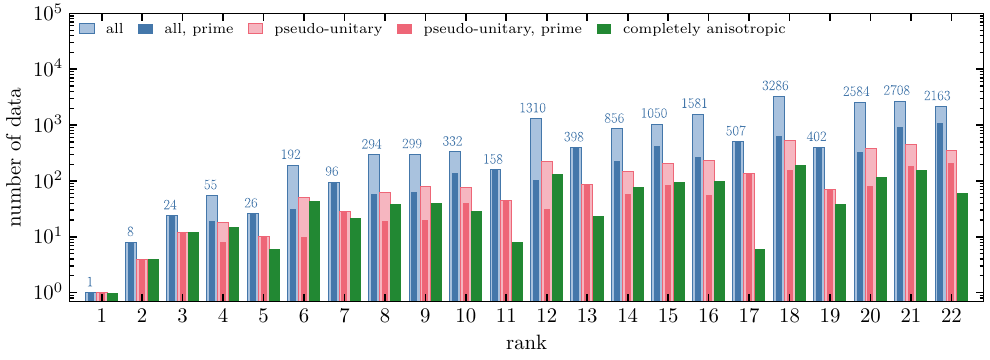}
        \caption{Modular data through rank 22.}
        \label{fig:data-counts-bosonic}
    \end{subfigure}\\[3mm]
    \begin{subfigure}[t]{\textwidth}
        \centering
        \includegraphics[width=\textwidth]{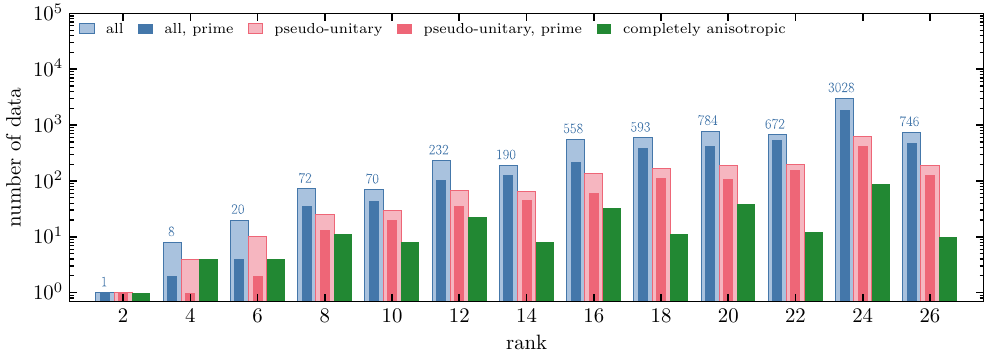}
        \caption{Supermodular data through rank 26.}
        \label{fig:data-counts-fermionic}
    \end{subfigure}
    \caption{Number of modular and supermodular data at each rank, with the total shown above the bars. The light bars count, respectively, all data and the subset which is pseudo-unitary, meaning that its quantum dimensions are positive and hence it could possibly correspond to a full unitary modular tensor category.  The dark inner bars count the subset of the data which are prime, i.e.\ do not split into a product of lower-rank examples.  Finally, the green bars count completely anisotropic data corresponding to terminal elements in a Witt class that do not admit any possible anyon condensation.}
    \label{fig:data-counts}
\end{figure}

A classification of modular categories through their modular data was initiated in~\cite{Rowell:2007dge}. Since the modular data defines a congruence representation of $\sl{\zee{}}$~\cite{ngCongruenceSubgroupsGeneralized2010,dongCongruencePropertyConformal2015}, Ng, Rowell, Wang and Wen reconstructed it from irreducible representations, reaching a partial classification at rank 12~\cite{ngReconstructionModularData2023,ngClassificationModularData2025}.\footnote{An alternative approach is the holomorphic modular bootstrap of rational conformal field theory, in which characters are constructed as solutions of modular linear differential equations and the modular data is read off from their $\sl{\zee{}}$ transformations~\cite{Mathur:1988na,Mathur:1988gt,Mukhi:2019xjy,Kaidi:2021ent}. It also constrains the central charge, but assumes the modular category is realized by an RCFT.} Fermionic topological orders were classified in the same way through rank 10~\cite{bondersonCongruenceSubgroupsSuperModular2018,choClassificationFermionicTopological2023}, and lower-rank results appeared in~\cite{Wen:2015qgn,lanTheory2+1DFermionic2016,Lan:2016seg}. In this work we extend and build on the above. Adapting the methods in~\cite{ngReconstructionModularData2023,ngClassificationModularData2025,choClassificationFermionicTopological2023}, we implement a numerical algorithm that scales far better at large rank and allows us to classify modular data up to rank 22 and supermodular data up to rank 26. This performance, however, comes at the cost of an exact certificate that every set of modular data at a given rank has been identified. Counts of the identified data at each rank are shown in \cref{fig:data-counts}, and the full counts are listed in \cref{tab:bosonic-counts,tab:fermionic-counts} of \cref{sec:results}. Even at lower rank, we identify 72 examples of modular data through rank 12 not included in~\cite{ngClassificationModularData2025} as well as 7 examples there which are not categorifiable. These are detailed in \cref{sec:closure} and \cref{sec:reconstruction}, respectively.


\subsection{Charting the Anyon Atlas}
\begin{figure}[ht!]
    \centering
    \includegraphics[width=.32\textwidth]{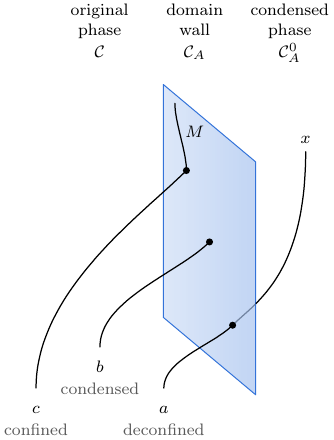}
    \includegraphics[width=.67\textwidth]{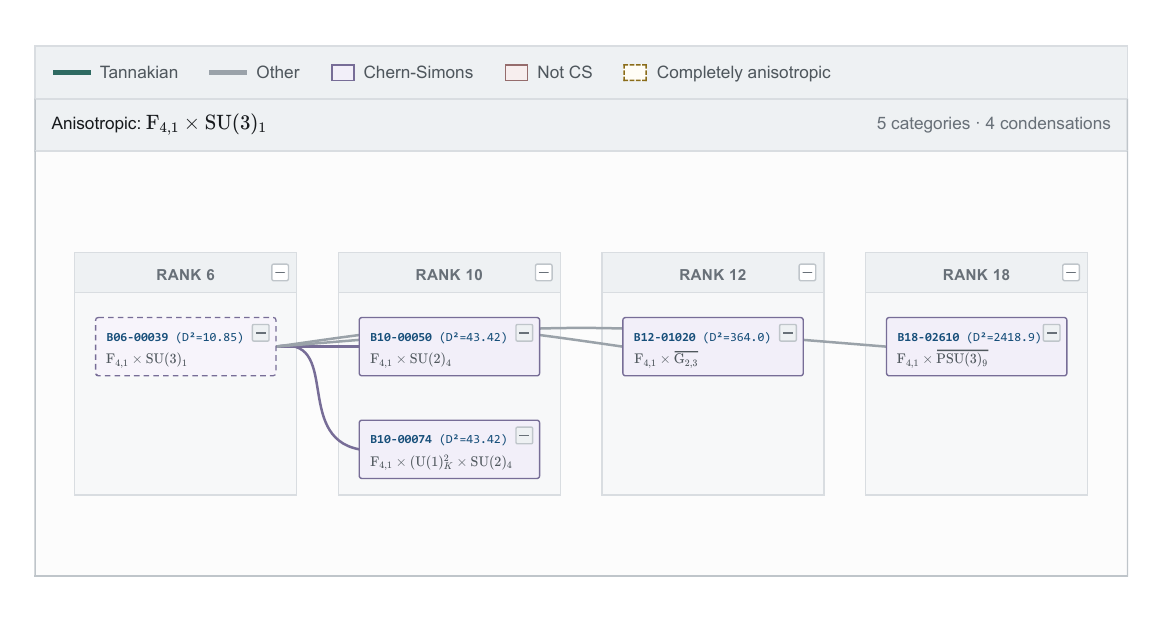}
    \caption{The Witt group is composed of equivalence classes of modular categories related by condensation and gauging, or, equivalently, 1- and 0-form gauging. Condensation between two phases, and its effect on simple lines, is illustrated schematically on the left. 
    Two phases related by condensation admit a gapped domain wall interface. Anyonic excitations in one phase may be deconfined, surviving as freely propagating excitations in the condensed phase; confined, unable to propagate into the condensed phase but surviving as excitations localized to the interface; or condensed, becoming identified with the vacuum in the condensed phase. In the database, we represent these condensation relations as directed multigraphs, with edges corresponding to condensations. Each graph is rooted in a unique completely anisotropic category, to which all Witt-equivalent categories condense. An example of these graphs is given on the right for the anisotropic $\text{F}_{4,1}\times \SU(3)_1$.}
    \label{fig:Witt}
\end{figure}

One might hope that a classification of modular categories by rank will provide insight into the structure of the space of 2+1d TQFTs.  Several types of relationships are of specific interest in our investigations:
\begin{itemize}
    \item Given two theories $\mathcal{C}_{1}$ and $\mathcal{C}_{2}$, of ranks $r_{1}$ and $r_{2}$, we can stack them to form a composite theory $\mathcal{C}_{1}\boxtimes \mathcal{C}_{2}$ of rank $r_{1}r_{2}$.  An MTC which does not split into a non-trivial product is referred to as \emph{prime}.
    \item Given any theory, one can construct a new one by gauging any non-anomalous 0-form and 1-form global symmetries.   This includes both invertible group-like symmetries and non-invertible symmetries. In the case of 1-form global symmetries, this operation is known as anyon condensation and is related to the study of local modules of the condensing algebra object \cite{Kirillov:2001, Bais:2002, Frohlich:2003, Bais:2008, kongAnyonCondensationTensor2021}.  A theory that admits no condensable anyons (gaugable 1-form symmetry) is referred to as (completely) \emph{anisotropic}.  
    \item Bosonic and fermionic theories are related by gauging the fermion parity symmetry $(-1)^{F}$.
\end{itemize}
Theories connected by sequences of gaugings and condensations form a Witt equivalence class. Equivalently, two theories are in the same Witt class when they can be separated by a gapped domain wall, up to stacking with invertible phases~\cite{davydovWittGroupNondegenerate2011,Kitaev:2011dxc,Lan:2014uaa}. Stacking makes the Witt classes into a group, the Witt group, in which the inverse of a class is its orientation reversal. Each Witt class contains a unique completely anisotropic theory, which admits no condensation, so the Witt group describes topological orders up to gauging, and its anisotropic representatives are the building blocks from which the other theories in a class are obtained. See \cref{fig:Witt}.

Using our classification results we initiate a deeper exploration of the space of 2+1d TQFTs and their relations. For each set of data we determine its decomposition into decoupled layers (prime factorization), its sets of anyons closed under fusion (fusion subcategories), the sets of bosonic anyons that can be condensed together (Tannakian subcategories), the candidate 0-form symmetries permuting anyons (permutation symmetries of $(S,T)$), and, for supermodular data, the bosonic theories obtained by gauging fermion parity (minimal modular extensions). For pseudo-unitary data we enumerate the candidate anyon condensations between entries, which determine the candidate Witt classes within the classified ranks, and we identify the data realized by Chern--Simons theories with compact connected gauge group, either directly or as condensations of parent theories up to rank 2500. 
The relations among the data also serve as checks on the completeness of the classification (see \cref{sec:closure}).

\subsection{Are All 2+1d TQFTs Related to Chern--Simons?}
\label{sec:introcs}

Most known modular categories come from Chern--Simons theories and quantum groups, either directly or via (possibly non-invertible) gauging or condensation.   Little is known about the possible theories that go beyond Chern--Simons. Over time, as knowledge and examples have accumulated, the field has developed several possible hypotheses about the centrality of Chern--Simons examples.  
\begin{itemize}
    \item \textbf{Conjecture:} \emph{Every unitary MTC is defined by a classical gauge theory.}  More sharply, this asserts that every unitary MTC can be built from a Chern--Simons theory based on a compact gauge group, possibly including gauging invertible 1-form and 0-form global symmetries. A conjecture of this form appeared in~\cite{mooreTamingConformalZoo1989}.  The interpretation in terms of gauge theory is natural.  Condensation of abelian anyons performs a quotient by a subgroup of the center of the gauge group, allowing non-simply connected possibilities.  Meanwhile, gauging invertible 0-form symmetries allows the possibility of disconnected gauge groups.

    This conjecture is known to be false.  The most direct low-rank counterexample is the MTC defined by the Drinfeld center $\mathcal{Z}(\text{Haagerup}_{3})$ of the Haagerup--Izumi category, at rank 12.  (See \cite{Teleman:2022Haagerup} for a detailed discussion.)  More generally, in our data we find 271 of the 3220 pseudo-unitary modular data up to rank 22, and 148 of the 1723 pseudo-unitary supermodular data up to rank 26, that have no Chern--Simons realization via any compact gauge group, including disconnected and non-simply connected ones.  A breakdown is shown in \cref{fig:cs_counts}.

\begin{figure}[!ht]
    \centering
    \includegraphics[width=.49\textwidth]{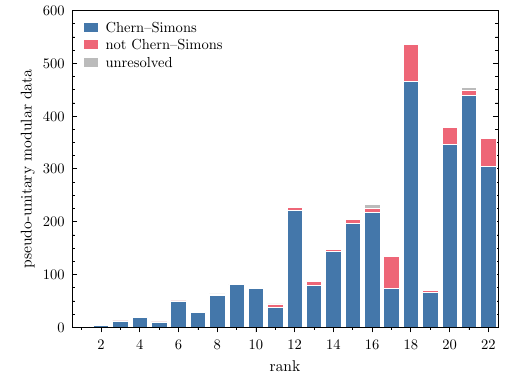}
    \includegraphics[width=.49\textwidth]{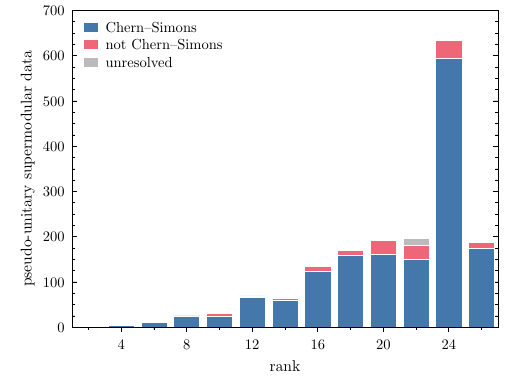}
    \caption{Counts of pseudo-unitary modular (left) and supermodular (right) data identified as Chern--Simons theories based on possibly disconnected and non-simply connected gauge groups, or proven not to be. Unresolved cases may be invertible gaugings of Chern--Simons theories at ranks greater than 2500.}
    \label{fig:cs_counts}
\end{figure}
    
    \item \textbf{Conjecture:} \emph{Every unitary MTC is Witt-equivalent to a Chern--Simons theory.}  This relaxes the above by allowing the 0-form and 1-form gaugings to be non-invertible.  A conjecture of this form appears in~\cite{davydovWittGroupNondegenerate2011,muegerModularCategories2012}. It removes the counterexample of $\mathcal{Z}(\text{Haagerup}_{3})$, for instance, since that example can be obtained by non-invertible 0-form gauging of ordinary $\mathbb{Z}_{3}$ gauge theory, i.e.\ $\mathcal{Z}(\mathbb{Z}_{3})$.  More broadly, since any Drinfeld center admits a topological boundary condition, all Drinfeld centers are Witt-equivalent to a trivial Chern--Simons theory.  It is worth emphasizing that in a modern physical interpretation this conjecture would assert that all MTCs are gauge theories of a kind, just not necessarily those based on classical groups.  Indeed, the string-net constructions of Drinfeld centers may all be viewed as a novel kind of lattice gauge theory, and in general non-invertible condensations of any symmetry define a kind of gauge theory.

    We provide the first substantial evidence that this conjecture is also false.  Specifically, we identify interesting prime, completely anisotropic data: one Galois orbit each at ranks 11 and 12, one each at ranks 18 and 22, and six orbits in the supermodular case, with no known connection by any sequence of invertible or non-invertible gauging operations to Chern--Simons theory. Moreover, for the orbits at ranks 11 and 12 we solve the pentagon and hexagon equations, with a method to be described in a separate paper, and obtain unitary, completely anisotropic modular tensor categories.  These are therefore counterexamples to the conjecture unless they come from a non-invertible anyon condensation in a Chern--Simons parent of very high rank.\footnote{Note, however, that in all 2108 non-invertible condensations of Chern--Simons theories in our data, we observe that the result is again a Chern--Simons theory. Yet this is also true for the 617 non-invertible condensations from data not identified as Chern--Simons.}  For instance, for the rank-11 categories, we show that a potential Chern--Simons parent theory must have rank larger than $10^4$, and the smallest identified candidate satisfying all known constraints has rank of order $10^{14}$ (see \cref{app:notCS}).

\end{itemize}

    We thus conjecture that the example at rank 11, as well as those at ranks 12, 18 and 22, and the analogous supermodular data we identify give the first examples of theories that are not Witt-equivalent to any Chern--Simons theory.  These examples would therefore be fundamentally beyond any notion of gauge theory.

\subsection{\texttt{AnyonAtlas}}

\begin{figure}[h!]
    \centering
    \includegraphics[width=\linewidth]{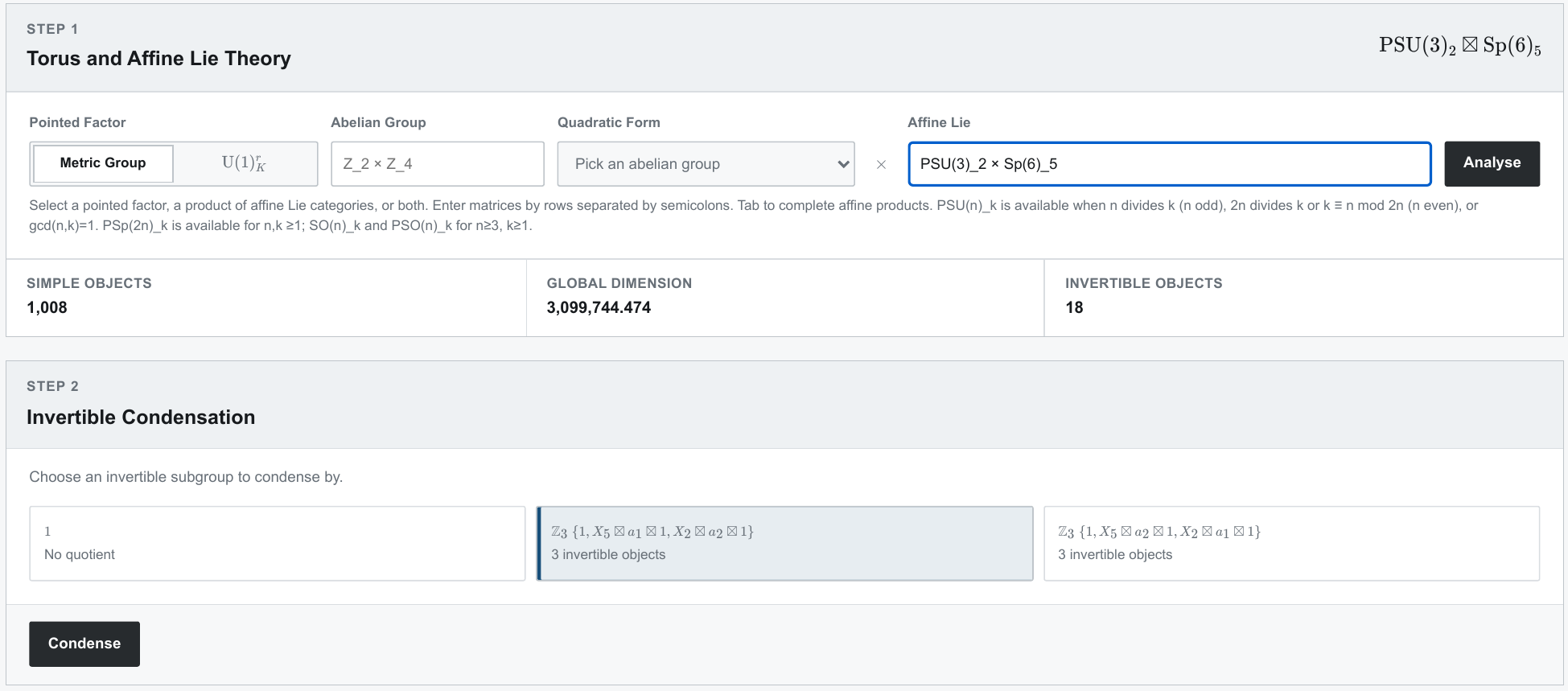}
    \caption{\href{http://www.AnyonAtlas.org}{\texttt{AnyonAtlas.org}} includes a calculator for the modular data of any Chern--Simons theory with a compact, connected gauge group, including spin theories. That is, it covers invertible anyon condensations of $\Uone$ and affine Lie modular categories. We compute the modular data of these condensates by the procedure developed in~\cite{Fuchs:1996dd}.}
    \label{fig:calculator}
\end{figure}

We have made the classified data available through the \texttt{AnyonAtlas} website 
\begin{center}
\href{http://www.anyonatlas.org}{\texttt{AnyonAtlas.org}}
\end{center}
Each entry has its own page with its exact modular data, quantum dimensions, twists and fusion rules, together with the derived properties described above. The catalog can be searched and filtered by rank, central charge, global dimension and properties such as pseudo-unitarity, primality and anisotropy. A separate tool identifies an entry from partial data such as its twists or quantum dimensions, and the condensation graph can be explored interactively. 
To facilitate greater investigation of the categories with a Chern--Simons realization, we include a calculator based on~\cite{Fuchs:1996dd} that allows one to extract the modular data of examples with compact and connected gauge group.  See Figure \ref{fig:calculator}.
All data can be downloaded in formats suited to Mathematica, GAP, Sage and Python. We hope that \texttt{AnyonAtlas} will serve as a resource for the community.\footnote{Complementary resources are the AnyonWiki~\cite{vercleyenAnyonWiki2023}, which lists low-rank multiplicity-free fusion rings and fusion categories together with explicit $F$-symbols~\cite{vercleyenLowRankFusion2023,vercleyenLowRankMultiplicityFreeComplex2026}, and the online database of vertex operator algebras and modular categories~\cite{gannonOnlineDatabaseVOA}, which catalogues MTCs arising from vertex operator algebras.} Future updates will include the $F$- and $R$-symbols of the categorified entries, along with additional features.

\subsection*{Organization}

This paper is organized as follows: In \cref{sec:modularcategories} we review the relevant facts about modular and supermodular categories. In \cref{sec:algorithm} we describe the classification algorithm in detail, and in \cref{sec:results} we present the resulting data and describe its structure. Finally, in \cref{sec:chernsimons} we relate the data to Chern--Simons theories with compact gauge group and discuss the structure of the Witt group that it reveals at low rank. 
In the appendices we review modular and supermodular categories (\cref{app:modular-tensor-categories}), bound the scaling of congruence representations of $\sl{\zee{}}$ (\cref{app:scaling}), derive a closed form for the higher central charges of Chern--Simons theories (\cref{app:higher-central-charges}), and analyze non-invertible condensations to the exotic anisotropic categories (\cref{app:notCS}).

\section{Modular and Supermodular Categories}
\label{sec:modularcategories}

In this section we review the familiar arguments for why topological orders are described by modular (and supermodular) categories. 
Then we detail the particular aspects of these categories that are central to our classification. 

\subsection{Topological Order and TQFT}
\label{sec:topological-order}

A gapped phase of a local quantum many-body system in 2+1d has intrinsic topological order when its ground state is long-range entangled, meaning that it cannot be connected to a product state by a finite-depth circuit of local unitaries, even when no symmetry is imposed~\cite{Wen:1989iv,Chen:2010gda}. Such a phase has a ground-state degeneracy on the torus that is stable under local perturbations~\cite{Wen:1989zg}, and its point-like excitations are anyons. An anyon type is a superselection sector of excitations, two excitations being of the same type when they are related by an operator acting near them. The local excitations, created from the ground state by local operators, form the trivial sector. A topological order is bosonic when all of its local excitations are bosons, and fermionic otherwise.\footnote{In general, long-range entanglement does not guarantee a TQFT description at long distances, as fracton phases show~\cite{vijayFractonTopologicalOrder2016,slagleQuantumFieldTheory2017}. In our analysis we assume that the long-distance descriptions of the topological orders we investigate are TQFTs.}

Kitaev argued that the anyons of a bosonic topological order are described by a unitary premodular, or ribbon, category~\cite{kitaevAnyonsExactlySolved2006}. Consistent fusion of anyons requires them to be labeled by the simple objects of a fusion category. In 2+1d, anyons can be braided, so the category is braided. The existence of antiparticles requires rigidity, and a spherical structure is needed to give a consistent meaning to pair creation and annihilation. Together these give a premodular category, whose defining properties are reviewed in \cref{app:modular-tensor-categories}. If, in addition, every non-trivial anyon can be detected remotely by an Aharonov--Bohm-type braiding experiment, the only anyon that braids trivially with all others is the vacuum, and the category is modular. A modular category $\mathcal M$ determines the chiral central charge $c_-$ modulo 8 (see \cref{sec:modulardata}), and two topological orders with the same $\mathcal M$ differ by stacking with copies of the invertible $\text{E}_{8,1}$ phase. A bosonic topological order is therefore labeled by the pair $(\mathcal M, c_-)$.

The same structure follows from TQFT. The Reshetikhin--Turaev construction assigns to a modular category $\mathcal M$ a 2+1d TQFT whose framed line operators are labeled by the simple objects of $\mathcal M$~\cite{reshetikhinInvariants3manifoldsLink1991,turaevModularCategories3Manifold1992,bakalovLecturesTensorCategories2000}. When $\mathcal M$ is the category of lines of a Chern--Simons theory, the construction reproduces that theory up to stacking with invertible phases~\cite{wittenQuantumFieldTheory1989}. Conversely, the line operators of a unitary, bosonic, once-extended 2+1d TQFT form a modular category. The Hilbert space on a torus has a basis labeled by the lines and carries a projective representation of the mapping class group $\sl{\zee{}}$, so the $S$ matrix is invertible. This is the TQFT counterpart of remote detectability. The correspondence between modular categories and once-extended 2+1d TQFTs is established in~\cite{bartlettModularCategoriesRepresentations2015a}. Since a topological order is expected to flow to a TQFT at long distances~\cite{rowellMathematicsTopologicalQuantum2018}, topological orders without symmetry are classified by modular categories together with the central charge. Topological orders with a finite symmetry group $G$ are described by $G$-crossed braided extensions of modular categories~\cite{barkeshliSymmetryFractionalizationDefects2019}, which we do not consider here.

In a fermionic topological order, we label anyon sectors up to the action of local operators that are even under fermion parity. The local fermion is then a distinguished anyon $\psi$ with fermionic statistics that braids trivially with all anyons. Since fermion parity is carried by $\psi$, the fusion spaces are ordinary vector spaces, and the associator, unitors and braiding are ordinary unitary maps. The pentagon and hexagon equations are therefore unchanged, and no Koszul signs are needed. With the same assumption of remote detectability as in the bosonic case, the anyons form a premodular category $\C$ whose Müger center is $\mathcal Z_2(\C) \simeq \sVec$, generated by $\psi$. Such categories are called supermodular. They were proposed to describe fermionic topological orders in~\cite{lanTheory2+1DFermionic2016} and are studied further in~\cite{bruillardFermionicModularCategories2017}.

The supermodular category $\C$ does not record how the theory depends on the spin structure. In 2+1d, fermion parity can be gauged, and the gauging can be undone~\cite{thorngrenAnomaliesBosonization2020}. Gauging $(-1)^F$ produces a bosonic topological order whose modular category $\mathcal M$ contains $\C$ and satisfies $D_{\mathcal M}^2 = 2D_\C^2$. Such an $\mathcal M$ is a minimal modular extension of $\C$. The existence of such extensions and the 16 choices are discussed in \cref{sec:supermodulardata}. The category $\C$ determines $c_-$ modulo $\frac12$, which corresponds to stacking with $p+ip$ superconductors, and each minimal modular extension fixes a lift to $c_-$ modulo 8. A fermionic topological order is therefore labeled by a triple
\begin{equation}
  (\C, \mathcal M, c_-) \sim (\mathcal M, \psi, c_-),
\end{equation}
where $\mathcal M$ is a minimal modular extension of a unitary supermodular category $\C$, or equivalently $\psi$ is a choice of fermion in $\mathcal M$. The pair $(\C, c_-)$ already fixes $\mathcal M$. We include $\mathcal M$ in the data because, given $\C$ and $c_-$, there is in general no explicit construction of the extension.

\subsection{Premodular Categories and Modular Data}
\label{sec:modulardata}

The defining properties of modular and supermodular categories can be found in, among others,~\cite{kitaevAnyonsExactlySolved2006,bakalovLecturesTensorCategories2000} for modular categories and~\cite{bruillardFermionicModularCategories2017} for supermodular categories. \Cref{app:modular-tensor-categories} reviews these definitions
and derives several identities relating fusion, braiding,
and modular data. Both modular and supermodular categories are examples of premodular categories: braided fusion categories equipped with a compatible ribbon structure. That is, there exist natural isomorphisms $\theta_X: X\rightarrow X$, called twists, satisfying the balancing condition
\begin{equation}\theta_{a\otimes b}
  = (\theta_a\otimes\theta_b)\circ\beta_{b,a}\circ\beta_{a,b}\end{equation}
and such that the induced pivotal structure is spherical. Here, $\beta_{a,b}:a\otimes b\to b\otimes a$ is the braiding. In the modular case, the category is required to have non-degenerate braiding, i.e.\ its Müger center $\mathcal Z_2(\C)$ is $\text{Vec}$, the trivial MTC with only the vacuum anyon. In the supermodular case, the Müger center is instead $\mathcal Z_2(\C) = \sVec$, generated by an invertible fermion $\psi$ ($\theta_\psi = -1$) that braids trivially with every object.

Much of the data of these categories, such as the $F$- and $R$-symbols, depends on choices of bases in the Hom spaces $V^{bc}_a = \Hom_\C(a,b\otimes c)$. The modular data $(S,T)$, by contrast, is independent of these choices and captures many of the categories' important features. This data will be the object of our classification.

The modular $T$ matrix is constructed from the twists as
\begin{equation}T = \operatorname{diag}(\theta_a)_{a\in\irr(\C)}\end{equation}
By Vafa's theorem~\cite{vafaClassificationConformalTheories1988}, the twists are roots of unity, $\theta_a = e^{2\pi i h_a}$, so $N=\operatorname{ord}(T)$ is finite. The modular $S$ matrix is given by evaluating the Hopf link. In any premodular category, the $S$ matrix is symmetric and yields the quantum dimensions $S_{1a} = d_a$. The preference for this normalization will be motivated in \cref{sec:representation}. The $S$ matrix satisfies two important properties. First,
\begin{equation}S_{a,b} = \sum_{c\in\irr(\C)}N_{a\bar b}^c\frac{\theta_c}{\theta_a\theta_b}d_c\end{equation}
which follows from applying balancing to its definition, and, second,
\begin{equation}\label{eq:protoverlinde}
      \frac{S_{ad}S_{bd}}{S_{1d}}=\sum_{c\in\irr(\C)}N_c^{ab}S_{cd}
\end{equation}
which follows from fusing the links. In the case where $S$ is invertible, and therefore unitary with normalization $s=\frac{S}{D}$, we have charge conjugation
\begin{equation}S^2 = D^2 C, ~~~~ C_{ab} = \delta_{a\bar b}\end{equation}
with $C^2 =1$ and $[C,T]=0$. Invertibility applied to \cref{eq:protoverlinde} yields the Verlinde formula,
\begin{equation}
  N_c^{ab} = \frac{1}{D^2}\sum_{d\in\irr(\C)} \frac{S_{ad}S_{bd}S_{\bar cd}}{S_{1d}}
\end{equation}
Moreover, $S$ is invertible if and only if $\C$ is modular; in the supermodular case, as we will see, one can instead work with an invertible $\hat S$ on the fermionic quotient.

The charge-conjugation matrix, which encodes duality, is particularly useful to us because it determines which second Frobenius--Schur indicators vanish:
\begin{equation}\nu_2(a) = \begin{cases}
0, & a\neq \bar a\\
\pm 1, & a = \bar a
\end{cases}\end{equation}
The indicators and their relation to modular data are reviewed
in \cref{app:fs-indicators}. With our braiding convention,
\begin{equation}\nu_2(a) = \theta_a R_1^{aa}\end{equation}
for $a = \bar a$.

For modular categories, $(s,T)$ furnish a projective representation of $\sl{\zee{}}$:\footnote{Throughout we will use the convention that $\mods = \left(\begin{smallmatrix}0&-1\\1&0\end{smallmatrix}\right)$, so that $(\rho(\mods)\rho(\modt))^3 = \rho(\mods)^2$.}
\begin{equation}
    s^4 = I,~~~(sT)^3 = \xi(\C)s^2
\end{equation}
The projective factor is the multiplicative central charge, given in terms of the modular data as
\begin{equation} \xi(\C)=\frac1D \sum_{c\in\irr(\C)} \theta_cd_c^2 = e^{2\pi i c_-/8}\end{equation}
The chiral central charge $c_-$ is rational, with $c_- \in \frac{2}{N}\mathbb Z$. Supermodular data is discussed in more detail in \cref{sec:supermodulardata}, while further aspects of the associated representation theory, leveraged heavily in the classification, are described in \cref{sec:representation}.

Given the above results, when $S$ is invertible, the modular data determines much of the corresponding categorical information. In fact, it was recently proven that the modular data is sufficient to determine the $R$-symbols in some basis~\cite{ngRecoveringRsymbolsModular2024}. The only missing categorical data, then, are the $F$-symbols. The known conditions on the modular data, including the above results, are not believed to be sufficient to guarantee that $F$-symbols satisfying the pentagon and hexagon equations exist. Examples of previously identified modular data in the classification through rank 12 that do not admit a categorification were found and are discussed later in the paper. Further, when a categorification exists, it need not be unique. The existence of such modular isotopes was demonstrated in~\cite{mignardModularCategoriesAre2021}, with the first known example at rank 49. Candidate supermodular isotopes can be identified in our data, since supermodular data impose weaker constraints, as we now discuss.

\subsection{Supermodular Categories}
\label{sec:supermodulardata}

Supermodular categories contain an invertible, transparent fermion $\psi$. By balancing, then, $\theta_{a\otimes \psi} = -\theta_a$. Simple objects therefore come in non-isomorphic pairs $a$ and $a\otimes \psi$, so the rank must be even. By the definition of $S$,
\begin{equation}S_{a\otimes \psi, b} = S_{a,b}\end{equation}
so, as noted above, $S$ is not invertible. Instead, one is led to consider $(\hat S, \hat T^2)$, where
\begin{equation} S = \begin{pmatrix}
    1 & 1\\ 1 & 1
  \end{pmatrix} \otimes \hat S, ~~~~~~~~ T = \begin{pmatrix}
    1 & 0\\ 0 & -1
  \end{pmatrix} \otimes \hat T\end{equation}
Only $\hat T^2$ is meaningful: the choice of square root $\hat T$, i.e.\ of a representative in each pair, does not affect the data. The pair $(\hat S,\hat T^2)$ is understood as the data of the fermionic quotient $\C/\psi$ identifying $a$ and $a\otimes \psi$. As in the modular case, we have $\hat S^2 = \frac12 D^2 C$, for $C$ the charge-conjugation matrix of $\C/\psi$, and $[C,\hat T^2]=0$. So, as shown in~\cite{bruillardFermionicModularCategories2017}, $(\hat S, \hat T^2)$ define a projective representation of the index-3 subgroup
\begin{equation}\Gamma_\theta = \left<\mods,\modt^2\right> <\sl{\zee{}}\end{equation}
As in the modular case, we may therefore work with a symmetric, invertible $\hat S$, unitary if one takes the normalization $\hat s = \frac{\sqrt{2} \,\hat S}{D}$. The Verlinde formula is then
\begin{equation}N^{ab}_c + N^{ab}_{c\otimes \psi} = \frac{2}{D^2} \sum_{[d]\in\irr(\C/\psi)} \frac{\hat S_{ad} \hat S_{bd} \hat S_{\bar cd}}{\hat S_{1d}}\end{equation}
Unlike in the modular case, $\hat S$ determines the fusion rules only up to a lift compatible with balancing:
\begin{equation}\hat S_{a,b} = \sum_{[c]\in\irr(\C/\psi)}(N^{a\bar b}_c - N^{a\bar b}_{c\otimes \psi})\frac{\theta_c}{\theta_a\theta_b}d_c\end{equation}
Here, the twists are the entries of any square root of $\hat T^2$. A single set of supermodular data can admit multiple fusion lifts; examples occur in our data, the first at rank 18. These provide candidates for supermodular isotopes. 

An important notion, on which our fermionic classification depends, is that of a (minimal) modular extension~\cite{bruillardFermionicModularCategories2017}. Given a supermodular category $\C$, a minimal modular extension of it is a modular category $\mathcal M$ containing $\C$ as a braided subcategory (the Müger centralizer of a distinguished fermion $\psi$) and satisfying $D_{\mathcal M}^2 = 2D_\C^2$. The objects of $\mathcal M$ split as $\C_0 \oplus \C_1$ with $\C_0 \simeq \C$. Simple objects in $\C_1$, unlike those in $\C_0$, have $\theta_{a\otimes \psi} = \theta_a$ and so may come either in pairs $a\otimes \psi \not\simeq a$ or as singlets $a\otimes \psi \simeq a$. There must be as many of these pairs or singlets as there are pairs in $\C_0$\footnote{For $a\in\C_x$, the monodromy (double braiding) is $M_{\psi,a} = \frac{S_{\psi,a}}{S_{1,a}} = (-1)^{x}$. Let $\C_0$ have $n$ pairs and $\C_1$ have $p$ pairs and $q$ singlets. The fusion matrix $N_\psi$ is an involution, since $\psi\otimes\psi\simeq 1$, and has eigenvalues $+1$ with multiplicity $n+p+q$ and $-1$ with multiplicity $n+p$. Since $S$ diagonalizes $N_\psi$, with eigenvalues $M_{\psi,a}$, comparing multiplicities yields $n = p+q$.}, so a minimal extension of a rank-$r$ category $\C$ has rank in $[\frac{3r}{2},2r]$. If a modular category $\mathcal M$ contains an invertible fermion $\psi$, it is a minimal extension of the $\psi$-neutral sector and is termed a spin modular category. Spin modular categories capture the complete data of fermionic topological orders, or spin TQFTs, beyond their anyon content alone, up to stacking with invertible $\text{E}_{8,1}$ phases. Passing to the minimal extension of $\C$ is understood as gauging $(-1)^F$ in the corresponding topological order, and the objects of $\C_1$ are identified with its fermion parity defects.

When a minimal extension exists, one can always construct all 16, preserving unitarity. In particular, if $\C$ is pseudo-unitary and admits at least one minimal extension $\mathcal M_0$, it possesses exactly 16 minimal pseudo-unitary extensions. These can be constructed from the minimal extensions $\mathcal S_\nu$ of $\sVec$\footnote{At rank 3, Ising-type: $\text{SU}(2)_2, ~\text{E}_{8,2},
~\text{Spin}(5)_1,~\text{Spin}(7)_1,~\text{Spin}(9)_1,~\text{Spin}(11)_1,~\text{Spin}(13)_1,~\text{Spin}(17)_1$; and at rank 4, pointed: $\text{SU}(4)_1,
~\text{Spin}(8)_1,~\text{Spin}(10)_1,~\text{Spin}(14)_1,~\text{Spin}(16)_1,~\text{Spin}(18)_1$ as well as the two theories $\Uone^2_K$ with $K=\pm\bigl( \begin{smallmatrix} 2 & 0 \\ 0 &  2 \end{smallmatrix} \bigr)$.} by condensation:
\begin{equation}\mathcal M_\nu = (\mathcal M_0\boxtimes \mathcal S_\nu)_{A_\nu}^0\end{equation}
for the condensable algebra $A_\nu = 1\oplus (\psi\boxtimes f)$, with fermion $f\in\mathcal S_\nu$. More details on condensation are provided in \cref{sec:condensation}. The existence of these extensions of pseudo-unitary $\C$ has been proven in~\cite{johnson-freydMinimalNondegenerateExtensions2023}, leaving open only whether non-pseudo-unitary $\C$ always admit minimal extensions. In our classification, as we are primarily interested in unitary categories, it is assumed that any $\C$ possesses a minimal extension.

\subsection{Congruence Property and Galois Automorphisms}
\label{sec:representation}

Modular data give not just projective representations of $\sl{\zee{}}$ but congruence representations~\cite{ngCongruenceSubgroupsGeneralized2010}. A representation $\rho:\sl{\zee{}}\rightarrow \mathrm{GL}(V)$ is said to be congruence if its kernel contains a principal congruence subgroup,
\begin{equation}\Gamma(n) = \ker(\sl{\zee{}}\rightarrow\sl{\zee{n}})\end{equation}
Equivalently, $\rho$ factors through $\sl{\zee{n}}$ for some $n$. The smallest such $n$ is the congruence level of $\rho$; for congruence representations of $\sl{\zee{}}$, it equals $\operatorname{ord}(\rho(\modt))$. 
Explicitly, the modular data $(S,T)$ is formed from a linear congruence representation as
\begin{equation}S = \frac{\rho(\mods)}{\rho(\mods)_{11}}, ~~~~~T = \frac{\rho(\modt)}{\rho(\modt)_{11}}\end{equation}
While for fixed $d\geq 2$ there are infinitely many inequivalent $d$-dimensional representations of $\sl{\zee{}}$, only finitely many of them are congruence. This fact allows us to proceed with a classification by finite enumeration of the possible modular data at fixed rank. 

It was similarly shown in~\cite{bondersonCongruenceSubgroupsSuperModular2018} that, for supermodular categories admitting minimal extensions, this congruence property continues to hold for fermion-neutral sectors, and therefore supermodular data form congruence representations of $\Gamma_\theta$. However, \cite{bondersonCongruenceSubgroupsSuperModular2018} notes that no analogous formula is known for the level of these representations. We provide one here: the level of a congruence representation $\rho$ of $\Gamma_\theta$ is
\begin{equation}l = \operatorname{lcm}(2\operatorname{ord}(\rho(\modt^2)), \operatorname{ord}(\rho(\mods\modt^2)))\end{equation}
\begin{proof} Let $n$ be the level of $\rho$, so that $\Gamma(n)\subseteq \ker\rho = K$. Clearly, $l|n$. $\modt:z\rightarrow z+1$ fixes the cusp $\infty$ and the cusp width is, therefore, the least $m$ such that $\modt^m\in K$. So, $m|2\operatorname{ord}(\rho(\modt^2))$. Similarly, $\mods\modt^2$ fixes the cusp $-1$ and so its width must divide $\operatorname{ord}(\rho(\mods\modt^2))$. The general level $g$, the lcm of the cusp widths, must therefore divide $l$. Then, by the generalized Wohlfahrt theorem~\cite[Prop.~3]{kimingLiftsProjectiveCongruence2011},
\begin{equation}\Gamma(2l) \subseteq \Gamma(2g)\subseteq K\end{equation}
So, $n|2l$, and we need only consider the elements of $\Gamma(l)/\Gamma(2l)$. For $A\in \Gamma(l)$, there exist $a,b,c,d\in \mathbb Z$ such that
\begin{equation}A = \begin{pmatrix}
  1+la & lb\\ lc & 1+ld
\end{pmatrix}\end{equation}
Since
\begin{equation}\det A = 1 = 1 + l(a+d) \pmod{2l}\end{equation}
we must have $a = d \pmod{2}$. Note that we have used that $l$ is even. So,
\begin{equation}A \mod 2l = I + l\begin{pmatrix}
  a & b\\c & a
\end{pmatrix}, ~~~~~a,b,c\in\{0,1\}\end{equation}
This quotient, $\Gamma(l)/\Gamma(2l)\simeq\zee{2}^3$, is generated by the images of $X=\modt^l$, $Y = \mods\modt^{-l}\mods^{-1}$, and $Z=(\mods\modt^2)^{-l}$, all of which lie in $\Gamma_\theta$. Then, by definition, $\rho(X)=\rho(Y)=\rho(Z)=I$, ensuring that $\Gamma(l) \subseteq K$. Hence, $n=l$. \end{proof}

Knowing the linear representation from which the fermionic data is constructed allows us to determine its central charge$\pmod{\frac12}$. Although the Gauss sums used in the modular case vanish, the central charge of a supermodular category can be inherited from its minimal extensions, each of which has a distinct $c_- \pmod8$. The linear representations offer a more direct route, since~\cite{choClassificationFermionicTopological2023}
\begin{equation}\rho(\modt^2) = e^{-2\pi i c_- /12} \hat T^2\end{equation}

Closely related to the congruence property is the fact that the unnormalized entries of $S$ are cyclotomic integers with conductor dividing $N=\operatorname{ord}(T)$~\cite{ngCongruenceSubgroupsGeneralized2010}:
\begin{equation}S_{ij} \in \zee{}[\zeta_N] \subset \mathbb Q(\zeta_N)\end{equation}
Besides providing an additional constraint, this property underlies our exact reconstruction of the data (\cref{sec:reconstruction}). Although the algorithm itself is numeric, the final modular data can be reconstructed with the knowledge that their entries must be cyclotomic integers. When $\varphi(N)>2$, $\zee{}[\zeta_N]$ is dense in $\mathbb C$, so this property alone is generally insufficient to guarantee that the data can be reconstructed exactly. One can, however, also use the fact that the full Galois action on $S$ is known, as described in~\cite{dongCongruencePropertyConformal2015}. There, it is shown that the matrices
\begin{equation}D_a = \rho(\mathfrak d_a), ~~~~\mathfrak d_a = \modt^a\mods\modt^b\mods\modt^a\mods^{-1} = \begin{pmatrix}
  a & 0\\0 & b
\end{pmatrix} \pmod{n}\end{equation}
for $a\in(\zee{}/n\zee{})^\times \simeq \operatorname{Gal}(\mathbb Q(\zeta_n)/\mathbb Q)$ and $ab = 1 \pmod{n}$, are signed permutation matrices implementing the Galois action on modular data. Namely, for $\sigma_a \in \operatorname{Gal}(\mathbb Q(\zeta_n)/\mathbb Q)$,
\begin{equation}\sigma_a \rho(\mods) = D_a \rho(\mods), ~~~~~ \sigma_{a^2} \rho(\modt) = D_a \rho(\modt) D_a^T\end{equation}

Integrality and the Galois action are less developed in the fermionic case, so we describe them here, assuming that a minimal extension exists. In~\cite{choClassificationFermionicTopological2023}, the matrices
\begin{equation}H_a =   \rho\left(h_a\right), ~~~~ h_a=(\modt^2)^{(1-a)/2}\mods(\modt^2\mods)^{b-1}(\modt^2)^{(1-a)/2}\mods^{-1}=\begin{pmatrix}
  a & 0\\0 & b
\end{pmatrix} \pmod{l}\end{equation}
for $ab = 1 \pmod{l}$, are recognized to agree with the bosonic Galois matrices and therefore required to be signed permutations. We can further require that they yield the correct Galois action. Given a minimal extension with linear $\sl{\zee{}}$ representation $\Phi$ of level $n$, there exists a subspace $W$ on which $\Phi$ restricts to a representation $\rho$ of $\Gamma_\theta$ of level $l|n$, i.e.,
\begin{equation}\Phi(\gamma)|_W = \rho(\gamma), ~~~~\gamma\in\Gamma_\theta\end{equation}
Then, for $A = a \pmod{l}$,
\begin{equation}D_A|_W = \Phi(h_A)|_W = \rho(h_A) = \rho(h_a) = H_a\end{equation}
so the Galois action on the $\Gamma_\theta$ representation is given by
\begin{equation}\sigma_a \rho(\mods) = H_a \rho(\mods), ~~~~~ \sigma_{a^2} \rho(\modt^2) = H_a \rho(\modt^2) H_a^T\end{equation}
for $a\in(\zee{}/l\zee{})^\times$. Finally, this same argument implies that the entries of $\hat S$ are cyclotomic integers in $\zee{}[\zeta_L]\subset \mathbb Q(\zeta_L)$ for $L = \operatorname{lcm}(2\operatorname{ord}(\hat T^2), \operatorname{pord}(\hat S\hat T^2))$, where $\operatorname{pord}$ denotes the projective order, the minimal $m$ such that $(\hat S\hat T^2)^m \propto I$.

We emphasize this because, starting at rank 12, there exist candidate supermodular data satisfying the conditions of~\cite{choClassificationFermionicTopological2023} but without categorification. This follows from the Galois action result above. These representations satisfy all previously known conditions (Verlinde, balancing, etc.) with $H_a$ signed permutations, but do not give the correct Galois action at level $l= \operatorname{lcm}(2\operatorname{ord}(\rho(\modt^2)),\operatorname{ord}(\rho(\mods\modt^2)))$. For example:
\begin{equation}
\rho(\mods)
=\frac12
\begin{pmatrix}
 \frac{i\sqrt2}{2} &  \frac{i\sqrt2}{2} &  \frac{i\sqrt2}{2} & -i &  \frac{i\sqrt2}{2} & -i \\[2mm]
 \frac{i\sqrt2}{2} & -\frac{i\sqrt2}{2} &  \frac{i\sqrt2}{2} & -1 & -\frac{i\sqrt2}{2} &  1 \\[2mm]
 \frac{i\sqrt2}{2} &  \frac{i\sqrt2}{2} &  \frac{i\sqrt2}{2} &  i &  \frac{i\sqrt2}{2} &  i \\[2mm]
 -i & -1 & i & 0 & 1 & 0 \\[2mm]
 \frac{i\sqrt2}{2} & -\frac{i\sqrt2}{2} & \frac{i\sqrt2}{2} & 1 & -\frac{i\sqrt2}{2} & -1 \\[2mm]
 -i & 1 & i & 0 & -1 & 0
\end{pmatrix}, ~~~~~~\rho(\modt^2)
=
\operatorname{diag}
\left(
\zeta_8^{6},
\zeta_8^{2},
\zeta_8^{6},
\zeta_8^{7},
\zeta_8^{2},
\zeta_8^{7}
\right)
\end{equation}
These cases emphasize that known conditions on modular data are almost certainly insufficient to guarantee categorification. Examples of this particular form do not occur in the bosonic case, at least through rank 22.

\section{Classification Algorithm}
\label{sec:algorithm}

The classification algorithm is inspired by and adapted from those implemented in~\cite{ngClassificationModularData2025} for modular and~\cite{choClassificationFermionicTopological2023} for supermodular categories. The exact methods used there, in principle, offer guarantees of exhaustiveness, but become intractable even in certain rank-12 cases and scale very poorly with rank. In lieu of this, we perform an analogous procedure numerically. This has the advantage of scaling far better, but at the cost of losing the guarantee of exhaustiveness. We return to this point below. The problem consists of enumerating candidate congruence representations, then searching for a simple-object basis satisfying the conditions described in the preceding section. This generally involves a numerical search over continuous changes of basis. No purely numerical search can guarantee that every suitable basis has been found. Instead, substantial effort was made to ensure the numerical search is sufficiently exhaustive for the classification to be made with reasonable confidence. These efforts are described throughout this section and in \cref{sec:closure}.

\begin{figure}[p]
    \centering
    \includegraphics{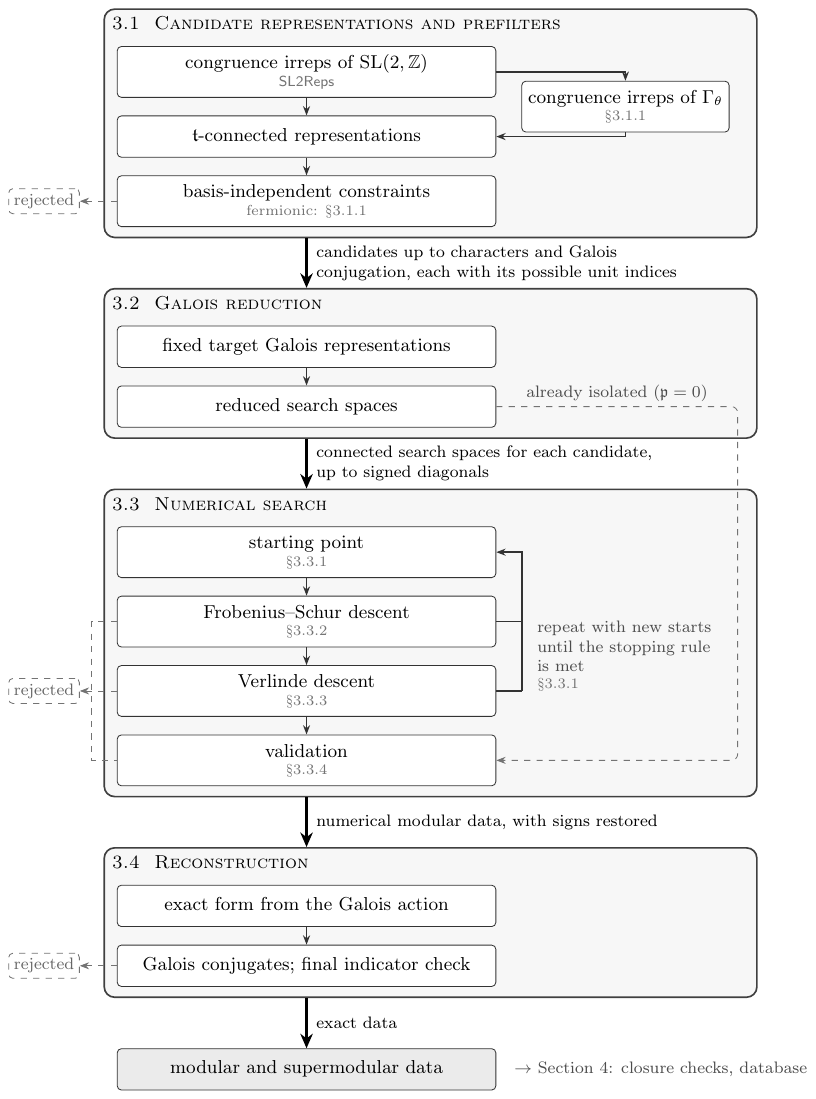}
    \caption{Flowchart of the modular and supermodular data classification algorithm.}
    \label{fig:algorithm}
\end{figure}

A flowchart of the classification algorithm is presented in \cref{fig:algorithm}. We begin by enumerating the relevant congruence representations of $\sl{\zee{}}$, $(\tilde\rho(\mods),\rho(\modt))$, computed from irreducibles provided by the \textsf{GAP} package \textsf{SL2Reps}~\cite{ngSymmetricRepresentations$textSL_2mathbbZ$2023}. In order to minimize the number of numeric searches that must be completed, these are then tested against as many practical basis-independent conditions as are known, referred to hereafter as the prefilter. The remaining candidate representations and choices of possible unit index, to which the later conditions are sensitive, then proceed to the numeric search looking for simple-object bases satisfying the rest of the conditions for modular data. These changes of basis must keep $\rho(\modt)$ diagonal and $\rho(\mods)$ symmetric. So, these searches are restricted to the continuous parameter space:
\begin{equation}\rho(\mods) = U\tilde\rho(\mods)U^T,~~~~~ U\in\mathrm{Stab}_{O(r)}(\rho(\modt)) \simeq \prod_i O(n_i)\end{equation}
for $n_i$ the multiplicities of the $\rho(\modt)$ eigenvalues. Throughout, a tilde denotes the initial basis produced by \textsf{SL2Reps}; since $U$ commutes with $\rho(\modt)$, $\tilde\rho(\modt) = \rho(\modt)$. This search space is reduced in dimension by the requirement that the Galois matrices $D_a$ be signed permutations. The numeric search then proceeds by descent on cost functions, first ensuring the correct second Frobenius--Schur indicators, then costing for integer fusion rules in a second descent. The numeric search is by far the most subtle portion of the algorithm, and the only one that is possibly inexhaustive.

The final stage of the algorithm leverages the algebraic properties of modular data, described in \cref{sec:representation}, to reconstruct the exact modular data reliably, even from numerical precision far lower than what was available. Then, the modular data is formed by normalizing these representations for a valid unit index $u$, and conjugating $\rho(\mods)$ by a signed diagonal matrix,
\begin{equation}S_{ij} = \sigma_i\sigma_j \frac{\rho(\mods)_{ij}}{\rho(\mods)_{uu}}, ~~~~~ \theta_{i}  = \frac{\rho(\modt)_{ii}}{\rho(\modt)_{uu}}\end{equation}
The signed diagonal restores signs that were quotiented out earlier to reduce the number of searches.

\subsection{Candidate Representations and Prefilters}
\label{sec:prefilter}

\begin{figure}[!ht]
    \centering
    \includegraphics[width=.49\textwidth]{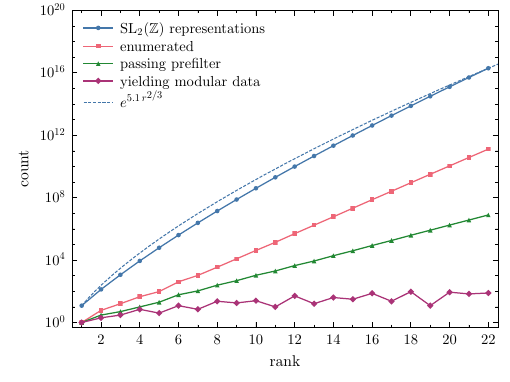}
    \includegraphics[width=.49\textwidth]{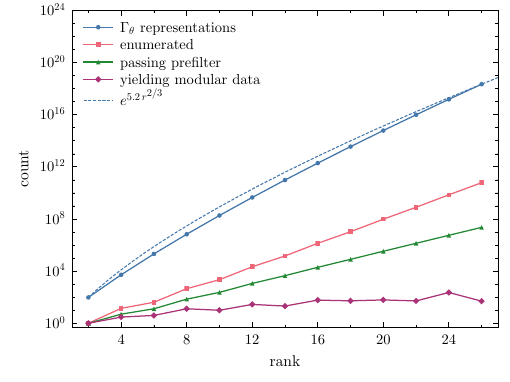}
    \caption{Counts of representations tested in each stage of the algorithm, for modular categories on the left and supermodular categories on the right. Enumerated representations are congruence representations with connected $\tilde\rho(\modt)$ (respectively $\tilde\rho(\modt^2)$) spectra, up to character tensoring. Those passing the prefilter are counted up to Galois conjugation as well, and those yielding modular data are the classes that also pass the search described in \cref{sec:search}. Finally, a representative fit of the scaling of the number of congruence representations is provided.}
    \label{fig:counts}
\end{figure}

Irreducible representations of $\sl{\zee{}}$ are generated by \textsf{SL2Reps} in a basis with symmetric unitary $\tilde\rho(\mods)$ and diagonal $\tilde\rho(\modt)$; such a basis always exists~\cite{ngSymmetricRepresentations$textSL_2mathbbZ$2023}. An irreducible congruence representation $\rho$ of level $n = \prod_i p_i^{k_i}$ is given by $\rho = \bigotimes_i \rho_i$ for $\rho_i$ of level $p_i^{k_i}$. In this way, all congruence representations of $\sl{\zee{}}$ can be constructed from the classified prime-power level irreducibles. The number of $r$-dimensional congruence representations is bounded by $N_\text{rep}(r) \leq \exp(r^{2/3+o(1)})$; this bound, derived in \cref{app:scaling}, is sharp and compared with the data in \cref{fig:counts}. None of the checks applied by the prefilter appear to reduce this scaling asymptotically, so this bound sets the scaling of the algorithm as implemented. This bound need not translate into one on the number of modular data, whose asymptotics are unknown: very few representations yield valid data, but each may yield several. 

To minimize the number of representations tested, only those whose irreducible summands are connected through shared $\modt$ eigenvalues are generated; that is, the graph with a vertex for each irreducible summand, and an edge between summands sharing a $\rho(\modt)$ eigenvalue, must be connected~\cite{bruillardClassificationModularCategories2015}. This must be the case as, if
\begin{equation}\tilde\rho = \tilde\rho_1 \oplus \tilde\rho_2\end{equation}
for any (possibly reducible) $\tilde\rho_{1,2}$ with non-overlapping $\modt$-spectra, no change of basis $U\in \prod_i O(n_i)$ will give $\rho(\mods)$ a column with no vanishing entries. There can therefore be no choice of basis and unit index yielding non-zero quantum dimensions $d_i$. Before the remaining checks, these representations are quotiented by the action of tensoring with characters. All but one of the checks applied are invariant under $\tilde\rho\sim \tilde\rho\otimes \chi$, so this quotient provides almost a 12-fold reduction in the tests needed and a 96-fold one in the fermionic case discussed below.

We pass now to the constraints applied by the prefilter. These involve the following $\tilde\rho(\modt)$ eigenvalue index sets: the indices of one-dimensional eigenspaces, on which any change of basis acts only by a sign,
\begin{equation}I_\text{ndeg} = \{i : \text{mult}(\tilde\rho(\modt)_{ii})=1 \}\end{equation}
for each eigenvalue $\tilde\theta$ of $\tilde\rho(\modt)$, the eigenspace index set
\begin{equation}I = \{i:\tilde\rho(\modt)_{ii} = \tilde\theta\}\end{equation}
and the set $I_\text{nonzero}$ of indices $i$ for which, for some admissible $U$, row $i$ of $U\tilde\rho(\mods)U^{-1}$ has no vanishing entries. The conditions tested on the quotiented, connected-$\modt$ representations are as follows. Most of these were applied, in related forms, in~\cite{ngReconstructionModularData2023,ngClassificationModularData2025}, with proofs provided for convenience and since several are modified.\footnote{Several additional constraints have been proposed. For example, $\operatorname{ord}(\tilde\rho(\modt)) \mid 12\, \operatorname{ord}\left(\frac{\tilde\rho(\modt)}{\tilde\rho(\modt)_{uu}}\right)$ follows from~\cite{dongCongruencePropertyConformal2015} and was included as a filtering constraint in~\cite{ngClassificationModularData2025}. These are not listed below, as they hold for all congruence representations of $\sl{\zee{}}$.} The proofs rely on the classification of prime-power level irreducible representations~\cite{nobsIrreduziblenDarstellungenGruppen1976,nobsIrreduziblenDarstellungenGruppen1976b,ngSymmetricRepresentations$textSL_2mathbbZ$2023,ngReconstructionModularData2023}, for which the possible dimensions are summarized in \cref{tab:dims} of \cref{app:scaling}. Constraints are listed in the order in which they are tested, chosen for computational efficiency.
\begin{enumerate}
\item If $\operatorname{pord}(\tilde\rho(\modt)) = p >3$ is prime, then $\tilde\rho$ is not a direct sum of a $\frac{p+1}{2}$-dimensional irrep and more than one 1-dimensional representation. Here $\operatorname{pord}$ is the projective order of $\tilde\rho(\modt)$, the minimal $n$ such that $\tilde\rho(\modt)^n \propto I$.
\begin{itemize}
  \item The proof is rather involved. See Proposition~3.22 of~\cite{ngReconstructionModularData2023}.
\end{itemize}
\item If $\operatorname{pord}(\tilde\rho(\modt)) = pq$ for primes $q > p >3$ such that $pq = 3\pmod4$, then $\operatorname{rank}(\tilde \rho) \neq \frac{p+q}{2} + 1$, and if $p>5$, $\operatorname{rank}(\tilde \rho) > \frac{p+q}{2} + 1$.
\begin{itemize}
  \item This is Proposition~3.11 of~\cite{ngReconstructionModularData2023}. Consider rank $r \leq \frac{p+q}{2} + 1$. Consider first the case of an irreducible subrepresentation $\rho$ of level $pq$, which must split as $\rho = \rho_1 \otimes \rho_2$. A level-$p$ irrep has dimension at least $\frac{p-1}{2}$, so $\frac{p-1}{2}\frac{q-1}{2} \leq \frac{p+q}{2}+1$. This requires $p=5$, $q=7$, and hence $\dim\rho = 6$. Level-$p$ irreps of least dimension have $\modt$-spectra consisting of distinct primitive $p$-th roots of unity. Hence $r = 7$ is impossible: no character could share a $\modt$ eigenvalue with $\rho$.\\
  Now, if there is no such irrep, there must be irreducible subrepresentations $\rho_1, \rho_2$ of levels $p$ and $q$. If either has least dimension,
  \begin{equation}
  \tilde \rho = \rho_1 \oplus \rho_2 \oplus \rho_3\end{equation}
  where $\rho_3$ is at most 2-dimensional. No $\rho_3$ of dimension at most 2 can share a $\modt$ eigenvalue with both of the other subrepresentations, as it would need a level divisible by $pq$. Similarly, $\rho_1, \rho_2$ share no $\modt$ eigenvalue, contradicting connectivity. Hence neither has least dimension, so $\dim \rho_1 = \frac{p+1}{2}$ and $\dim \rho_2 = \frac{q+1}{2}$, saturating the bound on $r$. We have $\rho_1(\mods)^2 = \left(\frac{-1}{p}\right) I$ and $\rho_2(\mods)^2 = \left(\frac{-1}{q}\right) I$ for representations of this dimension and level. Since $pq = 3\pmod4$, these signs are opposite, so $|\operatorname{Tr}\tilde\rho(\mods)^2| = \frac{q-p}{2}$ and there are $r-|\operatorname{Tr} \tilde\rho(\mods)^2| = p+1$ non-self-dual objects. This is a contradiction: dual objects share a twist, but $\tilde\rho(\modt)$ has only a single degenerate eigenvalue, of multiplicity 2, so at most two objects can be non-self-dual.
\end{itemize}
\item If $\tilde \rho = \rho\oplus n\chi$ for an irreducible $\rho$ with non-degenerate $\rho(\modt)$ and $n\geq 2$, then $(\rho(\mods)\chi(\mods)^{-1})^2 = I$.
\begin{itemize}
  \item Since $(\mods\modt)^3 = \mods^2$ is central in $\sl{\zee{}}$ and $\mods^4 = \id$, we have $\rho(\mods^2) = c I$ and $\chi(\mods^2) = k$ with $c,k=\pm1$; we want to show $c = k$. Let $d = \dim\rho \geq 2$ (if $d = 1$, $\rho(\modt) = \chi(\modt)$ forces $\rho = \chi$, as $\chi(\mods) = \chi(\modt)^{-3}$ for characters). By connectivity, $\chi(\modt)$ is an eigenvalue of $\rho(\modt)$, so $\tilde\rho(\modt)$ has $d-1$ one-dimensional eigenspaces and one $(n+1)$-dimensional eigenspace. In the simple-object basis, $U\tilde\rho(\mods^2)U^T = \kappa C$ for a single sign $\kappa$. The $d-1$ objects spanning one-dimensional eigenspaces are self-dual, so $\kappa = c$. On the $(n+1)$-dimensional eigenspace, $C = c\,U\tilde\rho(\mods^2)U^T$ therefore has eigenvalue $1$ with multiplicity one and $ck$ with multiplicity $n$. As $C$ permutes the $n+1$ objects in this eigenspace, $ck = -1$ would require $n$ disjoint transpositions among them, so $n\leq 1$. Hence $c = k$ for $n\geq 2$. 
\end{itemize}
\item There exists an index set $I$ such that $\tilde\rho(\mods)_{IJ} \neq 0$ for all $I\neq J$, and $\begin{cases}
  \tilde\rho(\mods)_{II}\neq 0, &\text{if}~|I| = 1\\
  \tilde\rho(\mods)_{II} \neq \lambda\,\id, &\text{if}~|I| >1
\end{cases}$.
\begin{itemize}
    \item This is a requirement for $I\cap I_\text{nonzero} \neq \varnothing$ so that there are suitable unit indices with non-vanishing quantum dimensions.
\end{itemize}
\item $\operatorname{Tr} \tilde \rho(\mods^2) \in \mathbb Z \setminus \{0\}$.
\begin{itemize}
    \item $\operatorname{Tr} \tilde \rho(\mods^2) = \pm \operatorname{Tr} C$, and the unit $1$ is self-dual.
\end{itemize}
\item $\operatorname{sgn}(\operatorname{Tr} \tilde\rho(\mods^2)) \operatorname{Tr}_{I}(\tilde\rho(\mods^2)) \geq 0$ for all $I$.
\begin{itemize}
    \item $\rho(\mods^2) = \pm C$ with a single overall sign, and $\operatorname{Tr}_I C \geq 0$ counts the self-dual objects in $I$. So, this sign must agree on each subtrace.
\end{itemize}
\item For some $I$ with $I\cap I_\text{nonzero} \neq \varnothing$, $|\operatorname{Tr}_{I}(\tilde\rho(\mods)^2)| >0$.
\begin{itemize}
    \item The unit $1$ is self-dual.
\end{itemize}
\item For all $\sigma_a\in\operatorname{Gal}(\mathbb Q(\zeta_n)/\mathbb Q)$, with $ab = 1\pmod n$, and all $i,j\in I_\text{ndeg}$, \begin{equation}\sigma_a(\tilde\rho(\mods)_{ij}) = (\tilde\rho(\modt)^a\tilde\rho(\mods)\tilde\rho(\modt)^b\tilde\rho(\mods)\tilde\rho(\modt)^a)_{ij}\end{equation}
\end{enumerate}
The final check is not generally character-tensoring invariant. Instead, if a representative fails there, the check is repeated for the other representations in its orbit. The index choices satisfying condition~7 yield the possible unit indices, over which the search later branches. Lastly, the representations passing all the conditions above are further quotiented by Galois conjugation. Representatives of the classes under simultaneous character tensoring and Galois conjugation are the candidates on which the numerical search is performed.

Various other algebraic tests are possible at the prefilter stage (see~\cite{ngClassificationModularData2025}), but are not implemented, as they reject few, if any, representations. This is at no cost in exactness; such representations were simply cheaper to reject in the search than with these conditions. 
In order to understand the scaling of the problem, counts of the representations before and after the prefiltering are presented in \cref{fig:counts}. A rough fit of the scaling derived in \cref{app:scaling} is given there, showing agreement with the data. Both the restricted enumeration and the remaining prefilter checks greatly reduce the number of representations that must be searched over. On the other hand, neither alters the scaling of the problem, so that a bottleneck is inevitably reached. Notably, the number of representations surviving the prefilter increasingly overshoots the number that actually yield modular data, which appears to grow far more slowly. Additional constraints, narrowing the objective onto these successful candidates, would greatly increase the rank at which this method remains tractable.

\subsubsection{Fermionic Prefiltering}

The fermionic prefilter proceeds analogously. First, irreducible representations of $\Gamma_\theta$ must be constructed from those of $\sl{\zee{}}$. Irreducibles of dimension $d$ are constructed by decomposing the restrictions of representations of dimensions $d$, $3d$, and $\frac{3d}{2}$ if $d$ is even. While this is possible algebraically at low rank, and an efficient procedure is detailed in~\cite{choClassificationFermionicTopological2023}, it becomes intractable well before the 13-dimensional representations needed at rank 26, which require decomposing 39-dimensional ones (recall that $d$-dimensional representations of $\Gamma_\theta$ produce supermodular data of rank $2d$). The decomposition is instead performed numerically. This is efficient, but it means the algebraic checks of the bosonic prefilter cannot be performed in the fermionic case. Once again, this does not reduce the exactness of the final result; it is merely a performance compromise.

As before, the prefilter forms all representations of a given dimension with connected $\modt^2$ graphs, quotiented by character tensoring. Unlike congruence representations of $\sl{\zee{}}$, those of $\Gamma_\theta$ need not admit a basis where $\rho(\mods)$ is symmetric and $\rho(\modt^2)$ is diagonal. The first example occurs at dimension 8, where the irreducible representation formed by
\begin{equation}
S_+
=
\frac{i}{2}
\begin{pmatrix}
0_4&F_4\\
\overline{F}_4&0_4
\end{pmatrix},
~~~~~F_4=
\begin{pmatrix}
1&1&1&1\\
1&i&-1&-i\\
1&-1&1&-1\\
1&-i&-1&i
\end{pmatrix}
\end{equation}
and $T_+^2
=
\operatorname{diag}\!\left(
\zeta_8,\zeta_8^3,\zeta_8^5,\zeta_8^7,
1,\zeta_4,-1,-\zeta_4
\right)$
does not admit a symmetric-diagonal basis. The prefilter only accepts representations that do admit such a basis. With only the numeric forms available, the prefiltering checks that can be performed are more limited. With the index sets now taken from the eigenspaces of $\tilde\rho(\modt^2)$, they are:
\begin{enumerate}
  \item For any irreducible representation $\rho$ with non-degenerate $\rho(\modt^2)$, $\tilde\rho = n\rho$ only if $n = 1$.
  \item There exists an index set $I$ such that $\tilde\rho(\mods)_{IJ} \neq 0$ for all $I\neq J$, and if $|I| = 1$, $\tilde\rho(\mods)_{II}\neq 0$, and if $|I| > 1$, $\tilde\rho(\mods)_{II} \neq \lambda\,\id$.
  \item $\operatorname{Tr}\tilde \rho(\mods^2) \in \mathbb Z \setminus \{0\}$.
  \item $\operatorname{sgn}(\operatorname{Tr} \tilde\rho(\mods^2)) \operatorname{Tr}_I(\tilde\rho(\mods^2)) \geq 0$ for all $I$.
  \item For some $I$ with $I\cap I_\text{nonzero} \neq \varnothing$, $|\operatorname{Tr}_I(\tilde\rho(\mods)^2)| >0$.
\end{enumerate}
The final check again yields the possible unit indices, and Galois quotienting is as in the bosonic case.

\subsection{Galois Reduction}
\label{sec:reduction}
For each candidate representation, we must now look for a basis satisfying the conditions of modular data, listed later in this section. This is generally a continuous problem, searching over changes of basis forming $\prod_i O(n_i)$, with $n_i$ the multiplicities of the $\rho(\modt)$ eigenvalues.
Before performing the gradient descent, we would like to reduce the dimension of the search space as much as possible. The primary way this is done is by fixing the form of the target signed-\hspace{0pt}permutation Galois matrices $\{D_a\}$,
\begin{equation}
    U\tilde D_a U^T = D_a, ~~~~~\tilde D_a = \tilde \rho(\modt^a\mods\modt^b\mods\modt^a\mods^{-1})
\end{equation}
for $a\in (\zee{}/n\zee{})^\times$, with $ab = 1\pmod n$ and $n=\operatorname{ord}(\rho(\modt))$. In the fermionic case, we instead fix $\{H_a\}$,
\begin{equation}
    U\tilde H_a U^T = H_a, ~~~~~\tilde H_a = \tilde \rho\left((\modt^2)^\frac{1-a}{2}\mods(\modt^2\mods)^{b-1}(\modt^2)^\frac{1-a}{2}\mods^{-1}\right)
\end{equation}
for $a\in (\zee{}/l\zee{})^\times$, with $ab = 1\pmod l$ and $l=\operatorname{lcm}(2\operatorname{ord}(\rho(\modt^2)), \operatorname{ord}(\rho(\mods\modt^2)))$. 
By the results of \cref{sec:representation}, the same reduction applies in the fermionic case. We use the notation of the bosonic case for the remainder of this section, but the fermionic case proceeds simply by substituting $\modt$ with $\modt^2$. Since $[U,\rho(\modt)] = 0$, the possible signed permutations can be restricted by the Galois action on $\rho(\modt)$:
\begin{equation}
    D_a \rho(\modt) D_a^T = \rho(\modt)^{a^2}
\end{equation}
That is, $(D_a)_{ij} \neq 0$ only if $\rho(\modt)_{jj} = \rho(\modt)_{ii}^{a^2}$. The number of possible target Galois families $\{D_a\}$ grows with $\operatorname{ord}(\rho(\modt))$. This in turn scales as
\begin{equation}
    \operatorname{ord}(\rho(\modt)) \leq \exp\left((1+o(1))\sqrt{2r\log r}\right)
\end{equation}
and the same bound holds for the level $l$ in the fermionic case. This bound is sharper than that given in~\cite{bruillardRankfinitenessModularCategories2015}, with its worst-case scaling achieved asymptotically, and none of the prefiltering conditions suggest a reduction in this. A proof is given in \cref{app:scaling}. Therefore, enumerating these target families must be done as efficiently as possible. The target families are formed only for the quotient $G = (\zee{}/n\zee{})^\times/{\sim}$, where $a\sim a'$ if $\tilde D_a = \tilde D_{a'}$.

We enumerate the possible signed-permutation representations $\{D_a\}$ of $G$ that could correspond to Galois actions as follows. Signed monomial representations decompose into transitive monomial representations on each $G$-orbit. A finite-dimensional representation is transitive monomial if and only if it is induced from a one-dimensional representation of a subgroup. On the basis lines $L_i=\mathbb R e_i$, $G$ acts by permutations. Then, the stabilizer $H = \operatorname{Stab}_G(L)$ of a basis line $L$ acts on the subspace spanned by $L$ as a character $\chi(h)\in\{\pm1\}$. The representation of $G$ on the full orbit is therefore the induced representation $\operatorname{Ind}_H^G\chi$. Since $G$ is abelian, the pair $(H,\chi)$ uniquely identifies the transitive monomial summand up to isomorphism and, since $\operatorname{Stab}_G(L) = \operatorname{Stab}_G(L')$ for lines $L, L'$ in the same orbit, does not depend on the choice of line $L$. Importantly, we need not distinguish isomorphic Galois representations. Isomorphisms of transitive monomial representations are signed permutations acting by conjugation. These factor as permutations and signed diagonal matrices, the former of which are equivalences of the modular data and therefore need not be considered. The signed diagonal matrices are quotiented here, as the signs they introduce can be restored at the end, minimizing the number of independent descents.

The possible transitive summands are further restricted by their action on $\rho(\modt)$. For each $g\in G$, let $a(g) \in (\zee{}/n\zee{})^\times$ label its representative, so $g = \tilde D_{a(g)}$, and let $\rho(\modt)_{ii}=e^{2\pi i q_i/n}$. The action of $g$ on the exponents is given by
\begin{equation}g\cdot q_i = a(g)^2 q_i\end{equation}
splitting the twists within an orbit where $G$ acts transitively. The Galois actions considered are, therefore, those for which the stabilizer $H$ of the line $L$ is a subgroup of the stabilizer of its exponent $q$:
\begin{equation}
    H \leq K_q = \{g\in G : a(g)^2 q = q \pmod{n}\}.
\end{equation}
Once again, $K_q$ does not depend on the choice of line representative $L$. With this, we assemble the possible Galois representations $\{D_a\}$, matched to the eigenvalue spectrum of $\rho(\modt)$.

These candidate Galois families must be reachable from the initial set $\{\tilde D_a\}$. The $U$ that realize them form the reduced search space, given by the intertwiner locus
\begin{equation}
    \mathcal U(\{D_a\}) = \left\{U\in\prod_i O(n_i): U\tilde D_a = D_a U~ \forall a \in (\zee{}/n\zee{})^\times\right\} \,.
\end{equation}
In order to compute it, we use the fact that, if non-empty, it is $U_0 \cdot Z$ for a single point $U_0$ and $Z$ the centralizer of $\{\tilde D_a\}$ in $\prod_i O(n_i)$. For a twist-orbit representative $q$, let $F_q$ be the real span of the basis vectors indexed by $\{i:q_i=q\}$, with $n_q=\dim_{\mathbb R}F_q$. Its complexification is the corresponding eigenspace of $\rho(\modt)$. The action of the stabilizer $K_q$ preserves $F_q$, so, restricted to $F_q$, both $\{\tilde D_a\}$ and $\{D_a\}$ yield representations of $K_q$. Since the allowed change of basis on this fiber is \(O(n_q)\), the source and target restrictions must be isomorphic as real \(K_q\)-representations. Because \(K_q\) is abelian, after complexification these representations decompose into characters. Matching orthonormal bases of the corresponding character spaces produces real orthogonal intertwiner blocks which are assembled across each orbit to give the intertwiner $U_0$.

If the $K_q$-representations on the $F_q$ have $k$ real isotypic summands, $Z$ has $2^k$ connected components. Representative points $U_A$ in each connected component of $\mathcal U(\{D_a\})$ are formed by reflections indexed by subsets $A$ of the set $J$ of real isotypic summands. Splitting $Z$ into connected components allows us to describe the descent procedure on a single one.
Before passing to the search, these connected components are quotiented by the identification $\tilde\rho(\mods)_A = U_A \tilde\rho(\mods) U_A^T \sim \tilde\rho(\mods)_0$ if they differ by conjugation by a signed diagonal. This completes the signed-diagonal quotienting, this time from those commuting with the $\{D_a\}$.

\subsection{Numerical Search}
\label{sec:search}

Having constructed the reduced search spaces, we now wish to find all points where $(\rho(\mods),\rho(\modt))$ yield modular data. One such condition is that the fusion rules
    \begin{equation} \tilde N_k^{ij} = \sum_{l} \frac{\rho(\mods)_{li}\rho(\mods)_{lj}\rho(\mods^{-1})_{lk}}{\rho(\mods)_{lu}}\end{equation}
be integers (signs have not yet been reintroduced, so they may still be negative). The set of points satisfying this condition cannot be positive-dimensional. It is not obvious, however, that this rules out sequences of integer fusion rules converging as $\rho(\mods)_{lu}\rightarrow 0$, and rank-finiteness alone is not sufficient to guarantee this. Even if no such sequences exist, if the $\tilde N_k^{ij}$ remain real as $\rho(\mods)_{lu}\rightarrow 0$, they will oscillate increasingly rapidly and, in general, should pass arbitrarily close to integers, so that a numerical evaluation need not be able to distinguish them from integers. This behavior complicates standard descent techniques applied to $\tilde N_k^{ij}$: in general, for no cost function will we be able to identify all minima, nor should we expect to identify all points within numerical precision of zero cost.

\begin{figure}[!ht]
    \centering
    \includegraphics[width=.49\textwidth]{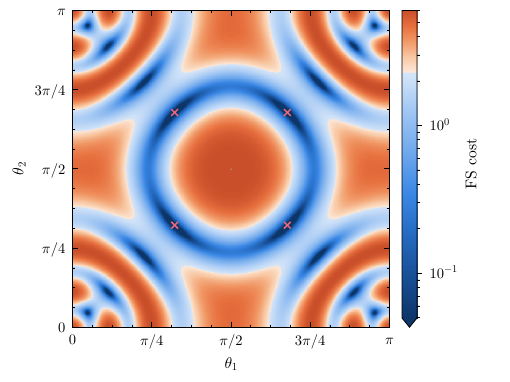}
    \includegraphics[width=.49\textwidth]{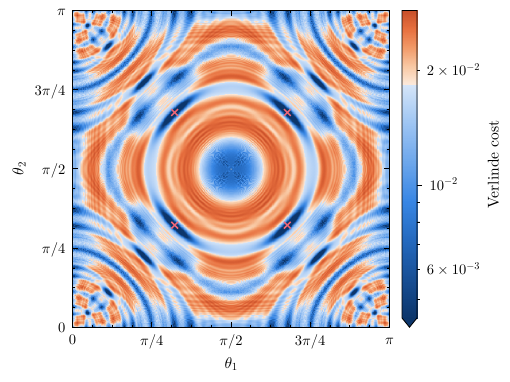}
    \caption{Representative cost functions from a candidate representation yielding modular data, evaluated on the Galois-reduced search spaces. Frobenius--Schur indicator costs are on the left, while Verlinde costs are on the right. Solution points are marked in red. The Frobenius--Schur cost is far smoother than the Verlinde cost and therefore behaves better under gradient descent. In this case, the Frobenius--Schur cost is sufficient to isolate the solutions, although this is not generally the case. It should be noted that this example was selected for visualization purposes and is relatively trivial: solution-bearing representations were tested with as much as 70-dimensional search spaces.}
    \label{fig:costs}
\end{figure}

The search addresses this, broadly, as follows. Initial $U$ within the search space are sampled using a scrambled Halton sequence. These are then costed first against the second Frobenius--Schur indicator conditions. Descent on these is far better behaved than on the fusion rules but is generally insufficient to isolate points. Instead, this descent is used to bring sampled starts closer to solution-bearing Verlinde basins. Not-necessarily-isolated points satisfying the indicator constraints are then used as starting points for a two-stage descent applied to the fusion rules. Each zero of this cost is then passed to the validation step. The search continues on a given component until the stopping condition below is satisfied. For some sense of the difference in the descent behavior on the indicator and Verlinde costs, an example from a representation yielding modular data is provided in \cref{fig:costs}. 

Each descent depends on a candidate representation, a connected component of the intertwiner locus $\mathcal U(\{D_a\})$ with an identified $U_A$, and a choice of unit index. Once the connected components and $U_A$ have been identified, the search space then depends only on the candidate representation. The Lie algebra of $Z$ is
\begin{equation}\mathfrak g = \{X : X^T = -X, [X,\rho(\modt)] = [X, \tilde D_a] = 0~\forall a \in (\zee{}/n\zee{})^\times\}\end{equation}
and we let $\mathfrak h = \{X\in\mathfrak g: [X,\tilde\rho(\mods)]=0\}$. The search is performed in $\mathfrak p = \mathfrak h^\perp \simeq \mathfrak g/\mathfrak h$, the tangent space of the compact homogeneous space constructed by quotienting directions preserving $\tilde\rho(\mods)$. In general, $\mathfrak p$ need not be a Lie subalgebra of $\mathfrak g$; for example, $\mathfrak{so}(3)/\mathfrak{so}(2) \simeq T_p S^2$. In practice, the basis $\{X_k\}$ of $\mathfrak p$ is formed starting from the standard block skew-symmetric basis for the Lie algebra of $\prod_i O(n_i)$. The $\{\tilde D_a\}$ commutation constraints are imposed by SVD to obtain a basis for $\mathfrak g$. Then, another round of SVD on $X\mapsto[X,\tilde\rho(\mods)]$ splits $\mathfrak g$, with the non-singular directions giving the search basis $\{X_k\}$ of $\mathfrak p$. If $\mathfrak p = 0$, the search space consists of isolated points, which are passed directly to the validation stage; otherwise, the descent is performed.

In the simplest case, the search space is a torus, meaning the $X_k$ commute and can be normalized to have integral frequencies. In that case, one has a chart
\begin{equation}U(\theta) = U_A \exp\left(\sum_k \theta_k X_k\right)\end{equation}
When this is not the case, a surjective chart can be formed as follows. Choose an orthogonal basis of $\mathfrak p$ with respect to the Frobenius inner product, normalized as $\operatorname{Tr}(X_i X_j^T) = 2\delta_{ij}$. $\mathfrak p$ is the tangent space of a compact normal homogeneous space, so by Hopf--Rinow, $\exp$ is surjective. A shortest lift $X$ has skew eigenvalues $\pm i\varphi_j$ with $|\varphi_j| \leq \pi$.
Therefore, $\left\lVert X \right\rVert_F^2 = \sum_{i,j} |X_{ij}|^2 = 2\sum_j \varphi_j^2 \leq r \pi^2$, where $r$ is the rank. So, any shortest lift $X = \sum_j c_j X_j$ has coefficients bounded by $|c_j| \leq \left\lVert c \right\rVert = \left\lVert X \right\rVert_F/\sqrt{2}\leq \pi \sqrt{\frac{r}{2}} = R$. One can thus form a surjective chart with shortest lifts in $[-R,R]^d$, where $d = \dim\mathfrak p$, as
\begin{equation}U(\theta) = U_A \exp\left(R\sum_k \sin(\theta_k)X_k\right), ~~~~~\theta \in [0,2\pi)^d\end{equation}
These global charts are used only for sampling, and not for descent updating.

\subsubsection{Sampling and Stopping Procedure}
Starting points for the descent are selected by an affine Owen-type scrambled Halton sequence~\cite{haltonEfficiencyCertainQuasirandom1960,owenRandomlyPermutedTmsNets1995}. These are chosen as they sample more uniformly than pseudorandom sequences for small numbers of samples. The $(j+1)$-th prime, $p_j$, is assigned to the $j$-th coordinate. Then, writing $n$ in base $p_j$ with digits $a_{j,k}$,
\begin{equation}n = \sum_{k\geq 0} a_{j,k}p_j^k\end{equation}
let
\begin{equation}\phi_{p_j}(n) = \sum_{k\geq 0} a_{j,k} p_j^{-k-1}\end{equation}
The $j$-th coordinate of the $s$-th sample point is given by \begin{equation}\frac{1}{2\pi}\theta_j^{(s)} = \phi_{p_j} (s+1)\pmod1\end{equation}
These sequences become linearly correlated in higher dimensions, so a scrambling procedure is used. For each coefficient $a_{j,k}$, the scramble determines integers $\alpha\in\{1,\dots,p_j-1\}$ and $\beta\in\{0,\dots,p_j-1\}$ by a hash depending on a seed, $j$, and $a_{j,l}$ for all $l< k$. Then, $a_{j,k}$ is substituted with $b_{j,k} = \alpha a_{j,k} + \beta \pmod{p_j}$.

Descents are performed in four streams with distinct scramble offset seeds. Each stream reaches a checkpoint whenever its number of samples reaches a power of two. The stopping condition is as follows. For each representation, connected component of the intertwiner locus, and choice of unit index, descents continue until two consecutive checkpoints, starting at 16, pass without finding additional data, so each stream performs at least 32 samples. If any such branch of a representation realizes modular data, all branches of that representation are revisited with eight streams using new seeds and at least 64 samples each. This revisit is cheap as very few representations yield data. The stopping condition is subtle, as this is the point where modular data could be missed. The procedure deliberately takes more samples than necessary in known cases to avoid this. In every case below rank 16, for example, all modular data are found with only four streams of no more than 32 samples.

\subsubsection{Frobenius--Schur Descent}

Having sampled starting points, we first precondition the descent on the second Frobenius--Schur indicators
\begin{equation}\nu_2(i) = \sum_l \frac{\rho(\mods\modt^2\mods)_{lu}\rho(\mods\modt^{-2}\mods^{-1})_{lu}\rho(\mods^{-1})_{li}}{\rho(\mods)_{lu}}\end{equation}
Although these are generally insufficient to isolate solutions, they have known values in $\{0,\pm1\}$ (as opposed to $\zee{}$ for Verlinde) and therefore have better-behaved and non-oscillatory costs. From a performance standpoint, the Frobenius--Schur descent rejects all but a tiny fraction of candidate representations, and its cost is much cheaper to evaluate than the Verlinde cost. On candidates that do reach a zero-cost endpoint of the Frobenius--Schur descent, the Verlinde descent is started from this point; this starting point is therefore closer to the Verlinde basins of valid modular data than the original Halton-sampled points.

Both $\nu_2$ and $N^{ij}_k$ depend on a choice of unit index. The search once again branches over the possible units identified by the prefilter, taken up to permutations that fix $\rho(\modt)$ and $\{D_a\}$. Further, since $\{D_a\}$ is fixed on the branch, the charge-conjugation matrix is known:
\begin{equation}D_{-1} = \rho(\mods^2)\end{equation}
either by the group relations or by noting $D_{-1}$ acts as complex conjugation. This holds similarly in the fermionic case. Unit candidates are then rejected if they are not fixed by $\operatorname{sgn}(\operatorname{Tr} D_{-1})\, D_{-1}$.

With $D_{-1}$ fixed, we can determine which indicators are non-zero. The Frobenius--Schur cost function is computed as
\begin{equation}R_\text{FS} = \left\{\nu_2(i)^2 - |(D_{-1})_{ii}|\right\}_{i=1}^r, ~~~~~C_\text{FS} = \left\lVert R_\text{FS} \right\rVert^2\end{equation}
The local updates in the search are Cayley updates by right multiplication. A step $\delta \in \mathbb R^d$ acts as $U\mapsto UQ(\delta)$, where
\begin{equation}Q(\delta) = \left(I-\frac12 \delta\cdot X\right)^{-1}\left(I+\frac12 \delta\cdot X\right)\end{equation}
The descent is performed in two stages: first, a large-step coordinate-pattern search moves from the start into a nearby deep basin. The purpose of this is to avoid nearby non-zero local minima. Then, the rest of the descent is performed by damped Gauss--Newton or Levenberg--Marquardt-style refinement. The $(2r\times d)$ Jacobian, $J$, is formed and
\begin{equation}(J^TJ+\lambda I)\,\delta = -J^T R_\text{FS}\end{equation}
is solved by $LDL^T$ decomposition. The damping parameter, $\lambda$, interpolates between gradient descent and Gauss--Newton: further from a basin, larger damping provides more robust updates close to gradient descent, while, within a basin, weaker damping offers the better performance of Gauss--Newton. Having reached a zero of the cost, the descent moves to Verlinde starting from that point, while a non-zero minimum rejects that start, and the search moves to the next.

Before turning to the Verlinde descent, we note that various other constraints could be incorporated as costs. For example, $\nu_n(a) \in \zee{}[\zeta_n]$~\cite{ngHigherFrobeniusSchurIndicators2007} and is real, so, in particular, $\nu_3,\nu_4,\nu_6\in\mathbb Z$. Although adding these constraints does, in general, reduce the dimension of the $\nu_2$ solution locus, they suffer from the same complications as Verlinde, having unfixed values and, hence, oscillatory costs. Any constraint added here would be most beneficial if it were known to take a fixed value.

\subsubsection{Verlinde Descent}

The rest of the descent costs the fusion rules
\begin{equation}\tilde N_k^{ij} = \sum_{l} \frac{\rho(\mods)_{li}\rho(\mods)_{lj}\rho(\mods^{-1})_{lk}}{\rho(\mods)_{lu}}\in\mathbb Z\end{equation}
These need not be non-negative, as signs have yet to be reintroduced. The unit row of $\rho(\mods)$ in the target basis must be either real or pure imaginary, determined by $D_{-1}$. If $\operatorname{Tr} D_{-1}>0$, the unit row is real, and otherwise imaginary. These cases are costed separately. The Verlinde search splits into a first descent to localize a deep basin, and then a second descent with fixed fusion rules into that basin.

By commutativity, the independent fusion coefficients are labeled by triples
\begin{equation}\mathcal A_u = \{(i,j,\bar k): 0< i \leq j \leq r, i,j \neq u, 0<k\leq r\}\end{equation}
Then, by Frobenius reciprocity (which holds in the fermionic case as well), this is further restricted to the set $\mathcal B_u$ of unordered multisets $\{i,j,\bar k\}$.\footnote{For triples $\alpha = (i,j,k)$ and $\alpha' = (i',j',k')$ with identical multisets in $\mathcal B_u$, their fusion rules are related by $N_\alpha = \epsilon_k\epsilon_{k'}N_{\alpha'}$.} Duals are again given by the charge-conjugation matrix
\begin{equation}(D_{-1})_{ij} = \epsilon_i \delta_{j\bar i}, ~~~~~\epsilon_i = \pm 1\end{equation}
Let $M=|\mathcal A_u|$, $m_\alpha$ the size of the orbit of $\alpha\in\mathcal B_u$, and $F_\alpha$ a representative of the fusion rules of that orbit. Then, with
\begin{equation}q_j^R = \operatorname{Im} \rho(\mods)_{uj}, ~~~~ q^I_j = \operatorname{Re} \rho(\mods)_{uj}\end{equation}
the first-stage cost function is\footnote{Since our primary aim in the descent is to isolate a finite number of points in the continuous search space, we need only cost as many fusion matrices as are needed for their combined spectrum to be non-degenerate. Although this is valid locally, it is not obvious how to determine in general which set of fusion matrices will produce a non-degenerate spectrum in the desired basis.}
\begin{equation}R_\text{wrap}^p = \left(\left\{\frac{q_j^p}{\sqrt{r}}\right\}_{j=1}^{r}, \left\{\sqrt{\frac{m_\alpha}{2M}}\frac{\sin(\pi \operatorname{Re} F_\alpha)}{\pi}, \sqrt{\frac{m_\alpha}{2M}}\operatorname{Im} F_\alpha\right\}_{\alpha \in \mathcal B_u}\right), ~~~~~ C_\text{wrap}^p(S) = \lVert R_\text{wrap}^p\rVert^2\end{equation}
for phases $p\in\{I,R\}$ determined by $D_{-1}$.

Damped Gauss--Newton descent is performed as in the indicator case, localizing a deep basin. Within such a basin, the Verlinde formula rounded to integers yields constant fusion rules:
\begin{equation}n_\alpha = \text{round}(\operatorname{Re} F_\alpha)\end{equation}
These $n_\alpha$ are then tested for associativity
\begin{equation}\sum_m N^{ij}_m N^{mk}_l = \sum_m N^{jk}_m N^{im}_l\end{equation}
before the second descent; a failure rejects the start. The second-stage cost function is then
\begin{equation}R_\text{fixed}^p = \left(\left\{\frac{q_j^p}{\sqrt{r}}\right\}_{j=1}^{r}, \left\{\sqrt{\frac{m_\alpha}{2M}}(\operatorname{Re} F_\alpha - n_\alpha), \sqrt{\frac{m_\alpha}{2M}}\operatorname{Im} F_\alpha\right\}_{\alpha \in \mathcal B_u}\right), ~~~~~ C_\text{fixed}^p(S) = \lVert R_\text{fixed}^p\rVert^2\end{equation}
A zero of this cost function located within numerical tolerance is isolated and is passed to the final validation step. A non-zero minimum rejects the start.

\subsubsection{Validation}
Isolated points need only be tested for the remaining conditions on modular data. For points already isolated by the Galois reduction, the fusion rules
are tested. Then, we reintroduce the signs quotiented earlier to form the modular data:
\begin{equation}S_{ij} = \sigma_i \sigma_j \frac{\rho(\mods)_{ij}}{\rho(\mods)_{uu}}, ~~~~~\theta_i = \frac{\rho(\modt)_{ii}}{\rho(\modt)_{uu}}\end{equation}
with $\sigma_i = \pm 1$ and $\sigma_u = 1$. In the bosonic case, we verify:
\begin{enumerate}
  \item $\nu_1(i) = \delta_{iu}$ and $|\nu_2(i)| = |(D_{-1})_{ii}|$ (the latter is tested again because the Verlinde descent may move off the indicator solution locus).
  \item $N_k^{ij} = \sigma_i\sigma_j\sigma_k \tilde N_k^{ij} \in \zee{\geq 0}$.
  \item $C_{ij} = (S^2)_{ij}/D^2 = N^{ij}_u$.
  \item $S_{ij} = \sum_k N_{k}^{i\bar j} \frac{\theta_k}{\theta_i\theta_j}d_k$.
\end{enumerate}
Far more conditions than these are tested, in particular multiple categorifiability constraints on fusion rings, but we list only those that rejected candidate data.

The fermionic case is somewhat more subtle. One has
\begin{equation}N_k^{ij} + N_{k\otimes \psi}^{ij} = \sigma_i\sigma_j\sigma_k \tilde N_k^{ij} \in \zee{\geq 0}\end{equation}
So the fusion rules are not fully determined. Instead, we must solve for a fusion lift compatible with balancing,
\begin{equation}\hat S_{ij} = \sum_{[k]\in\irr(\C/\psi)} (N_k^{i\bar j} - N_{k\otimes \psi}^{i\bar j}) \frac{\theta_k}{\theta_i\theta_j}d_k\end{equation}
In large-rank cases, these lifts become non-trivial to determine, and they are neither guaranteed to exist nor necessarily unique. 

\subsection{Reconstruction}
\label{sec:reconstruction}

The final step is to leverage the Galois transformations and integrality of $S$ to reconstruct its exact algebraic form. $T$ is straightforward to reconstruct since its entries are roots of unity. The Galois action on $\rho(\mods)$ is
\begin{equation}\sigma_a(\rho(\mods)_{ij}) = \epsilon_a(i) \rho(\mods)_{\hat\sigma_a(i)j}\end{equation}
So, on $S$ (normalized so that $S_{uu}=1$)
\begin{equation}\sigma_a (S_{ij}) = \frac{\epsilon_a(i)}{\epsilon_a(u)}\frac{S_{\hat\sigma_a(i)j}}{S_{\hat\sigma_a(u)u}}\end{equation}
The requirement that these be algebraic integers in $\mathbb Q(\zeta_N)$ is then expressed as a $\varphi(N)\times \varphi(N)$ Vandermonde-type linear system in each entry. The fermionic case proceeds identically, substituting $L = \operatorname{lcm}(2\operatorname{ord}(\hat T^2), \operatorname{pord}(\hat S\hat T^2))$ for $N$.

Having recovered the exact forms, we can generate the Galois-conjugate data that had earlier been quotiented out. Then, a final condition is tested on the data, implemented here as it is not Galois-invariant. For
\begin{equation}\nu_n(a) = \frac{1}{D^2} \sum_{i,j} N_{a}^{ij}d_i d_j \left(\frac{\theta_i}{\theta_j}\right)^n\end{equation}
these indicators must satisfy $|\nu_n(a)| \leq \dim \Hom_\C(1, a^{\otimes n})$~\cite{ngHigherFrobeniusSchurIndicators2007}. This constraint is not implemented in the existing rank-12 classification~\cite{ngClassificationModularData2025}, and rejects two of its identified data at rank 8 and five at rank 12. As a result of this, we do not expect that the known conditions are sufficient to guarantee categorification, although we have sought to test as many as possible.

An analogous constraint is applied in the fermionic case. For $n\geq 3$, the bosonic formula for $\nu_n(a)$, evaluated on $\C/\psi$, does not yield the Frobenius--Schur indicators; for odd $n$ it computes $\nu_n(a)-\nu_n(a\otimes\psi)$. It can nonetheless be bounded by $\dim \Hom_\C(1, a^{\otimes n}) + \dim \Hom_\C(\psi, a^{\otimes n})$, analogously to the bosonic case. For $n=2$, however, it does equal $\nu_2(a)$, since $\Hom_\C(\psi,a\otimes a)=0$ (otherwise $\bar a\simeq a\otimes\psi$, contradicting $\theta_{a\otimes\psi}=-\theta_a$). This justifies the Frobenius--Schur descent in the fermionic case.

\section{Classification and Mapping of Data through Ranks 22 and 26}
\label{sec:results}

The algorithm above was used to identify modular data up to rank 22 and supermodular data up to rank 26. The search was performed on a dual-socket AMD EPYC 9654 machine with 192 physical cores and 2.27~TB of RAM, provided by \href{https://projectnumina.ai}{Project Numina}. The total wall time of the classification, for both bosonic and fermionic data, was approximately two weeks. This data has been made available at \href{https://anyonatlas.org}{\texttt{AnyonAtlas.org}}. The precise counts of the identified data are provided in \cref{tab:bosonic-counts,tab:fermionic-counts}.

\begin{table}[htbp]
\centering
\begin{tabular}{r||rr|rr|r}
Rank & Total & Prime & Pseudo-unitary & Prime & Completely anisotropic \\
\hline
1 & 1 & 1 & 1 & 1 & 1 \\
2 & 8 & 8 & 4 & 4 & 4 \\
3 & 24 & 24 & 12 & 12 & 12 \\
4 & 55 & 19 & 18 & 8 & 15 \\
5 & 26 & 26 & 10 & 10 & 6 \\
6 & 192 & 32 & 50 & 10 & 44 \\
7 & 96 & 96 & 28 & 28 & 22 \\
8 & 294 & 58 & 63 & 19 & 38 \\
9 & 299 & 64 & 81 & 20 & 41 \\
10 & 332 & 140 & 76 & 40 & 29 \\
11 & 158 & 158 & 44 & 44 & 8 \\
12 & 1310 & 102 & 227 & 31 & 134 \\
13 & 398 & 398 & 86 & 86 & 24 \\
14 & 856 & 232 & 150 & 58 & 78 \\
15 & 1050 & 426 & 204 & 84 & 96 \\
16 & 1581 & 272 & 232 & 57 & 99 \\
17 & 507 & 507 & 137 & 137 & 6 \\
18 & 3286 & 638 & 536 & 160 & 190 \\
19 & 402 & 402 & 70 & 70 & 38 \\
20 & 2584 & 334 & 380 & 82 & 118 \\
21 & 2708 & 916 & 454 & 182 & 156 \\
22 & 2165 & 1109 & 357 & 213 & 60 \\
\hline
Total & 18332 & 5962 & 3220 & 1356 & 1219 \\
\end{tabular}
\caption{Modular data identified through rank 22.}
\label{tab:bosonic-counts}
\end{table}

\begin{table}[htbp]
\centering
\begin{tabular}{r||rr|rr|r}
Rank & Total & Prime & Pseudo-unitary & Prime & Completely anisotropic \\
\hline
2  & 1    & 1    & 1    & 1   & 1  \\
4  & 8    & 8    & 4    & 4   & 4  \\
6  & 20   & 20   & 10   & 10  & 4  \\
8  & 72   & 36   & 25   & 15  & 11 \\
10 & 70   & 70   & 30   & 30  & 8  \\
12 & 232  & 80   & 69   & 31  & 23 \\
14 & 190  & 190  & 66   & 66  & 8  \\
16 & 558  & 176  & 135  & 61  & 33 \\
18 & 593  & 384  & 170  & 116 & 11 \\
20 & 784  & 256  & 192  & 80  & 39 \\
22 & 672  & 672  & 196  & 196 & 12 \\
24 & 3028 & 1054 & 637  & 281 & 88 \\
26 & 746  & 746  & 188  & 188 & 10 \\
\hline
Total & 6974 & 3693 & 1723 & 1079 & 252 \\
\end{tabular}
\caption{Supermodular data identified through rank 26.}
\label{tab:fermionic-counts}
\end{table}

\subsection{Closure}
\label{sec:closure}

We obtain all data found in the known classifications of modular data through rank 12~\cite{ngClassificationModularData2025}, of integral modular data through rank 13~\cite{alekseyevClassificationIntegralModular2026}, and of supermodular data through rank 10~\cite{choClassificationFermionicTopological2023}, apart from the modular data rejected by the Frobenius--Schur bound of \cref{sec:reconstruction}. In addition, we identify 72 sets of modular data through rank 12, satisfying all known conditions, that do not appear in~\cite{ngClassificationModularData2025}. These comprise 24 non-pseudo-unitary modular data at ranks 8, 11, and 12 that are Galois conjugate to data appearing in that classification, and 48 data at rank 12, forming 9 Galois orbits, that arise from representations absent from their list. These missing cases are not among the representations highlighted there as unresolved.

Despite this apparent completeness at low ranks, numeric searches alone cannot guarantee that all solutions have been identified. In addition to the stopping conditions of the search itself, we perform various closure checks to test exhaustiveness. The search does not treat the data predicted by these checks differently from any other data, so the checks provide an unbiased probe of its completeness. The closure conditions are as follows:
\begin{enumerate}
  \item Primes: For $\mathcal M \subset \C$ a braided subcategory, if $\mathcal M$ has non-degenerate braiding, it must be that $\C = \mathcal M \boxtimes \mathcal M'$ with both $\mathcal M$ and $\mathcal M'$ modular. In the supermodular case, if $\mathcal M$ is a supermodular subcategory, one has $\C = \mathcal M \boxtimes_{\sVec} \mathcal M'$ with $\mathcal M'$ also supermodular. This is used to determine the prime factorizations of the data. Each such prime factor is verified to be present in the data.\footnote{It should be noted that in neither modular nor supermodular categories is this decomposition unique, either at the level of their data or at the level of the categories. The database lists alternative factorizations when they exist.}
  \item Products: Having computed the prime modular and supermodular data above, we verify that all products of these are contained in the data.
  \item Split Fermionic: A supermodular category is termed split if it is of the form $\C = \mathcal M \boxtimes \sVec$ with $\mathcal M$ modular. As described in~\cite{choClassificationFermionicTopological2023}, these split categories are produced from $\Gamma_\theta$ representations extending to representations of $\sl{\zee{}}$. In these cases, the corresponding modular factor is verified to be contained in the data.
  \item Minimal Extensions: The classification procedure assumes that a minimal extension exists; we require it to be present in the data whenever its rank is within the classified range. In particular, these extensions come at two ranks in $[\frac{3r}{2},2r]$, with $r = \operatorname{rank}(\C)$. If $\C$ is pseudo-unitary, it has exactly 16 pseudo-unitary minimal extensions and 16 non-pseudo-unitary ones; each set has eight at each of the two (not necessarily distinct) ranks. Finally, if extensions are found at rank $R$, the other half must be found at $R' = \frac{7r}{2} - R$, providing a sharper rank bound.
  \item Fermion-Neutral Sectors: Conversely, a spin modular category, having an invertible object $\psi$ with $\psi\otimes\psi= 1$ and $\theta_\psi = -1$, must have its fermion-neutral sector present in the supermodular data.
  \item Chern--Simons Closure: As will be discussed later, the modular data of Chern--Simons theories with a torus or simply connected gauge group (and any of their products) have known forms. These are all verified to exist within the data up to the ranks investigated.
  \item Tannakian Condensation: The existence of a Tannakian subcategory $\operatorname{Rep}(G)$ and its corresponding condensable algebra $A=\operatorname{Fun}(G)$ can be guaranteed from the modular data alone. For each data containing such an $A$, the corresponding condensate is verified to be present in the data whenever its rank is within the classified range. Anyon condensation and Tannakian subcategories are discussed in more detail in \cref{sec:database}.
\end{enumerate}
Although sporadic, these checks are quite stringent. Since completeness is least certain at high rank, the ``bottom-up'' checks are particularly valuable. Product closure essentially guarantees that no non-prime data are missed. Non-prime data are often the hardest to find, as they typically have very large search spaces. For example, one non-prime data set at rank 16 has a $60$-dimensional search space; had the search missed it, product closure would have flagged the failure. Similarly, minimal-extension closure effectively guarantees that spin modular categories are not missed, since the supermodular classification at the corresponding lower rank is much easier: that same rank-16 data would have been flagged as an extension of a rank-8 supermodular category whose 4-dimensional $\Gamma_\theta$ representation has only a 6-dimensional search space. Lastly, metric groups provide closure in important cases: at rank 16, for example, $\Uone_K^2$ with $K=\bigl( \begin{smallmatrix} 0 & 4 \\ 4 & 0 \end{smallmatrix} \bigr)$, a prime non-spin data set with a 36-dimensional search space, would have been flagged had the search missed it.

\subsection{Classification Database}
\label{sec:database}

\href{http://www.anyonatlas.org}{\texttt{AnyonAtlas.org}} also contains several derived properties of each data set, some of which we describe here. First, we compute the fusion subcategories of every data set, each of which may or may not admit a categorification. Of these, the Tannakian subcategories, $\operatorname{Rep}(G)$ for some finite group $G$, are of particular interest. By Deligne's theorem~\cite{deligneCategoriesTensorielles2002}, every symmetric fusion subcategory (one with $\frac{S_{ij}S_{11}}{S_{i1}S_{j1}} = 1$ for all its simple objects $i,j$) is super-Tannakian, and is Tannakian if moreover $\theta_i > 0$ for all its simple objects. On the other hand, fusion rings alone need not identify $G$. In all cases encountered, the rest of the modular data was sufficient to identify $G$ uniquely, but this is not known to hold in general. Relatedly, we construct the symmetry group $\operatorname{Sym} \irr (\C)$ of the modular data, consisting of the permutations of simple objects that fix $S$ and $T$. This is of interest as there is a homomorphism
\begin{equation}\operatorname{Aut}_\text{br}(\C)\rightarrow \operatorname{Sym} \irr (\C)\end{equation}
given by the action of a braided autoequivalence on simple objects. Braided autoequivalences are themselves of interest because they control the possible gaugings of 0-form $G$ symmetries of $\C$~\cite{barkeshliSymmetryFractionalizationDefects2019} or, equivalently, the categories containing $\operatorname{Rep}(G)$ whose condensation yields $\C$~\cite{drinfeldBraidedFusionCategories2009}. However, this homomorphism is, in general, neither injective nor surjective.

For the fermionic data, we also compute the minimal extensions when these appear within the data. For split $\C = \mathcal M \boxtimes \sVec$, the extensions are simply $\mathcal M\boxtimes\mathcal S_\nu$, but for primitive (non-split) supermodular data, there is no known general construction for the data of their minimal extensions. In addition, we compute the fusion lifts of every supermodular entry. A fusion lift must exist for the data to be categorifiable, but it need not be unique. Several entries have multiple fusion lifts which, if each admits a categorification, would have distinct minimal extensions. Their minimal extensions, however, lie outside the range of our bosonic data: the first supermodular data with multiple fusion lifts appear at rank 18, and their minimal extensions have rank at least 27.

Lastly, we identify, to the extent possible, all data realized by a Chern--Simons theory with a compact gauge group. When connected, these gauge groups are of the form
\begin{equation}G = \frac{T\times \tilde G}{\Gamma}\end{equation}
where $T \simeq \Uone^r$ is a torus, $\tilde G$ is a compact, simply connected, semisimple Lie group, and $\Gamma$ is a finite subgroup of the center of $T\times\tilde G$. For an integral, symmetric, non-degenerate matrix $K$, the category corresponding to $\Uone^r_K$ is the metric group category $\C(A,q)$, with abelian group $A = \zee{}^r/K\zee{}^r$ and quadratic form $q$. The simple factors of $\tilde G$ yield the affine Lie algebra categories $\C(\mathfrak g, k)$. All theories with gauge group of the form $T\times \tilde G$ are easily identified in the database. Quotients of the gauge group by center subgroups are realized categorically by condensation of $\operatorname{Rep}(\Gamma)$ subcategories~\cite{mooreTamingConformalZoo1989}, which we turn to now.

\subsubsection{Anyon Condensations}
\label{sec:condensation}

The most substantial derived data in the database are the possible condensations between its entries. Anyon condensation is detailed in~\cite{kongAnyonCondensationTensor2021}; the terminology used
here is collected in \cref{tab:condensation-glossary}. Physically, condensation corresponds to 1-form symmetry gauging and equivariantization to 0-form gauging. An important qualification here is that, beyond Tannakian cases, the modular data is insufficient to guarantee that a candidate algebra is condensable. In particular, associativity and commutativity depend on the $F$- and $R$-symbols. The modular data is also generally insufficient to compute the data of the condensate $\C_A^0$. Assuming our data are complete, however, each candidate algebra can be matched to its condensate within the data. Whenever the candidate condensate's rank lies within the classified range, the conditions below were sufficient to determine it uniquely.

For each set of pseudo-unitary modular and supermodular data, we enumerate the finite set of possible algebra objects
\begin{equation}A = \bigoplus_i n_i X_i\end{equation}
satisfying the necessary conditions:
\begin{enumerate}
  \item $n_1 = 1$, $n_i >0 \implies \theta_i = 1$, $n_i = n_{\bar i}$.
  \item $n_i \leq d_i - \delta(d_i)$ for $\delta(x) = 1$ if $x\notin \zee{}$ and $0$ otherwise~\cite{chatterjeeHolographicTheoryContinuous2023}.
  \item $H_A = \{X_i\in \operatorname{Inv}(\C): n_i > 0\}$ is a subgroup of the group of invertible objects $\operatorname{Inv}(\C)$ such that $n_{gi} = n_i$ for all $g\in H_A$.
  \item If $n_i,n_j > 0$, then there exists $k$ such that $N^{ij}_k, n_k >0$.
  \item There exists a symmetric, positive semidefinite $Z\in M_r(\mathbb Z_{\geq 0})$ such that $Z_{1i} = n_i$, $[Z,S] = [Z,T] = 0$, and $Z_{ij} \leq \sum_k N^{i\bar j}_k n_k$.
\end{enumerate}
These possible algebra objects are then matched to their condensates $\C_A^0$ by the following conditions:
\begin{enumerate}
  \item $D_\C^2 = d_A^2 D_{\C_A^0}^2$, for $d_A = \sum_i n_i d_i$.
  \item The set of twists of $\C_A^0$ is a subset of those of $\C$.
  \item The higher central charges agree, $\xi_n(\C) = \xi_n(\C^0_A)$ for $\operatorname{gcd}(n, \operatorname{ord}(T_\C)) = 1$. The higher central charges are computed as \begin{equation}\xi_n(\C) = \frac{\tau_n(\C)}{|\tau_n(\C)|}, \qquad \tau_n(\C) = \sum_c\theta_c^nd_c^2\end{equation}
  and provide a coarse invariant by which condensations are grouped in the database. They are described in greater detail in~\cite{ngHigherGaussSums2019,ngHigherCentralCharges2022,kaidiHigherCentralCharges2022}. In the supermodular case, when a minimal extension is known, the higher central charges are inherited up to a 16th root of unity.
  \item $\operatorname{rank}(\C_A^0) = r' \leq \frac{D_\C^2}{d_A^2}$ and $r' \leq \sum_i s_i$ for $s_i = \sum_j n_j N^{ij}_i$. Note, however, that one can have $r' > r$, so the condensate may lie beyond the classified rank. The few instances of this in our data are left unresolved.
  \item There exists a branching matrix $B\in M_{r'\times r}(\zee{\geq 0})$ such that $BS_\C = d_A\, S_{\C_A^0} B$ (equivalently, $B s_\C = s_{\C_A^0} B$ for the unitary $s = S/D$), $BT_\C = T_{\C_A^0} B$, and $Z = B^T B$.
\end{enumerate}
We also construct the associated NIM-reps~\cite{ostrikModuleCategoriesWeak2001} in most cases, but this becomes difficult in general, so it is not used to reject candidate algebras. In this way, we identify all candidate condensations between pseudo-unitary modular and supermodular data in the database. Condensation is still well-defined in non-pseudo-unitary cases, but several restrictions, such as $\theta_a = 1$, no longer apply, and the unitary categories are of greatest physical interest. As with modular data, which may lack a categorification, these candidate algebras may not be condensable. Several consistency checks are performed, including transitivity: for condensations $\mathcal A \rightarrow \mathcal B$ and $\mathcal B \rightarrow \C$, there must exist a condensation $\mathcal A \rightarrow \C$. 

In addition to identifying condensations within the data, we also attempt to identify all $\operatorname{Rep}(\Gamma)$ condensations of Chern--Simons theories. There is no a priori bound on the rank of a Chern--Simons theory with gauge group $T\times\tilde G$ whose $\operatorname{Rep}(\Gamma)$ condensation yields a given $(T\times\tilde G)/\Gamma$ theory, so we enumerate parent theories up to rank 2500. For each data set, the database lists the identified condensation realizations with lowest-rank parents.  Including these condensations, only 383 of 3220 pseudo-unitary modular data and 241 of 1723 pseudo-unitary supermodular data do not have an identified compact connected Chern--Simons realization up to rank 2500. Some of these may still be Chern--Simons via condensations of higher-rank parents: in one case, the lowest-rank invertible 0-form gauging we found has rank 1936. 

To facilitate the investigation of Chern--Simons modular categories, we provide at \href{http://www.AnyonAtlas.org/chern-simons}{\texttt{AnyonAtlas.org}} a calculator for computing the modular data of Chern--Simons theories with compact connected gauge groups at ranks far exceeding our classification. This is shown in \cref{fig:calculator}. The necessary condensations are computed using the procedure developed in~\cite{Fuchs:1996dd}.

For the remaining data, we also consider gaugings of 0-form $H$-symmetries in Chern--Simons theories with connected gauge groups. As described in~\cite{barkeshliSymmetryFractionalizationDefects2019}, an $H$-gauging yields a modular category with a $\text{Rep}(H)$ subcategory which, when condensed, yields the original category. Unlike condensations, these have sharp rank bounds, and so it is merely a matter of identifying all the remaining data having $\text{Rep}(H)$ subcategories condensing to connected Chern--Simons. These cases are included as they describe the most general notion of gauge theories: all Chern--Simons theories with compact disconnected gauge group $G$ having $\pi_0(G) = H$ are described by $H$-gaugings of this form. In particular, $A_4$ and $S_4$ Dijkgraaf--Witten theories are included as Chern--Simons, despite seemingly not admitting a description in terms of some connected gauge group. Henceforth we describe as Chern--Simons all categories that can be constructed in this manner by invertible condensations and $H$-gaugings.

Almost all the remaining pseudo-unitary data can be proven not to be Chern--Simons. As noted, ruling out $H$-gaugings is possible by finite enumeration, so we need only consider invertible condensations. In~\cite{gannonAlgebraicNumberFields2019}, the dimension fields $\mathbb Q(D^2)$ of each $\C(\mathfrak g, k)$ are identified, and take the form $\mathbb Q_n = \mathbb Q(\cos(\frac{2\pi}{n}))$ in all but four exceptional cases. For a candidate data $\mathcal M$ to be realizable as Chern--Simons, $\mathbb Q(D^2)$ must be some compositum of these dimension fields. Since condensation by invertibles preserves the dimension field, $\mathcal M$ must have an invertible 0-form gauging of the form
\begin{equation}\C(A,q) \boxtimes X_0 \boxtimes X_1\end{equation}
where $X_0$ is any product of the non-pointed affine Lie categories with global dimension field $\mathbb Q$\footnote{These are $\SU(2)_4,~ \SU(3)_3,~\SU(4)_2,~\Spin(2n+1)_1$ for $n=2,\ldots,9$, $\Spin(2n+1)_2,~\Spin(2n)_2,~ \Sp(8)_1$. These are drawn from Figure 2 in~\cite{gannonAlgebraicNumberFields2019}, with several typos corrected.} and $X_1$ is any product of the finitely many affine Lie categories whose compositum of dimension fields yields $\mathbb Q(D^2)$. Further constraints follow from requiring $\mathcal M$ to be a condensate of a category of this form. For each decomposition of $\mathbb Q(D^2)$ as a compositum of the affine Lie dimension fields, consider one of the affine Lie categories, $\C$, realizing a field in that compositum. Then, for modular categories, the Müger centralizer of the pointed subcategory $\C_\text{pt}$ is the adjoint subcategory~\cite{etingofTensorCategories2015}:
\begin{equation}\C_\text{ad} = (\C_\text{pt})' = \left\langle X\otimes X^* \;:\; X\in \irr(\mathcal C)\right\rangle\end{equation} 
and for supermodular categories, $\C_\text{ad} \subseteq (\C_\text{pt})'$, with equality if $\psi\in \C_\text{ad}$. For each non-invertible $Y\in \C_\text{ad}$, we test whether $\mathcal M$ has a simple object $X$ such that
\begin{equation}d_Y/d_X \in \zee{>0}, ~~~~\theta_X = \theta_Y~ \text{or}~ \overline{\theta_Y}\end{equation}
with the conjugate accounting for orientation reversals. If $\mathcal M$ is obtained from $\mathcal E \boxtimes \C$ by condensing invertible objects, for some $\mathcal E$, then $1_{\mathcal E}\boxtimes Y$, being in the centralizer of $(\mathcal E \boxtimes \C)_\text{pt}$, must appear in $\mathcal M$, with the degree of its splitting given by $d_Y/d_X \in \zee{>0}$. For $\mathcal M$ to be realizable as a Chern--Simons category, these constraints must be satisfied for at least one decomposition of $\mathbb Q(D^2)$ and choice of corresponding affine Lie categories. 

The counts of pseudo-unitary data identified as Chern--Simons under our general definition above and those proven not to be are presented in \cref{fig:cs_counts}. By this argument, all but 20 of the bosonic, and 23 of the fermionic, data without identified Chern--Simons realizations are provably not Chern--Simons. In particular, this argument resolves all of the possibly non-Chern--Simons data identified in~\cite{ModularTensorCategories} and~\cite{ngClassificationModularData2025}.

\subsection{Structure Across Rank}

\begin{figure}[!ht]
    \centering
    \includegraphics[width=.49\textwidth]{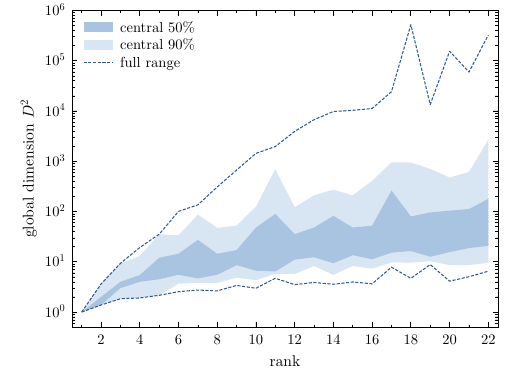}
    \includegraphics[width=.49\textwidth]{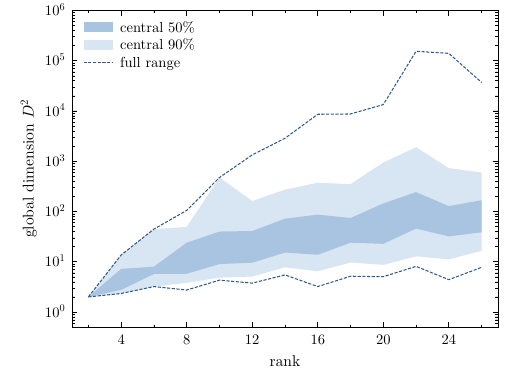}
    \caption{Global dimensions of the identified modular (left) and supermodular (right) data.}
    \label{fig:globaldim}
\end{figure}

\begin{figure}[!ht]
    \centering
    \includegraphics[width=.49\textwidth]{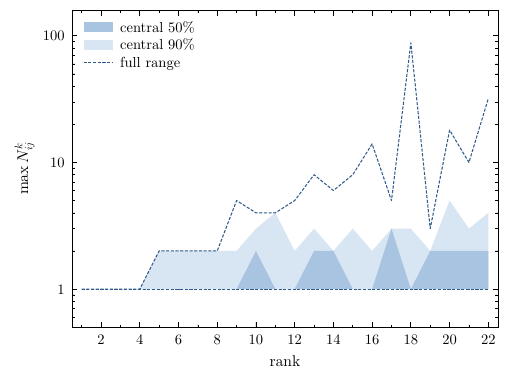}
    \includegraphics[width=.49\textwidth]{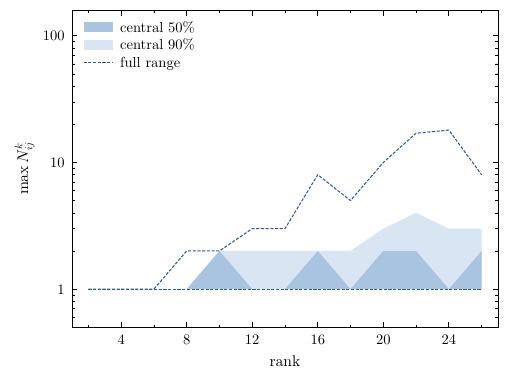}
    \caption{Largest fusion coefficients in the identified modular (left) and supermodular (right) data. At almost every rank, at least half the data are multiplicity-free.}
    \label{fig:Nmax}
\end{figure}

In this subsection, we look briefly at how important parameters of the identified data scale with rank. The scaling of the global dimension $D^2$ is presented in \cref{fig:globaldim}. The data suggest that $D^2$ grows more slowly than $e^r$, but no explicit upper bound is known, as emphasized in~\cite{bruillardRankfinitenessModularCategories2015}. Among pseudo-unitary data, $D^2 \geq r$, with equality exactly for pointed categories, all of which are (abelian) Chern--Simons theories. The data with the largest $D^2$, on the other hand, are generally not Chern--Simons. Notably, the bosonic outlier at rank 18 (\href{http://www.anyonatlas.org/data/B18-02829}{\texttt{B18-02829}} in the database) is not a Chern--Simons theory, but appears to be a non-invertible 0-form gauging of $\Uone_K^2$ with $K = \left(\begin{smallmatrix}
  -2 & 1 \\1 & 2
\end{smallmatrix}\right)$. The largest fusion coefficients in each set of data are presented in \cref{fig:Nmax}. As with $D^2$, the data with the largest fusion coefficients generally do not correspond to Chern--Simons theories. The rank-18 outlier is again visible, having largest fusion coefficient 88. For both $D^2$ and $\max N^{ij}_k$, a practical bound would substantially simplify the search by excluding the regions where the Verlinde cost becomes highly oscillatory.

\begin{figure}[!ht]
    \centering
    \includegraphics[width=.49\textwidth]{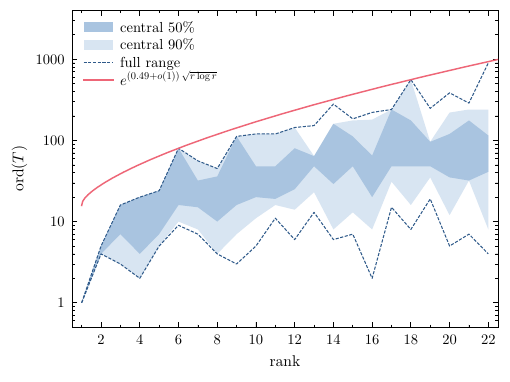}
    \includegraphics[width=.49\textwidth]{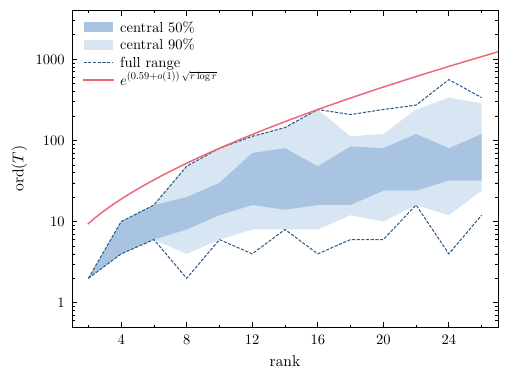}
    \caption{$\operatorname{ord}(T)$ in the identified modular (left) and supermodular (right) data. Fits of the scaling upper bound presented in \cref{app:scaling} are also shown.}
    \label{fig:ordT}
\end{figure}

The scaling of $\operatorname{ord}(T)$ within our data is presented in \cref{fig:ordT}. A fit of the bound derived in \cref{app:scaling} of $\operatorname{ord}(\rho(\modt)) \leq \exp\left((1+o(1))\sqrt{2r\log r}\right)$ is also presented, suggesting that $\operatorname{ord}(T)$ scales in modular categories as it does in general congruence representations of $\sl{\zee{}}$, possibly with a smaller exponent factor.

\section{Chern--Simons and the Witt Group}
\label{sec:chernsimons}

Having constructed the condensation graphs of pseudo-unitary modular data, truncated at the classified ranks, we can study the structure of the Witt group. The Witt group was first introduced in the study of centers of fusion categories, $\mathcal Z(\mathcal F)$, or, equivalently, modular categories with Lagrangian algebras---algebras condensing to $\text{Vec}$~\cite{davydovWittGroupNondegenerate2011}. It was observed that $\mathcal W$, the quotient of the monoid of modular categories by the submonoid of Drinfeld centers, has a group structure analogous to that of the classical Witt group of quadratic forms over a field. The elements of $\mathcal W$ are equivalence classes $[\C]$, where $\C \sim \mathcal M$ if $\C \boxtimes \mathcal M^\text{rev} \simeq \mathcal Z(\mathcal F)$ for some fusion category $\mathcal F$. In more physical terms we can rephrase this as follows: $\C \sim \mathcal M$ precisely when there is a topological domain wall between $\C$ and $\mathcal{M}$.  (See \cref{fig:Witt}.)  The connection between the two definitions is then the standard folding trick, i.e.\ observing that a topological interface between $\C$ and $\mathcal{M}$ is equivalent to a topological boundary of $\C \boxtimes \mathcal M^\text{rev}$, together with the observation that Drinfeld centers always themselves admit a topological boundary.

For our purposes, the Witt group is of interest because its classes consist of categories related by sequences of condensations, equivalently of 0- and 1-form gaugings. A useful observation to organize Witt equivalence classes is that they are directed.  Specifically, in each equivalence class there is a unique modular tensor category in which no anyon condensation is possible~\cite{davydovWittGroupNondegenerate2011}.  This representative is referred to as \emph{anisotropic}.  In a fixed Witt class every MTC admits an anyon condensation directly to the anisotropic representative.   In the Witt-trivial case, the anisotropic representative is the trivial theory with only the vacuum anyon, and the corresponding algebras are Lagrangian.   Thus understanding the full structure of the Witt group is equivalent to studying anisotropic MTCs, and whole equivalence classes are then reduced to the study of various gaugings. 

Our understanding of the structure of the Witt group is, on the whole, rather limited. Our primary tools for constructing modular categories are metric groups, affine Lie algebras, and quantum groups, together with invertible gaugings of 0- and 1-form global symmetries.  These together make up the categories we call Chern--Simons, 
by which we mean gauge theories with any compact gauge group.  Little is known about how much of $\mathcal W$ is and is not represented by these examples. Chern--Simons categories are unitary, and so can only cover $\mathcal W_U \leq \mathcal W_\text{pu} < \mathcal W$, where $\mathcal W_U$ is the unitary Witt group and $\mathcal W_\text{pu}$ is the pseudo-unitary Witt group. Pseudo-unitarity, meaning positive quantum dimensions, is not known to be sufficient for unitarity, which requires a Hom-basis where the $F$-symbols are unitary. (However, no counterexamples are known.) A pseudo-unitary category may be Witt-equivalent to a unitary one, but they must, in that case, share a unitary anisotropic representative.

As described in \cref{sec:introcs}, it is known that not all unitary MTCs are literally given by Chern--Simons theories. For example, $\mathcal Z(\text{Haagerup}_3)$~\cite{izumiStructureSectorsAssociatedLongo2000,hongEXOTICMODULARTENSOR2008,evansExoticnessRealisabilityTwisted2011} is known to not be Chern--Simons.  A weaker statement, however, was proposed in~\cite{davydovWittGroupNondegenerate2011,muegerModularCategories2012}:

\textbf{Conjecture:} Every unitary Witt equivalence class is represented by a Chern--Simons theory.

Our data provide strong evidence against this conjecture. Several candidate anisotropic data that are not Chern--Simons theories were noted in~\cite{ngClassificationModularData2025}.  Although one of the rank-12 data there is ruled out, as it does not satisfy $|\nu_3(a)|\leq \dim \Hom_\C(1,a^{\otimes 3})$, individual Galois orbits of potentially exotic anisotropic data remain at ranks 11 and 12. By the argument of \cref{sec:condensation}, these are not themselves Chern--Simons. We identify two more Galois orbits of prime, exotic, anisotropic data in the modular case, at ranks 18 and 22, as well as six such orbits in the supermodular case. 

To obtain a counterexample to the conjecture, we must show that one of these exotic data admits a unitary categorification and that the resulting category cannot be obtained by condensation from any Chern--Simons theory. For context, the modular data of one of the pair of rank-11 exotics (\href{http://www.anyonatlas.org/data/B11-00119}{\texttt{B11-00119}} in the database) is
\begin{equation}
T=\operatorname{diag}\!\left(
1,\,
\E{\frac{3}{35}},\,
\E{\frac{13}{35}},\,
\E{\frac{33}{35}},\,
\E{\frac45},\,
1,\,
\E{\frac17},\,
\E{\frac27},\,
\E{\frac47},\,
\E{\frac25},\,
\E{\frac45}
\right)
\end{equation}
\begin{equation}
S=
\left(
\begin{array}{rrrrrrrrrrr}
1 & a & a & a & b & c & d & d & d & e & f \\
a & g & h & -i & d & a & j & -k & -l & 0 & -d \\
a & h & -i & g & d & a & -k & -l & j & 0 & -d \\
a & -i & g & h & d & a & -l & j & -k & 0 & -d \\
b & d & d & d & -c & -f & -a & -a & -a & e & 1 \\
c & a & a & a & -f & 1 & d & d & d & -e & -b \\
d & j & -k & -l & -a & d & -h & i & -g & 0 & a \\
d & -k & -l & j & -a & d & i & -g & -h & 0 & a \\
d & -l & j & -k & -a & d & -g & -h & i & 0 & a \\
e & 0 & 0 & 0 & e & -e & 0 & 0 & 0 & -e & e \\
f & -d & -d & -d & 1 & -b & a & a & a & e & -c
\end{array}
\right)
\end{equation}
for $c_m^n:=\zeta_m^n+\zeta_m^{-n}$, $\zeta_m = e^{2\pi i/m}$, and
\begin{equation}
\begin{aligned}
a={}&2+c_{35}^{4}+c_{35}^{6}-c_{35}^{8}+c_{35}^{9}
      -c_{35}^{12}-c_{5}^{2}-c_{35}^{17}\\
b={}&2c_{35}^{2}+2c_{35}^{3}+c_{35}^{4}+c_{35}^{6}
      +c_{35}^{9}+2c_{35}^{13}-3c_{5}^{2}\\
c={}&b-c_{5}^{2}\\
d={}&c+1\\
e={}&2+3c_{35}^{2}+3c_{35}^{3}+2c_{35}^{4}+2c_{35}^{6}
      +2c_{35}^{9}+3c_{35}^{13}-4c_{5}^{2}\\
f={}&e+c_{5}^{1}\\
g={}&3c_{35}^{1}+2c_{35}^{3}+c_{35}^{4}+2c_{35}^{6}
      -c_{35}^{11}-2c_{35}^{13}-3c_{7}^{3}-2c_{35}^{17}\\
h={}&2c_{35}^{2}-2c_{35}^{3}+2c_{35}^{4}+c_{35}^{9}
      -3c_{7}^{2}+3c_{35}^{11}-2c_{35}^{12}-c_{35}^{16}\\
i={}&c_{35}^{1}+2c_{35}^{2}+3c_{7}^{1}-c_{35}^{6}
      +2c_{35}^{8}-2c_{35}^{9}-2c_{35}^{13}-3c_{35}^{16}\\
j={}&c_{35}^{2}+2c_{7}^{1}-2c_{35}^{6}-c_{35}^{9}
      -2c_{35}^{13}-c_{7}^{3}-2c_{35}^{16}\\
k={}&2c_{35}^{1}+c_{35}^{6}-c_{7}^{2}-2c_{35}^{11}
      -c_{35}^{13}-2c_{7}^{3}-2c_{35}^{17}\\
l={}&-c_{35}^{3}+c_{35}^{4}-c_{7}^{1}-2c_{7}^{2}
      +2c_{35}^{11}-2c_{35}^{12}-2c_{35}^{16}
\end{aligned}
\end{equation}
Categorification requires solving the pentagon equations, which in general is notoriously difficult. For a sense of the challenge this poses, see~\cite{vercleyenLowRankFusion2023,vercleyenLowRankMultiplicityFreeComplex2026}, which classify low-rank multiplicity-free fusion rings and solve the pentagon equations for their categorifications, and~\cite{maurerFsymbolsRsymbolsDrinfeld2026}, where the authors exploit properties specific to centers of known fusion categories. The lowest-rank exotic cases, at ranks 11 and 12, have the most $F$-symbols of any modular data at their ranks, with largest fusion coefficients 4 and 5, respectively. We have obtained complete solutions for the $F$-symbols of both exotic orbits, at ranks 11 and 12.  The details will be presented in a subsequent publication. In both cases, this yields a unitary anisotropic modular category that is not Chern--Simons. The remaining gap is whether these exotics are nonetheless condensations from Chern--Simons theories by algebras with support on non-invertible simples. While this appears exceedingly unlikely, the machinery does not yet exist to rule this out directly.  We have, however, confirmed to high rank that no such non-invertible condensation exists.  For instance, in the rank-11 case, the rank of any putative Chern--Simons parent must be at least $10^{4}$, and the smallest candidate identified has rank of order $10^{14}$.  Further details are provided in \cref{app:notCS}.

A final observation which, if true, would close this remaining gap is the following:\\[5pt]
\textbf{Question:} Are non-invertible 1-form gaugings of Chern--Simons theories with compact gauge groups themselves equivalent to Chern--Simons theories?\\[5pt]
That is, every category obtained by condensing an algebra supported on non-invertibles in a Chern--Simons category would also be obtainable by condensing an algebra supported on invertibles in a possibly different Chern--Simons category. Once again, there is no particular reason to expect that this would be the case, but in all 2108 such condensations we identify in our data (across both modular and supermodular data), the condensate is itself a Chern--Simons theory. Of course, at low rank almost all the data are realized by Chern--Simons, and of the 617 non-invertible condensations from data not identified as Chern--Simons, all of them also yield Chern--Simons theories. That being said, many higher-rank examples were tested, and all were consistent with this. For simple gauge groups, non-invertible condensations of Chern--Simons theories are in some sense exceptional since, as shown in~\cite{gannonExoticQuantumSubgroups2023}, for fixed $\mathfrak g$ there is a finite level $k$ above which $\C(\mathfrak g,k)$ has no condensable algebras with support on non-invertibles. That alone does not explain this pattern, however, so it would be intriguing if some version of the conjecture were shown to hold at arbitrary rank. 

\FloatBarrier


\vspace{5pt}
\noindent\textbf{Acknowledgements.} We thank Lea  Bottini, Clement Delcamp, Nicholas Holfester and Anton Kapustin for many illuminating discussions. We are grateful to \href{https://projectnumina.ai}{Project Numina} for computing support and collaboration on related topics. JBD is supported by a Four-Year Fellowship (4YF) from the University of British Columbia. CC and RT are supported by the Simons Collaboration on Global Categorical Symmetries and the US Department of Energy Grant 5-29073.

\appendix
\crefalias{section}{appendix}

\section{Review of Modular and Supermodular Categories}
\label{app:modular-tensor-categories}

In this appendix we review the basic definitions of modular tensor categories (MTCs) and supermodular categories (SMCs). These categories describe the fusion and braiding of topological line operators in 2+1d TQFTs and of anyonic excitations in topologically ordered phases. The modular data are derived from the underlying categorical data. Further reviews of this material can be found, for example, in
\cite{kitaevAnyonsExactlySolved2006,bakalovLecturesTensorCategories2000}.

For reference, we collect terminology used in this appendix and the
main text in two glossary tables below. \Cref{tab:mtc-glossary} relates basic
categorical notions to their physical interpretation in a TQFT.
\Cref{tab:condensation-glossary} collects definitions used in
our discussion of anyon condensation in
\cref{sec:condensation}~\cite{kongAnyonCondensationTensor2021,davydovWittGroupNondegenerate2011}.
As in \cref{sec:condensation}, the definitions in \Cref{tab:condensation-glossary} are applicable in pseudo-unitary categories with their positive spherical structures.

\begingroup
\small
\renewcommand{\arraystretch}{1.4}
\setlength{\LTcapwidth}{\linewidth}
\begin{longtable}{|p{\dimexpr0.34\linewidth-2\tabcolsep-1.5\arrayrulewidth\relax}|p{\dimexpr0.66\linewidth-2\tabcolsep-1.5\arrayrulewidth\relax}|}
\caption{Categorical terminology and its physical interpretation in a TQFT.}
\label{tab:mtc-glossary}\\
\hline
\textbf{Mathematical term} & \textbf{Physics interpretation} \\
\hline
\endfirsthead
\hline
\textbf{Mathematical term} & \textbf{Physics interpretation} \\
\hline
\endhead
Simple object $a$ & Labels a simple topological line operator, or an anyon superselection sector in a topologically ordered phase. \\
\hline
Morphism & A map between objects, represented in the TQFT by
an operator at a junction of lines. Fusion and splitting
junctions are examples. \\
\hline
$V_c^{ab}:=\operatorname{Hom}_{\mathcal C}(c,a\otimes b)$ & The vector space of junction operators splitting a $c$-line into an $a$-line and a $b$-line. In a unitary TQFT, it can be identified with the Hilbert space of a sphere with three outgoing punctures labelled $a,b,\bar c$. \\
\hline
Dual object $\bar a$ & The same line with reversed orientation, corresponding to the antiparticle sector in a topologically ordered phase. \\
\hline
Invertible object $a$ & A line whose fusion with its oppositely oriented partner gives only the vacuum: $a\otimes\bar a\cong1$. In a topologically ordered phase, this is an abelian anyon sector. \\
\hline
Categorical (quantum) dimension $d_a$ &
The amplitude of a closed loop formed by an $a$-line. In a unitary theory, it also
controls the growth of state spaces under repeated fusion. \\
\hline
Associator, represented by $F$ & Relates different associative groupings in successive fusion, such as $(ab)c$ and $a(bc)$. In a unitary TQFT, the corresponding fusion trees specify different bases of the same Hilbert space, with the endpoints fixed. $F$ relates the expansion coefficients in these bases, leaving the physical state unchanged.\\
\hline
Braiding $R$ & Exchanges two line endpoints on a spatial surface by braiding the corresponding lines in bulk spacetime. In a unitary TQFT, this is realized as a unitary linear map between the Hilbert spaces associated with the two endpoint configurations. \\
\hline
Double braiding (monodromy) $M$ & A full winding of one line around another in bulk spacetime, returning the endpoints to their original positions. In a unitary TQFT, its action is a linear operator on the original Hilbert space. \\
\hline
Ribbon twist $\theta_a$ & The scalar acquired under a full twist of the framing of an $a$-line. In a topologically ordered phase, this gives the topological-spin phase of the corresponding anyon sector. \\
\hline
\end{longtable}
\endgroup
\begingroup
\small
\renewcommand{\arraystretch}{1.4}
\setlength{\LTcapwidth}{\linewidth}
\begin{longtable}{|p{\dimexpr0.34\linewidth-2\tabcolsep-1.5\arrayrulewidth\relax}|p{\dimexpr0.66\linewidth-2\tabcolsep-1.5\arrayrulewidth\relax}|}
\caption{Terminology for anyon condensation.}
\label{tab:condensation-glossary}\\
\hline
\textbf{Mathematical term} & \textbf{Definition} \\
\hline
\endfirsthead
\hline
\textbf{Mathematical term} & \textbf{Definition} \\
\hline
\endhead
Algebra object $(A,m,\eta)$ & An object $A$ equipped with an associative multiplication $m:A\otimes A\to A$ and a unit $\eta:1\to A$. These maps are additional data beyond the decomposition $A=\bigoplus_i n_iX_i$. \\
\hline
Condensable algebra & A connected, commutative, separable algebra
object. The vacuum line of the condensed TQFT is represented by $A$.
The decomposition $A=\bigoplus_i n_iX_i$ expresses this vacuum
in terms of line sectors of the original TQFT. \\
\hline
$A$-module; $\mathcal C_A$ & An object equipped with a consistent
action of $A$; $\mathcal C_A$ denotes the category of these modules.
For condensable $A$, simple modules label topological lines running
along the condensation interface separating the original and
condensed TQFTs. \\
\hline
Local $A$-module; $\mathcal C_A^0$ & An $A$-module whose action is unchanged when the acting $A$-line is taken once around it; $\mathcal C_A^0$ denotes the category of these modules. For condensable $A$, simple local modules label the deconfined bulk lines of the condensed TQFT. \\
\hline
Tannakian subcategory $\operatorname{Rep}(G)$ & A symmetric fusion
subcategory braided-equivalent to the category of finite-dimensional
representations of a finite group $G$. Its regular algebra
$\operatorname{Fun}(G)$ is condensable. \\
\hline
Branching matrix $B$ & For a simple local $A$-module $M_\alpha$, $B_{\alpha i}$ is the multiplicity of $X_i$ in its underlying object in $\mathcal C$. The condensed vacuum is the module $A$, so $B_{1i}=n_i$. \\
\hline
Witt-equivalent MTCs & Two MTCs $\mathcal C$ and $\mathcal D$
are Witt-equivalent if there exist condensable algebras
$A\in\mathcal C$ and $B\in\mathcal D$ such that
$\mathcal C_A^0\simeq\mathcal D_B^0$ as braided fusion categories.
Thus, suitable condensations in the two theories give equivalent
categories of deconfined bulk lines. \\
\hline
\end{longtable}
\endgroup

\subsection{Fusion and Braiding}
\label{app:fusion-duality-dimensions}
\label{app:associativity-braiding-twists}

An MTC $\mathcal C$ has finitely many simple objects, and every object is a finite direct sum of simple objects. In the TQFT description, a simple object $a$ labels a topological line operator. When the TQFT describes a topologically ordered phase, the same label specifies an anyon superselection sector.

Maps between objects, also called \textit{morphisms}, form complex vector spaces. The tensor product describes fusion of the corresponding line operators:
\begin{align}
a \otimes b \cong \bigoplus_{c \in \operatorname{Irr}(\mathcal{C})} N_{ab}^c c\,, \qquad (\mathcal N_a)_b{}^c=N_{ab}^{c}.\label{eq:fusion}
\end{align}
with $N_{ab}^c$ being the non-negative integer fusion coefficients. We use $\mathcal{N}_a$ to denote the fusion matrix, with row index $b$ and column index $c$. Simple object labels range over a set we denote $\operatorname{Irr}(\mathcal{C})$. We denote the unique \textit{vacuum} object as $1$, where $a \otimes 1 \cong a$ for each object $a$. A splitting junction is represented diagrammatically as
\begin{align}
    \vcenter{\hbox{\StraightFusionTree}}\in V_c^{ab}\,. \label{eq:straight-fusion-tree}
\end{align}
The splitting and fusion spaces are the complex vector spaces
\begin{align}
    V_c^{ab}:=\operatorname{Hom}_{\mathcal C}(c,a\otimes b),
    \qquad
    V_{ab}^{c}:=\operatorname{Hom}_{\mathcal C}(a\otimes b,c),
    \qquad
    \dim V_c^{ab}=\dim V_{ab}^{c}=N_{ab}^{c}\,.
\end{align}
Diagrams such as \cref{eq:straight-fusion-tree} are to be understood as compositions of maps, read from bottom to top. Thus $V_c^{ab}$ is a \textit{splitting space}, while $V_{ab}^{c}$ is the oppositely oriented \textit{fusion space}.

In a unitary TQFT, a three-dimensional ball containing a splitting junction determines a state on its boundary sphere. The three incident lines meet the boundary at three punctures. With all puncture labels taken in the outgoing convention,
\begin{align}
    \mathcal H(S^2;a,b,\bar c)
    \cong\operatorname{Hom}_{\mathcal C}(1,a\otimes b\otimes\bar c)
    \cong V_c^{ab}\,.
    \label{eq:splitting-sphere-state}
\end{align}
The label $\bar c$ accounts for the line entering the junction. Thus the space of splitting junctions is naturally identified with the Hilbert space assigned to this three-punctured sphere.

An MTC has duality data that furnish a \textit{rigid} structure: each simple object $a$ has a dual object
\begin{align}
    a^\vee=\bar a\,, \label{eq:duality}
\end{align}
together with \textit{coevaluation} and \textit{evaluation} maps
\begin{align}
    \operatorname{coev}_a:1\longrightarrow a\otimes\bar a,
    \qquad
    \operatorname{ev}_a:\bar a\otimes a\longrightarrow1\,.
\end{align}
Diagrammatically,
\begin{align}
    \operatorname{coev}_a=
    \vcenter{\hbox{\CoevaluationCup}}\in V_1^{a\bar a}\,,
    \qquad
    \operatorname{ev}_a=
    \vcenter{\hbox{\EvaluationCap}}\in V_{\bar a a}^{1}\,.
    \label{eq:coev-ev-maps}
\end{align}
Arrows indicate line orientation; reversing the orientation of an $a$-line gives a $\bar a$-line. The coevaluation and evaluation maps are represented by creation and annihilation junctions for lines labelled by $a$ and $\bar a$. The rigid structure demands that they compose consistently. The consistency conditions are
\begin{align}
    \vcenter{\hbox{\LeftSnakeWithIdentities}}
    &=\vcenter{\hbox{\LeftSnakeA}}
    =\vcenter{\hbox{\StraightOrientedLine{a}}}\,,
    &
    \vcenter{\hbox{\RightSnakeWithIdentities}}
    &=\vcenter{\hbox{\RightSnakeDualA}}
    =\vcenter{\hbox{\StraightOrientedLine{\bar a}}}\,.
    \label{eq:snake-identities}
\end{align}
It is a requirement of the MTC that $a \otimes \bar{a} \cong 1 + \dotsm $, specifically with the vacuum appearing with $N_{a\bar{a}}^1=1$. Where $a \otimes \bar{a} \cong 1$ we refer to $a$ as invertible, and conversely if there are other objects in the fusion then $a$ is referred to as non-invertible. The invertible simple objects form an abelian group under $\otimes$. In a topologically ordered phase, they label the abelian anyon sectors.

In $G_k$ Chern--Simons theory with compact, simply connected,
simple gauge group $G$ and positive integer $k$, the simple lines are
Wilson lines labelled by the integrable highest weights at level
$k$.\footnote{More general
compact connected gauge groups are discussed in \cref{sec:database}.
Negative level $k$ gives the orientation-reversed theory.}
The invertible simple lines generate the 1-form symmetry of the TQFT.

Taking the dual twice gives an object isomorphic to the original,
\begin{align}
    \bar{\bar{a}} \cong a\,. \label{eq:pivotal}
\end{align}
A \textit{pivotal structure} chooses these identifications in a manner that is consistent with tensor products and with maps between objects. A pivotal structure gives two ways to close a map $f:a\to a$ into a loop, yielding scalars that are referred to as its left and right traces. The pivotal structure is \textit{spherical} when
\begin{align}
    \operatorname{tr}_L(f)=\operatorname{tr}_R(f)
\end{align}
for every such map $f$. Graphically, this means that closed planar loop diagrams have a consistent evaluation when drawn on the sphere $S^2$. With the spherical pivotal structure, closing the identity map of the line $a$ in either way gives its quantum dimension $d_a$:
\begin{align}
    d_a=\operatorname{tr}_L(\operatorname{id}_a)
       =\operatorname{tr}_R(\operatorname{id}_a),
    \qquad d_a=d_{\bar a}\,.
\end{align}
\begin{align}
    \vcenter{\hbox{\QuantumDimensionLoop{a}}}
    =\vcenter{\hbox{\ReverseQuantumDimensionLoop{a}}}=d_a\,.
\end{align}
In the TQFT path integral on $S^3$, inserting the closed $a$-line
shown above multiplies the result by $d_a$.

Consistency under \cref{eq:fusion} leads to the property that the quantum dimensions realize a homomorphism of the fusion ring
\begin{align}
    d_a d_b
    &=\vcenter{\hbox{\NestedDimensionLoops{a}{b}}}
    =\sum_{c,\alpha}
      \vcenter{\hbox{\ResolvedDimensionLoops{a}{b}{c}{\alpha}}}
    =\sum_cN_{ab}^{c}d_c\,.
    \label{eq:fusion-dimension}
\end{align}
In the intermediate diagram, $\alpha$ labels a chosen basis of $V_c^{ab}$ and $\alpha_*$ its composition-dual basis in $V_{ab}^c$, normalized so that fusion after splitting gives $\operatorname{id}_c$ for matching indices and zero otherwise.
The total quantum dimension is $D$, and the global dimension is $D^2$:
\begin{align}
    D=\sqrt{\sum_{a\in\operatorname{Irr}(\mathcal C)}d_a^2},
    \qquad \dim(\mathcal C)=D^2\,. \label{eq:total-quantum-dimension}
\end{align}
An MTC is \textit{pseudo-unitary} if its spherical structure can be
chosen so that $d_a>0$ for every simple object $a$. This positive
spherical structure is unique, and we use it in the pseudo-unitary
case~\cite{etingofFusionCategories2017}.
Since $a\otimes\bar a$ contains the vacuum and $d_{\bar a}=d_a$,
\cref{eq:fusion-dimension} implies $d_a\geq1$, with equality
if and only if $a$ is invertible.

The fusion rules also determine another notion of dimension,
the \textit{Frobenius--Perron dimension}. For every simple object
in any MTC, whether or not it is pseudo-unitary,
\begin{align}
    f_a=\operatorname{FPdim}(a)>0\,.
\end{align}
The $f_a$ form the unique strictly positive vector
$\mathbf f=(f_b)_{b\in\operatorname{Irr}(\mathcal C)}$, normalized
by $f_1=1$, satisfying
\begin{align}
    \mathcal N_a\mathbf f=f_a\mathbf f
    \qquad\text{for every }a\,.
\end{align}
The Frobenius--Perron dimension of the category is
\begin{align}
    \operatorname{FPdim}(\mathcal C)=\sum_a f_a^2\,.
\end{align}

In the pseudo-unitary case, the categorical and Frobenius--Perron
dimensions agree:
\begin{align}
    d_a=f_a,\qquad D^2=\operatorname{FPdim}(\mathcal C)\,.
\end{align}
Every unitary MTC is pseudo-unitary. Unitarity additionally
requires an adjoint operation on maps, taking $f:X\to Y$ to
$f^\dagger:Y\to X$ and inducing positive-definite inner products
on the fusion and splitting spaces.

An MTC specifies a distinguished linear isomorphism that relates distinct associative orderings of fusion
\begin{align}
    F^{abc}_d:
    \bigoplus_{e\in\operatorname{Irr}(\mathcal C)}V_d^{ec}\otimes V_e^{ab}
    &\longrightarrow
    \bigoplus_{f\in\operatorname{Irr}(\mathcal C)}V_d^{af}\otimes V_f^{bc}\,.
    \label{eq:f-map}
\end{align}
Choosing bases for the splitting spaces represents these maps
by matrices $\bigl[F^{abc}_d\bigr]$, called the
\textit{$F$-symbols}. Diagrammatically,
\begin{align}
    \vcenter{\hbox{\LeftAssociatedFusionTree}}
    &=\sum_{f,\mu,\nu}
      \bigl[F^{abc}_d\bigr]_{(e,\alpha,\beta),(f,\mu,\nu)}
      \vcenter{\hbox{\RightAssociatedFusionTree}}\,.
    \label{eq:f-move}
\end{align}
The multiplicity indices range over
\begin{align}
    \alpha=1,\ldots,N_{ec}^d,\quad
    \beta=1,\ldots,N_{ab}^e,\quad
    \mu=1,\ldots,N_{af}^d,\quad
    \nu=1,\ldots,N_{bc}^f\,.
\end{align}
We write the source basis index first and the target basis
index second.

In a unitary TQFT, the two splitting trees in \cref{eq:f-move}
specify bases of the same Hilbert space
$\mathcal H(S^2;a,b,c,\bar d)$. The labelled punctures remain fixed.
The matrix $[F_d^{abc}]$ relates the expansion coefficients of a
given physical state in these bases.

The bases of the individual splitting spaces $V_c^{ab}$ are not
intrinsic data of the MTC. Changing these bases changes the
$F$-symbols, but can be viewed as a gauge redundancy of the MTC.

For four objects, different sequences of $F$-moves relating
the same two associative groupings must give the same map.
This consistency requirement is expressed by the
\textit{pentagon equations}:
\begin{align}
    \vcenter{\hbox{\PentagonCoherenceDiagram}}\,.
    \label{eq:pentagon}
\end{align}
In components,
\begin{align}
\begin{aligned}
&\sum_{\lambda}
[F_x^{ecd}]_{(f,\alpha,\beta),(g,\lambda,\rho)}
[F_x^{abg}]_{(e,\lambda,\gamma),(h,\mu,\nu)}
\\
&\quad=
\sum_{k,\sigma,\tau,\kappa}
[F_f^{abc}]_{(e,\beta,\gamma),(k,\sigma,\tau)}
[F_x^{akd}]_{(f,\alpha,\sigma),(h,\mu,\kappa)}
[F_h^{bcd}]_{(k,\kappa,\tau),(g,\nu,\rho)}\,.
\end{aligned}
\label{eq:pentagon-components}
\end{align}
Here the source multiplicity indices $\alpha,\beta,\gamma$ label bases of
$V_x^{fd},V_f^{ec},V_e^{ab}$, respectively, while the target indices
$\mu,\nu,\rho$ label bases of $V_x^{ah},V_h^{bg},V_g^{cd}$.
The equation holds for every choice of these indices and the external and internal simple-object labels. Summations run over the allowed intermediate channels and vertex multiplicities.

The simple objects, their fusion and splitting spaces, duality data, and consistent maps $F$ define the underlying fusion category. An MTC additionally has a braiding structure. Braiding adds linear maps relating different orderings of line endpoints:
\begin{align}
    R^{ab}_c:V_c^{ab}&\longrightarrow V_c^{ba}\,,
    \label{eq:r-map}
\end{align}
Choosing bases gives matrices $\bigl[R_c^{ab}\bigr]$, called the
\textit{$R$-symbols}. Diagrammatically,
\begin{align}
    \vcenter{\hbox{\BraidedSplittingTree}}
    &=\sum_{\beta}
      \bigl[R_c^{ab}\bigr]_{\alpha\beta}
      \vcenter{\hbox{\ReorderedSplittingTree}}\,.
    \label{eq:r-move}
\end{align}
where $\alpha$ and $\beta$ again run over the associated fusion
multiplicities. That the $a$-line crosses over the $b$-line from
the left is meaningful and corresponds to the ordering of the
labels in $R_c^{ab}$. Like the $F$-symbols, the $R$-symbols depend
on the bases chosen for the splitting spaces and the choice is
likewise a gauge redundancy of the MTC.

In a unitary MTC, the $F$- and $R$-maps are unitary, so in orthonormal bases
\begin{align}
    [F_d^{abc}]^\dagger[F_d^{abc}]=I,
    \qquad [R_c^{ab}]^\dagger[R_c^{ab}]=I\,.
\end{align}

A \textit{double-braiding} is the composition of maps
\begin{align}
    M_c^{ab}:=R^{ba}_c\circ R^{ab}_c:
    V_c^{ab}&\longrightarrow V_c^{ab}\,,
    \label{eq:monodromy-map}
\end{align}
enacting
\begin{align}
    \vcenter{\hbox{\DoubleBraidedSplittingTree}}
    &=\sum_{\gamma}
      \bigl[M_c^{a,b}\bigr]_{\alpha\gamma}
      \vcenter{\hbox{\StraightSplittingTree}}\,.
    \label{eq:double-braid}
\end{align}
In a unitary TQFT, $R_c^{ab}$ is realized as a unitary linear map between the Hilbert spaces assigned to the two labelled puncture configurations obtained by exchanging $a$ and $b$. The exchange is represented by braiding the corresponding lines in 2+1d spacetime. The double braid $M_c^{ab}$ returns the labelled punctures to their original positions and acts on the original Hilbert space.

Two lines $a$ and $b$ are said to be \textit{transparent} to one another when
\begin{align}
    M_c^{ab}=\operatorname{id}_{V_c^{ab}}
    \qquad\text{for every fusion channel }c\,.
\end{align}
The lines transparent to every other line form the \textit{M\"uger center} $\mathcal Z_2(\mathcal C)$. Braiding is \textit{non-degenerate} when the vacuum is its only transparent simple line.

The consistency of sequences of fusing and braiding operations is ensured by the \textit{hexagon equations}
\begin{align}
    \vcenter{\hbox{\HexagonCoherenceDiagram}}\,.
    \label{eq:hexagon}
\end{align}
In components,
\begin{align}
\begin{aligned}
&\sum_{\lambda,\kappa}
[R_e^{ab}]_{\beta\lambda}
[F_d^{bac}]_{(e,\alpha,\lambda),(g,\mu,\kappa)}
[R_g^{ac}]_{\kappa\nu}
\\
&\quad=
\sum_{f,\sigma,\tau,\eta}
[F_d^{abc}]_{(e,\alpha,\beta),(f,\sigma,\tau)}
[R_d^{af}]_{\sigma\eta}
[F_d^{bca}]_{(f,\eta,\tau),(g,\mu,\nu)}\,.
\end{aligned}
\label{eq:hexagon-components}
\end{align}
Here the source indices $\alpha,\beta$ label bases of
$V_d^{ec},V_e^{ab}$, while the target indices $\mu,\nu$
label bases of $V_d^{bg},V_g^{ca}$.
The displayed diagram gives one hexagon equation. There is another hexagon obtained by reversing every crossing, replacing each $R_z^{xy}$ by $(R_z^{yx})^{-1}$ while leaving the $F$-symbols unchanged. Both hexagon equations must hold for all allowed simple-object labels and multiplicity indices.

The \textit{ribbon structure} assigns a twist map $\theta_X:X\to X$ to each object, compatible with braiding and duality. On a simple object $a$, the twist map is multiplication by a nonzero scalar $\theta_a$:
\begin{align}
    \theta_X&:X \rightarrow X\,,
    &
    \StandardTwistEquation\,.
    \label{eq:topological-twist}
\end{align}
The lines carry a \textit{framing}, a continuous choice of normal
direction along the line, which can be represented by thickening
the line into a ribbon. The twisted and untwisted diagrams of \cref{eq:topological-twist} are
distinct as framed diagrams, with $\theta_a$ relating them. For a precise exposition of ribbon category graphical calculus,
see~\cite{bakalovLecturesTensorCategories2000}.

The scalar $\theta_a$ is the topological twist of the line labelled by $a$ and, in a topologically ordered phase, of the corresponding anyon sector. The modular $T$ matrix is given entirely by the topological twists:
\begin{align}
    T_{ab}=\delta_{ab}\theta_a,\qquad
    \theta_1=1,\qquad \theta_{\bar a}=\theta_a\,.
\end{align}
The twists are roots of unity~\cite{vafaClassificationConformalTheories1988}, so $T$ has finite order $N=\operatorname{ord}(T)$.
Anyon sectors with $\theta_a=1$ are called bosonic, while those with $\theta_a=-1$ are called fermionic.

Compatibility with braiding is expressed by \textit{balancing}, which in a splitting channel gives the following identity:
\begin{align}
    \theta_c\,\operatorname{id}_{V_c^{ab}}
    &=\theta_a\theta_b\,M_c^{ab}\,,
    \label{eq:balancing}
\end{align}
which is an equality of linear maps on $V_c^{ab}$. Graphically,
\begingroup
\makeatletter
\let\LocalBalancingLabel\ltx@label
\def\ltx@label#1{\LocalBalancingLabel{#1}\LocalAliasLabel{eq:graphical-balancing}}
\makeatother
\begin{align}
    \theta_c\,\vcenter{\hbox{\BalancingStraightChannel}}
    =\vcenter{\hbox{\SimultaneouslyTwistedPair[1]}}
    =\vcenter{\hbox{\DoubleBraidWithIndividualTwists[1]}}
    =\theta_a\theta_b\,\vcenter{\hbox{\BalancingBraidedChannel}}\,.
    \label{eq:channel-balancing}
\end{align}
\endgroup

A fusion category with compatible braiding and ribbon structures, satisfying balancing, is called \textit{premodular}. A premodular category is an MTC when its braiding is furthermore non-degenerate, having only the vacuum line in its M\"uger center $\mathcal Z_2(\mathcal C)$.

\subsection{Frobenius--Schur Indicators}
\label{app:fs-indicators}

A pivotal fusion category possesses a sequence of invariants known as the \textit{Frobenius--Schur indicators}~\cite{ngHigherFrobeniusSchurIndicators2007}. Rigidity together with the pivotal structure defines a cyclic rotation operator
\begin{align}
    E_X^{(n)}:\operatorname{Hom}(1,X^{\otimes n})
    \longrightarrow\operatorname{Hom}(1,X^{\otimes n})\,,
    \label{eq:higher-fs-rotation}
\end{align}
obtained by moving one external $X$-leg around the diagram to the opposite side.
Taking the trace of this rotation gives the indicator:
\begin{align}
    \nu_n(X)=\operatorname{Tr}_{\operatorname{Hom}(1,X^{\otimes n})}E_X^{(n)}\,.
    \label{eq:higher-fs-definition}
\end{align}
Applying the cyclic rotation $n$ times gives the identity:
\begin{align}
    \left(E_X^{(n)}\right)^n=\mathrm{id}\,.
    \label{eq:higher-fs-rotation-order}
\end{align}
Hence the eigenvalues of $E_X^{(n)}$ are $n$-th roots of unity, and their trace is a cyclotomic integer,
\begin{align}
    \nu_n(X)\in\mathbb Z[\zeta_n]\,.
    \label{eq:higher-fs-cyclotomic}
\end{align}
Since each eigenvalue has absolute value one, the same
argument gives
\begin{align}
    |\nu_n(X)|\leq
    \dim\operatorname{Hom}(1,X^{\otimes n})\,.
\end{align}
In a braided category with a ribbon structure compatible with the pivotal structure, the cyclic rotation for $n=2$ can be implemented by an exchange followed by a twist. On the vacuum splitting space
\begin{align}
    V_1^{aa}=\operatorname{Hom}(1,a\otimes a)\,,
    \label{eq:nu2-vacuum-space}
\end{align}
this gives
\begin{align}
    E_a^{(2)}=\theta_aR_1^{aa}\,.
    \label{eq:nu2-rotation}
\end{align}
The second Frobenius--Schur indicator is therefore~\cite{kitaevAnyonsExactlySolved2006,ngHigherFrobeniusSchurIndicators2007}
\begin{align}
    \nu_2(a)=\operatorname{Tr}_{V_1^{aa}}\!\left(\theta_a R_1^{aa}\right)\,.
    \label{eq:nu2-trace}
\end{align}
For a simple object, $\dim V_1^{aa}=N_{aa}^{1}=\delta_{a\bar a}$. Thus $\nu_2(a)=0$ when $a\ne\bar a$, because the trace is over the zero vector space. When $a=\bar a$, the space is one-dimensional and the operation acts by the scalar $\nu_2(a)=\theta_aR_1^{aa}$:
\begin{align}
    \vcenter{\hbox{\FSVertex{2}}}
    =\theta_a\,\vcenter{\hbox{\FSVertex{1}}}
    =\nu_2(a)\,\vcenter{\hbox{\FSVertex{0}}}\,.
    \label{eq:nu2-braid-twist}
\end{align}
Applying this operation twice gives a double braid and a twist on each line. By the balancing relation of \cref{eq:balancing}, these can be replaced by a twist of the vacuum line, which is trivial:
\begin{align}
    \vcenter{\hbox{\FSSquare}}
    =\vcenter{\hbox{\FSVacuumTwist}}
    =\vcenter{\hbox{\FSVertex{0}}}\,.
    \label{eq:nu2-square}
\end{align}
Consequently, $(\theta_aR_1^{aa})^2=1$ in the self-dual case, and
\begin{align}
    \nu_2(a)=\begin{cases}
        0,& a\ne\bar a,\\
        +1\ \text{or}\ -1,& a=\bar a.
    \end{cases}
    \label{eq:nu2-values}
\end{align}

In a modular category, the Verlinde formula
\cref{eq:appendix-verlinde} and $d_i=S_{1i}$ allow these
indicators to be computed entirely from the modular
data~\cite{ngCongruenceSubgroupsGeneralized2010},
\begin{align}
    \nu_n(a)=\frac{1}{D^2}
    \sum_{i,j\in\operatorname{Irr}(\mathcal C)}N_{ij}^{a}d_i d_j
    \left(\frac{\theta_i}{\theta_j}\right)^n\,.
    \label{eq:higher-fs-modular-data}
\end{align}
This formula allows the constraints on $\nu_2$, together with
arithmetic constraints and dimension bounds on higher indicators,
to be checked directly on candidate modular data, as discussed
in \cref{sec:search,sec:reconstruction}.

\subsection{Modular Data}
\label{app:modular-data}

The modular $S$ matrix is derived from the ribbon and braiding data~\cite{bakalovLecturesTensorCategories2000}. Specifically
\begin{align}
    S_{ab}
    &=\vcenter{\hbox{\HopfLinkAmplitude{a}{b}}}\nonumber\\
    &=\sum_{c\in\operatorname{Irr}(\mathcal C)}d_c\,
      \operatorname{Tr}_{V_c^{a\bar b}}\!\left(M_c^{a,\bar b}\right)\nonumber\\
    &=\sum_{c\in\operatorname{Irr}(\mathcal C)}
      N_{a\bar b}^{c}\frac{\theta_c}{\theta_a\theta_b}d_c\,.
    \label{eq:s-matrix}
\end{align}
In the final equality we used the balancing condition \cref{eq:channel-balancing}. The diagram is the oriented \textit{Hopf link}. A vacuum component can be removed from the Hopf link, leaving only one loop:
\begin{align}
    \vcenter{\hbox{\HopfLinkAmplitude[gray!75]{1}{a}}}
    &=\vcenter{\hbox{\MatchedHopfUnknot{a}}}=d_a\,.
\end{align}
Therefore
\begin{align}
    S_{1a}=S_{a1}=d_a,\qquad S_{11}=1\,.
    \label{eq:s-vacuum-row}
\end{align}
The $S$-matrix is symmetric:
\begin{align}
    S_{ab}=S_{ba}\,.
\end{align}
The two components of the Hopf link can be exchanged by an ambient isotopy
preserving their orientations and framings. This interchanges the
labels $a$ and $b$ without changing the link amplitude.

A closed $a$ line encircling a simple $x$ line defines an endomorphism of $x$, hence a scalar multiple of $\operatorname{id}_x$. Closing the $x$ line gives the Hopf-link amplitude $S_{ax}$, while closing $\operatorname{id}_x$ gives $d_x=S_{1x}$. Therefore

\begin{align}
    \vcenter{\hbox{\EncirclingLine{a}{x}}}
    =\frac{S_{ax}}{S_{1x}}
      \vcenter{\hbox{\EncirclingIdentityLine{.90}{x}}}\,.
    \label{eq:s-encircling}
\end{align}
Two loops labelled $a,b$ encircling the same $x$ line can be evaluated either by applying the encircling identity twice or by fusing the loops:
\begin{align}
    \frac{S_{ax}S_{bx}}{S_{1x}^{\,2}}
      \vcenter{\hbox{\EncirclingIdentityLine{1.65}{x}}}
    =\vcenter{\hbox{\TwoEncirclingLoops{a}{b}{x}}}
    =\sum_cN_{ab}^{c}\,
      \vcenter{\hbox{\EncirclingLine{c}{x}}}\,.
\end{align}
Equating the two evaluations gives
\begin{align}
    \frac{S_{ax}S_{bx}}{S_{1x}}
    =\sum_cN_{ab}^{c}S_{cx}\,.
    \label{eq:s-fused-link}
\end{align}
In terms of the fusion matrices $(\mathcal N_a)_b{}^c=N_{ab}^{c}$, this identity reads
\begin{align}
    \mathcal N_aS=S\Lambda_a,
    \qquad (\Lambda_a)_{xx}=\frac{S_{ax}}{S_{1x}}\,,
    \label{eq:s-fusion-diagonalization}
\end{align}
where $\Lambda_a$ is diagonal. For an MTC, $S$ is invertible, so
\begin{align}
    \mathcal N_a=S\Lambda_aS^{-1}\,.
\end{align}
Thus the columns of $S$ simultaneously diagonalize the fusion matrices.

The Hopf-link matrix also satisfies
\begin{align}
    S^2=D^2C,\qquad C_{ab}=\delta_{a\bar b}\,,
    \label{eq:appendix-s-square}
\end{align}
where $C$ is the charge-conjugation matrix. Consequently,
\begin{align}
    S^{-1}=\frac1{D^2}SC,\qquad
    (S^{-1})_{xc}=\frac1{D^2}S_{\bar c x}\,,
\end{align}
where we used the symmetry of $S$ in the second equality.
Taking the $b,c$ matrix element of
$\mathcal N_a=S\Lambda_aS^{-1}$ gives the Verlinde formula:
\begin{align}
    N_{ab}^{c}=\frac1{D^2}\sum_x
    \frac{S_{ax}S_{bx}S_{\bar c x}}{S_{1x}}\,.
    \label{eq:appendix-verlinde}
\end{align}

The modular data also describe the action of the torus mapping
class group in the associated unitary 2+1d TQFT. For the theory defined on a solid torus, a Hilbert space is assigned to the boundary $\mathcal{H}(T^2)$. There is a distinguished basis for $\mathcal{H}(T^2)$ labelled by the simple objects: the state $|a\rangle$ is obtained by inserting an $a$ line along the core of the solid torus. The mapping class group is
$\sl{\mathbb Z}$, whose generators
\begin{align}
    \mathfrak s=\begin{pmatrix}0&-1\\1&0\end{pmatrix},
    \qquad
    \mathfrak t=\begin{pmatrix}1&1\\0&1\end{pmatrix}
\end{align}
are represented projectively in this basis by
\begin{align}
    \mathfrak s\longmapsto s:=\frac{S}{D},
    \qquad
    \mathfrak t\longmapsto T\,.
\end{align}
They satisfy
\begin{align}
    s^2=C,\qquad
    (sT)^3=\xi\,C,\qquad
    \xi=\frac1D\sum_a\theta_a d_a^2\,,
\end{align}
where $\xi$ is the multiplicative central charge of the MTC. The first relation implements charge conjugation:
$\mathfrak s^2=-I$ reverses the core line, replacing $a$ by $\bar a$.

For any MTC, including those that are not pseudo-unitary, the normalized modular $s$ and $T$ matrices
are unitary~\cite{etingofFusionCategories2017}:
\begin{align}
    s^\dagger s=I,\qquad T^\dagger T=I\,.
\end{align}

\subsection{Supermodular Categories}
\label{app:supermodular-categories}

The fusion and braiding of excitations in a fermionic topological
phase are described by a \emph{supermodular category} (SMC)~\cite{bruillardFermionicModularCategories2017}.
These are premodular categories whose M\"uger center is
\begin{align}
    \mathcal Z_2(\mathcal C)\simeq\sVec,
\end{align}
where $\sVec$ is a category consisting of two simple objects: the vacuum $\mathbf1$
and an invertible fermion $\psi$, satisfying
\begin{align}
    \psi\otimes\psi\cong\mathbf1\,, \qquad \theta_\psi=-1\,, \qquad d_\psi=1\,.
\end{align}
The fermion is transparent to every simple object $a\in\mathcal C$,
\begin{align}
    M_{a\otimes\psi}^{a\psi} = \operatorname{id}_{V_{a\otimes\psi}^{a\psi}}\,.
\end{align}
Because $\psi$ is invertible, $a\otimes\psi$ is simple, and the splitting space $V_{a\otimes\psi}^{a\psi}$ is one-dimensional. Applying the balancing relation \cref{eq:balancing} to this channel therefore gives
\begin{align}
    \theta_{a\otimes\psi} = \theta_a\theta_\psi = -\theta_a.
\end{align}
Since twists are nonzero and preserved under isomorphism, $a$ and $a\otimes\psi$ cannot be isomorphic. Fusion with $\psi$ therefore partitions the simple objects into two-element equivalence classes under $a\sim a\otimes\psi$:
\begin{align}
    [a]:=\{a,a\otimes\psi\}.
\end{align}
It follows that the rank of a supermodular category is even.

The transparency of $\psi$ also constrains the $S$-matrix. From its definition as the Hopf-link invariant \cref{eq:s-matrix}, we have
\begin{align}
        S_{\psi b} &= d_{\psi\otimes\bar b}\, \operatorname{Tr}_{V_{\psi\otimes\bar b}^{\psi\bar b}} \left(M_{\psi\otimes\bar b}^{\psi,\bar b}\right)\\
        &=d_\psi d_b\\
        &=S_{\mathbf1b}.
\end{align}
Applying the identity \cref{eq:s-fused-link} then gives
\begin{align}
    S_{a\otimes\psi,b}
    &=\sum_c N_{a\psi}^{c}S_{cb}\\
    &=\frac{S_{ab}S_{\psi b}}{S_{\mathbf1b}}\\
    &=S_{ab}.
\end{align}
Hence, the two members of every class $[a]$ have identical rows and, by symmetry, identical columns in $S$. Therefore $S$ is not invertible. Choosing one representative from each of the $n$ classes, we then order the simple objects as
\begin{align}
    \bigl(a_0,\ldots,a_{n-1},a_0\otimes\psi,\ldots,a_{n-1}\otimes\psi\bigr),
    \qquad a_0=\mathbf1\,. \label{eq:super-ordering}
\end{align}
Defining the $n\times n$ matrices
\begin{align}
    \widehat S_{ij}:=S_{a_i a_j},
    \qquad
    \widehat T_{ij}:=\delta_{ij}\theta_{a_i}\,,
\end{align}
we then have
\begin{align}
    S=
    \begin{pmatrix}
        \widehat S&\widehat S\\
        \widehat S&\widehat S
    \end{pmatrix}
    \,,
    \qquad
    T=
    \begin{pmatrix}
        \widehat T&0\\
        0&-\widehat T
    \end{pmatrix}\,.
\end{align}
Replacing a representative $a_i$ by $a_i\otimes\psi$ leaves $\widehat S$ unchanged and reverses the sign of $\widehat T_{ii}$. Thus the reduced modular data independent of these choices are
\begin{align}
    \left(\widehat S,\widehat T^2\right)\,.
\end{align}

Where a fermionic topological phase admits a spin TQFT description, its supermodular category $\mathcal C$ describes the fusion and braiding of topological line operators. Gauging fermion parity $(-1)^F$ yields a bosonic TQFT whose topological line operators form a modular tensor category $\mathcal M$. The original local fermion $\psi$ now carries the nontrivial $\mathbb Z_2$ gauge charge. The additional anyons are fermion parity fluxes, characterized by monodromy $-1$ with $\psi$~\cite{bhardwajStateSumConstructions2017}.

A \emph{minimal modular extension} of $\mathcal C$ is a modular category $\mathcal M$ containing $\mathcal C$ as a braided subcategory, with $\mathcal C$ precisely the sector having trivial monodromy with $\psi$. The pair $(\mathcal M,\psi)$ is called a \emph{spin modular category}~\cite{bruillardFermionicModularCategories2017}. Fermion condensation of $\psi$ in the bosonic theory recovers the spin TQFT, making $\psi$ a local fermion~\cite{bhardwajStateSumConstructions2017,gaiottoSpinTQFTsFermionic2016}.

Every pseudo-unitary supermodular category admits a minimal modular
extension~\cite{johnson-freydMinimalNondegenerateExtensions2023}.
For the more general supermodular categories considered here,
we assume the existence of such an extension and derive its
consequences.

Let $S^{\mathcal M}$ be the unnormalized $S$-matrix of the extension $\mathcal M$. Applying \cref{eq:s-fused-link} to $\psi\otimes\psi\cong\mathbf1$ gives
\begin{align}
    \frac{(S^{\mathcal M}_{\psi x})^2}{d_x}=d_x,
    \qquad
    \mu_x:=\frac{S^{\mathcal M}_{\psi x}}{d_x}\in\{+1,-1\}\,.
\end{align}
Since $\psi$ is self-dual and invertible, double braiding acts on the one-dimensional splitting space $V_{\psi\otimes x}^{\psi x}$ by the scalar $\mu_x$:
\begin{align}
    M_{\psi\otimes x}^{\psi x}
    =\mu_x\operatorname{id}_{V_{\psi\otimes x}^{\psi x}}\,.
\end{align}
The simple objects of $\mathcal{M}$ therefore split into two sectors,
\begin{align}
    \mathcal M=\mathcal C_0\oplus\mathcal C_1,
    \qquad
    \mu_x=(-1)^\alpha\quad\text{for }x\in\mathcal C_\alpha,
    \qquad
    \mathcal C_0=\mathcal C\,.
\end{align}
The objects of $\mathcal C_1$ are the fermion parity fluxes.

Consider a splitting junction in $V_c^{ab}$, with its multiplicity label suppressed. Moving a $\psi$-loop past the junction and evaluating it on either side gives
\begin{align}
    \mu_c\,\vcenter{\hbox{\FermionGradingPlainTree}}
    &=\vcenter{\hbox{\FermionGradingLoopTree{0}}}
    =\vcenter{\hbox{\FermionGradingLoopTree{1}}}
    =\mu_a\mu_b\,\vcenter{\hbox{\FermionGradingPlainTree}}\,.
    \label{eq:fermion-fusion-grading}
\end{align}
Thus every fusion channel $c$ of $a\otimes b$ satisfies
\begin{align}
    \mu_c=\mu_a\mu_b\,.
\end{align}
If $a\in\mathcal C_\alpha$ and $b\in\mathcal C_\beta$, then $c\in\mathcal C_{\alpha+\beta}$, with addition modulo two.

The sums of squared quantum dimensions over the simple objects
of $\mathcal C_0$ and $\mathcal C_1$ are equal. Since $\mathcal M$ is modular, its $S$-matrix satisfies \cref{eq:appendix-s-square}. Taking the $\mathbf1,\psi$ matrix element gives
\begin{align}
    0
    &=\left(\left(S^{\mathcal M}\right)^2\right)_{\mathbf1\psi}\\
    &=\sum_{x\in\irr(\mathcal M)}
        S^{\mathcal M}_{\mathbf1x}S^{\mathcal M}_{x\psi}\\
    &=\sum_{x\in\irr(\mathcal M)}\mu_xd_x^2\,.
\end{align}
The first equality follows because $\bar\psi\cong\psi\not\cong\mathbf1$. Splitting the sum into the two sectors yields
\begin{align}
    \sum_{x\in\irr(\mathcal C_0)}d_x^2
    =\sum_{x\in\irr(\mathcal C_1)}d_x^2,
    \qquad
    D_{\mathcal M}^{\,2}=2D_{\mathcal C}^{\,2}\,.
\end{align}
Balancing in the unique $\psi\otimes x$ channel gives
\begin{align}
    \theta_{\psi\otimes x}=-\mu_x\theta_x\,.
\end{align}
For $x\in\mathcal C_0$, the objects $x$ and $\psi\otimes x$ have opposite twists and are therefore non-isomorphic. For $x\in\mathcal C_1$, they have the same twist and may be isomorphic.

We can now bound the rank of $\mathcal M$ by counting how fusion with $\psi$ permutes its simple objects. Since $\psi\otimes\psi\cong\mathbf1$, this permutation exchanges two objects or fixes an object:
\begin{align}
    x\longleftrightarrow\psi\otimes x,
    \qquad\text{or}\qquad
    \psi\otimes x\cong x\,.
\end{align}
The $2n$ simple objects of $\mathcal C_0$ form $n$ exchanged pairs. Let $\mathcal C_1$ contain $p$ exchanged pairs and $q$ fixed objects. Each exchanged pair contributes one $+1$ eigenvector of the fusion matrix $\mathcal N_\psi$, given by the sum of the two corresponding basis vectors. Each fixed object contributes another $+1$ eigenvector. The $+1$ eigenspace therefore has dimension $n+p+q$.

By the fusion-matrix diagonalization \cref{eq:s-fusion-diagonalization}, the eigenvalues of $\mathcal N_\psi$ are also
\begin{align}
    \frac{S^{\mathcal M}_{\psi x}}{S^{\mathcal M}_{\mathbf1x}}
    =\mu_x\,.
\end{align}
These equal $+1$ precisely for the $2n$ simple objects of $\mathcal C_0$. Equating the two counts gives
\begin{align}
    n+p+q=2n,
    \qquad
    p+q=n\,.
\end{align}
Consequently,
\begin{align}
    \operatorname{rank}(\mathcal M)=2n+2p+q=3n+p\,.
\end{align}
Since $0\leq p\leq n$, writing $r=\operatorname{rank}(\mathcal C)=2n$ gives~\cite{bruillard2020classification}
\begin{align}
    \frac{3r}{2}\leq\operatorname{rank}(\mathcal M)\leq2r\,.
\end{align}
The reduced modular data are obtained by restricting the modular data of $\mathcal M$ to an invariant subspace.
Let us extend the ordering used in \cref{eq:super-ordering} to include the remaining simple objects of $\mathcal{M}$, placing those belonging to $\mathcal C_1$ at the end:
\begin{align}
    \bigl(a_0,\ldots,a_{n-1},a_0\otimes\psi,\ldots,a_{n-1}\otimes\psi, b_1, \ldots b_{\mathrm{rank}(\mathcal{C}_1)} \bigr),
    \qquad a_0=\mathbf1\,.
\end{align}
Applying \cref{eq:s-fused-link} to fusion with $\psi$ gives
\begin{align}
    S^{\mathcal M}_{a\otimes\psi,x}
    =\mu_xS^{\mathcal M}_{ax},
\end{align}
so the full matrix has the block form
\begin{align}
    S^{\mathcal M}
    =\begin{pmatrix}
        \widehat S&\widehat S&B\\
        \widehat S&\widehat S&-B\\
        B^{\mathsf T}&-B^{\mathsf T}&E
    \end{pmatrix},
\end{align}
for some matrices $B,E$. Consequently,
\begin{align}
    S^{\mathcal M}
    \begin{pmatrix}v\\v\\0\end{pmatrix}
    =\begin{pmatrix}
        2\widehat S v\\
        2\widehat S v\\
        0
    \end{pmatrix}.
\end{align}
Also, $\theta_{a\otimes\psi}^{\,2}=\theta_a^2$, so
\begin{align}
    \left(T^{\mathcal M}\right)^2
    \begin{pmatrix}v\\v\\0\end{pmatrix}
    =\begin{pmatrix}
        \widehat T^2v\\
        \widehat T^2v\\
        0
    \end{pmatrix}.
\end{align}
Thus the subspace
\begin{align}
    W=\left\{
        \frac1{\sqrt2}\begin{pmatrix}v\\v\\0\end{pmatrix}
        :v\in\mathbb C^n
    \right\}
\end{align}
is preserved by $s^{\mathcal M}=S^{\mathcal M}/D_{\mathcal M}$ and $(T^{\mathcal M})^2$. In these normalized coordinates, choosing $D_{\mathcal M}=\sqrt2D_{\mathcal C}$, their restrictions are
\begin{align}
    s^{\mathcal M}\Big|_W=\widehat s
    :=\frac{\sqrt2\,\widehat S}{D_{\mathcal C}},
    \qquad
    \left(T^{\mathcal M}\right)^2\Big|_W=\widehat T^2\,.
\end{align}
The square of $s^{\mathcal M}$ acts by charge conjugation, which
preserves $W$ and sends $[a]$ to $[\bar a]$. Its restriction to $W$
is the charge-conjugation matrix $C$ on the fermion pairs:
\begin{align}
    \widehat s^{\,2}=C,
    \qquad
    C_{[a][b]}=\delta_{[a],[\bar b]},
    \qquad
    \widehat S^{\,2}=\frac{D_{\mathcal C}^{\,2}}2C\,. \label{eq:super-charge-conjugation}
\end{align}
Since $C^2=I$, both $\widehat S$ and $\widehat s$ are invertible.

The matrices $s^{\mathcal M}$ and $T^{\mathcal M}$ define a projective representation of $\sl{\mathbb Z}$. Since $W$ is preserved by $s^{\mathcal M}$ and $(T^{\mathcal M})^2$, their restrictions define a projective representation of
\begin{align}
    \Gamma_\theta=\langle\mathfrak s,\mathfrak t^2\rangle
    <\sl{\mathbb Z},
    \qquad
    \mathfrak s\longmapsto\widehat s,
    \qquad
    \mathfrak t^2\longmapsto\widehat T^2\,.
\end{align}

In the corresponding spin TQFT, $\Gamma_\theta$ is the subgroup of the torus mapping class group preserving the spin structure with antiperiodic boundary conditions around both cycles~\cite{bondersonCongruenceSubgroupsSuperModular2018}.

The reduced matrices determine fusion coefficients for the equivalence classes:
\begin{align}
    \widehat N_{[a][b]}^{[c]}:=N_{ab}^{c}+N_{ab}^{c\otimes\psi}\,.
\end{align}
Writing $\widehat S_{[a][b]}:=S_{ab}$ and grouping
\cref{eq:s-fused-link} by equivalence class gives
\begin{align}
    \frac{\widehat S_{[a][x]}\widehat S_{[b][x]}}
         {\widehat S_{[\mathbf1][x]}}
    =\sum_{[c]}\widehat N_{[a][b]}^{[c]}\widehat S_{[c][x]}\,.
\end{align}
Using \cref{eq:super-charge-conjugation} to multiply both sides by $\widehat S^{-1}=2\widehat SC/D_{\mathcal C}^{\,2}$ yields the reduced Verlinde formula~\cite{bruillardFermionicModularCategories2017}
\begin{align}
    \widehat N_{[a][b]}^{[c]}
    =\frac{2}{D_{\mathcal C}^{\,2}}\sum_{[x]}
    \frac{\widehat S_{[a][x]}\widehat S_{[b][x]}\widehat S_{[\bar c][x]}}
         {\widehat S_{[\mathbf1][x]}}\,.
\end{align}
Choose representatives $a,b$ of the input classes and one
representative $c$ of each output class. Grouping the balancing
expression in \cref{eq:s-matrix} by equivalence class, and using
$\theta_{c\otimes\psi}=-\theta_c$, gives
\begin{align}
    \widehat S_{[a][b]}
    =\sum_{[c]}\bigl(N_{a\bar b}^{c}-N_{a\bar b}^{c\otimes\psi}\bigr)
    \frac{\theta_c}{\theta_a\theta_b}d_c\,.
\end{align}
The reduced Verlinde formula determines the sums
$N_{ab}^{c}+N_{ab}^{c\otimes\psi}$.
A splitting into individual nonnegative integer coefficients
$N_{ab}^{c}$ and $N_{ab}^{c\otimes\psi}$ must also satisfy
balancing, associativity, and the vacuum and duality relations.

\section{Scaling of Congruence Representations of \texorpdfstring{$\sl{\zee{}}$}{SL(2,Z)}}
\label{app:scaling}

A couple of brief proofs are given here for the scaling relations stated for congruence representations of $\sl{\zee{}}$. Both bounds are sharp for congruence representations, but modular data may scale more moderately (or more dramatically, in the case of the number of modular data, although \cref{fig:counts} suggests otherwise). Both results leverage the classification of prime-power level representations in~\cite{nobsIrreduziblenDarstellungenGruppen1976,nobsIrreduziblenDarstellungenGruppen1976b}. 
A useful fact is that the irreducible, congruence, prime-power level representations of $\sl{\zee{}}$ occur at exactly the dimensions listed in \cref{tab:dims}.

\begin{table}[ht]
\centering
\begin{tabular}{ c|c } 
  prime-power level & irrep. dimensions\\
 \hline
 1 & 1\\
 2 & 1, 2\\
 3, 4 & 1, 2, 3\\
 8 & 2, 3, 4, 6\\
 $2^\lambda,\quad \lambda\geq 4$ & \( 3\cdot2^{\lambda-4},\quad 3\cdot2^{\lambda-3},\quad 3\cdot2^{\lambda-2},\quad 2^{\lambda-1},\quad 3\cdot2^{\lambda-1}\)\\
 $p,\quad p\ge5$ odd & $\frac{1}{2}(p-1),\quad\frac{1}{2}(p+1),\quad p-1,\quad p,\quad p+1$\\
 $p^\lambda,\quad p$ odd, \(\lambda\ge2\)	& \( \frac{1}2p^{\lambda-2}(p^2-1),\quad p^{\lambda-1}(p-1),\quad p^{\lambda-1}(p+1)\)
\end{tabular}
 \caption{Dimensions of irreducible prime-power level representations of $\sl{\zee{}}$.}
 \label{tab:dims}
\end{table}

\textbf{Lemma:} The number $N_\text{rep}(r)$ of isomorphism classes of $r$-dimensional congruence representations satisfies
\begin{equation}N_\text{rep}(r) \leq \exp\left(r^{2/3 + o(1)}\right)\end{equation}
\begin{proof}
The irreducible prime-power level representations $\pi$ of $\sl{\zee{q}}$, with $q = p^\lambda$, satisfy
\begin{equation}\operatorname{dim}\pi \geq c q\end{equation} where $c = \frac{3}{16}$ is constant. Then, let $b(q)$ be the number of irreducible representations of level $q$. We have
\begin{equation}b(q) c^2 p^{2\lambda} \leq \sum_{\substack{\pi\in\irr(\sl{\zee{q}})\\\text{level}~q}}(\dim \pi)^2 \leq |\sl{\zee{q}}|\end{equation} 
Then, we have $|\sl{\zee{p^\lambda}}|  = p^{3\lambda}(1-p^{-2})$, so that 
\begin{equation}b(q) \leq c^{-2}q\end{equation}
Any irreducible congruence representation $\rho$ of $\sl{\zee{}}$ of level $N = \prod_{i=1}^k q_i$, with $q_i = p_i^{\lambda_i}$ for distinct primes $p_i$, factors as a tensor product $\bigotimes_{i=1}^k \pi_i$ of irreducible representations $\pi_i$ of level $q_i$. So, 
\begin{equation}\operatorname{dim} \rho = \prod_{i=1}^k \operatorname{dim} \pi_i \geq c^k N~~~~\text{and} ~~~~ b(N) \leq c^{-2k}N\end{equation}
Then, the number of factors is $k = O(\frac{\log N}{\log \log N})$, so that 
\begin{equation}b(N) \leq N^{1+o(1)}\end{equation}
Summing over the irreducibles of level $N\leq d^{1+o(1)}$, the number of irreducible representations of dimension $\leq d$ is
\begin{equation}A(d) \leq d^{2+o(1)}\end{equation}
Then, considering all congruence representations, the generating function $F(z)$ satisfies
\begin{equation}F(z)\coloneq \sum_{r\geq 0} N_\text{rep}(r) z^r = \prod_{n\geq 1} (1-z^n)^{-a_n}\end{equation}
for $a_n = A(n) - A(n-1)$ the number of irreducible representations of dimension $n$. Then $N_\text{rep}(r) \leq e^{tr} F(e^{-t})$ for any $t>0$, and
\begin{equation}\log F(e^{-t}) = \sum_{k \geq 1} \frac{1}{k} S(tk)\end{equation}
where
\begin{equation}S(u) = \sum_{n\geq 1} e^{-un}(A(n) - A(n-1)) \leq u^{-2-o(1)}\end{equation}
Finally, we have
\begin{equation}\log F(e^{-t}) \leq t^{-2 - o(1)}\end{equation}
from which the result follows, taking $t = r^{-1/(3+o(1))}$.
\end{proof}

\textbf{Lemma:} For a congruence representation $\rho$ of $\sl{\zee{}}$ of dimension $r$,
\begin{equation}\operatorname{ord}(\rho(\modt)) \leq \exp\left((1+o(1))\sqrt{2r\log r}\right)\end{equation}
\begin{proof} 
For a representation $\rho = \bigoplus_{j=1}^s \rho_j$ (with each $\rho_j$ irreducible) of level $N = \prod_{p|N} p^{\lambda_p}$, each $\rho_j$ decomposes as $\bigotimes \rho_{j,p}$, each of level $p^{e_{j,p}}$. Further, $\lambda_p = \max_j e_{j,p}$, and we have an upper bound
\begin{equation}\log\operatorname{ord}\rho(\modt) \leq \sum_{p|N}\lambda_p \log p\end{equation}
We first exclude the contributions from small primes. By \cref{tab:dims}, $p^{\lambda_p} \leq \frac{16}{3}r$, so
\begin{equation}\sum_{\substack{p\leq \log r\\\lambda_p > 0}} \lambda_p \log p \leq O((\log r)^2)\end{equation}
These therefore contribute negligibly to the desired bound for $p\leq \log r$. Instead, for $p>\log r$, each prime-power level irreducible is of dimension $d_{j,p} \geq \frac12(1+o(1)) p^{e_{j,p}}$. So,
\begin{equation}\dim \rho_j \geq \prod_{p> \log r} d_{j,p} \geq \sum_{\substack{p> \log r\\e_{j,p}>0}} \frac12(1+o(1)) p^{e_{j,p}}\end{equation}
and, summing over $j$,
\begin{equation}r\geq \frac12(1+o(1))\sum_{\substack{p> \log r\\\lambda_p>0}}  p^{\lambda_p}\end{equation}
Fix $x \geq e$. If $p^{\lambda_p} \geq x$, then
\begin{equation}\frac{\log p^{\lambda_p}}{p^{\lambda_p}} \leq \frac{\log x}{x}\eqcolon t\end{equation}
Then, for $B = (2+o(1))r$,
\begin{equation}\sum_{p > \log r} \lambda_p \log p \leq t B + \sum_{\substack{p > \log r,~\lambda_p > 0\\ p^{\lambda_p} \leq x}} (\lambda_p \log p - t p^{\lambda_p})\end{equation}
One would expect $\operatorname{ord}(T)$ to be dominated by primes with $\lambda_p = 1$, and this ends up being the case. Namely, for $\lambda_p  = 1$,
\begin{equation}\sum_{p\leq x}(\log p - tp) = x - t \frac{x^2}{2\log x} + o(x) = \frac{x}{2} + o(x)\end{equation}
while for $\lambda_p \geq 2$,
\begin{equation}\sum_{\substack{\lambda_p \geq 2\\p^{\lambda_p}\leq x}} \lambda_p \log p = O(\sqrt{x}(\log x)^2) = o(x)\end{equation}
Hence,
\begin{equation}\sum_{\lambda_p > 0} \lambda_p \log p \leq B \frac{\log x}{x} + \frac{x}{2} + o(x)\end{equation}
and the result follows, for $x = \sqrt{B \log B}$.

\end{proof}

\section{Higher Central Charges of Chern--Simons Theories}
\label{app:higher-central-charges}

The higher central charges of a modular category $\C$ are the phases of its higher Gauss sums,
\begin{equation}
  \xi_n(\C) = \frac{\tau_n(\C)}{|\tau_n(\C)|}, \qquad \tau_n(\C) = \sum_{a\in\irr(\C)} \theta_a^n\, d_a^2 ,
  \label{eq:hcc-def}
\end{equation}
defined for $n$ coprime to $\operatorname{ord}(T)$ \cite{ngHigherGaussSums2019}. Throughout this appendix $e(x)\equiv\exp(2\pi i x)$. At $n=1$, $\xi_1 = e(c_-/8)$ with $c_-$ the chiral central charge. Like $\xi_1$, the higher central charges are invariants of the Witt class of $\C$ \cite{ngHigherCentralCharges2022}. They are unchanged by anyon condensation and multiply under stacking. A theory with $\xi_n\neq1$ for some $n$ admits no gapped boundary even when $c_-\equiv0\pmod 8$ \cite{kaidiHigherCentralCharges2022}. In the main text they are used to match condensations in \cref{sec:condensation} and to constrain Chern--Simons realizations of the exotic data in \cref{app:notCS}.

A closed formula for the higher central charges of $G_k$ Chern--Simons theory with general $G$ does not appear in the literature. Ng, Schopieray and Wang posed this as Question 7.1 of \cite{ngHigherGaussSums2019}. In this appendix we derive one from a theorem of \cite{ngHigherCentralCharges2022} and the Galois action on affine characters \cite{Coste:1993af}. For $G$ simple and simply connected, with category $\C(\mathfrak g,k)$ and chiral central charge $c_- = k\dim\mathfrak g/(k+h^\vee)$,
\begin{equation}
  \boxed{\;\xi_n\big(\C(\mathfrak g,k)\big) = s_n \left(\frac{d}{n}\right) (-1)^{\frac{(n-1)|\Delta_+|}{2}}\, e\!\left(\frac{\bar n\, c_-}{8} + \frac{n(\bar n^2-1)(\dim\mathfrak g - c_-)}{24}\right)\;}
  \label{eq:hcc-result}
\end{equation}
Here $n>0$ is coprime to $2(k+h^\vee)\,|P/Q^\vee|$, and $\bar n$ is an inverse of $n$ modulo $8(k+h^\vee)\,|P/Q^\vee|$, where $P$ is the weight lattice of $\mathfrak g$, $Q^\vee$ its coroot lattice, and $|P/Q^\vee|$ the order of the finite group $P/Q^\vee$, equal to the determinant of the Gram matrix of the coroots (for simply laced $\mathfrak g$, to the order of the center of $G$). The Jacobi symbol $\left(\frac{d}{n}\right)$ is evaluated at $d = (k+h^\vee)^r\,|P/Q^\vee|$, with $r$ the rank of $\mathfrak g$, and $|\Delta_+|$ is the number of positive roots. The sign $s_n$ is the sign of a product over the positive roots $\Delta_+$,
\begin{equation}
  s_n = \operatorname{sgn}\prod_{\alpha>0}\sin\!\left(\frac{\pi\bar n(\rho,\alpha)}{k+h^\vee}\right) ,
  \label{eq:hcc-weyl-sign}
\end{equation}
with $\rho$ the Weyl vector and $(\cdot,\cdot)$ the invariant form normalized so that long roots have length squared 2. \Cref{table:hcc-data} lists the remaining data for each $\mathfrak g$. Since $\xi_n$ depends only on the class of $n$ modulo $\operatorname{ord}(T)$, and each class coprime to $\operatorname{ord}(T)$ contains integers coprime to $2(k+h^\vee)|P/Q^\vee|$, \cref{eq:hcc-result} determines $\xi_n$ for each admissible $n$.

\begin{table}[!ht]
\centering
\begin{tabular}{c|cccc}
  $\mathfrak g$ & $h^\vee$ & $\dim\mathfrak g$ & $|\Delta_+|$ & $|P/Q^\vee|$ \\
  \hline
  $\mathfrak{su}(N)$, $N\geq2$ & $N$ & $N^2-1$ & $N(N-1)/2$ & $N$ \\
  $\mathfrak{so}(2r+1)$, $r\geq2$ & $2r-1$ & $r(2r+1)$ & $r^2$ & $4$ \\
  $\mathfrak{sp}(2r)$, $r\geq1$ & $r+1$ & $r(2r+1)$ & $r^2$ & $2^r$ \\
  $\mathfrak{so}(2r)$, $r\geq3$ & $2r-2$ & $r(2r-1)$ & $r(r-1)$ & $4$ \\
  $\mathfrak g_2$ & $4$ & $14$ & $6$ & $3$ \\
  $\mathfrak f_4$ & $9$ & $52$ & $24$ & $4$ \\
  $\mathfrak e_6$ & $12$ & $78$ & $36$ & $3$ \\
  $\mathfrak e_7$ & $18$ & $133$ & $63$ & $2$ \\
  $\mathfrak e_8$ & $30$ & $248$ & $120$ & $1$
\end{tabular}
\caption{The data entering \cref{eq:hcc-result} for each simple $\mathfrak g$.}
\label{table:hcc-data}
\end{table}

\Cref{eq:hcc-result} follows from a general result of Ng, Rowell, Wang and Zhang \cite{ngHigherCentralCharges2022}. The modular data of $\C(\mathfrak g,k)$ lie in the cyclotomic field generated by $\zeta = e\big(1/(8(k+h^\vee)|P/Q^\vee|)\big)$, and $\operatorname{ord}(T)$ divides $8(k+h^\vee)|P/Q^\vee|$. A Galois automorphism $\sigma$ of this field acts on the modular data by a signed permutation $\hat\sigma$ of the anyons, $\sigma(s_{ab}) = \epsilon_\sigma(a)\, s_{\hat\sigma(a),b}$ with $s = S/D$ \cite{Coste:1993af}. For $n$ coprime to $2(k+h^\vee)|P/Q^\vee|$, let $\sigma$ be the automorphism of this field with $\sigma(\zeta) = \zeta^{\bar n}$, so that $\sigma^{-1}$ acts on the twists as $\theta_a\mapsto\theta_a^n$. Theorem 4.1 of \cite{ngHigherCentralCharges2022} then reads\footnote{Theorem, equation and remark numbers refer to the arXiv version of \cite{ngHigherCentralCharges2022}.}
\begin{equation}
  \xi_n = \operatorname{sgn}\sigma(D)\;\, \sigma(\xi_1)\;\, \theta_{\hat\sigma(1)}^{\,n} .
  \label{eq:hcc-nrwz}
\end{equation}
The theorem is stated there in terms of the Frobenius--Schur exponent, which equals $\operatorname{ord}(T)$ for a modular category, and with the last factor expressed through a cube root of $\xi_1$; that factor equals $\theta_{\hat\sigma(1)}^{\,n}$ by their Eq.~(4.14). In a pseudo-unitary theory $\operatorname{sgn}\sigma(D) = \epsilon_\sigma(1)$, the sign with which $\sigma$ carries the vacuum row of $s$ to the row of $\hat\sigma(1)$ (their Remark 4.2). Beyond $c_-$, $\xi_n$ is therefore fixed by the twist of the Galois image of the vacuum and by the sign $\epsilon_\sigma(1)$. For $\C(\mathfrak g,k)$ both follow from the Kac--Peterson formula.

The anyons of $\C(\mathfrak g,k)$ are labeled by integrable weights $\lambda$, and the shifted weights $\lambda+\rho$ run over the weights $\beta\in P$ in the open alcove $\{\beta : 0 < (\beta,\alpha) < k+h^\vee \text{ for all } \alpha>0\}$. The vacuum corresponds to $\rho$. The twists are $\theta_\lambda = e\big(((\lambda+\rho,\lambda+\rho) - (\rho,\rho))/2(k+h^\vee)\big)$, and the Kac--Peterson formula gives
\begin{equation}
  s_{\lambda\mu} = x \sum_{w\in W} \det(w)\, e\!\left(-\frac{(w(\lambda+\rho), \mu+\rho)}{k+h^\vee}\right), \qquad x = \frac{i^{|\Delta_+|}}{\sqrt d} ,
  \label{eq:hcc-kac-peterson}
\end{equation}
with $W$ the Weyl group. Acting with $\sigma$ multiplies the exponents by $\bar n$, which is the same as replacing the shifted weight $\lambda+\rho$ by $\bar n(\lambda+\rho)$. The rescaled weight generally lies outside the alcove, but the sum over $W$ is antisymmetric under Weyl reflections and invariant under translations by $(k+h^\vee)Q^\vee$, so it can be folded back into the alcove by an element of the affine Weyl group $W\ltimes (k+h^\vee)Q^\vee$, which contributes the sign $\det(\omega)$ of its Weyl part $\omega$. Together with the conjugation of the prefactor $x$, this gives $\hat\sigma$ and $\epsilon_\sigma(\lambda) = \det(\omega)\,\sigma(x)/x$ (Lemma 3.9 of \cite{gannonExoticQuantumSubgroups2023}). The Galois image of the vacuum is therefore the anyon at $\bar n\rho$, folded back into the alcove:
\begin{equation}
  \hat\sigma(1) + \rho = \omega(\bar n\rho) + (k+h^\vee)\gamma, \qquad \omega\in W,\ \gamma\in Q^\vee .
  \label{eq:hcc-partner}
\end{equation}

The twist of $\hat\sigma(1)$ is unaffected by the folding. The norm is $W$-invariant, and a translation by $(k+h^\vee)\gamma$ changes $(\beta,\beta)/2(k+h^\vee)$ by the integer $(\beta,\gamma)+(k+h^\vee)(\gamma,\gamma)/2$, because $Q^\vee$ is even and dual to $P$. Using $(\rho,\rho) = h^\vee\dim\mathfrak g/12$ and $h^\vee\dim\mathfrak g/(k+h^\vee) = \dim\mathfrak g - c_-$,
\begin{equation}
  \theta_{\hat\sigma(1)}^{\,n} = e\!\left(\frac{n(\bar n^2-1)(\rho,\rho)}{2(k+h^\vee)}\right) = e\!\left(\frac{n(\bar n^2-1)(\dim\mathfrak g - c_-)}{24}\right) .
  \label{eq:hcc-partner-twist}
\end{equation}

The sign $\epsilon_\sigma(1) = \det(\omega)\,\sigma(x)/x$ has two factors. The parity of the folding, $\det(\omega)$, can be read off without finding $\omega$ from $F(\beta) = \prod_{\alpha>0}\sin\big(\pi(\beta,\alpha)/(k+h^\vee)\big)$, the numerator of the quantum dimensions. $F$ changes sign under Weyl reflections, is invariant under translations by $(k+h^\vee)Q^\vee$ (a translation by $(k+h^\vee)\gamma$ multiplies it by $(-1)^{(\gamma,2\rho)} = 1$), and is positive inside the alcove, so evaluating it on both sides of \cref{eq:hcc-partner} gives $\det(\omega) = \operatorname{sgn}F(\bar n\rho) = s_n$. The conjugation of the prefactor gives $\sigma(i^{|\Delta_+|})/i^{|\Delta_+|} = (-1)^{(\bar n-1)|\Delta_+|/2}$ and $\sigma(\sqrt d)/\sqrt d = \left(\frac{d}{\bar n}\right)$, since $\sigma$ acts on $\mathbb Q(\sqrt d)$ through its quadratic character. Both factors take the same value at $n$ as at $\bar n$, because $n\bar n\equiv1$ modulo 4 and modulo the conductor of $\mathbb Q(\sqrt d)$, both of which divide $8(k+h^\vee)|P/Q^\vee|$. Hence
\begin{equation}
  \operatorname{sgn}\sigma(D) = s_n \left(\frac{d}{n}\right) (-1)^{\frac{(n-1)|\Delta_+|}{2}} .
  \label{eq:hcc-sign}
\end{equation}
Finally $\sigma(\xi_1) = e(\bar n c_-/8)$, and substituting this, \cref{eq:hcc-partner-twist} and \cref{eq:hcc-sign} into \cref{eq:hcc-nrwz} gives \cref{eq:hcc-result}. The modulus $8(k+h^\vee)|P/Q^\vee|$ makes $e(\bar n c_-/8)$ well defined. Since $|P/Q^\vee|\,\rho\in Q^\vee$, changing $\bar n$ by a multiple of the modulus moves $\bar n\rho$ by an element of $(k+h^\vee)Q^\vee$, and its image in the alcove is unchanged.

We compared \cref{eq:hcc-result} with a direct evaluation of \cref{eq:hcc-def} over the alcove for the simple $\mathfrak g$ of rank at most 7, together with $\mathfrak{su}(9)$ and $\mathfrak e_8$, at levels up to 12 for $\mathfrak{su}(2)$ and $\mathfrak g_2$ and up to 2 to 5 for the algebras of rank 7 and 8 (2{,}904 cases), and with modular data from an independent implementation of the Kac--Peterson formula that constructs the Weyl group explicitly (622 cases, reaching $\mathfrak e_7$ at level 1 and $\mathfrak e_8$ at level 2). In all cases the two agree. The formula also satisfies the sum rules implied by 20 conformal embeddings and by level-rank duality of $\SU(N)_k$.

\section{Non-Invertible Condensations to Exotic Anisotropic Categories}
\label{app:notCS}

We expect the following holds for the exotic modular categories $\mathcal M$ we identified:

\textbf{Conjecture:} $\mathcal M$ is not Witt-equivalent to any Chern--Simons category of the form
\begin{equation}\C = \mathcal P \boxtimes \C(\mathfrak g_1, k_1) \boxtimes \cdots \boxtimes \C(\mathfrak g_n, k_n) \end{equation}
for $\mathcal P$ pointed. 

In all cases, the exotic $\mathcal M$ is anisotropic, simple, and non-pointed. In particular, then, it possesses property S, as defined in~\cite{davydovWittGroupNondegenerate2011}. Anisotropy reduces the statement to $\mathcal M\neq \C_{A_\text{max}}^0$, and by the argument of \cref{sec:condensation} we may assume that $A_\text{max}$ has support on non-invertibles. In the following we focus only on the rank-11 exotics for simplicity.

Although the known constraints on this form of condensation are insufficient to rule out all possible $\C$, they are nonetheless very restrictive. For general condensations we require:
\begin{enumerate}
  \item $\xi_n(\mathcal M) = \xi_n(\C)$ for $\operatorname{gcd}(n,\operatorname{ord}(T_\C)) = 1$.
  \item $\mathbb Q(\{d\}_{\mathcal M}) \subset \mathbb Q(\{d\}_{\C})$. The dimension fields for affine Lie categories are classified in~\cite{gannonAlgebraicNumberFields2019}.
  \item For each simple $x\in\mathcal M$, there must be a simple $y\in\C$ such that $\theta_x = \theta_y$ and $d_x \leq d_y$. We may further require that $d_y / d_x$ be totally positive~\cite{gannonExoticQuantumSubgroups2023}.
\end{enumerate}
Although no single $\C(\mathfrak g,k)$ has higher central charges compatible with those of $\mathcal M$, one can always engineer a product that does. Similarly, the other conditions become less and less constraining as one extends to higher-rank $\C$. 

Some headway can be made by leveraging property S as follows. Consider the homomorphism
\begin{equation}S: \mathcal W \rightarrow s\mathcal W,~~~~~ [\mathcal D] \mapsto [\mathcal D\boxtimes \sVec]\end{equation}
By~\cite{davydovStructureWittGroup2011}, the super-Witt group has a well-understood structure
\begin{equation}s\mathcal W = s\mathcal W_\text{pt} \oplus s\mathcal W_2 \oplus s\mathcal W_\infty\end{equation}
That is, it is generated by pointed categories and by $s$-simple categories $\mathcal S$ having either order 2 ($\overline{\mathcal S} := \mathcal S^\text{rev} \simeq \mathcal S$) or infinite order. By virtue of $\mathcal M$ having property S, $\mathcal M \boxtimes \sVec$ is $s$-simple and in this case has infinite order. Then $\mathcal M$ can be Witt-equivalent to $\C$ only if $\mathcal M$ or $\overline{\mathcal M}$ lies within the $s$-simple decomposition of a single non-pointed affine Lie factor, $[\C(\mathfrak g_i, k_i)\boxtimes \sVec]$. Now, let $\mathcal A_i$ be the anisotropic representative of $\C(\mathfrak g_i, k_i)$. If $\mathcal A_i$ is non-spin, $\mathcal A_i \boxtimes \sVec$ is anisotropic, and we immediately have that $\mathcal M$ (or $\overline{\mathcal M}$, though for simplicity we omit this) must be a subcategory of $\mathcal A_i$. On the other hand, if $\mathcal A_i$ is spin, $\mathcal A_i \boxtimes \sVec$ has maximal condensable algebra $A = 1 \oplus (\psi \boxtimes f)$. Then, the anisotropic representative $(\mathcal A_i \boxtimes \sVec)_A^0$ is precisely the Müger centralizer $\psi' = \mathcal B_i$ of the fermion $\psi \in \mathcal A_i$. As before, then, $\mathcal M \subset \mathcal B_i \subsetneq \mathcal A_i$.

Applying this to the constraints above, we must have that $\mathcal M \subset \mathcal A_i = (\C(\mathfrak g_i, k_i))_{A_\text{max}}^0$ for some $i$ in the product. We need only consider $A_\text{max}$ having support on non-invertibles, since the subcategories of $(\C(\mathfrak g_i, k_i))_R^0$ for $R$ supported on invertibles are fully classified in~\cite{schopierayPrimeDecompositionModular2018}. Similarly, by the higher central charge observation earlier (\cref{app:higher-central-charges}), $\mathcal M = \mathcal A_i$ is immediately ruled out. The remaining constraints are then:
\begin{enumerate}
  \item $\mathbb Q(\{d\}_{\mathcal M}) \subset \mathbb Q_{\kappa_i}$, for $\mathbb Q_n = \mathbb Q(\cos(\frac{2\pi}{n}))$ (bar a few exceptions, see~\cite{gannonAlgebraicNumberFields2019}). In our case, this implies that $35|\kappa_i = k_i + h_i^\vee$.
  \item As before, for each simple $x\in\mathcal M$, there must be a simple $y\in\C(\mathfrak g_i, k_i)$ such that $\theta_x = \theta_y$ and $d_x \leq d_y$ with $d_y / d_x$ totally positive.
  \item It is known that, for fixed $\mathfrak g$, there is a finite level $k$ above which $\C(\mathfrak g,k)$ has condensable algebras supported only on invertible objects. This bound scales roughly as $\mathcal O(n^3)$ in the rank $n$ of $\mathfrak g$~\cite{gannonExoticQuantumSubgroups2023}. In our case, then, we must have $k_i$ below this bound, and level-rank dualities can be used to further rule out certain $k$ below it.
\end{enumerate}
While restrictive, these constraints are insufficient to conclusively rule out $\mathcal M$ as Witt-equivalent to Chern--Simons. Infinite families can be constructed satisfying all the above constraints. The lowest-rank example we identify is, adopting the notation used in the database,
\begin{equation}\mathcal R_t = \text{U}(1)_K^2 \boxtimes \text{G}_{2,1}^{\boxtimes (1+20t)} \boxtimes \Sp(136)_1 \boxtimes \Spin(66)_6 \boxtimes \Sp(50)_9\end{equation}
for $t \in \zee{\geq 0}$ and $K = \left(\begin{smallmatrix}
  2 & 1\\1 & 4
\end{smallmatrix}\right)$. $\Sp(50)_9$ is the factor whose $s$-simple decomposition might contain $\mathcal M$. No subproduct is Witt-trivial, so this cannot be trivially reduced to a simpler product. At $t=0$ this has rank $5.44 \times 10^{14}$ and, although not proven to be the lowest-rank example,\footnote{We also exhaustively tested any possible Chern--Simons candidate up to rank $10^4$, guaranteeing at least that as a lower bound.} suggests that the existing constraints are at least sufficient to exclude any remotely low-rank candidates.


\printbibliography

\end{document}

%% file: figures/tikz-diagrams/rigidity-diagrams.tex
\input{figures/tikz-diagrams/diagram-style.tex}
\usetikzlibrary{calc,decorations.markings,positioning}

\newcommand{\RigidityCenteredArrow}{%
  \fill[black]
    (-1.75pt,-1.25pt) -- (2.05pt,0pt) -- (-1.75pt,1.25pt) -- cycle;%
}

\tikzset{
  appendix snake oriented/.style={
    appendix strand,
    postaction={decorate},
    decoration={
      markings,
      mark=at position 0.08 with {\RigidityCenteredArrow},
      mark=at position 0.50 with {\RigidityCenteredArrow},
      mark=at position 0.92 with {\RigidityCenteredArrow}
    }
  },
  appendix identity oriented/.style={
    appendix strand,
    postaction={decorate},
    decoration={
      markings,
      mark=at position 0.50 with {\RigidityCenteredArrow}
    }
  }
}

\newcommand{\LeftSnakeA}{%
  \begin{tikzpicture}[baseline=-0.52ex, x=0.54cm, y=0.54cm]
    \path[use as bounding box] (-1.00,-1.55) rectangle (1.00,1.55);
    \coordinate (LSBottom) at (0.68,-1.22);
    \coordinate (LSRightCap) at (0.68,0.34);
    \coordinate (LSMiddleCap) at (0,0.34);
    \coordinate (LSMiddleCup) at (0,-0.34);
    \coordinate (LSLeftCup) at (-0.68,-0.34);
    \coordinate (LSTop) at (-0.68,1.22);
    \draw[appendix snake oriented]
      (LSBottom) -- (LSRightCap)
      .. controls (0.68,0.78) and (0,0.78) .. (LSMiddleCap)
      -- (LSMiddleCup)
      .. controls (0,-0.78) and (-0.68,-0.78) .. (LSLeftCup)
      -- (LSTop);
    \node[appendix label, above=0pt of LSTop] {$a$};
    \node[appendix label, below=0pt of LSBottom] {$a$};
    \node[appendix label] at (0.25,0) {$\bar a$};
  \end{tikzpicture}%
}

\newcommand{\RightSnakeDualA}{%
  \begin{tikzpicture}[baseline=-0.52ex, x=0.54cm, y=0.54cm]
    \path[use as bounding box] (-1.00,-1.55) rectangle (1.00,1.55);
    \coordinate (RSBottom) at (-0.68,-1.22);
    \coordinate (RSLeftCap) at (-0.68,0.34);
    \coordinate (RSMiddleCap) at (0,0.34);
    \coordinate (RSMiddleCup) at (0,-0.34);
    \coordinate (RSRightCup) at (0.68,-0.34);
    \coordinate (RSTop) at (0.68,1.22);
    \draw[appendix snake oriented]
      (RSBottom) -- (RSLeftCap)
      .. controls (-0.68,0.78) and (0,0.78) .. (RSMiddleCap)
      -- (RSMiddleCup)
      .. controls (0,-0.78) and (0.68,-0.78) .. (RSRightCup)
      -- (RSTop);
    \node[appendix label, above=0pt of RSTop] {$\bar a$};
    \node[appendix label, below=0pt of RSBottom] {$\bar a$};
    \node[appendix label] at (-0.25,0) {$a$};
  \end{tikzpicture}%
}

\newcommand{\StraightOrientedLine}[1]{%
  \begin{tikzpicture}[baseline=-0.52ex, x=0.54cm, y=0.54cm]
    \path[use as bounding box] (-0.42,-1.55) rectangle (0.42,1.55);
    \coordinate (RIBottom) at (0,-1.22);
    \coordinate (RITop) at (0,1.22);
    \draw[appendix identity oriented] (RIBottom) -- (RITop);
    \node[appendix label, above=0pt of RITop] {$#1$};
    \node[appendix label, below=0pt of RIBottom] {$#1$};
  \end{tikzpicture}%
}

\tikzset{appendix unit strand/.style={
  appendix strand,draw=gray!75,postaction={decorate},
  decoration={markings,mark=at position .50 with {
    \fill[gray!75] (-1.75pt,-1.25pt)--(2.05pt,0pt)--(-1.75pt,1.25pt)--cycle;
  }}
}}
\newcommand{\CoevaluationCup}{%
  \begin{tikzpicture}[baseline=-.52ex,x=.54cm,y=.54cm]
    \draw[appendix unit strand] (-.45,-1.15)--(-.45,-.4875);
    \draw[appendix strand,postaction={decorate},decoration={markings,
      mark=at position .15 with {\RigidityCenteredArrow},
      mark=at position .85 with {\RigidityCenteredArrow}}]
      (0,.7)--(0,0).. controls (0,-.65) and (-.9,-.65)..(-.9,0)--(-.9,.7);
    \node[appendix vertex] at (-.45,-.4875) {};
    \node[appendix label,above=1pt] at (-.9,.7) {$a$};
    \node[appendix label,above=1pt] at (0,.7) {$\bar a$};
    \node[appendix label,below=1pt,text=gray!75] at (-.45,-1.15) {$1$};
  \end{tikzpicture}%
}
\newcommand{\EvaluationCap}{%
  \begin{tikzpicture}[baseline=-.52ex,x=.54cm,y=.54cm]
    \draw[appendix unit strand] (-.45,.4875)--(-.45,1.15);
    \draw[appendix strand,postaction={decorate},decoration={markings,
      mark=at position .15 with {\RigidityCenteredArrow},
      mark=at position .85 with {\RigidityCenteredArrow}}]
      (0,-.7)--(0,0).. controls (0,.65) and (-.9,.65)..(-.9,0)--(-.9,-.7);
    \node[appendix vertex] at (-.45,.4875) {};
    \node[appendix label,below=1pt] at (-.9,-.7) {$\bar a$};
    \node[appendix label,below=1pt] at (0,-.7) {$a$};
    \node[appendix label,above=1pt,text=gray!75] at (-.45,1.15) {$1$};
  \end{tikzpicture}%
}
\newcommand{\SnakeWithIdentities}[3]{%
  \begin{tikzpicture}[baseline=-.52ex,x=.54cm,y=.54cm]
    \path[use as bounding box] (-1,-1.55) rectangle (1,1.55);
    \begin{scope}[xscale=#1]
      \draw[appendix unit strand] (.68,-1.08)
        .. controls (.12,-1.08) and (-.34,-1.08) .. (-.34,-.67);
      \draw[appendix unit strand] (.34,.67)
        .. controls (.34,1.08) and (-.12,1.08) .. (-.68,1.08);
      \node[appendix label,below=1pt,text=gray!75] at (.05,-1.04) {$1$};
      \node[appendix label,above=1pt,text=gray!75] at (-.05,1.04) {$1$};
      \draw[appendix snake oriented] (.68,-1.22)--(.68,.34)
        .. controls (.68,.78) and (0,.78)..(0,.34)--(0,-.34)
        .. controls (0,-.78) and (-.68,-.78)..(-.68,-.34)--(-.68,1.22);
      \node[appendix vertex] at (-.34,-.67) {};
      \node[appendix vertex] at (.34,.67) {};
      \node[appendix vertex] at (.68,-1.08) {};
      \node[appendix vertex] at (-.68,1.08) {};
      \node[appendix label,above=0pt] at (-.68,1.22) {$#2$};
      \node[appendix label,below=0pt] at (.68,-1.22) {$#2$};
      \node[appendix label] at (.25,0) {$#3$};
    \end{scope}
  \end{tikzpicture}%
}
\newcommand{\LeftSnakeWithIdentities}{\SnakeWithIdentities{1}{a}{\bar a}}
\newcommand{\RightSnakeWithIdentities}{\SnakeWithIdentities{-1}{\bar a}{a}}

%% file: figures/tikz-diagrams/diagram-style.tex
\tikzset{
  appendix strand/.style={
    draw=black,
    line width=0.75pt,
    line cap=round,
    line join=round
  },
  appendix crossing/.style={
    appendix strand,
    preaction={draw=white, line width=3.2pt}
  },
  appendix vertex/.style={circle, fill=black, inner sep=1.25pt},
  appendix label/.style={font=\scriptsize, inner sep=1pt},
  appendix channel/.style={font=\scriptsize, inner sep=0pt},
  appendix basis/.style={font=\scriptsize, inner sep=0pt}
}

%% file: figures/tikz-diagrams/quantum-dimension-diagrams.tex
\input{figures/tikz-diagrams/diagram-style.tex}
\usetikzlibrary{arrows.meta,decorations.markings,positioning}

\newcommand{\AppendixCenteredArrow}{%
  \fill[black]
    (-1.75pt,-1.25pt) -- (2.05pt,0pt) -- (-1.75pt,1.25pt) -- cycle;%
}

\tikzset{
  appendix loop oriented/.style={
    appendix strand,
    postaction={decorate},
    decoration={
      markings,
      mark=at position 0.17 with
        {\AppendixCenteredArrow},
      mark=at position 0.67 with
        {\AppendixCenteredArrow}
    }
  },
  appendix edge oriented/.style={
    appendix strand,
    postaction={decorate},
    decoration={
      markings,
      mark=at position 0.30 with
        {\AppendixCenteredArrow}
    }
  }
}

\newcommand{\QuantumDimensionLoop}[1]{%
  \begin{tikzpicture}[baseline=-0.55ex, x=0.72cm, y=0.72cm]
    \path[use as bounding box] (-0.92,-0.72) rectangle (0.72,0.72);
    \coordinate (LoopCenter) at (0,0);
    \coordinate (LoopLabelAnchor) at (-0.50,0);
    \draw[appendix loop oriented] (LoopCenter) circle[radius=0.50];
    \node[appendix label, left=2pt of LoopLabelAnchor] {$#1$};
  \end{tikzpicture}%
}

\newcommand{\NestedDimensionLoops}[2]{%
  \begin{tikzpicture}[baseline=-0.55ex, x=0.76cm, y=0.76cm]
    \path[use as bounding box] (-0.92,-0.98) rectangle (1.18,0.98);
    \coordinate (NestedCenter) at (0,0);
    \coordinate (OuterRight) at (0.72,0);
    \draw[appendix loop oriented]
      (NestedCenter) ellipse[x radius=0.72,y radius=0.86];
    \draw[appendix loop oriented]
      (NestedCenter) ellipse[x radius=0.52,y radius=0.66];
    \node[appendix label] at (NestedCenter) {$#1$};
    \node[appendix label, right=1.5pt of OuterRight] {$#2$};
  \end{tikzpicture}%
}

\newcommand{\ResolvedDimensionLoops}[4]{%
  \begin{tikzpicture}[baseline=-0.55ex, x=0.86cm, y=0.86cm]
    \path[use as bounding box] (-1.10,-0.98) rectangle (1.30,0.98);
    \coordinate (ResolvedUpper) at (0.50,0.45);
    \coordinate (ResolvedLower) at (0.50,-0.45);
    \coordinate (ResolvedALabel) at (0.02,0);
    \coordinate (ResolvedBLabel) at (-0.61,0);
    \coordinate (ResolvedCLabel) at (0.37,0);

    \draw[appendix edge oriented]
      (ResolvedUpper)
      .. controls (0.14,0.69) and (-0.19,0.43) .. (-0.19,0)
      .. controls (-0.19,-0.43) and (0.14,-0.69) .. (ResolvedLower);
    \draw[appendix edge oriented]
      (ResolvedUpper)
      .. controls (0.78,0.82) and (-0.87,0.91) .. (-0.91,0)
      .. controls (-0.87,-0.91) and (0.78,-0.82) .. (ResolvedLower);
    \draw[appendix edge oriented]
      (ResolvedLower) -- (ResolvedUpper);

    \node[appendix label] at (ResolvedALabel) {$#1$};
    \node[appendix label] at (ResolvedBLabel) {$#2$};
    \node[appendix label] at (ResolvedCLabel) {$#3$};
    \node[appendix vertex] (ResolvedUpperDot) at (ResolvedUpper) {};
    \node[appendix vertex] (ResolvedLowerDot) at (ResolvedLower) {};
    \node[appendix basis, right=3.5pt of ResolvedUpperDot] {$#4$};
    \node[appendix basis, right=3.5pt of ResolvedLowerDot] {${#4}_*$};
  \end{tikzpicture}%
}

\newcommand{\ReverseQuantumDimensionLoop}[1]{%
  \begin{tikzpicture}[baseline=-.55ex,x=.72cm,y=.72cm]
    \path[use as bounding box] (-.92,-.72) rectangle (.72,.72);
    \draw[appendix loop oriented] (.5,0) arc[start angle=0,end angle=-360,radius=.5];
    \node[appendix label,left=2pt] at (-.5,0) {$#1$};
  \end{tikzpicture}%
}

%% file: figures/tikz-diagrams/f-move-diagrams.tex
\input{figures/tikz-diagrams/diagram-style.tex}
\usetikzlibrary{positioning}

\newcommand{\LeftAssociatedFusionTree}{%
  \begin{tikzpicture}[baseline=-0.52ex, x=0.58cm, y=0.54cm]
    \path[use as bounding box] (-1.48,-1.10) rectangle (1.48,2.42);
    \coordinate (LaTop) at (-1.20,2.12);
    \coordinate (LbTop) at (-0.16,2.12);
    \coordinate (LcTop) at (1.18,2.12);
    \coordinate (Lupper) at (-0.64,1.08);
    \coordinate (Llower) at (0,0);
    \coordinate (Lbottom) at (0,-0.82);
    \draw[appendix strand] (LaTop)
      .. controls (-1.20,1.70) and (-0.88,1.27) .. (Lupper);
    \draw[appendix strand] (LbTop)
      .. controls (-0.16,1.70) and (-0.39,1.27) .. (Lupper);
    \draw[appendix strand] (Lupper)
      .. controls (-0.58,0.75) and (-0.25,0.22)
      .. coordinate[pos=0.62] (LeMid) (Llower);
    \draw[appendix strand] (LcTop)
      .. controls (1.18,1.28) and (0.63,0.36) .. (Llower);
    \draw[appendix strand] (Llower) -- (Lbottom);
    \node[appendix vertex] (LupperDot) at (Lupper) {};
    \node[appendix vertex] (LlowerDot) at (Llower) {};
    \node[appendix label, above=0pt of LaTop] {$a$};
    \node[appendix label, above=0pt of LbTop] {$b$};
    \node[appendix label, above=0pt of LcTop] {$c$};
    \node[appendix basis, left=3.5pt of LupperDot] {$\beta$};
    \node[appendix channel, left=4.5pt of LeMid] {$e$};
    \node[appendix basis, right=3.5pt of LlowerDot] {$\alpha$};
    \node[appendix label, below=0pt of Lbottom] {$d$};
  \end{tikzpicture}%
}

\newcommand{\RightAssociatedFusionTree}{%
  \begin{tikzpicture}[baseline=-0.52ex, x=0.58cm, y=0.54cm]
    \path[use as bounding box] (-1.48,-1.10) rectangle (1.48,2.42);
    \coordinate (RaTop) at (-1.18,2.12);
    \coordinate (RbTop) at (0.16,2.12);
    \coordinate (RcTop) at (1.20,2.12);
    \coordinate (Rupper) at (0.64,1.08);
    \coordinate (Rlower) at (0,0);
    \coordinate (Rbottom) at (0,-0.82);
    \draw[appendix strand] (RbTop)
      .. controls (0.16,1.70) and (0.39,1.27) .. (Rupper);
    \draw[appendix strand] (RcTop)
      .. controls (1.20,1.70) and (0.88,1.27) .. (Rupper);
    \draw[appendix strand] (Rupper)
      .. controls (0.58,0.75) and (0.25,0.22)
      .. coordinate[pos=0.62] (RfMid) (Rlower);
    \draw[appendix strand] (RaTop)
      .. controls (-1.18,1.28) and (-0.63,0.36) .. (Rlower);
    \draw[appendix strand] (Rlower) -- (Rbottom);
    \node[appendix vertex] (RupperDot) at (Rupper) {};
    \node[appendix vertex] (RlowerDot) at (Rlower) {};
    \node[appendix label, above=0pt of RaTop] {$a$};
    \node[appendix label, above=0pt of RbTop] {$b$};
    \node[appendix label, above=0pt of RcTop] {$c$};
    \node[appendix basis, right=3.5pt of RupperDot] {$\nu$};
    \node[appendix channel, right=4.5pt of RfMid] {$f$};
    \node[appendix basis, left=3.5pt of RlowerDot] {$\mu$};
    \node[appendix label, below=0pt of Rbottom] {$d$};
  \end{tikzpicture}%
}

%% file: figures/tikz-diagrams/pentagon-diagrams.tex
\input{figures/tikz-diagrams/diagram-style.tex}
\usetikzlibrary{arrows.meta,calc,positioning}

\newcommand{\PentagonChannelAt}[4]{%
  \coordinate (#1Label) at ($(#1)!5pt!#3:(#2)$);
  \node[appendix channel] at (#1Label) {$#4$};
}

\newcommand{\PentagonExternalLabels}[5]{%
  \node[appendix label, above=0pt of PTopA] {$#1$};
  \node[appendix label, above=0pt of PTopB] {$#2$};
  \node[appendix label, above=0pt of PTopC] {$#3$};
  \node[appendix label, above=0pt of PTopD] {$#4$};
  \node[appendix label, below=0pt of PBottom] {$#5$};
}

\newcommand{\PentagonLeftComb}{%
  \begin{tikzpicture}[baseline=-0.52ex, x=0.42cm, y=0.40cm]
    \path[use as bounding box] (-1.62,-1.05) rectangle (1.62,2.50);
    \coordinate (PTopA) at (-1.38,2.18);
    \coordinate (PTopB) at (-0.54,2.18);
    \coordinate (PTopC) at (0.34,2.18);
    \coordinate (PTopD) at (1.38,2.18);
    \coordinate (PVertexOne) at (-0.96,1.34);
    \coordinate (PVertexTwo) at (-0.43,0.58);
    \coordinate (PVertexThree) at (0,-0.18);
    \coordinate (PBottom) at (0,-0.78);
    \draw[appendix strand] (PTopA)
      .. controls (-1.38,1.84) and (-1.15,1.52) .. (PVertexOne);
    \draw[appendix strand] (PTopB)
      .. controls (-0.54,1.84) and (-0.75,1.52) .. (PVertexOne);
    \draw[appendix strand] (PVertexOne)
      .. controls (-0.88,1.06) and (-0.58,0.76)
      .. coordinate[pos=0.50] (PEdgeE)
         coordinate[pos=0.60] (PEdgeEForward) (PVertexTwo);
    \draw[appendix strand] (PTopC)
      .. controls (0.34,1.53) and (-0.04,0.82) .. (PVertexTwo);
    \draw[appendix strand] (PVertexTwo)
      .. controls (-0.39,0.29) and (-0.16,0.02)
      .. coordinate[pos=0.50] (PEdgeF)
         coordinate[pos=0.60] (PEdgeFForward) (PVertexThree);
    \draw[appendix strand] (PTopD)
      .. controls (1.38,1.30) and (0.58,0.18) .. (PVertexThree);
    \draw[appendix strand] (PVertexThree) -- (PBottom);
    \node[appendix vertex] at (PVertexOne) {};
    \node[appendix vertex] at (PVertexTwo) {};
    \node[appendix vertex] at (PVertexThree) {};
    \PentagonChannelAt{PEdgeE}{PEdgeEForward}{-90}{e}
    \PentagonChannelAt{PEdgeF}{PEdgeFForward}{-90}{f}
    \PentagonExternalLabels{a}{b}{c}{d}{x}
  \end{tikzpicture}%
}

\newcommand{\PentagonPairTree}{%
  \begin{tikzpicture}[baseline=-0.52ex, x=0.42cm, y=0.40cm]
    \path[use as bounding box] (-1.62,-1.05) rectangle (1.62,2.50);
    \coordinate (PTopA) at (-1.38,2.18);
    \coordinate (PTopB) at (-0.46,2.18);
    \coordinate (PTopC) at (0.46,2.18);
    \coordinate (PTopD) at (1.38,2.18);
    \coordinate (PVertexOne) at (-0.90,1.18);
    \coordinate (PVertexTwo) at (0.90,1.18);
    \coordinate (PVertexThree) at (0,-0.18);
    \coordinate (PBottom) at (0,-0.78);
    \draw[appendix strand] (PTopA)
      .. controls (-1.38,1.76) and (-1.14,1.38) .. (PVertexOne);
    \draw[appendix strand] (PTopB)
      .. controls (-0.46,1.76) and (-0.66,1.38) .. (PVertexOne);
    \draw[appendix strand] (PTopC)
      .. controls (0.46,1.76) and (0.66,1.38) .. (PVertexTwo);
    \draw[appendix strand] (PTopD)
      .. controls (1.38,1.76) and (1.14,1.38) .. (PVertexTwo);
    \draw[appendix strand] (PVertexOne)
      .. controls (-0.77,0.72) and (-0.30,0.14)
      .. coordinate[pos=0.50] (PEdgeE)
         coordinate[pos=0.60] (PEdgeEForward) (PVertexThree);
    \draw[appendix strand] (PVertexTwo)
      .. controls (0.77,0.72) and (0.30,0.14)
      .. coordinate[pos=0.50] (PEdgeG)
         coordinate[pos=0.60] (PEdgeGForward) (PVertexThree);
    \draw[appendix strand] (PVertexThree) -- (PBottom);
    \node[appendix vertex] at (PVertexOne) {};
    \node[appendix vertex] at (PVertexTwo) {};
    \node[appendix vertex] at (PVertexThree) {};
    \PentagonChannelAt{PEdgeE}{PEdgeEForward}{-90}{e}
    \PentagonChannelAt{PEdgeG}{PEdgeGForward}{90}{g}
    \PentagonExternalLabels{a}{b}{c}{d}{x}
  \end{tikzpicture}%
}

\newcommand{\PentagonLowerLeftTree}{%
  \begin{tikzpicture}[baseline=-0.52ex, x=0.42cm, y=0.40cm]
    \path[use as bounding box] (-1.62,-1.05) rectangle (1.62,2.50);
    \coordinate (PTopA) at (-1.38,2.18);
    \coordinate (PTopB) at (-0.42,2.18);
    \coordinate (PTopC) at (0.46,2.18);
    \coordinate (PTopD) at (1.38,2.18);
    \coordinate (PVertexOne) at (0.02,1.34);
    \coordinate (PVertexTwo) at (-0.48,0.58);
    \coordinate (PVertexThree) at (0,-0.18);
    \coordinate (PBottom) at (0,-0.78);
    \draw[appendix strand] (PTopB)
      .. controls (-0.42,1.84) and (-0.20,1.52) .. (PVertexOne);
    \draw[appendix strand] (PTopC)
      .. controls (0.46,1.84) and (0.24,1.52) .. (PVertexOne);
    \draw[appendix strand] (PVertexOne)
      .. controls (-0.04,1.06) and (-0.34,0.76)
      .. coordinate[pos=0.50] (PEdgeK)
         coordinate[pos=0.60] (PEdgeKForward) (PVertexTwo);
    \draw[appendix strand] (PTopA)
      .. controls (-1.38,1.46) and (-0.86,0.78) .. (PVertexTwo);
    \draw[appendix strand] (PVertexTwo)
      .. controls (-0.43,0.29) and (-0.17,0.02)
      .. coordinate[pos=0.50] (PEdgeF)
         coordinate[pos=0.60] (PEdgeFForward) (PVertexThree);
    \draw[appendix strand] (PTopD)
      .. controls (1.38,1.30) and (0.58,0.18) .. (PVertexThree);
    \draw[appendix strand] (PVertexThree) -- (PBottom);
    \node[appendix vertex] at (PVertexOne) {};
    \node[appendix vertex] at (PVertexTwo) {};
    \node[appendix vertex] at (PVertexThree) {};
    \PentagonChannelAt{PEdgeK}{PEdgeKForward}{90}{k}
    \PentagonChannelAt{PEdgeF}{PEdgeFForward}{-90}{f}
    \PentagonExternalLabels{a}{b}{c}{d}{x}
  \end{tikzpicture}%
}

\newcommand{\PentagonLowerRightTree}{%
  \begin{tikzpicture}[baseline=-0.52ex, x=0.42cm, y=0.40cm]
    \path[use as bounding box] (-1.62,-1.05) rectangle (1.62,2.50);
    \coordinate (PTopA) at (-1.38,2.18);
    \coordinate (PTopB) at (-0.46,2.18);
    \coordinate (PTopC) at (0.42,2.18);
    \coordinate (PTopD) at (1.38,2.18);
    \coordinate (PVertexOne) at (-0.02,1.34);
    \coordinate (PVertexTwo) at (0.48,0.58);
    \coordinate (PVertexThree) at (0,-0.18);
    \coordinate (PBottom) at (0,-0.78);
    \draw[appendix strand] (PTopB)
      .. controls (-0.46,1.84) and (-0.24,1.52) .. (PVertexOne);
    \draw[appendix strand] (PTopC)
      .. controls (0.42,1.84) and (0.20,1.52) .. (PVertexOne);
    \draw[appendix strand] (PVertexOne)
      .. controls (0.04,1.06) and (0.34,0.76)
      .. coordinate[pos=0.50] (PEdgeK)
         coordinate[pos=0.60] (PEdgeKForward) (PVertexTwo);
    \draw[appendix strand] (PTopD)
      .. controls (1.38,1.46) and (0.86,0.78) .. (PVertexTwo);
    \draw[appendix strand] (PVertexTwo)
      .. controls (0.43,0.29) and (0.17,0.02)
      .. coordinate[pos=0.50] (PEdgeH)
         coordinate[pos=0.60] (PEdgeHForward) (PVertexThree);
    \draw[appendix strand] (PTopA)
      .. controls (-1.38,1.30) and (-0.58,0.18) .. (PVertexThree);
    \draw[appendix strand] (PVertexThree) -- (PBottom);
    \node[appendix vertex] at (PVertexOne) {};
    \node[appendix vertex] at (PVertexTwo) {};
    \node[appendix vertex] at (PVertexThree) {};
    \PentagonChannelAt{PEdgeK}{PEdgeKForward}{-90}{k}
    \PentagonChannelAt{PEdgeH}{PEdgeHForward}{90}{h}
    \PentagonExternalLabels{a}{b}{c}{d}{x}
  \end{tikzpicture}%
}

\newcommand{\PentagonRightComb}{%
  \begin{tikzpicture}[baseline=-0.52ex, x=0.42cm, y=0.40cm]
    \path[use as bounding box] (-1.62,-1.05) rectangle (1.62,2.50);
    \coordinate (PTopA) at (-1.38,2.18);
    \coordinate (PTopB) at (-0.34,2.18);
    \coordinate (PTopC) at (0.54,2.18);
    \coordinate (PTopD) at (1.38,2.18);
    \coordinate (PVertexOne) at (0.96,1.34);
    \coordinate (PVertexTwo) at (0.43,0.58);
    \coordinate (PVertexThree) at (0,-0.18);
    \coordinate (PBottom) at (0,-0.78);
    \draw[appendix strand] (PTopC)
      .. controls (0.54,1.84) and (0.75,1.52) .. (PVertexOne);
    \draw[appendix strand] (PTopD)
      .. controls (1.38,1.84) and (1.15,1.52) .. (PVertexOne);
    \draw[appendix strand] (PVertexOne)
      .. controls (0.88,1.06) and (0.58,0.76)
      .. coordinate[pos=0.50] (PEdgeG)
         coordinate[pos=0.60] (PEdgeGForward) (PVertexTwo);
    \draw[appendix strand] (PTopB)
      .. controls (-0.34,1.53) and (0.04,0.82) .. (PVertexTwo);
    \draw[appendix strand] (PVertexTwo)
      .. controls (0.39,0.29) and (0.16,0.02)
      .. coordinate[pos=0.50] (PEdgeH)
         coordinate[pos=0.60] (PEdgeHForward) (PVertexThree);
    \draw[appendix strand] (PTopA)
      .. controls (-1.38,1.30) and (-0.58,0.18) .. (PVertexThree);
    \draw[appendix strand] (PVertexThree) -- (PBottom);
    \node[appendix vertex] at (PVertexOne) {};
    \node[appendix vertex] at (PVertexTwo) {};
    \node[appendix vertex] at (PVertexThree) {};
    \PentagonChannelAt{PEdgeG}{PEdgeGForward}{90}{g}
    \PentagonChannelAt{PEdgeH}{PEdgeHForward}{90}{h}
    \PentagonExternalLabels{a}{b}{c}{d}{x}
  \end{tikzpicture}%
}

\newcommand{\PentagonEdge}[6]{%
  \coordinate (Pentagon#1#2Start) at
    ($(Pentagon#1.center)!1.05cm!(Pentagon#2.center)$);
  \coordinate (Pentagon#1#2End) at
    ($(Pentagon#2.center)!1.05cm!(Pentagon#1.center)$);
  \coordinate (Pentagon#1#2Mid) at
    ($(Pentagon#1.center)!0.50!(Pentagon#2.center)$);
  \coordinate (Pentagon#1#2LabelAnchor) at
    ($(Pentagon#1#2Mid)+(#3:0.18cm)$);
  \draw[appendix map] (Pentagon#1#2Start) -- (Pentagon#1#2End);
  \node[appendix map label, anchor=#4, yshift=#5]
    at (Pentagon#1#2LabelAnchor) {#6};
}

\newcommand{\PentagonCoherenceDiagram}{%
  \begin{tikzpicture}[
    appendix map/.style={
      draw=black,
      line width=0.70pt,
      -{Stealth[length=4.2pt,width=3.3pt]}
    },
    appendix map label/.style={
      font=\small,
      inner sep=0pt
    }
  ]
    \node[inner sep=0pt] (PentagonA) at (162:2.55cm)
      {\PentagonLeftComb};
    \node[inner sep=0pt] (PentagonB) at (90:2.55cm)
      {\PentagonPairTree};
    \node[inner sep=0pt] (PentagonC) at (18:2.55cm)
      {\PentagonRightComb};
    \node[inner sep=0pt] (PentagonD) at (-54:2.55cm)
      {\PentagonLowerRightTree};
    \node[inner sep=0pt] (PentagonE) at (-126:2.55cm)
      {\PentagonLowerLeftTree};

    \PentagonEdge{A}{B}{126}{south east}{0pt}{$F_x^{ecd}$}
    \PentagonEdge{B}{C}{54}{south west}{0pt}{$F_x^{abg}$}
    \PentagonEdge{A}{E}{-162}{north east}{2pt}{$F_f^{abc}$}
    \PentagonEdge{E}{D}{-90}{north}{0pt}{$F_x^{akd}$}
    \PentagonEdge{D}{C}{-18}{north west}{2pt}{$F_h^{bcd}$}
  \end{tikzpicture}%
}

%% file: figures/tikz-diagrams/r-move-diagrams.tex
\input{figures/tikz-diagrams/diagram-style.tex}
\usetikzlibrary{positioning}

\newcommand{\BraidedSplittingTree}{%
  \begin{tikzpicture}[baseline=-0.52ex, x=0.58cm, y=0.54cm]
    \path[use as bounding box] (-1.15,-1.10) rectangle (1.15,2.25);
    \coordinate (BleftTop) at (-0.60,1.98);
    \coordinate (BrightTop) at (0.60,1.98);
    \coordinate (Bvertex) at (0,-0.35);
    \coordinate (BleftStart) at (-0.60,0.58);
    \coordinate (BrightStart) at (0.60,0.58);
    \coordinate (Bbottom) at (0,-0.92);
    \draw[appendix strand] (Bvertex)
      .. controls (-0.16,-0.12) and (-0.60,0.30) .. (BleftStart)
      .. controls (-0.60,0.98) and (0.60,1.26) .. (0.60,1.66)
      -- (BrightTop);
    \draw[appendix strand] (Bvertex)
      .. controls (0.16,-0.12) and (0.60,0.30) .. (BrightStart)
      .. controls (0.60,0.98) and (-0.60,1.26) .. (-0.60,1.66)
      -- (BleftTop);
    \draw[appendix crossing] (-0.31000,0.93500)
      .. controls (-0.12200,1.06020) and (0.12200,1.17980)
      .. (0.31000,1.30500);
    \draw[appendix strand] (Bbottom) -- (Bvertex);
    \node[appendix vertex] (BvertexDot) at (Bvertex) {};
    \node[appendix label, above=0pt of BleftTop] {$b$};
    \node[appendix label, above=0pt of BrightTop] {$a$};
    \node[appendix basis, left=3.5pt of BvertexDot] {$\alpha$};
    \node[appendix label, below=0pt of Bbottom] {$c$};
  \end{tikzpicture}%
}

\newcommand{\ReorderedSplittingTree}{%
  \begin{tikzpicture}[baseline=-0.52ex, x=0.58cm, y=0.54cm]
    \path[use as bounding box] (-1.15,-1.10) rectangle (1.15,2.25);
    \coordinate (SleftTop) at (-0.72,1.92);
    \coordinate (SrightTop) at (0.72,1.92);
    \coordinate (Svertex) at (0,-0.35);
    \coordinate (Sbottom) at (0,-0.92);
    \draw[appendix strand] (Svertex)
      .. controls (-0.24,0.02) and (-0.72,1.10) .. (SleftTop);
    \draw[appendix strand] (Svertex)
      .. controls (0.24,0.02) and (0.72,1.10) .. (SrightTop);
    \draw[appendix strand] (Sbottom) -- (Svertex);
    \node[appendix vertex] (SvertexDot) at (Svertex) {};
    \node[appendix label, above=0pt of SleftTop] {$b$};
    \node[appendix label, above=0pt of SrightTop] {$a$};
    \node[appendix basis, right=3.5pt of SvertexDot] {$\beta$};
    \node[appendix label, below=0pt of Sbottom] {$c$};
  \end{tikzpicture}%
}

%% file: figures/tikz-diagrams/monodromy-diagrams.tex
\input{figures/tikz-diagrams/diagram-style.tex}
\usetikzlibrary{positioning}

\newcommand{\DoubleBraidedSplittingTree}{%
  \begin{tikzpicture}[baseline=-0.52ex, x=0.58cm, y=0.54cm]
    \path[use as bounding box] (-1.02,-1.10) rectangle (1.02,2.35);
    \coordinate (DleftTop) at (-0.56,2.02);
    \coordinate (DrightTop) at (0.56,2.02);
    \coordinate (Dvertex) at (0,-0.42);
    \coordinate (Dbottom) at (0,-0.96);
    \draw[appendix strand] (Dvertex)
      .. controls (-0.15,-0.24) and (-0.56,-0.02) .. (-0.56,0.25)
      .. controls (-0.56,0.57) and (0.56,0.78) .. (0.56,1.10)
      .. controls (0.56,1.42) and (-0.56,1.63) .. (-0.56,1.95)
      -- (DleftTop);
    \draw[appendix strand] (Dvertex)
      .. controls (0.15,-0.24) and (0.56,-0.02) .. (0.56,0.25)
      .. controls (0.56,0.57) and (-0.56,0.78) .. (-0.56,1.10)
      .. controls (-0.56,1.42) and (0.56,1.63) .. (0.56,1.95)
      -- (DrightTop);
    \draw[appendix crossing] (-0.28934,0.53062)
      .. controls (-0.11386,0.62858) and (0.11386,0.72142)
      .. (0.28934,0.81938);
    \draw[appendix crossing] (-0.28934,1.38062)
      .. controls (-0.11386,1.47858) and (0.11386,1.57142)
      .. (0.28934,1.66938);
    \draw[appendix strand] (Dbottom) -- (Dvertex);
    \node[appendix vertex] (DvertexDot) at (Dvertex) {};
    \node[appendix label, above=0pt of DleftTop] {$a$};
    \node[appendix label, above=0pt of DrightTop] {$b$};
    \node[appendix basis, left=3.5pt of DvertexDot] {$\alpha$};
    \node[appendix label, below=0pt of Dbottom] {$c$};
  \end{tikzpicture}%
}

\newcommand{\StraightSplittingTree}{%
  \begin{tikzpicture}[baseline=-0.52ex, x=0.58cm, y=0.54cm]
    \path[use as bounding box] (-1.02,-1.10) rectangle (1.02,2.35);
    \coordinate (TleftTop) at (-0.72,2.02);
    \coordinate (TrightTop) at (0.72,2.02);
    \coordinate (Tvertex) at (0,-0.42);
    \coordinate (Tbottom) at (0,-0.96);
    \draw[appendix strand] (Tvertex)
      .. controls (-0.24,-0.02) and (-0.72,1.16) .. (TleftTop);
    \draw[appendix strand] (Tvertex)
      .. controls (0.24,-0.02) and (0.72,1.16) .. (TrightTop);
    \draw[appendix strand] (Tbottom) -- (Tvertex);
    \node[appendix vertex] (TvertexDot) at (Tvertex) {};
    \node[appendix label, above=0pt of TleftTop] {$a$};
    \node[appendix label, above=0pt of TrightTop] {$b$};
    \node[appendix basis, right=3.5pt of TvertexDot] {$\gamma$};
    \node[appendix label, below=0pt of Tbottom] {$c$};
  \end{tikzpicture}%
}

%% file: figures/tikz-diagrams/hexagon-diagrams.tex
\input{figures/tikz-diagrams/diagram-style.tex}
\usetikzlibrary{arrows.meta,calc,positioning}

\newcommand{\HexagonChannelAt}[4]{%
  \coordinate (#1Label) at ($(#1)!5pt!#3:(#2)$);
  \node[appendix channel] at (#1Label) {$#4$};
}

\newcommand{\HexagonLeftTree}[5]{%
  \begin{tikzpicture}[baseline=-0.52ex, x=0.48cm, y=0.45cm]
    \path[use as bounding box] (-1.48,-1.10) rectangle (1.48,2.42);
    \coordinate (HLaTop) at (-1.20,2.12);
    \coordinate (HLbTop) at (-0.16,2.12);
    \coordinate (HLcTop) at (1.18,2.12);
    \coordinate (HLupper) at (-0.64,1.08);
    \coordinate (HLlower) at (0,0);
    \coordinate (HLbottom) at (0,-0.82);
    \draw[appendix strand] (HLaTop)
      .. controls (-1.20,1.70) and (-0.88,1.27) .. (HLupper);
    \draw[appendix strand] (HLbTop)
      .. controls (-0.16,1.70) and (-0.39,1.27) .. (HLupper);
    \draw[appendix strand] (HLupper)
      .. controls (-0.58,0.75) and (-0.25,0.22)
      .. coordinate[pos=0.50] (HLEdge)
         coordinate[pos=0.60] (HLEdgeForward) (HLlower);
    \draw[appendix strand] (HLcTop)
      .. controls (1.18,1.28) and (0.63,0.36) .. (HLlower);
    \draw[appendix strand] (HLlower) -- (HLbottom);
    \node[appendix vertex] at (HLupper) {};
    \node[appendix vertex] at (HLlower) {};
    \node[appendix label, above=0pt of HLaTop] {$#1$};
    \node[appendix label, above=0pt of HLbTop] {$#2$};
    \node[appendix label, above=0pt of HLcTop] {$#3$};
    \HexagonChannelAt{HLEdge}{HLEdgeForward}{-90}{#4}
    \node[appendix label, below=0pt of HLbottom] {$#5$};
  \end{tikzpicture}%
}

\newcommand{\HexagonRightTree}[5]{%
  \begin{tikzpicture}[baseline=-0.52ex, x=0.48cm, y=0.45cm]
    \path[use as bounding box] (-1.48,-1.10) rectangle (1.48,2.42);
    \coordinate (HRaTop) at (-1.18,2.12);
    \coordinate (HRbTop) at (0.16,2.12);
    \coordinate (HRcTop) at (1.20,2.12);
    \coordinate (HRupper) at (0.64,1.08);
    \coordinate (HRlower) at (0,0);
    \coordinate (HRbottom) at (0,-0.82);
    \draw[appendix strand] (HRbTop)
      .. controls (0.16,1.70) and (0.39,1.27) .. (HRupper);
    \draw[appendix strand] (HRcTop)
      .. controls (1.20,1.70) and (0.88,1.27) .. (HRupper);
    \draw[appendix strand] (HRupper)
      .. controls (0.58,0.75) and (0.25,0.22)
      .. coordinate[pos=0.50] (HREdge)
         coordinate[pos=0.60] (HREdgeForward) (HRlower);
    \draw[appendix strand] (HRaTop)
      .. controls (-1.18,1.28) and (-0.63,0.36) .. (HRlower);
    \draw[appendix strand] (HRlower) -- (HRbottom);
    \node[appendix vertex] at (HRupper) {};
    \node[appendix vertex] at (HRlower) {};
    \node[appendix label, above=0pt of HRaTop] {$#1$};
    \node[appendix label, above=0pt of HRbTop] {$#2$};
    \node[appendix label, above=0pt of HRcTop] {$#3$};
    \HexagonChannelAt{HREdge}{HREdgeForward}{90}{#4}
    \node[appendix label, below=0pt of HRbottom] {$#5$};
  \end{tikzpicture}%
}

\newcommand{\HexagonSourceBraidedTree}{%
  \begin{tikzpicture}[baseline=-0.52ex, x=0.48cm, y=0.45cm]
    \path[use as bounding box] (-1.48,-1.10) rectangle (1.48,2.42);
    \coordinate (HSbTop) at (-1.20,2.12);
    \coordinate (HScTop) at (0,2.12);
    \coordinate (HSaTop) at (1.20,2.12);
    \coordinate (HSupper) at (-0.64,1.02);
    \coordinate (HSlower) at (0,0);
    \coordinate (HSbottom) at (0,-0.82);
    \coordinate (HSbTurn) at (-0.28,1.32);
    \coordinate (HSaTurnOne) at (-0.94,1.34);
    \coordinate (HSaTurnTwo) at (-0.28,1.60);
    \coordinate (HScTurn) at (0.78,1.45);
    \draw[appendix strand] (HSupper)
      .. controls (-0.48,1.18) and (-0.30,1.20) .. (HSbTurn)
      .. controls (-0.40,1.60) and (-0.95,1.80) .. (HSbTop);
    \draw[appendix strand] (HSlower)
      .. controls (0.45,0.42) and (0.85,1.05) .. (HScTurn)
      .. controls (0.70,1.74) and (0.25,1.92) .. (HScTop);
    \draw[appendix strand] (HSupper)
      .. controls (-0.82,1.12) and (-0.98,1.20) .. (HSaTurnOne)
      .. controls (-0.82,1.56) and (-0.48,1.56) .. (HSaTurnTwo)
      .. controls (0.15,1.66) and (0.78,1.84) .. (HSaTop);
    \draw[appendix crossing] (HSaTurnOne)
      .. controls (-0.82,1.56) and (-0.48,1.56) .. (HSaTurnTwo);
    \draw[appendix crossing] (HSaTurnTwo)
      .. controls (0.15,1.66) and (0.78,1.84) .. (HSaTop);
    \draw[appendix strand] (HSupper)
      .. controls (-0.57,0.70) and (-0.25,0.22)
      .. coordinate[pos=0.50] (HSEdge)
         coordinate[pos=0.60] (HSEdgeForward) (HSlower);
    \draw[appendix strand] (HSbottom) -- (HSlower);
    \node[appendix vertex] at (HSupper) {};
    \node[appendix vertex] at (HSlower) {};
    \node[appendix label, above=0pt of HSbTop] {$b$};
    \node[appendix label, above=0pt of HScTop] {$c$};
    \node[appendix label, above=0pt of HSaTop] {$a$};
    \HexagonChannelAt{HSEdge}{HSEdgeForward}{-90}{e}
    \node[appendix label, below=0pt of HSbottom] {$d$};
  \end{tikzpicture}%
}

\newcommand{\HexagonTopLeftBraidedTree}{%
  \begin{tikzpicture}[baseline=-0.52ex, x=0.48cm, y=0.45cm]
    \path[use as bounding box] (-1.48,-1.10) rectangle (1.48,2.42);
    \coordinate (HTLbTop) at (-1.20,2.12);
    \coordinate (HTLcTop) at (0,2.12);
    \coordinate (HTLaTop) at (1.20,2.12);
    \coordinate (HTLupper) at (-0.64,1.08);
    \coordinate (HTLlower) at (0,0);
    \coordinate (HTLbottom) at (0,-0.82);
    \coordinate (HTLaTurn) at (0.05,1.52);
    \coordinate (HTLcTurn) at (0.78,1.42);
    \draw[appendix strand] (HTLupper)
      .. controls (-0.86,1.34) and (-1.16,1.74) .. (HTLbTop);
    \draw[appendix strand] (HTLlower)
      .. controls (0.45,0.45) and (0.85,1.00) .. (HTLcTurn)
      .. controls (0.72,1.72) and (0.22,1.88) .. (HTLcTop);
    \draw[appendix strand] (HTLupper)
      .. controls (-0.42,1.28) and (-0.08,1.38) .. (HTLaTurn)
      .. controls (0.30,1.78) and (0.90,1.88) .. (HTLaTop);
    \draw[appendix crossing] (HTLaTurn)
      .. controls (0.30,1.78) and (0.90,1.88) .. (HTLaTop);
    \draw[appendix strand] (HTLupper)
      .. controls (-0.58,0.75) and (-0.25,0.22)
      .. coordinate[pos=0.50] (HTLEdge)
         coordinate[pos=0.60] (HTLEdgeForward) (HTLlower);
    \draw[appendix strand] (HTLbottom) -- (HTLlower);
    \node[appendix vertex] at (HTLupper) {};
    \node[appendix vertex] at (HTLlower) {};
    \node[appendix label, above=0pt of HTLbTop] {$b$};
    \node[appendix label, above=0pt of HTLcTop] {$c$};
    \node[appendix label, above=0pt of HTLaTop] {$a$};
    \HexagonChannelAt{HTLEdge}{HTLEdgeForward}{-90}{e}
    \node[appendix label, below=0pt of HTLbottom] {$d$};
  \end{tikzpicture}%
}

\newcommand{\HexagonTopRightBraidedTree}{%
  \begin{tikzpicture}[baseline=-0.52ex, x=0.48cm, y=0.45cm]
    \path[use as bounding box] (-1.48,-1.10) rectangle (1.48,2.42);
    \coordinate (HTRbTop) at (-1.20,2.12);
    \coordinate (HTRcTop) at (0,2.12);
    \coordinate (HTRaTop) at (1.20,2.12);
    \coordinate (HTRupper) at (0.64,1.08);
    \coordinate (HTRlower) at (0,0);
    \coordinate (HTRbottom) at (0,-0.82);
    \coordinate (HTRaTurn) at (0.31,1.60);
    \coordinate (HTRcTurn) at (0.95,1.60);
    \draw[appendix strand] (HTRlower)
      .. controls (-0.24,0.36) and (-0.77,1.28) .. (HTRbTop);
    \draw[appendix strand] (HTRupper)
      .. controls (0.88,1.28) and (1.02,1.45) .. (HTRcTurn)
      .. controls (0.84,1.85) and (0.22,1.90) .. (HTRcTop);
    \draw[appendix strand] (HTRupper)
      .. controls (0.40,1.28) and (0.24,1.45) .. (HTRaTurn)
      .. controls (0.45,1.85) and (1.00,1.90) .. (HTRaTop);
    \draw[appendix crossing] (HTRaTurn)
      .. controls (0.45,1.85) and (1.00,1.90) .. (HTRaTop);
    \draw[appendix strand] (HTRupper)
      .. controls (0.58,0.75) and (0.25,0.22)
      .. coordinate[pos=0.50] (HTREdge)
         coordinate[pos=0.60] (HTREdgeForward) (HTRlower);
    \draw[appendix strand] (HTRbottom) -- (HTRlower);
    \node[appendix vertex] at (HTRupper) {};
    \node[appendix vertex] at (HTRlower) {};
    \node[appendix label, above=0pt of HTRbTop] {$b$};
    \node[appendix label, above=0pt of HTRcTop] {$c$};
    \node[appendix label, above=0pt of HTRaTop] {$a$};
    \HexagonChannelAt{HTREdge}{HTREdgeForward}{90}{g}
    \node[appendix label, below=0pt of HTRbottom] {$d$};
  \end{tikzpicture}%
}

\newcommand{\HexagonBottomLeftBraidedTree}{%
  \begin{tikzpicture}[baseline=-0.52ex, x=0.48cm, y=0.45cm]
    \path[use as bounding box] (-1.48,-1.10) rectangle (1.48,2.42);
    \coordinate (HBLbTop) at (-1.20,2.12);
    \coordinate (HBLcTop) at (0,2.12);
    \coordinate (HBLaTop) at (1.20,2.12);
    \coordinate (HBLupper) at (-0.60,1.18);
    \coordinate (HBLlower) at (0,0);
    \coordinate (HBLbottom) at (0,-0.82);
    \coordinate (HBLfAbove) at (-0.42,0.95);
    \coordinate (HBLfBend) at (0.38,0.48);
    \coordinate (HBLaBend) at (-0.38,0.48);
    \coordinate (HBLaAbove) at (0.45,0.95);
    \draw[appendix strand] (HBLbTop)
      .. controls (-1.20,1.72) and (-0.88,1.36) .. (HBLupper);
    \draw[appendix strand] (HBLcTop)
      .. controls (0,1.72) and (-0.30,1.36) .. (HBLupper);
    \draw[appendix strand] (HBLupper)
      .. controls (-0.54,1.08) and (-0.48,1.00)
      .. (HBLfAbove)
      .. controls (-0.20,0.82) and (0.38,0.72) .. (HBLfBend)
      .. controls (0.38,0.28) and (0.12,0.15) .. (HBLlower);
    \draw[appendix strand] (HBLlower)
      .. controls (-0.12,0.15) and (-0.38,0.28) .. (HBLaBend)
      .. controls (-0.38,0.72) and (0.20,0.82) .. (HBLaAbove)
      .. controls (0.66,1.08) and (1.02,1.72) .. (HBLaTop);
    \draw[appendix crossing] (HBLaBend)
      .. controls (-0.38,0.72) and (0.20,0.82) .. (HBLaAbove);
    \draw[appendix strand] (HBLbottom) -- (HBLlower);
    \node[appendix vertex] at (HBLupper) {};
    \node[appendix vertex] at (HBLlower) {};
    \node[appendix label, above=0pt of HBLbTop] {$b$};
    \node[appendix label, above=0pt of HBLcTop] {$c$};
    \node[appendix label, above=0pt of HBLaTop] {$a$};
    \coordinate (HBLfLabel) at ($(HBLupper)+(0:9pt)$);
    \node[appendix channel] at (HBLfLabel) {$f$};
    \node[appendix label, below=0pt of HBLbottom] {$d$};
  \end{tikzpicture}%
}

\newcommand{\HexagonRadius}{3.20cm}
\newcommand{\HexagonLabelClearance}{0.22cm}

\newcommand{\HexagonEdge}[8]{%
  \coordinate (Hexagon#1#2Start) at
    ($(Hexagon#1.center)!#3!(Hexagon#2.center)$);
  \coordinate (Hexagon#1#2End) at
    ($(Hexagon#2.center)!#4!(Hexagon#1.center)$);
  \coordinate (Hexagon#1#2Mid) at
    ($(Hexagon#1.center)!0.50!(Hexagon#2.center)$);
  \coordinate (Hexagon#1#2LabelAnchor) at
    ($(Hexagon#1#2Mid)+(#5:\HexagonLabelClearance)$);
  \draw[appendix map] (Hexagon#1#2Start) -- (Hexagon#1#2End);
  \node[appendix map label, anchor=#6, yshift=#7]
    at (Hexagon#1#2LabelAnchor) {#8};
}

\newcommand{\HexagonCoherenceDiagram}{%
  \begin{tikzpicture}[
    appendix map/.style={
      draw=black,
      line width=0.70pt,
      -{Stealth[length=4.2pt,width=3.3pt]}
    },
    appendix map label/.style={
      font=\small,
      inner sep=0pt
    }
  ]
    \node[inner sep=0pt] (HexagonA) at (180:\HexagonRadius)
      {\HexagonSourceBraidedTree};
    \node[inner sep=0pt] (HexagonB) at (120:\HexagonRadius)
      {\HexagonTopLeftBraidedTree};
    \node[inner sep=0pt] (HexagonC) at (60:\HexagonRadius)
      {\HexagonTopRightBraidedTree};
    \node[inner sep=0pt] (HexagonD) at (0:\HexagonRadius)
      {\HexagonRightTree{b}{c}{a}{g}{d}};
    \node[inner sep=0pt] (HexagonE) at (-60:\HexagonRadius)
      {\HexagonLeftTree{b}{c}{a}{f}{d}};
    \node[inner sep=0pt] (HexagonF) at (-120:\HexagonRadius)
      {\HexagonBottomLeftBraidedTree};

    \HexagonEdge{A}{B}{1.18cm}{1.05cm}{150}{south east}{0pt}{$R_e^{ab}$}
    \HexagonEdge{B}{C}{1.115cm}{1.115cm}{90}{south}{0pt}{$F_d^{bac}$}
    \HexagonEdge{C}{D}{1.05cm}{1.18cm}{30}{south west}{0pt}{$R_g^{ac}$}
    \HexagonEdge{A}{F}{1.05cm}{1.18cm}{-150}{north east}{0pt}{$F_d^{abc}$}
    \HexagonEdge{F}{E}{1.115cm}{1.115cm}{-90}{north}{0pt}{$R_d^{af}$}
    \HexagonEdge{E}{D}{1.18cm}{1.05cm}{-30}{north west}{0pt}{$F_d^{bca}$}
  \end{tikzpicture}%
}

%% file: figures/tikz-diagrams/ribbon-twist-diagrams.tex
\input{figures/tikz-diagrams/diagram-style.tex}
\input{figures/tikz-diagrams/ribbon-geometry.tex}
\newcommand{\StandardTwistSelfLoop}[1]{%
  \begin{tikzpicture}[baseline=-.52ex,x=.48cm,y=.48cm]
    \path[use as bounding box] (-2.15,-2.4) rectangle (.4,2.4);
    \DrawRibbonCenter{appendix strand}
    \DrawRibbonCenterOverpass{appendix crossing}
    \node[appendix label,above=2pt] at (0,2) {$#1$};
    \node[appendix label,below=2pt] at (0,-2) {$#1$};
  \end{tikzpicture}%
}
\newcommand{\StandardStraightLine}[1]{%
  \begin{tikzpicture}[baseline=-.52ex,x=.48cm,y=.48cm]
    \path[use as bounding box] (-.43,-2.4) rectangle (.43,2.4);
    \draw[appendix strand] (0,-2) -- (0,2);
    \node[appendix label,above=2pt] at (0,2) {$#1$};
    \node[appendix label,below=2pt] at (0,-2) {$#1$};
  \end{tikzpicture}%
}
\newcommand{\StandardTwistEquation}{%
  \vcenter{\hbox{\StandardTwistSelfLoop{a}}}
  =\theta_a\,\vcenter{\hbox{\StandardStraightLine{a}}}%
}

%% file: figures/tikz-diagrams/ribbon-geometry.tex
\newcommand{\DrawRibbonCenter}[1]{%
    \draw[#1] (0.000000,-2.000000)
      .. controls (0.000000,-1.750000) and (-0.093000,-1.512210) .. (-0.210000,-1.255800)
      .. controls (-0.231667,-1.208317) and (-0.253125,-1.161971) .. (-0.274375,-1.116741)
      .. controls (-0.295625,-1.071510) and (-0.316667,-1.027396) .. (-0.337500,-0.984375)
      .. controls (-0.358333,-0.941354) and (-0.378958,-0.899427) .. (-0.399375,-0.858572)
      .. controls (-0.419792,-0.817717) and (-0.440000,-0.777933) .. (-0.460000,-0.739200)
      .. controls (-0.480000,-0.700467) and (-0.499792,-0.662783) .. (-0.519375,-0.626128)
      .. controls (-0.538958,-0.589473) and (-0.558333,-0.553846) .. (-0.577500,-0.519225)
      .. controls (-0.596667,-0.484604) and (-0.615625,-0.450990) .. (-0.634375,-0.418359)
      .. controls (-0.653125,-0.385729) and (-0.671667,-0.354083) .. (-0.690000,-0.323400)
      .. controls (-0.708333,-0.292717) and (-0.726458,-0.262996) .. (-0.744375,-0.234216)
      .. controls (-0.762292,-0.205435) and (-0.780000,-0.177596) .. (-0.797500,-0.150675)
      .. controls (-0.815000,-0.123754) and (-0.832292,-0.097752) .. (-0.849375,-0.072647)
      .. controls (-0.866458,-0.047542) and (-0.883333,-0.023333) .. (-0.900000,0.000000)
      .. controls (-0.916667,0.023333) and (-0.933125,0.045792) .. (-0.949375,0.067397)
      .. controls (-0.965625,0.089002) and (-0.981667,0.109754) .. (-0.997500,0.129675)
      .. controls (-1.013333,0.149596) and (-1.028958,0.168685) .. (-1.044375,0.186966)
      .. controls (-1.059792,0.205246) and (-1.075000,0.222717) .. (-1.090000,0.239400)
      .. controls (-1.105000,0.256083) and (-1.119792,0.271979) .. (-1.134375,0.287109)
      .. controls (-1.148958,0.302240) and (-1.163333,0.316604) .. (-1.177500,0.330225)
      .. controls (-1.191667,0.343846) and (-1.205625,0.356723) .. (-1.219375,0.368878)
      .. controls (-1.233125,0.381033) and (-1.246667,0.392467) .. (-1.260000,0.403200)
      .. controls (-1.273333,0.413933) and (-1.286458,0.423967) .. (-1.299375,0.433322)
      .. controls (-1.312292,0.442677) and (-1.325000,0.451354) .. (-1.337500,0.459375)
      .. controls (-1.350000,0.467396) and (-1.362292,0.474760) .. (-1.374375,0.481491)
      .. controls (-1.386458,0.488221) and (-1.398333,0.494317) .. (-1.410000,0.499800)
      .. controls (-1.421667,0.505283) and (-1.433125,0.510154) .. (-1.444375,0.514434)
      .. controls (-1.455625,0.518715) and (-1.466667,0.522404) .. (-1.477500,0.525525)
      .. controls (-1.488333,0.528646) and (-1.498958,0.531198) .. (-1.509375,0.533203)
      .. controls (-1.519792,0.535208) and (-1.530000,0.536667) .. (-1.540000,0.537600)
      .. controls (-1.550000,0.538533) and (-1.559792,0.538942) .. (-1.569375,0.538847)
      .. controls (-1.578958,0.538752) and (-1.588333,0.538154) .. (-1.597500,0.537075)
      .. controls (-1.606667,0.535996) and (-1.615625,0.534435) .. (-1.624375,0.532416)
      .. controls (-1.633125,0.530396) and (-1.641667,0.527917) .. (-1.650000,0.525000)
      .. controls (-1.658333,0.522083) and (-1.666458,0.518729) .. (-1.674375,0.514959)
      .. controls (-1.682292,0.511190) and (-1.690000,0.507004) .. (-1.697500,0.502425)
      .. controls (-1.705000,0.497846) and (-1.712292,0.492873) .. (-1.719375,0.487528)
      .. controls (-1.726458,0.482183) and (-1.733333,0.476467) .. (-1.740000,0.470400)
      .. controls (-1.746667,0.464333) and (-1.753125,0.457917) .. (-1.759375,0.451172)
      .. controls (-1.765625,0.444427) and (-1.771667,0.437354) .. (-1.777500,0.429975)
      .. controls (-1.783333,0.422596) and (-1.788958,0.414910) .. (-1.794375,0.406941)
      .. controls (-1.799792,0.398971) and (-1.805000,0.390717) .. (-1.810000,0.382200)
      .. controls (-1.815000,0.373683) and (-1.819792,0.364904) .. (-1.824375,0.355884)
      .. controls (-1.828958,0.346865) and (-1.833333,0.337604) .. (-1.837500,0.328125)
      .. controls (-1.841667,0.318646) and (-1.845625,0.308948) .. (-1.849375,0.299053)
      .. controls (-1.853125,0.289158) and (-1.856667,0.279067) .. (-1.860000,0.268800)
      .. controls (-1.863333,0.258533) and (-1.866458,0.248092) .. (-1.869375,0.237497)
      .. controls (-1.872292,0.226902) and (-1.875000,0.216154) .. (-1.877500,0.205275)
      .. controls (-1.880000,0.194396) and (-1.882292,0.183385) .. (-1.884375,0.172266)
      .. controls (-1.886458,0.161146) and (-1.888333,0.149917) .. (-1.890000,0.138600)
      .. controls (-1.891667,0.127283) and (-1.893125,0.115879) .. (-1.894375,0.104409)
      .. controls (-1.895625,0.092940) and (-1.896667,0.081404) .. (-1.897500,0.069825)
      .. controls (-1.898333,0.058246) and (-1.898958,0.046623) .. (-1.899375,0.034978)
      .. controls (-1.899792,0.023333) and (-1.900000,0.011667) .. (-1.900000,0.000000)
      .. controls (-1.900000,-0.011667) and (-1.899792,-0.023333) .. (-1.899375,-0.034978)
      .. controls (-1.898958,-0.046623) and (-1.898333,-0.058246) .. (-1.897500,-0.069825)
      .. controls (-1.896667,-0.081404) and (-1.895625,-0.092940) .. (-1.894375,-0.104409)
      .. controls (-1.893125,-0.115879) and (-1.891667,-0.127283) .. (-1.890000,-0.138600)
      .. controls (-1.888333,-0.149917) and (-1.886458,-0.161146) .. (-1.884375,-0.172266)
      .. controls (-1.882292,-0.183385) and (-1.880000,-0.194396) .. (-1.877500,-0.205275)
      .. controls (-1.875000,-0.216154) and (-1.872292,-0.226902) .. (-1.869375,-0.237497)
      .. controls (-1.866458,-0.248092) and (-1.863333,-0.258533) .. (-1.860000,-0.268800)
      .. controls (-1.856667,-0.279067) and (-1.853125,-0.289158) .. (-1.849375,-0.299053)
      .. controls (-1.845625,-0.308948) and (-1.841667,-0.318646) .. (-1.837500,-0.328125)
      .. controls (-1.833333,-0.337604) and (-1.828958,-0.346865) .. (-1.824375,-0.355884)
      .. controls (-1.819792,-0.364904) and (-1.815000,-0.373683) .. (-1.810000,-0.382200)
      .. controls (-1.805000,-0.390717) and (-1.799792,-0.398971) .. (-1.794375,-0.406941)
      .. controls (-1.788958,-0.414910) and (-1.783333,-0.422596) .. (-1.777500,-0.429975)
      .. controls (-1.771667,-0.437354) and (-1.765625,-0.444427) .. (-1.759375,-0.451172)
      .. controls (-1.753125,-0.457917) and (-1.746667,-0.464333) .. (-1.740000,-0.470400)
      .. controls (-1.733333,-0.476467) and (-1.726458,-0.482183) .. (-1.719375,-0.487528)
      .. controls (-1.712292,-0.492873) and (-1.705000,-0.497846) .. (-1.697500,-0.502425)
      .. controls (-1.690000,-0.507004) and (-1.682292,-0.511190) .. (-1.674375,-0.514959)
      .. controls (-1.666458,-0.518729) and (-1.658333,-0.522083) .. (-1.650000,-0.525000)
      .. controls (-1.641667,-0.527917) and (-1.633125,-0.530396) .. (-1.624375,-0.532416)
      .. controls (-1.615625,-0.534435) and (-1.606667,-0.535996) .. (-1.597500,-0.537075)
      .. controls (-1.588333,-0.538154) and (-1.578958,-0.538752) .. (-1.569375,-0.538847)
      .. controls (-1.559792,-0.538942) and (-1.550000,-0.538533) .. (-1.540000,-0.537600)
      .. controls (-1.530000,-0.536667) and (-1.519792,-0.535208) .. (-1.509375,-0.533203)
      .. controls (-1.498958,-0.531198) and (-1.488333,-0.528646) .. (-1.477500,-0.525525)
      .. controls (-1.466667,-0.522404) and (-1.455625,-0.518715) .. (-1.444375,-0.514434)
      .. controls (-1.433125,-0.510154) and (-1.421667,-0.505283) .. (-1.410000,-0.499800)
      .. controls (-1.398333,-0.494317) and (-1.386458,-0.488221) .. (-1.374375,-0.481491)
      .. controls (-1.362292,-0.474760) and (-1.350000,-0.467396) .. (-1.337500,-0.459375)
      .. controls (-1.325000,-0.451354) and (-1.312292,-0.442677) .. (-1.299375,-0.433322)
      .. controls (-1.286458,-0.423967) and (-1.273333,-0.413933) .. (-1.260000,-0.403200)
      .. controls (-1.246667,-0.392467) and (-1.233125,-0.381033) .. (-1.219375,-0.368878)
      .. controls (-1.205625,-0.356723) and (-1.191667,-0.343846) .. (-1.177500,-0.330225)
      .. controls (-1.163333,-0.316604) and (-1.148958,-0.302240) .. (-1.134375,-0.287109)
      .. controls (-1.119792,-0.271979) and (-1.105000,-0.256083) .. (-1.090000,-0.239400)
      .. controls (-1.075000,-0.222717) and (-1.059792,-0.205246) .. (-1.044375,-0.186966)
      .. controls (-1.028958,-0.168685) and (-1.013333,-0.149596) .. (-0.997500,-0.129675)
      .. controls (-0.981667,-0.109754) and (-0.965625,-0.089002) .. (-0.949375,-0.067397)
      .. controls (-0.933125,-0.045792) and (-0.916667,-0.023333) .. (-0.900000,0.000000)
      .. controls (-0.883333,0.023333) and (-0.866458,0.047542) .. (-0.849375,0.072647)
      .. controls (-0.832292,0.097752) and (-0.815000,0.123754) .. (-0.797500,0.150675)
      .. controls (-0.780000,0.177596) and (-0.762292,0.205435) .. (-0.744375,0.234216)
      .. controls (-0.726458,0.262996) and (-0.708333,0.292717) .. (-0.690000,0.323400)
      .. controls (-0.671667,0.354083) and (-0.653125,0.385729) .. (-0.634375,0.418359)
      .. controls (-0.615625,0.450990) and (-0.596667,0.484604) .. (-0.577500,0.519225)
      .. controls (-0.558333,0.553846) and (-0.538958,0.589473) .. (-0.519375,0.626128)
      .. controls (-0.499792,0.662783) and (-0.480000,0.700467) .. (-0.460000,0.739200)
      .. controls (-0.440000,0.777933) and (-0.419792,0.817717) .. (-0.399375,0.858572)
      .. controls (-0.378958,0.899427) and (-0.358333,0.941354) .. (-0.337500,0.984375)
      .. controls (-0.316667,1.027396) and (-0.295625,1.071510) .. (-0.274375,1.116741)
      .. controls (-0.253125,1.161971) and (-0.231667,1.208317) .. (-0.210000,1.255800)
      .. controls (-0.093000,1.512210) and (0.000000,1.750000) .. (0.000000,2.000000);
}
\newcommand{\DrawRibbonCenterOverpass}[1]{%
    \draw[#1] (-1.777500,-0.429975)
      .. controls (-1.771667,-0.437354) and (-1.765625,-0.444427) .. (-1.759375,-0.451172)
      .. controls (-1.753125,-0.457917) and (-1.746667,-0.464333) .. (-1.740000,-0.470400)
      .. controls (-1.733333,-0.476467) and (-1.726458,-0.482183) .. (-1.719375,-0.487528)
      .. controls (-1.712292,-0.492873) and (-1.705000,-0.497846) .. (-1.697500,-0.502425)
      .. controls (-1.690000,-0.507004) and (-1.682292,-0.511190) .. (-1.674375,-0.514959)
      .. controls (-1.666458,-0.518729) and (-1.658333,-0.522083) .. (-1.650000,-0.525000)
      .. controls (-1.641667,-0.527917) and (-1.633125,-0.530396) .. (-1.624375,-0.532416)
      .. controls (-1.615625,-0.534435) and (-1.606667,-0.535996) .. (-1.597500,-0.537075)
      .. controls (-1.588333,-0.538154) and (-1.578958,-0.538752) .. (-1.569375,-0.538847)
      .. controls (-1.559792,-0.538942) and (-1.550000,-0.538533) .. (-1.540000,-0.537600)
      .. controls (-1.530000,-0.536667) and (-1.519792,-0.535208) .. (-1.509375,-0.533203)
      .. controls (-1.498958,-0.531198) and (-1.488333,-0.528646) .. (-1.477500,-0.525525)
      .. controls (-1.466667,-0.522404) and (-1.455625,-0.518715) .. (-1.444375,-0.514434)
      .. controls (-1.433125,-0.510154) and (-1.421667,-0.505283) .. (-1.410000,-0.499800)
      .. controls (-1.398333,-0.494317) and (-1.386458,-0.488221) .. (-1.374375,-0.481491)
      .. controls (-1.362292,-0.474760) and (-1.350000,-0.467396) .. (-1.337500,-0.459375)
      .. controls (-1.325000,-0.451354) and (-1.312292,-0.442677) .. (-1.299375,-0.433322)
      .. controls (-1.286458,-0.423967) and (-1.273333,-0.413933) .. (-1.260000,-0.403200)
      .. controls (-1.246667,-0.392467) and (-1.233125,-0.381033) .. (-1.219375,-0.368878)
      .. controls (-1.205625,-0.356723) and (-1.191667,-0.343846) .. (-1.177500,-0.330225)
      .. controls (-1.163333,-0.316604) and (-1.148958,-0.302240) .. (-1.134375,-0.287109)
      .. controls (-1.119792,-0.271979) and (-1.105000,-0.256083) .. (-1.090000,-0.239400)
      .. controls (-1.075000,-0.222717) and (-1.059792,-0.205246) .. (-1.044375,-0.186966)
      .. controls (-1.028958,-0.168685) and (-1.013333,-0.149596) .. (-0.997500,-0.129675)
      .. controls (-0.981667,-0.109754) and (-0.965625,-0.089002) .. (-0.949375,-0.067397)
      .. controls (-0.933125,-0.045792) and (-0.916667,-0.023333) .. (-0.900000,0.000000)
      .. controls (-0.883333,0.023333) and (-0.866458,0.047542) .. (-0.849375,0.072647)
      .. controls (-0.832292,0.097752) and (-0.815000,0.123754) .. (-0.797500,0.150675)
      .. controls (-0.780000,0.177596) and (-0.762292,0.205435) .. (-0.744375,0.234216)
      .. controls (-0.726458,0.262996) and (-0.708333,0.292717) .. (-0.690000,0.323400)
      .. controls (-0.671667,0.354083) and (-0.653125,0.385729) .. (-0.634375,0.418359)
      .. controls (-0.615625,0.450990) and (-0.596667,0.484604) .. (-0.577500,0.519225)
      .. controls (-0.558333,0.553846) and (-0.538958,0.589473) .. (-0.519375,0.626128)
      .. controls (-0.499792,0.662783) and (-0.480000,0.700467) .. (-0.460000,0.739200)
      .. controls (-0.440000,0.777933) and (-0.419792,0.817717) .. (-0.399375,0.858572)
      .. controls (-0.378958,0.899427) and (-0.358333,0.941354) .. (-0.337500,0.984375)
      .. controls (-0.316667,1.027396) and (-0.295625,1.071510) .. (-0.274375,1.116741)
      .. controls (-0.253125,1.161971) and (-0.231667,1.208317) .. (-0.210000,1.255800)
;
}
\newcommand{\DrawRibbonLeft}[1]{%
    \draw[#1] (-0.220000,-2.000000)
      .. controls (-0.220000,-1.750000) and (-0.294929,-1.599635) .. (-0.410148,-1.347128)
      .. controls (-0.431485,-1.300367) and (-0.452601,-1.254761) .. (-0.473494,-1.210291)
      .. controls (-0.494387,-1.165820) and (-0.515058,-1.122483) .. (-0.535505,-1.080261)
      .. controls (-0.555951,-1.038039) and (-0.576174,-0.996931) .. (-0.596170,-0.956917)
      .. controls (-0.616166,-0.916903) and (-0.635937,-0.877982) .. (-0.655479,-0.840136)
      .. controls (-0.675021,-0.802289) and (-0.694335,-0.765516) .. (-0.713418,-0.729797)
      .. controls (-0.732501,-0.694078) and (-0.751354,-0.659413) .. (-0.769973,-0.625781)
      .. controls (-0.788592,-0.592150) and (-0.806977,-0.559552) .. (-0.825126,-0.527969)
      .. controls (-0.843274,-0.496385) and (-0.861186,-0.465816) .. (-0.878856,-0.436242)
      .. controls (-0.896527,-0.406668) and (-0.913957,-0.378088) .. (-0.931141,-0.350484)
      .. controls (-0.948326,-0.322880) and (-0.965265,-0.296252) .. (-0.981953,-0.270579)
      .. controls (-0.998641,-0.244907) and (-1.015079,-0.220192) .. (-1.031259,-0.196413)
      .. controls (-1.047439,-0.172635) and (-1.063363,-0.149795) .. (-1.079021,-0.127872)
      .. controls (-1.094680,-0.105950) and (-1.110074,-0.084947) .. (-1.125195,-0.064843)
      .. controls (-1.140315,-0.044739) and (-1.155163,-0.025535) .. (-1.169726,-0.007212)
      .. controls (-1.184289,0.011111) and (-1.198569,0.028553) .. (-1.212552,0.045133)
      .. controls (-1.226535,0.061713) and (-1.240222,0.077432) .. (-1.253598,0.092309)
      .. controls (-1.266974,0.107186) and (-1.280038,0.121221) .. (-1.292775,0.134435)
      .. controls (-1.305511,0.147649) and (-1.317919,0.160042) .. (-1.329978,0.171636)
      .. controls (-1.342037,0.183231) and (-1.353748,0.194027) .. (-1.365086,0.204050)
      .. controls (-1.376424,0.214073) and (-1.387389,0.223322) .. (-1.397955,0.231828)
      .. controls (-1.408521,0.240333) and (-1.418687,0.248094) .. (-1.428423,0.255146)
      .. controls (-1.438159,0.262198) and (-1.447466,0.268540) .. (-1.456311,0.274216)
      .. controls (-1.465155,0.279891) and (-1.473538,0.284899) .. (-1.481426,0.289293)
      .. controls (-1.489314,0.293686) and (-1.496708,0.297465) .. (-1.503579,0.300695)
      .. controls (-1.510451,0.303924) and (-1.516799,0.306604) .. (-1.522606,0.308814)
      .. controls (-1.528413,0.311023) and (-1.533678,0.312762) .. (-1.538400,0.314122)
      .. controls (-1.543122,0.315483) and (-1.547301,0.316465) .. (-1.550961,0.317169)
      .. controls (-1.554622,0.317874) and (-1.557763,0.318302) .. (-1.560444,0.318552)
      .. controls (-1.563126,0.318802) and (-1.565346,0.318876) .. (-1.567199,0.318858)
      .. controls (-1.569052,0.318839) and (-1.570538,0.318730) .. (-1.571778,0.318584)
      .. controls (-1.573017,0.318438) and (-1.574012,0.318256) .. (-1.574893,0.318053)
      .. controls (-1.575774,0.317849) and (-1.576543,0.317624) .. (-1.577323,0.317351)
      .. controls (-1.578103,0.317078) and (-1.578894,0.316757) .. (-1.579791,0.316330)
      .. controls (-1.580688,0.315902) and (-1.581691,0.315369) .. (-1.582857,0.314657)
      .. controls (-1.584023,0.313945) and (-1.585352,0.313053) .. (-1.586863,0.311913)
      .. controls (-1.588374,0.310773) and (-1.590067,0.309383) .. (-1.591931,0.307687)
      .. controls (-1.593795,0.305991) and (-1.595830,0.303988) .. (-1.598005,0.301640)
      .. controls (-1.600181,0.299292) and (-1.602497,0.296600) .. (-1.604913,0.293543)
      .. controls (-1.607330,0.290486) and (-1.609847,0.287065) .. (-1.612422,0.283276)
      .. controls (-1.614996,0.279487) and (-1.617629,0.275332) .. (-1.620279,0.270818)
      .. controls (-1.622929,0.266304) and (-1.625596,0.261433) .. (-1.628244,0.256222)
      .. controls (-1.630892,0.251011) and (-1.633520,0.245461) .. (-1.636098,0.239597)
      .. controls (-1.638676,0.233732) and (-1.641203,0.227553) .. (-1.643654,0.221087)
      .. controls (-1.646104,0.214622) and (-1.648477,0.207870) .. (-1.650753,0.200863)
      .. controls (-1.653028,0.193855) and (-1.655205,0.186592) .. (-1.657266,0.179105)
      .. controls (-1.659327,0.171618) and (-1.661272,0.163907) .. (-1.663088,0.156004)
      .. controls (-1.664904,0.148101) and (-1.666591,0.140007) .. (-1.668137,0.131753)
      .. controls (-1.669684,0.123499) and (-1.671090,0.115086) .. (-1.672348,0.106545)
      .. controls (-1.673606,0.098005) and (-1.674715,0.089337) .. (-1.675670,0.080574)
      .. controls (-1.676625,0.071812) and (-1.677425,0.062954) .. (-1.678068,0.054033)
      .. controls (-1.678710,0.045112) and (-1.679193,0.036127) .. (-1.679516,0.027111)
      .. controls (-1.679838,0.018095) and (-1.680000,0.009048) .. (-1.680000,0.000000)
      .. controls (-1.680000,-0.009048) and (-1.679838,-0.018095) .. (-1.679516,-0.027111)
      .. controls (-1.679193,-0.036127) and (-1.678710,-0.045112) .. (-1.678068,-0.054033)
      .. controls (-1.677425,-0.062954) and (-1.676625,-0.071812) .. (-1.675670,-0.080574)
      .. controls (-1.674715,-0.089337) and (-1.673606,-0.098005) .. (-1.672348,-0.106545)
      .. controls (-1.671090,-0.115086) and (-1.669684,-0.123499) .. (-1.668137,-0.131753)
      .. controls (-1.666591,-0.140007) and (-1.664904,-0.148101) .. (-1.663088,-0.156004)
      .. controls (-1.661272,-0.163907) and (-1.659327,-0.171618) .. (-1.657266,-0.179105)
      .. controls (-1.655205,-0.186592) and (-1.653028,-0.193855) .. (-1.650753,-0.200863)
      .. controls (-1.648477,-0.207870) and (-1.646104,-0.214622) .. (-1.643654,-0.221087)
      .. controls (-1.641203,-0.227553) and (-1.638676,-0.233732) .. (-1.636098,-0.239597)
      .. controls (-1.633520,-0.245461) and (-1.630892,-0.251011) .. (-1.628244,-0.256222)
      .. controls (-1.625596,-0.261433) and (-1.622929,-0.266304) .. (-1.620279,-0.270818)
      .. controls (-1.617629,-0.275332) and (-1.614996,-0.279487) .. (-1.612422,-0.283276)
      .. controls (-1.609847,-0.287065) and (-1.607330,-0.290486) .. (-1.604913,-0.293543)
      .. controls (-1.602497,-0.296600) and (-1.600181,-0.299292) .. (-1.598005,-0.301640)
      .. controls (-1.595830,-0.303988) and (-1.593795,-0.305991) .. (-1.591931,-0.307687)
      .. controls (-1.590067,-0.309383) and (-1.588374,-0.310773) .. (-1.586863,-0.311913)
      .. controls (-1.585352,-0.313053) and (-1.584023,-0.313945) .. (-1.582857,-0.314657)
      .. controls (-1.581691,-0.315369) and (-1.580688,-0.315902) .. (-1.579791,-0.316330)
      .. controls (-1.578894,-0.316757) and (-1.578103,-0.317078) .. (-1.577323,-0.317351)
      .. controls (-1.576543,-0.317624) and (-1.575774,-0.317849) .. (-1.574893,-0.318053)
      .. controls (-1.574012,-0.318256) and (-1.573017,-0.318438) .. (-1.571778,-0.318584)
      .. controls (-1.570538,-0.318730) and (-1.569052,-0.318839) .. (-1.567199,-0.318858)
      .. controls (-1.565346,-0.318876) and (-1.563126,-0.318802) .. (-1.560444,-0.318552)
      .. controls (-1.557763,-0.318302) and (-1.554622,-0.317874) .. (-1.550961,-0.317169)
      .. controls (-1.547301,-0.316465) and (-1.543122,-0.315483) .. (-1.538400,-0.314122)
      .. controls (-1.533678,-0.312762) and (-1.528413,-0.311023) .. (-1.522606,-0.308814)
      .. controls (-1.516799,-0.306604) and (-1.510451,-0.303924) .. (-1.503579,-0.300695)
      .. controls (-1.496708,-0.297465) and (-1.489314,-0.293686) .. (-1.481426,-0.289293)
      .. controls (-1.473538,-0.284899) and (-1.465155,-0.279891) .. (-1.456311,-0.274216)
      .. controls (-1.447466,-0.268540) and (-1.438159,-0.262198) .. (-1.428423,-0.255146)
      .. controls (-1.418687,-0.248094) and (-1.408521,-0.240333) .. (-1.397955,-0.231828)
      .. controls (-1.387389,-0.223322) and (-1.376424,-0.214073) .. (-1.365086,-0.204050)
      .. controls (-1.353748,-0.194027) and (-1.342037,-0.183231) .. (-1.329978,-0.171636)
      .. controls (-1.317919,-0.160042) and (-1.305511,-0.147649) .. (-1.292775,-0.134435)
      .. controls (-1.280038,-0.121221) and (-1.266974,-0.107186) .. (-1.253598,-0.092309)
      .. controls (-1.240222,-0.077432) and (-1.226535,-0.061713) .. (-1.212552,-0.045133)
      .. controls (-1.198569,-0.028553) and (-1.184289,-0.011111) .. (-1.169726,0.007212)
      .. controls (-1.155163,0.025535) and (-1.140315,0.044739) .. (-1.125195,0.064843)
      .. controls (-1.110074,0.084947) and (-1.094680,0.105950) .. (-1.079021,0.127872)
      .. controls (-1.063363,0.149795) and (-1.047439,0.172635) .. (-1.031259,0.196413)
      .. controls (-1.015079,0.220192) and (-0.998641,0.244907) .. (-0.981953,0.270579)
      .. controls (-0.965265,0.296252) and (-0.948326,0.322880) .. (-0.931141,0.350484)
      .. controls (-0.913957,0.378088) and (-0.896527,0.406668) .. (-0.878856,0.436242)
      .. controls (-0.861186,0.465816) and (-0.843274,0.496385) .. (-0.825126,0.527969)
      .. controls (-0.806977,0.559552) and (-0.788592,0.592150) .. (-0.769973,0.625781)
      .. controls (-0.751354,0.659413) and (-0.732501,0.694078) .. (-0.713418,0.729797)
      .. controls (-0.694335,0.765516) and (-0.675021,0.802289) .. (-0.655479,0.840136)
      .. controls (-0.635937,0.877982) and (-0.616166,0.916903) .. (-0.596170,0.956917)
      .. controls (-0.576174,0.996931) and (-0.555951,1.038039) .. (-0.535505,1.080261)
      .. controls (-0.515058,1.122483) and (-0.494387,1.165820) .. (-0.473494,1.210291)
      .. controls (-0.452601,1.254761) and (-0.431485,1.300367) .. (-0.410148,1.347128)
      .. controls (-0.294929,1.599635) and (-0.220000,1.750000) .. (-0.220000,2.000000);
}
\newcommand{\DrawRibbonLeftOverpass}[1]{%
    \draw[#1] (-1.604913,-0.293543)
      .. controls (-1.602497,-0.296600) and (-1.600181,-0.299292) .. (-1.598005,-0.301640)
      .. controls (-1.595830,-0.303988) and (-1.593795,-0.305991) .. (-1.591931,-0.307687)
      .. controls (-1.590067,-0.309383) and (-1.588374,-0.310773) .. (-1.586863,-0.311913)
      .. controls (-1.585352,-0.313053) and (-1.584023,-0.313945) .. (-1.582857,-0.314657)
      .. controls (-1.581691,-0.315369) and (-1.580688,-0.315902) .. (-1.579791,-0.316330)
      .. controls (-1.578894,-0.316757) and (-1.578103,-0.317078) .. (-1.577323,-0.317351)
      .. controls (-1.576543,-0.317624) and (-1.575774,-0.317849) .. (-1.574893,-0.318053)
      .. controls (-1.574012,-0.318256) and (-1.573017,-0.318438) .. (-1.571778,-0.318584)
      .. controls (-1.570538,-0.318730) and (-1.569052,-0.318839) .. (-1.567199,-0.318858)
      .. controls (-1.565346,-0.318876) and (-1.563126,-0.318802) .. (-1.560444,-0.318552)
      .. controls (-1.557763,-0.318302) and (-1.554622,-0.317874) .. (-1.550961,-0.317169)
      .. controls (-1.547301,-0.316465) and (-1.543122,-0.315483) .. (-1.538400,-0.314122)
      .. controls (-1.533678,-0.312762) and (-1.528413,-0.311023) .. (-1.522606,-0.308814)
      .. controls (-1.516799,-0.306604) and (-1.510451,-0.303924) .. (-1.503579,-0.300695)
      .. controls (-1.496708,-0.297465) and (-1.489314,-0.293686) .. (-1.481426,-0.289293)
      .. controls (-1.473538,-0.284899) and (-1.465155,-0.279891) .. (-1.456311,-0.274216)
      .. controls (-1.447466,-0.268540) and (-1.438159,-0.262198) .. (-1.428423,-0.255146)
      .. controls (-1.418687,-0.248094) and (-1.408521,-0.240333) .. (-1.397955,-0.231828)
      .. controls (-1.387389,-0.223322) and (-1.376424,-0.214073) .. (-1.365086,-0.204050)
      .. controls (-1.353748,-0.194027) and (-1.342037,-0.183231) .. (-1.329978,-0.171636)
      .. controls (-1.317919,-0.160042) and (-1.305511,-0.147649) .. (-1.292775,-0.134435)
      .. controls (-1.280038,-0.121221) and (-1.266974,-0.107186) .. (-1.253598,-0.092309)
      .. controls (-1.240222,-0.077432) and (-1.226535,-0.061713) .. (-1.212552,-0.045133)
      .. controls (-1.198569,-0.028553) and (-1.184289,-0.011111) .. (-1.169726,0.007212)
      .. controls (-1.155163,0.025535) and (-1.140315,0.044739) .. (-1.125195,0.064843)
      .. controls (-1.110074,0.084947) and (-1.094680,0.105950) .. (-1.079021,0.127872)
      .. controls (-1.063363,0.149795) and (-1.047439,0.172635) .. (-1.031259,0.196413)
      .. controls (-1.015079,0.220192) and (-0.998641,0.244907) .. (-0.981953,0.270579)
      .. controls (-0.965265,0.296252) and (-0.948326,0.322880) .. (-0.931141,0.350484)
      .. controls (-0.913957,0.378088) and (-0.896527,0.406668) .. (-0.878856,0.436242)
      .. controls (-0.861186,0.465816) and (-0.843274,0.496385) .. (-0.825126,0.527969)
      .. controls (-0.806977,0.559552) and (-0.788592,0.592150) .. (-0.769973,0.625781)
      .. controls (-0.751354,0.659413) and (-0.732501,0.694078) .. (-0.713418,0.729797)
      .. controls (-0.694335,0.765516) and (-0.675021,0.802289) .. (-0.655479,0.840136)
      .. controls (-0.635937,0.877982) and (-0.616166,0.916903) .. (-0.596170,0.956917)
      .. controls (-0.576174,0.996931) and (-0.555951,1.038039) .. (-0.535505,1.080261)
      .. controls (-0.515058,1.122483) and (-0.494387,1.165820) .. (-0.473494,1.210291)
      .. controls (-0.452601,1.254761) and (-0.431485,1.300367) .. (-0.410148,1.347128)
;
}
\newcommand{\DrawRibbonRight}[1]{%
    \draw[#1] (0.220000,-2.000000)
      .. controls (0.220000,-1.750000) and (0.108929,-1.424785) .. (-0.009852,-1.164472)
      .. controls (-0.031848,-1.116266) and (-0.053649,-1.069180) .. (-0.075256,-1.023191)
      .. controls (-0.096863,-0.977201) and (-0.118275,-0.932308) .. (-0.139495,-0.888489)
      .. controls (-0.160715,-0.844669) and (-0.181743,-0.801923) .. (-0.202580,-0.760227)
      .. controls (-0.223417,-0.718531) and (-0.244063,-0.677884) .. (-0.264521,-0.638264)
      .. controls (-0.284979,-0.598644) and (-0.305248,-0.560050) .. (-0.325332,-0.522459)
      .. controls (-0.345415,-0.484868) and (-0.365313,-0.448279) .. (-0.385027,-0.412669)
      .. controls (-0.404742,-0.377059) and (-0.424273,-0.342427) .. (-0.443624,-0.308750)
      .. controls (-0.462976,-0.275073) and (-0.482148,-0.242350) .. (-0.501144,-0.210558)
      .. controls (-0.520140,-0.178766) and (-0.538960,-0.147904) .. (-0.557609,-0.117947)
      .. controls (-0.576258,-0.087991) and (-0.594735,-0.058940) .. (-0.613047,-0.030771)
      .. controls (-0.631359,-0.002601) and (-0.649505,0.024688) .. (-0.667491,0.051120)
      .. controls (-0.685477,0.077552) and (-0.703304,0.103128) .. (-0.720979,0.127872)
      .. controls (-0.738653,0.152617) and (-0.756176,0.176530) .. (-0.773555,0.199637)
      .. controls (-0.790935,0.222744) and (-0.808171,0.245044) .. (-0.825274,0.266562)
      .. controls (-0.842377,0.288081) and (-0.859348,0.308818) .. (-0.876198,0.328798)
      .. controls (-0.893048,0.348778) and (-0.909778,0.368001) .. (-0.926402,0.386491)
      .. controls (-0.943026,0.404981) and (-0.959545,0.422737) .. (-0.975975,0.439784)
      .. controls (-0.992405,0.456830) and (-1.008748,0.473167) .. (-1.025022,0.488814)
      .. controls (-1.041296,0.504461) and (-1.057502,0.519419) .. (-1.073664,0.533706)
      .. controls (-1.089826,0.547994) and (-1.105944,0.561611) .. (-1.122045,0.574572)
      .. controls (-1.138146,0.587534) and (-1.154230,0.599839) .. (-1.170327,0.611498)
      .. controls (-1.186424,0.623156) and (-1.202534,0.634168) .. (-1.218689,0.644534)
      .. controls (-1.234845,0.654901) and (-1.251046,0.664622) .. (-1.267324,0.673689)
      .. controls (-1.283602,0.682755) and (-1.299959,0.691168) .. (-1.316421,0.698905)
      .. controls (-1.332883,0.706642) and (-1.349451,0.713704) .. (-1.366144,0.720055)
      .. controls (-1.382837,0.726406) and (-1.399655,0.732047) .. (-1.416600,0.736928)
      .. controls (-1.433544,0.741809) and (-1.450616,0.745931) .. (-1.467789,0.749237)
      .. controls (-1.484962,0.752543) and (-1.502237,0.755032) .. (-1.519556,0.756648)
      .. controls (-1.536874,0.758264) and (-1.554237,0.759007) .. (-1.571551,0.758836)
      .. controls (-1.588865,0.758665) and (-1.606129,0.757578) .. (-1.623222,0.755566)
      .. controls (-1.640316,0.753554) and (-1.657238,0.750615) .. (-1.673857,0.746779)
      .. controls (-1.690476,0.742942) and (-1.706790,0.738209) .. (-1.722677,0.732649)
      .. controls (-1.738564,0.727088) and (-1.754023,0.720702) .. (-1.768959,0.713589)
      .. controls (-1.783896,0.706477) and (-1.798309,0.698640) .. (-1.812143,0.690193)
      .. controls (-1.825977,0.681747) and (-1.839231,0.672692) .. (-1.851887,0.663143)
      .. controls (-1.864542,0.653594) and (-1.876600,0.643550) .. (-1.888069,0.633113)
      .. controls (-1.899538,0.622676) and (-1.910420,0.611846) .. (-1.920745,0.600704)
      .. controls (-1.931069,0.589562) and (-1.940837,0.578109) .. (-1.950087,0.566407)
      .. controls (-1.959337,0.554706) and (-1.968070,0.542756) .. (-1.976328,0.530605)
      .. controls (-1.984587,0.518454) and (-1.992371,0.506102) .. (-1.999721,0.493582)
      .. controls (-2.007071,0.481063) and (-2.013987,0.468376) .. (-2.020506,0.455547)
      .. controls (-2.027025,0.442718) and (-2.033146,0.429747) .. (-2.038902,0.416653)
      .. controls (-2.044657,0.403560) and (-2.050047,0.390343) .. (-2.055096,0.377019)
      .. controls (-2.060146,0.363695) and (-2.064856,0.350263) .. (-2.069247,0.336737)
      .. controls (-2.073639,0.323212) and (-2.077712,0.309591) .. (-2.081484,0.295889)
      .. controls (-2.085257,0.282187) and (-2.088728,0.268402) .. (-2.091912,0.254546)
      .. controls (-2.095096,0.240691) and (-2.097992,0.226764) .. (-2.100613,0.212779)
      .. controls (-2.103233,0.198793) and (-2.105577,0.184748) .. (-2.107652,0.170655)
      .. controls (-2.109728,0.156562) and (-2.111535,0.142421) .. (-2.113080,0.128244)
      .. controls (-2.114625,0.114067) and (-2.115908,0.099854) .. (-2.116932,0.085617)
      .. controls (-2.117957,0.071380) and (-2.118724,0.057119) .. (-2.119234,0.042845)
      .. controls (-2.119745,0.028571) and (-2.120000,0.014286) .. (-2.120000,0.000000)
      .. controls (-2.120000,-0.014286) and (-2.119745,-0.028571) .. (-2.119234,-0.042845)
      .. controls (-2.118724,-0.057119) and (-2.117957,-0.071380) .. (-2.116932,-0.085617)
      .. controls (-2.115908,-0.099854) and (-2.114625,-0.114067) .. (-2.113080,-0.128244)
      .. controls (-2.111535,-0.142421) and (-2.109728,-0.156562) .. (-2.107652,-0.170655)
      .. controls (-2.105577,-0.184748) and (-2.103233,-0.198793) .. (-2.100613,-0.212779)
      .. controls (-2.097992,-0.226764) and (-2.095096,-0.240691) .. (-2.091912,-0.254546)
      .. controls (-2.088728,-0.268402) and (-2.085257,-0.282187) .. (-2.081484,-0.295889)
      .. controls (-2.077712,-0.309591) and (-2.073639,-0.323212) .. (-2.069247,-0.336737)
      .. controls (-2.064856,-0.350263) and (-2.060146,-0.363695) .. (-2.055096,-0.377019)
      .. controls (-2.050047,-0.390343) and (-2.044657,-0.403560) .. (-2.038902,-0.416653)
      .. controls (-2.033146,-0.429747) and (-2.027025,-0.442718) .. (-2.020506,-0.455547)
      .. controls (-2.013987,-0.468376) and (-2.007071,-0.481063) .. (-1.999721,-0.493582)
      .. controls (-1.992371,-0.506102) and (-1.984587,-0.518454) .. (-1.976328,-0.530605)
      .. controls (-1.968070,-0.542756) and (-1.959337,-0.554706) .. (-1.950087,-0.566407)
      .. controls (-1.940837,-0.578109) and (-1.931069,-0.589562) .. (-1.920745,-0.600704)
      .. controls (-1.910420,-0.611846) and (-1.899538,-0.622676) .. (-1.888069,-0.633113)
      .. controls (-1.876600,-0.643550) and (-1.864542,-0.653594) .. (-1.851887,-0.663143)
      .. controls (-1.839231,-0.672692) and (-1.825977,-0.681747) .. (-1.812143,-0.690193)
      .. controls (-1.798309,-0.698640) and (-1.783896,-0.706477) .. (-1.768959,-0.713589)
      .. controls (-1.754023,-0.720702) and (-1.738564,-0.727088) .. (-1.722677,-0.732649)
      .. controls (-1.706790,-0.738209) and (-1.690476,-0.742942) .. (-1.673857,-0.746779)
      .. controls (-1.657238,-0.750615) and (-1.640316,-0.753554) .. (-1.623222,-0.755566)
      .. controls (-1.606129,-0.757578) and (-1.588865,-0.758665) .. (-1.571551,-0.758836)
      .. controls (-1.554237,-0.759007) and (-1.536874,-0.758264) .. (-1.519556,-0.756648)
      .. controls (-1.502237,-0.755032) and (-1.484962,-0.752543) .. (-1.467789,-0.749237)
      .. controls (-1.450616,-0.745931) and (-1.433544,-0.741809) .. (-1.416600,-0.736928)
      .. controls (-1.399655,-0.732047) and (-1.382837,-0.726406) .. (-1.366144,-0.720055)
      .. controls (-1.349451,-0.713704) and (-1.332883,-0.706642) .. (-1.316421,-0.698905)
      .. controls (-1.299959,-0.691168) and (-1.283602,-0.682755) .. (-1.267324,-0.673689)
      .. controls (-1.251046,-0.664622) and (-1.234845,-0.654901) .. (-1.218689,-0.644534)
      .. controls (-1.202534,-0.634168) and (-1.186424,-0.623156) .. (-1.170327,-0.611498)
      .. controls (-1.154230,-0.599839) and (-1.138146,-0.587534) .. (-1.122045,-0.574572)
      .. controls (-1.105944,-0.561611) and (-1.089826,-0.547994) .. (-1.073664,-0.533706)
      .. controls (-1.057502,-0.519419) and (-1.041296,-0.504461) .. (-1.025022,-0.488814)
      .. controls (-1.008748,-0.473167) and (-0.992405,-0.456830) .. (-0.975975,-0.439784)
      .. controls (-0.959545,-0.422737) and (-0.943026,-0.404981) .. (-0.926402,-0.386491)
      .. controls (-0.909778,-0.368001) and (-0.893048,-0.348778) .. (-0.876198,-0.328798)
      .. controls (-0.859348,-0.308818) and (-0.842377,-0.288081) .. (-0.825274,-0.266562)
      .. controls (-0.808171,-0.245044) and (-0.790935,-0.222744) .. (-0.773555,-0.199637)
      .. controls (-0.756176,-0.176530) and (-0.738653,-0.152617) .. (-0.720979,-0.127872)
      .. controls (-0.703304,-0.103128) and (-0.685477,-0.077552) .. (-0.667491,-0.051120)
      .. controls (-0.649505,-0.024688) and (-0.631359,0.002601) .. (-0.613047,0.030771)
      .. controls (-0.594735,0.058940) and (-0.576258,0.087991) .. (-0.557609,0.117947)
      .. controls (-0.538960,0.147904) and (-0.520140,0.178766) .. (-0.501144,0.210558)
      .. controls (-0.482148,0.242350) and (-0.462976,0.275073) .. (-0.443624,0.308750)
      .. controls (-0.424273,0.342427) and (-0.404742,0.377059) .. (-0.385027,0.412669)
      .. controls (-0.365313,0.448279) and (-0.345415,0.484868) .. (-0.325332,0.522459)
      .. controls (-0.305248,0.560050) and (-0.284979,0.598644) .. (-0.264521,0.638264)
      .. controls (-0.244063,0.677884) and (-0.223417,0.718531) .. (-0.202580,0.760227)
      .. controls (-0.181743,0.801923) and (-0.160715,0.844669) .. (-0.139495,0.888489)
      .. controls (-0.118275,0.932308) and (-0.096863,0.977201) .. (-0.075256,1.023191)
      .. controls (-0.053649,1.069180) and (-0.031848,1.116266) .. (-0.009852,1.164472)
      .. controls (0.108929,1.424785) and (0.220000,1.750000) .. (0.220000,2.000000);
}
\newcommand{\DrawRibbonRightOverpass}[1]{%
    \draw[#1] (-1.950087,-0.566407)
      .. controls (-1.940837,-0.578109) and (-1.931069,-0.589562) .. (-1.920745,-0.600704)
      .. controls (-1.910420,-0.611846) and (-1.899538,-0.622676) .. (-1.888069,-0.633113)
      .. controls (-1.876600,-0.643550) and (-1.864542,-0.653594) .. (-1.851887,-0.663143)
      .. controls (-1.839231,-0.672692) and (-1.825977,-0.681747) .. (-1.812143,-0.690193)
      .. controls (-1.798309,-0.698640) and (-1.783896,-0.706477) .. (-1.768959,-0.713589)
      .. controls (-1.754023,-0.720702) and (-1.738564,-0.727088) .. (-1.722677,-0.732649)
      .. controls (-1.706790,-0.738209) and (-1.690476,-0.742942) .. (-1.673857,-0.746779)
      .. controls (-1.657238,-0.750615) and (-1.640316,-0.753554) .. (-1.623222,-0.755566)
      .. controls (-1.606129,-0.757578) and (-1.588865,-0.758665) .. (-1.571551,-0.758836)
      .. controls (-1.554237,-0.759007) and (-1.536874,-0.758264) .. (-1.519556,-0.756648)
      .. controls (-1.502237,-0.755032) and (-1.484962,-0.752543) .. (-1.467789,-0.749237)
      .. controls (-1.450616,-0.745931) and (-1.433544,-0.741809) .. (-1.416600,-0.736928)
      .. controls (-1.399655,-0.732047) and (-1.382837,-0.726406) .. (-1.366144,-0.720055)
      .. controls (-1.349451,-0.713704) and (-1.332883,-0.706642) .. (-1.316421,-0.698905)
      .. controls (-1.299959,-0.691168) and (-1.283602,-0.682755) .. (-1.267324,-0.673689)
      .. controls (-1.251046,-0.664622) and (-1.234845,-0.654901) .. (-1.218689,-0.644534)
      .. controls (-1.202534,-0.634168) and (-1.186424,-0.623156) .. (-1.170327,-0.611498)
      .. controls (-1.154230,-0.599839) and (-1.138146,-0.587534) .. (-1.122045,-0.574572)
      .. controls (-1.105944,-0.561611) and (-1.089826,-0.547994) .. (-1.073664,-0.533706)
      .. controls (-1.057502,-0.519419) and (-1.041296,-0.504461) .. (-1.025022,-0.488814)
      .. controls (-1.008748,-0.473167) and (-0.992405,-0.456830) .. (-0.975975,-0.439784)
      .. controls (-0.959545,-0.422737) and (-0.943026,-0.404981) .. (-0.926402,-0.386491)
      .. controls (-0.909778,-0.368001) and (-0.893048,-0.348778) .. (-0.876198,-0.328798)
      .. controls (-0.859348,-0.308818) and (-0.842377,-0.288081) .. (-0.825274,-0.266562)
      .. controls (-0.808171,-0.245044) and (-0.790935,-0.222744) .. (-0.773555,-0.199637)
      .. controls (-0.756176,-0.176530) and (-0.738653,-0.152617) .. (-0.720979,-0.127872)
      .. controls (-0.703304,-0.103128) and (-0.685477,-0.077552) .. (-0.667491,-0.051120)
      .. controls (-0.649505,-0.024688) and (-0.631359,0.002601) .. (-0.613047,0.030771)
      .. controls (-0.594735,0.058940) and (-0.576258,0.087991) .. (-0.557609,0.117947)
      .. controls (-0.538960,0.147904) and (-0.520140,0.178766) .. (-0.501144,0.210558)
      .. controls (-0.482148,0.242350) and (-0.462976,0.275073) .. (-0.443624,0.308750)
      .. controls (-0.424273,0.342427) and (-0.404742,0.377059) .. (-0.385027,0.412669)
      .. controls (-0.365313,0.448279) and (-0.345415,0.484868) .. (-0.325332,0.522459)
      .. controls (-0.305248,0.560050) and (-0.284979,0.598644) .. (-0.264521,0.638264)
      .. controls (-0.244063,0.677884) and (-0.223417,0.718531) .. (-0.202580,0.760227)
      .. controls (-0.181743,0.801923) and (-0.160715,0.844669) .. (-0.139495,0.888489)
      .. controls (-0.118275,0.932308) and (-0.096863,0.977201) .. (-0.075256,1.023191)
      .. controls (-0.053649,1.069180) and (-0.031848,1.116266) .. (-0.009852,1.164472)
;
}

%% file: figures/tikz-diagrams/balancing-isotopy-diagrams.tex
\input{figures/tikz-diagrams/balancing-diagrams.tex}

\newcommand{\SimultaneouslyTwistedPair}[1][0]{%
  \begin{tikzpicture}[baseline=-.65ex,x=.55cm,y=.55cm]
    \path[use as bounding box] (-2.25,-3) rectangle (.55,2.4);
    \begin{scope}[shift={(0,.367816092)},yscale=.816091954]
    \DrawRibbonLeft{appendix strand}
    \DrawRibbonRight{appendix strand}
    \DrawRibbonLeftOverpass{appendix crossing}
    \DrawRibbonRightOverpass{appendix crossing}
    \node[appendix label,above=2pt] at (-.30,2) {$a$};
    \node[appendix label,above=2pt] at (.30,2) {$b$};
    \ifnum#1=1
      \draw[appendix strand] (0,-3.697887324) -- (0,-2.35);
      \draw[appendix strand] (0,-2.35)
        .. controls (-.15,-2.23) and (-.22,-2.2) .. (-.22,-2);
      \draw[appendix strand] (0,-2.35)
        .. controls (.15,-2.23) and (.22,-2.2) .. (.22,-2);
      \node[appendix vertex] at (0,-2.35) {};
      \node[appendix label,below=2pt] at (0,-3.697887324) {$c$};
    \else
      \node[appendix label,below=3pt] at (-.22,-2) {$a$};
      \node[appendix label,below=3pt] at (.22,-2) {$b$};
    \fi
    \end{scope}
  \end{tikzpicture}%
}

\newcommand{\DrawBalancingBraid}{%
  \draw[appendix strand] (-.65,-.1)
    .. controls (-.65,.2) and (.65,.4) .. (.65,.7)
    .. controls (.65,1) and (-.65,1.2) .. (-.65,1.5) -- (-.65,2);
  \draw[appendix strand] (.65,-.1)
    .. controls (.65,.2) and (-.65,.4) .. (-.65,.7)
    .. controls (-.65,1) and (.65,1.2) .. (.65,1.5) -- (.65,2);
  \draw[appendix crossing] (-.3358375,.16383)
    .. controls (-.1321625,.25617) and (.1321625,.34383) .. (.3358375,.43617);
  \draw[appendix crossing] (-.3358375,.96383)
    .. controls (-.1321625,1.05617) and (.1321625,1.14383) .. (.3358375,1.23617);
}
\newcommand{\BalancingChannelRoot}{%
  \draw[appendix strand] (0,-3.697887324) -- (0,-2.35);
  \draw[appendix strand] (0,-2.35)
    .. controls (-.25,-2.28) and (-.65,-2.3) .. (-.65,-2.1);
  \draw[appendix strand] (0,-2.35)
    .. controls (.25,-2.28) and (.65,-2.3) .. (.65,-2.1);
  \node[appendix vertex] at (0,-2.35) {};
  \node[appendix label,below=2pt] at (0,-3.697887324) {$c$};
}
\newcommand{\BalancingTopLabels}{%
  \node[appendix label,above=2pt] at (-.65,2) {$a$};
  \node[appendix label,above=2pt] at (.65,2) {$b$};
}
\newcommand{\DoubleBraidWithIndividualTwists}[1][0]{%
  \begin{tikzpicture}[baseline=-.65ex,x=.55cm,y=.55cm]
    \path[use as bounding box] (-1.7,-3) rectangle (.85,2.4);
    \begin{scope}[shift={(0,.367816092)},yscale=.816091954]
    \foreach \x in {-.65,.65}{
      \begin{scope}[shift={(\x,-1.1)},scale=.5]
        \DrawRibbonCenter{appendix strand}
        \DrawRibbonCenterOverpass{appendix crossing}
      \end{scope}
    }
    \DrawBalancingBraid
    \BalancingTopLabels
    \ifnum#1=1
      \BalancingChannelRoot
    \else
      \node[appendix label,below=3pt] at (-.65,-2.1) {$a$};
      \node[appendix label,below=3pt] at (.65,-2.1) {$b$};
    \fi
    \end{scope}
  \end{tikzpicture}%
}
\newcommand{\BalancingStraightChannel}{%
  \begin{tikzpicture}[baseline=-.65ex,x=.55cm,y=.55cm]
    \path[use as bounding box] (-.85,-3) rectangle (.85,2.4);
    \begin{scope}[shift={(0,.367816092)},yscale=.816091954]
    \BalancingChannelRoot
    \draw[appendix strand] (-.65,-2.1) -- (-.65,2);
    \draw[appendix strand] (.65,-2.1) -- (.65,2);
    \BalancingTopLabels
    \end{scope}
  \end{tikzpicture}%
}
\newcommand{\BalancingBraidedChannel}{%
  \begin{tikzpicture}[baseline=-.65ex,x=.55cm,y=.55cm]
    \path[use as bounding box] (-.85,-3) rectangle (.85,2.4);
    \begin{scope}[shift={(0,.367816092)},yscale=.816091954]
    \BalancingChannelRoot
    \draw[appendix strand] (-.65,-2.1) -- (-.65,-.1);
    \draw[appendix strand] (.65,-2.1) -- (.65,-.1);
    \DrawBalancingBraid
    \BalancingTopLabels
    \end{scope}
  \end{tikzpicture}%
}

%% file: figures/tikz-diagrams/balancing-diagrams.tex
\input{figures/tikz-diagrams/diagram-style.tex}
\newcommand{\StraightFusionTree}{%
  \begin{tikzpicture}[baseline=-0.52ex, x=0.54cm, y=0.54cm]
    \path[use as bounding box] (-0.85,-1.25) rectangle (0.85,2.20);
    \draw[appendix strand] (0,-1.05) -- (0,-0.52);
    \draw[appendix strand] (0,-0.52)
      .. controls (-0.14,-0.40) and (-0.52,-0.20) .. (-0.52,0.20)
      -- (-0.52,1.92);
    \draw[appendix strand] (0,-0.52)
      .. controls (0.14,-0.40) and (0.52,-0.20) .. (0.52,0.20)
      -- (0.52,1.92);
    \node[appendix vertex] at (0,-0.52) {};
    \node[appendix label, above] at (-0.52,1.92) {$a$};
    \node[appendix label, above] at (0.52,1.92) {$b$};
    \node[appendix label, below] at (0,-1.05) {$c$};
  \end{tikzpicture}%
}

%% file: figures/tikz-diagrams/frobenius-schur-diagrams.tex
\input{figures/tikz-diagrams/diagram-style.tex}
\tikzset{
  appendix fs vacuum/.style={appendix strand,draw=gray!75,line width=.75pt},
  appendix fs vacuum crossing/.style={appendix fs vacuum,
    preaction={draw=white,line width=3.2pt}}
}

\newcommand{\FSRoot}{%
  \draw[appendix fs vacuum] (0,-3.697887324)--(0,-2.35);
  \draw[appendix strand] (0,-2.35)
    .. controls (-.25,-2.28) and (-.65,-2.3) .. (-.65,-2.1);
  \draw[appendix strand] (0,-2.35)
    .. controls (.25,-2.28) and (.65,-2.3) .. (.65,-2.1);
  \node[appendix vertex] at (0,-2.35) {};
  \node[appendix label,below=2pt,text=gray!75] at (0,-3.697887324) {$1$};
}
\newcommand{\FSTops}{%
  \node[appendix label,above=2pt] at (-.65,2) {$a$};
  \node[appendix label,above=2pt] at (.65,2) {$a$};
}
\newcommand{\FSCrossing}{%
  \draw[appendix strand] (-.65,-.1)
    .. controls (-.65,.2) and (.65,.4) .. (.65,.7);
  \draw[appendix strand] (.65,-.1)
    .. controls (.65,.2) and (-.65,.4) .. (-.65,.7);
  \draw[appendix crossing] (-.3358375,.16383)
    .. controls (-.1321625,.25617) and (.1321625,.34383)
    .. (.3358375,.43617);
}
\newcommand{\FSVertex}[1]{%
  \begin{tikzpicture}[baseline=-.52ex,x=.50cm,y=.50cm]
    \ifnum#1=2
      \path[use as bounding box] (-1.9,-3) rectangle (.88,2.4);
    \else
      \path[use as bounding box] (-.88,-3) rectangle (.88,2.4);
    \fi
    \begin{scope}[shift={(0,.367816092)},yscale=.816091954]
    \FSRoot
    \ifnum#1=0
      \draw[appendix strand] (-.65,-2.1)--(-.65,2);
      \draw[appendix strand] (.65,-2.1)--(.65,2);
    \else
      \draw[appendix strand] (-.65,-2.1)--(-.65,-1.1);
      \draw[appendix strand] (.65,-2.1)--(.65,-1.1);
      \begin{scope}[shift={(0,-1)}]\FSCrossing\end{scope}
      \draw[appendix strand] (.65,-.3)--(.65,2);
      \ifnum#1=1
        \draw[appendix strand] (-.65,-.3)--(-.65,2);
      \else
        \begin{scope}[shift={(-.65,.85)},scale=.575]
          \DrawRibbonCenter{appendix strand}
          \DrawRibbonCenterOverpass{appendix crossing}
        \end{scope}
      \fi
    \fi
    \FSTops
    \end{scope}
  \end{tikzpicture}%
}
\newcommand{\FSSquare}{%
  \begin{tikzpicture}[baseline=-.52ex,x=.50cm,y=.50cm]
    \path[use as bounding box] (-1.7,-3) rectangle (.88,2.4);
    \begin{scope}[shift={(0,.367816092)},yscale=.816091954]
    \FSRoot
    \foreach \x in {-.65,.65}{
      \begin{scope}[shift={(\x,-1.1)},scale=.5]
        \DrawRibbonCenter{appendix strand}
        \DrawRibbonCenterOverpass{appendix crossing}
      \end{scope}
    }
    \FSCrossing
    \begin{scope}[shift={(0,.8)}]\FSCrossing\end{scope}
    \draw[appendix strand] (-.65,1.5)--(-.65,2);
    \draw[appendix strand] (.65,1.5)--(.65,2);
    \FSTops
    \end{scope}
  \end{tikzpicture}%
}
\newcommand{\FSVacuumTwist}{%
  \begin{tikzpicture}[baseline=-.52ex,x=.50cm,y=.50cm]
    \path[use as bounding box] (-1.25,-3) rectangle (.88,2.4);
    \begin{scope}[shift={(0,.367816092)},yscale=.816091954]
    \begin{scope}[shift={(0,-1.1)},scale=.5]
      \DrawRibbonCenter{appendix fs vacuum}
      \DrawRibbonCenterOverpass{appendix fs vacuum crossing}
    \end{scope}
    \draw[appendix fs vacuum] (0,-3.697887324)--(0,-2.1);
    \draw[appendix fs vacuum] (0,-.1)--(0,.15);
    \node[appendix vertex] at (0,.15) {};
    \draw[appendix strand] (0,.15)
      .. controls (-.25,.25) and (-.65,.4) .. (-.65,.65)--(-.65,2);
    \draw[appendix strand] (0,.15)
      .. controls (.25,.25) and (.65,.4) .. (.65,.65)--(.65,2);
    \node[appendix label,below=2pt,text=gray!75] at (0,-3.697887324) {$1$};
    \FSTops
    \end{scope}
  \end{tikzpicture}%
}

%% file: figures/tikz-diagrams/s-matrix-diagrams.tex
\input{figures/tikz-diagrams/diagram-style.tex}
\usetikzlibrary{calc,decorations.markings,positioning}

\newcommand{\HopfCenteredArrow}[1][black]{%
  \fill[#1]
    (-1.75pt,-1.25pt) -- (2.05pt,0pt) -- (-1.75pt,1.25pt) -- cycle;%
}
\newcommand{\HopfReversedArrow}[1][black]{%
  \fill[#1]
    (1.75pt,-1.25pt) -- (-2.05pt,0pt) -- (1.75pt,1.25pt) -- cycle;%
}

\tikzset{
  appendix hopf left oriented/.style={
    appendix strand,
    postaction={decorate},
    decoration={
      markings,
      mark=at position 0.30 with {\HopfReversedArrow},
      mark=at position 0.70 with {\HopfReversedArrow}
    }
  },
  appendix hopf right oriented/.style={
    appendix strand,
    postaction={decorate},
    decoration={
      markings,
      mark=at position 0.20 with {\HopfCenteredArrow},
      mark=at position 0.80 with {\HopfCenteredArrow}
    }
  }
}

\newcommand{\HopfLinkAmplitude}[3][black]{%
  \begin{tikzpicture}[baseline=-0.55ex, x=0.82cm, y=0.82cm]
    \path[use as bounding box] (-1.52,-0.86) rectangle (1.52,0.86);
    \coordinate (HopfLeftCenter) at (-0.36,0);
    \coordinate (HopfRightCenter) at (0.36,0);
    \coordinate (HopfLeftLabel) at (-0.98,0);
    \coordinate (HopfRightLabel) at (0.98,0);

    \draw[appendix strand,draw=#1,postaction={decorate},
      decoration={markings,
        mark=at position .30 with {\HopfReversedArrow[#1]},
        mark=at position .70 with {\HopfReversedArrow[#1]}}]
      (HopfLeftCenter) circle[radius=0.62];
    \draw[appendix hopf right oriented]
      (HopfRightCenter) circle[radius=0.62];
    \draw[appendix crossing,draw=#1]
      ($(HopfLeftCenter)+(44:0.62)$)
      arc[start angle=44,end angle=65,radius=0.62];
    \draw[appendix crossing]
      ($(HopfRightCenter)+(224:0.62)$)
      arc[start angle=224,end angle=245,radius=0.62];

    \node[appendix label, left=2pt of HopfLeftLabel,text=#1] {$#2$};
    \node[appendix label, right=2pt of HopfRightLabel] {$#3$};
  \end{tikzpicture}%
}

\newcommand{\MatchedHopfUnknot}[1]{%
  \begin{tikzpicture}[baseline=-0.55ex,x=0.82cm,y=0.82cm]
    \path[use as bounding box] (-0.86,-0.86) rectangle (1.16,0.86);
    \coordinate (MatchedLoopLabel) at (0.62,0);
    \draw[appendix hopf right oriented] (0,0) circle[radius=0.62];
    \node[appendix label,right=2pt of MatchedLoopLabel] {$#1$};
  \end{tikzpicture}%
}

\newcommand{\EncirclingLoop}[2]{%
  \draw[appendix hopf left oriented] (-.267096774,#1) circle[radius=.46];
  \node[appendix label,left=2pt] at (-.727096774,#1) {$#2$};
}
\newcommand{\EncirclingOverpasses}[1]{%
  \draw[appendix crossing] (0,{#1+.311612903})--(0,{#1+.437741935});
  \draw[appendix crossing]
    ($(-.267096774,#1)+(295:.46)$)
    arc[start angle=295,end angle=316,radius=.46];
}
\newcommand{\EncirclingVertical}[2]{%
  \draw[appendix strand,postaction={decorate},
    decoration={markings,mark=at position .5 with {\HopfCenteredArrow}}]
    (0,-#1)--(0,#1);
  \node[appendix label,below=0pt] at (0,-#1) {$#2$};
  \node[appendix label,above=0pt] at (0,#1) {$#2$};
}
\newcommand{\EncirclingLine}[2]{%
  \begin{tikzpicture}[baseline=-.55ex,x=.82cm,y=.82cm]
    \path[use as bounding box] (-1.10,-1.16) rectangle (.38,1.16);
    \EncirclingLoop{0}{#1}
    \EncirclingVertical{.90}{#2}
    \EncirclingOverpasses{0}
  \end{tikzpicture}%
}
\newcommand{\TwoEncirclingLoops}[3]{%
  \begin{tikzpicture}[baseline=-.55ex,x=.82cm,y=.82cm]
    \path[use as bounding box] (-1.10,-1.93) rectangle (.38,1.93);
    \EncirclingLoop{.72}{#1}
    \EncirclingLoop{-.72}{#2}
    \EncirclingVertical{1.65}{#3}
    \EncirclingOverpasses{.72}
    \EncirclingOverpasses{-.72}
  \end{tikzpicture}%
}
\newcommand{\EncirclingIdentityLine}[2]{%
  \begin{tikzpicture}[baseline=-.55ex,x=.82cm,y=.82cm]
    \path[use as bounding box] (-.35,{-#1-.26}) rectangle (.35,{#1+.26});
    \EncirclingVertical{#1}{#2}
  \end{tikzpicture}%
}

%% file: figures/tikz-diagrams/fermion-grading-diagrams.tex
\input{figures/tikz-diagrams/diagram-style.tex}
\usetikzlibrary{decorations.markings}

\newcommand{\FermionGradingArrow}{%
  \fill (-1.75pt,-1.25pt)--(2.05pt,0pt)--(-1.75pt,1.25pt)--cycle;%
}
\tikzset{
  appendix fermion grading oriented/.style={
    postaction={decorate},
    decoration={markings,mark=at position .72 with {\FermionGradingArrow}}
  }
}

\newcommand{\FermionGradingJunction}{%
  \draw[appendix crossing,appendix fermion grading oriented]
    (0,-1.25)--(0,0);
  \draw[appendix crossing,appendix fermion grading oriented]
    (0,0) .. controls (-.30,.23) and (-.72,.25) .. (-.72,.58)
    -- (-.72,1.65);
  \draw[appendix crossing,appendix fermion grading oriented]
    (0,0) .. controls (.30,.23) and (.72,.25) .. (.72,.58)
    -- (.72,1.65);
  \node[appendix vertex] at (0,0) {};
  \node[appendix label,above=2pt] at (-.72,1.65) {$a$};
  \node[appendix label,above=2pt] at (.72,1.65) {$b$};
  \node[appendix label,below=2pt] at (0,-1.25) {$c$};
}

\newcommand{\FermionGradingLoopTree}[1]{%
  \begin{tikzpicture}[baseline=0pt,x=.68cm,y=.68cm]
    \path[use as bounding box] (-1.70,-1.70) rectangle (1.30,2.05);
    \ifnum#1=0
      \def\GradingLoopY{-.65}\def\GradingLoopRX{.64}\def\GradingLoopRY{.24}%
    \else
      \def\GradingLoopY{1.00}\def\GradingLoopRX{1.16}\def\GradingLoopRY{.33}%
    \fi
    \draw[appendix strand] (-\GradingLoopRX,\GradingLoopY)
      arc[start angle=180,end angle=0,
          x radius=\GradingLoopRX,y radius=\GradingLoopRY];
    \FermionGradingJunction
    \draw[appendix crossing,appendix fermion grading oriented]
      (\GradingLoopRX,\GradingLoopY)
      arc[start angle=0,end angle=-180,
          x radius=\GradingLoopRX,y radius=\GradingLoopRY];
    \node[appendix label,left=2pt] at (-\GradingLoopRX,\GradingLoopY) {$\psi$};
  \end{tikzpicture}%
}

\newcommand{\FermionGradingPlainTree}{%
  \begin{tikzpicture}[baseline=0pt,x=.68cm,y=.68cm]
    \path[use as bounding box] (-1.03,-1.70) rectangle (1.03,2.05);
    \FermionGradingJunction
  \end{tikzpicture}%
}

%% file: mtc-bib.bib
@misc{Teleman:2022Haagerup,
    author = "Teleman, Constantin",
    title = "{The Haagerup TQFT is not a gauge theory}",
    year = "2022",
    note = "Unpublished note",
    url = "https://math.berkeley.edu/~teleman/math/Haagerup.pdf"
}

@article{Frohlich:2003,
    author        = {Fr{\"o}hlich, J{\"u}rg and Fuchs, J{\"u}rgen and Runkel, Ingo and Schweigert, Christoph},
    title         = {Correspondences of ribbon categories},
    journal       = {Adv. Math.},
    volume        = {199},
    pages         = {192--329},
    year          = {2006},
    doi           = {10.1016/j.aim.2005.04.007},
    eprint        = {math/0309465},
    archivePrefix = {arXiv},
    primaryClass  = {math.CT}
}

@article{Kirillov:2001,
    author        = {Kirillov, Alexander, Jr. and Ostrik, Viktor},
    title         = {On a {$q$}-Analogue of the {McKay} Correspondence and the {ADE} Classification of {$\widehat{\mathfrak{sl}}_2$} Conformal Field Theories},
    journal       = {Adv. Math.},
    volume        = {171},
    pages         = {183--227},
    year          = {2002},
    doi           = {10.1006/aima.2002.2072},
    eprint        = {math/0101219},
    archivePrefix = {arXiv},
    primaryClass  = {math.QA}
}

@article{Bais:2002,
    author        = {Bais, F. A. and Schroers, B. J. and Slingerland, J. K.},
    title         = {Broken Quantum Symmetry and Confinement Phases in Planar Physics},
    journal       = {Phys. Rev. Lett.},
    volume        = {89},
    pages         = {181601},
    year          = {2002},
    doi           = {10.1103/PhysRevLett.89.181601},
    eprint        = {hep-th/0205117},
    archivePrefix = {arXiv},
    primaryClass  = {hep-th}
}

@article{Bais:2008,
    author        = {Bais, F. A. and Slingerland, J. K.},
    title         = {Condensate induced transitions between topologically ordered phases},
    journal       = {Phys. Rev. B},
    volume        = {79},
    pages         = {045316},
    year          = {2009},
    doi           = {10.1103/PhysRevB.79.045316},
    eprint        = {0808.0627},
    archivePrefix = {arXiv},
    primaryClass  = {cond-mat.mes-hall}
}

@article{alekseyevClassificationIntegralModular2026,
  title = {Classification of Integral Modular Data up to Rank 13},
  author = {Alekseyev, Max A. and Bruns, Winfried and Palcoux, Sebastien and Petrov, Fedor V.},
  date = {2026-08-01},
  journaltitle = {J. Math. Phys.},
  volume = {67},
  number = {8},
  eprint = {2302.01613},
  eprinttype = {arXiv},
  eprintclass = {math.QA},
  pages = {081701},
  doi = {10.1063/5.0334158}
}

@article{Andersen:2022xmz,
  title = {Non-{{Abelian}} Braiding of Graph Vertices in a Superconducting Processor},
  author = {Andersen, T. I. and Lensky, Y. D. and Kechedzhi, K. and Drozdov, I. K. and Bengtsson, A. and Hong, S. and Morvan, A. and Mi, X. and Opremcak, A. and Acharya, R. and Allen, R. and Ansmann, M. and Arute, F. and Arya, K. and Asfaw, A. and Atalaya, J. and Babbush, R. and Bacon, D. and Bardin, J. C. and Bortoli, G. and Bourassa, A. and Bovaird, J. and Brill, L. and Broughton, M. and Buckley, B. B. and Buell, D. A. and Burger, T. and Burkett, B. and Bushnell, N. and Chen, Z. and Chiaro, B. and Chik, D. and Chou, C. and Cogan, J. and Collins, R. and Conner, P. and Courtney, W. and Crook, A. L. and Curtin, B. and Debroy, D. M. and Del Toro Barba, A. and Demura, S. and Dunsworth, A. and Eppens, D. and Erickson, C. and Faoro, L. and Farhi, E. and Fatemi, R. and Ferreira, V. S. and Burgos, L. F. and Forati, E. and Fowler, A. G. and Foxen, B. and Giang, W. and Gidney, C. and Gilboa, D. and Giustina, M. and Gosula, R. and Dau, A. G. and Gross, J. A. and Habegger, S. and Hamilton, M. C. and Hansen, M. and Harrigan, M. P. and Harrington, S. D. and Heu, P. and Hilton, J. and Hoffmann, M. R. and Huang, T. and Huff, A. and Huggins, W. J. and Ioffe, L. B. and Isakov, S. V. and Iveland, J. and Jeffrey, E. and Jiang, Z. and Jones, C. and Juhas, P. and Kafri, D. and Khattar, T. and Khezri, M. and Kieferová, M. and Kim, S. and Kitaev, A. and Klimov, P. V. and Klots, A. R. and Korotkov, A. N. and Kostritsa, F. and Kreikebaum, J. M. and Landhuis, D. and Laptev, P. and Lau, K.-M. and Laws, L. and Lee, J. and Lee, K. W. and Lester, B. J. and Lill, A. T. and Liu, W. and Locharla, A. and Lucero, E. and Malone, F. D. and Martin, O. and McClean, J. R. and McCourt, T. and McEwen, M. and Miao, K. C. and Mieszala, A. and Mohseni, M. and Montazeri, S. and Mount, E. and Movassagh, R. and Mruczkiewicz, W. and Naaman, O. and Neeley, M. and Neill, C. and Nersisyan, A. and Newman, M. and Ng, J. H. and Nguyen, A. and Nguyen, M. and Niu, M. Y. and O’Brien, T. E. and Omonije, S. and Petukhov, A. and Potter, R. and Pryadko, L. P. and Quintana, C. and Rocque, C. and Rubin, N. C. and Saei, N. and Sank, D. and Sankaragomathi, K. and Satzinger, K. J. and Schurkus, H. F. and Schuster, C. and Shearn, M. J. and Shorter, A. and Shutty, N. and Shvarts, V. and Skruzny, J. and Smith, W. C. and Somma, R. and Sterling, G. and Strain, D. and Szalay, M. and Torres, A. and Vidal, G. and Villalonga, B. and Heidweiller, C. V. and White, T. and Woo, B. W. K. and Xing, C. and Yao, Z. J. and Yeh, P. and Yoo, J. and Young, G. and Zalcman, A. and Zhang, Y. and Zhu, N. and Zobrist, N. and Neven, H. and Boixo, S. and Megrant, A. and Kelly, J. and Chen, Y. and Smelyanskiy, V. and Kim, E.-A. and Aleiner, I. and Roushan, P. and {Google Quantum AI and Collaborators}},
  date = {2023},
  journaltitle = {Nature},
  volume = {618},
  number = {7964},
  eprint = {2210.10255},
  eprinttype = {arXiv},
  eprintclass = {quant-ph},
  pages = {264--269},
  doi = {10.1038/s41586-023-05954-4}
}

@article{Apruzzi:2021nmk,
  title = {Symmetry {{TFTs}} from String Theory},
  author = {Apruzzi, Fabio and Bonetti, Federico and García Etxebarria, Iñaki and Hosseini, Saghar S. and Schafer-Nameki, Sakura},
  date = {2023},
  journaltitle = {Commun. Math. Phys.},
  volume = {402},
  number = {1},
  eprint = {2112.02092},
  eprinttype = {arXiv},
  eprintclass = {hep-th},
  pages = {895--949},
  doi = {10.1007/s00220-023-04737-2}
}

@article{Arovas:1984qr,
  title = {Fractional Statistics and the Quantum {{Hall}} Effect},
  author = {Arovas, D. and Schrieffer, J. R. and Wilczek, Frank},
  date = {1984},
  journaltitle = {Phys. Rev. Lett.},
  volume = {53},
  pages = {722--723},
  doi = {10.1103/PhysRevLett.53.722}
}

@book{bakalovLecturesTensorCategories2000,
  author    = "Bakalov, Bojko and Kirillov, Alexander A.",
  title     = "Lectures on tensor categories and modular functors",
  publisher = "American Mathematical Society",
  year      = "2001",
  series    = "University lecture series",
  number    = "21",
  URL       = "https://ci.nii.ac.jp/ncid/BA4965790X"
}

@article{Banerjee:2017ebi,
  title = {Observation of Half-Integer Thermal {{Hall}} Conductance},
  author = {Banerjee, Mitali and Heiblum, Moty and Umansky, Vladimir and Feldman, Dima E. and Oreg, Yuval and Stern, Ady},
  date = {2018},
  journaltitle = {Nature},
  volume = {559},
  number = {7713},
  eprint = {1710.00492},
  eprinttype = {arXiv},
  eprintclass = {cond-mat.mes-hall},
  pages = {205--210},
  doi = {10.1038/s41586-018-0184-1}
}

@article{barkeshliSymmetryFractionalizationDefects2019,
  title = {Symmetry {{Fractionalization}}, {{Defects}}, and {{Gauging}} of {{Topological Phases}}},
  author = {Barkeshli, Maissam and Bonderson, Parsa and Cheng, Meng and Wang, Zhenghan},
  date = {2019-09-20},
  journaltitle = {Phys. Rev. B},
  volume = {100},
  number = {11},
  eprint = {1410.4540},
  eprinttype = {arXiv},
  eprintclass = {cond-mat.str-el},
  pages = {115147},
  doi = {10.1103/PhysRevB.100.115147}
}

@online{bartlettModularCategoriesRepresentations2015a,
  title = {Modular Categories as Representations of the 3-Dimensional Bordism 2-Category},
  author = {Bartlett, Bruce and Douglas, Christopher L. and Schommer-Pries, Christopher J. and Vicary, Jamie},
  date = {2015-09-22},
  eprint = {1509.06811},
  eprinttype = {arXiv},
  eprintclass = {math.AT},
  doi = {10.48550/arXiv.1509.06811},
  pubstate = {prepublished}
}

@article{Bartolomei:2020qfr,
  title = {Fractional Statistics in Anyon Collisions},
  author = {Bartolomei, H. and Kumar, M. and Bisognin, R. and Marguerite, A. and Berroir, J.-M. and Bocquillon, E. and Plaçais, B. and Cavanna, A. and Dong, Q. and Gennser, U. and Jin, Y. and Fève, G.},
  date = {2020},
  journaltitle = {Science},
  volume = {368},
  number = {6487},
  eprint = {2006.13157},
  eprinttype = {arXiv},
  eprintclass = {cond-mat.mes-hall},
  pages = {173--177},
  doi = {10.1126/science.aaz5601}
}

@article{bhardwajStateSumConstructions2017,
  title = {State Sum Constructions of Spin-{{TFTs}} and String Net Constructions of Fermionic Phases of Matter},
  author = {Bhardwaj, Lakshya and Gaiotto, Davide and Kapustin, Anton},
  date = {2017-04},
  journaltitle = {J. High Energ. Phys.},
  volume = {2017},
  number = {4},
  eprint = {1605.01640},
  eprinttype = {arXiv},
  eprintclass = {cond-mat.str-el},
  pages = {96},
  doi = {10.1007/JHEP04(2017)096}
}

@article{bondersonCongruenceSubgroupsSuperModular2018,
  title = {Congruence Subgroups and Super-Modular Categories},
  author = {Bonderson, Parsa and Rowell, Eric and Wang, Zhenghan and Zhang, Qing},
  date = {2018-07},
  journaltitle = {Pacific J. Math.},
  volume = {296},
  number = {2},
  pages = {257--270},
  eprint = {1704.02041},
  eprinttype = {arXiv},
  eprintclass = {math.QA},
  doi = {10.2140/pjm.2018.296.257}
}

@article{bruillard2020classification,
  title = {Classification of Super-Modular Categories by Rank},
  author = {Bruillard, Paul and Galindo, César and Ng, Siu-Hung and Plavnik, Julia Y and Rowell, Eric C and Wang, Zhenghan},
  date = {2020},
  journaltitle = {Algebr. Represent. Theory},
  volume = {23},
  number = {3},
  eprint = {1705.05293},
  eprinttype = {arXiv},
  eprintclass = {math.QA},
  pages = {795--809},
  doi = {10.1007/s10468-019-09873-9}
}

@article{bruillardClassificationModularCategories2015,
  title = {On Classification of Modular Categories by Rank},
  author = {Bruillard, Paul and Ng, Siu-Hung and Rowell, Eric C. and Wang, Zhenghan},
  date = {2016},
  journaltitle = {Int. Math. Res. Not.},
  volume = {2016},
  number = {24},
  eprint = {1507.05139},
  eprinttype = {arXiv},
  eprintclass = {math.QA},
  pages = {7546--7588},
  doi = {10.1093/imrn/rnw020},
  langid = {english}
}

@article{bruillardFermionicModularCategories2017,
  title = {Fermionic {{Modular Categories}} and the 16-Fold {{Way}}},
  author = {Bruillard, Paul and Galindo, Cesar and Hagge, Tobias and Ng, Siu-Hung and Plavnik, Julia Yael and Rowell, Eric C. and Wang, Zhenghan},
  date = {2017-04-01},
  journaltitle = {J. Math. Phys.},
  volume = {58},
  number = {4},
  eprint = {1603.09294},
  eprinttype = {arXiv},
  eprintclass = {math.QA},
  pages = {041704},
  doi = {10.1063/1.4982048}
}

@article{bruillardRankfinitenessModularCategories2015,
  title = {Rank-Finiteness for Modular Categories},
  author = {Bruillard, Paul and Ng, Siu-Hung and Rowell, Eric C. and Wang, Zhenghan},
  date = {2016},
  journaltitle = {J. Amer. Math. Soc.},
  volume = {29},
  number = {3},
  eprint = {1310.7050},
  eprinttype = {arXiv},
  eprintclass = {math.QA},
  pages = {857--881},
  doi = {10.1090/jams/842}
}

@article{Cai:2023signatures,
  title = {Signatures of Fractional Quantum Anomalous {{Hall}} States in Twisted {{MoTe}}{\textsubscript{2}}},
  author = {Cai, Jiaqi and Anderson, Eric and Wang, Chong and Zhang, Xiaowei and Liu, Xiaoyu and Holtzmann, William and Zhang, Yinong and Fan, Fengren and Taniguchi, Takashi and Watanabe, Kenji and Ran, Ying and Cao, Ting and Fu, Liang and Xiao, Di and Yao, Wang and Xu, Xiaodong},
  date = {2023},
  journaltitle = {Nature},
  volume = {622},
  eprint = {2304.08470},
  eprinttype = {arXiv},
  eprintclass = {cond-mat.mes-hall},
  pages = {63--68},
  doi = {10.1038/s41586-023-06289-w}
}

@article{chatterjeeHolographicTheoryContinuous2023,
  title = {Holographic Theory for Continuous Phase Transitions -- the Emergence and Symmetry Protection of Gaplessness},
  author = {Chatterjee, Arkya and Wen, Xiao-Gang},
  date = {2023-08-02},
  journaltitle = {Phys. Rev. B},
  volume = {108},
  number = {7},
  eprint = {2205.06244},
  eprinttype = {arXiv},
  eprintclass = {cond-mat.str-el},
  pages = {075105},
  doi = {10.1103/PhysRevB.108.075105}
}

@article{Chen:2010gda,
  title = {Local Unitary Transformation, Long-Range Quantum Entanglement, Wave Function Renormalization, and Topological Order},
  author = {Chen, Xie and Gu, Zheng-Cheng and Wen, Xiao-Gang},
  date = {2010},
  journaltitle = {Phys. Rev. B},
  volume = {82},
  eprint = {1004.3835},
  eprinttype = {arXiv},
  eprintclass = {cond-mat.str-el},
  pages = {155138},
  doi = {10.1103/PhysRevB.82.155138}
}

@article{choClassificationFermionicTopological2023,
  title = {Classification of {{Fermionic Topological Orders}} from {{Congruence Representations}}},
  author = {Cho, Gil Young and Kim, Hee-cheol and Seo, Donghae and You, Minyoung},
  date = {2023-09-01},
  journaltitle = {Phys. Rev. B},
  volume = {108},
  number = {11},
  eprint = {2210.03681},
  eprinttype = {arXiv},
  eprintclass = {cond-mat.str-el},
  pages = {115103},
  doi = {10.1103/PhysRevB.108.115103}
}

@article{Coste:1993af,
  title = {Remarks on {{Galois}} Symmetry in Rational Conformal Field Theories},
  author = {Coste, A. and Gannon, T.},
  date = {1994},
  journaltitle = {Phys. Lett. B},
  volume = {323},
  pages = {316--321},
  doi = {10.1016/0370-2693(94)91226-2}
}

@article{davydovStructureWittGroup2011,
  title = {On the Structure of the {{Witt}} Group of Braided Fusion Categories},
  author = {Davydov, Alexei and Nikshych, Dmitri and Ostrik, Victor},
  date = {2013},
  journaltitle = {Selecta Math. (N.S.)},
  volume = {19},
  number = {1},
  eprint = {1109.5558},
  eprinttype = {arXiv},
  eprintclass = {math.QA},
  pages = {237--269},
  doi = {10.1007/s00029-012-0093-3}
}

@article{davydovWittGroupNondegenerate2011,
  title = {The {{Witt}} Group of Non-Degenerate Braided Fusion Categories},
  author = {Davydov, Alexei and Müger, Michael and Nikshych, Dmitri and Ostrik, Victor},
  date = {2013},
  journaltitle = {J. Reine Angew. Math.},
  volume = {677},
  eprint = {1009.2117},
  eprinttype = {arXiv},
  eprintclass = {math.QA},
  pages = {135--177},
  doi = {10.1515/crelle.2012.014}
}

@article{deligneCategoriesTensorielles2002,
  title = {Catégories tensorielles},
  author = {Deligne, P.},
  date = {2002},
  journaltitle = {Mosc. Math. J.},
  volume = {2},
  number = {2},
  pages = {227--248},
  doi = {10.17323/1609-4514-2002-2-2-227-248},
  langid = {french}
}

@article{dijkgraafTopologicalGaugeTheories1990,
  title = {Topological Gauge Theories and Group Cohomology},
  author = {Dijkgraaf, Robbert and Witten, Edward},
  date = {1990-01},
  journaltitle = {Commun. Math. Phys.},
  volume = {129},
  number = {2},
  pages = {393--429},
  publisher = {Springer},
  doi = {10.1007/BF02096988},
  langid = {english}
}

@article{dongCongruencePropertyConformal2015,
  title = {Congruence {{Property In Conformal Field Theory}}},
  author = {Dong, Chongying and Lin, Xingjun and Ng, Siu-Hung},
  date = {2015-11-04},
  journaltitle = {Algebra Number Theory},
  volume = {9},
  number = {9},
  eprint = {1201.6644},
  eprinttype = {arXiv},
  eprintclass = {math.QA},
  pages = {2121--2166},
  doi = {10.2140/ant.2015.9.2121}
}

@article{drinfeldBraidedFusionCategories2009,
  title = {On Braided Fusion Categories {{I}}},
  author = {Drinfeld, Vladimir and Gelaki, Shlomo and Nikshych, Dmitri and Ostrik, Victor},
  date = {2010},
  journaltitle = {Selecta Math. (N.S.)},
  volume = {16},
  number = {1},
  eprint = {0906.0620},
  eprinttype = {arXiv},
  eprintclass = {math.QA},
  pages = {1--119},
  doi = {10.1007/s00029-010-0017-z},
  langid = {english}
}

@article{etingofFusionCategories2017,
  title = {On Fusion Categories},
  author = {Etingof, Pavel and Nikshych, Dmitri and Ostrik, Viktor},
  date = {2005},
  journaltitle = {Ann. of Math.},
  volume = {162},
  number = {2},
  eprint = {math/0203060},
  eprinttype = {arXiv},
  pages = {581--642},
  doi = {10.4007/annals.2005.162.581}
}

@book{etingofTensorCategories2015,
  title = {Tensor Categories},
  author = {Etingof, Pavel and Gelaki, Shlomo and Nikshych, Dmitri and Ostrik, Victor},
  date = {2015},
  series = {Mathematical {{Surveys}} and {{Monographs}}},
  number = {205},
  publisher = {American Mathematical Society},
  location = {Providence, Rhode Island},
  doi = {10.1090/surv/205},
  langid = {english},
  pagetotal = {343}
}

@article{evansExoticnessRealisabilityTwisted2011,
  title = {The Exoticness and Realisability of Twisted {{Haagerup-Izumi}} Modular Data},
  author = {Evans, David E. and Gannon, Terry},
  date = {2011-10},
  journaltitle = {Commun. Math. Phys.},
  volume = {307},
  number = {2},
  eprint = {1006.1326},
  eprinttype = {arXiv},
  eprintclass = {math.QA},
  pages = {463--512},
  doi = {10.1007/s00220-011-1329-3},
  langid = {english}
}

@article{Freed:2022qnc,
  title = {Topological Symmetry in Quantum Field Theory},
  author = {Freed, Daniel S. and Moore, Gregory W. and Teleman, Constantin},
  date = {2024},
  journaltitle = {Quantum Topol.},
  volume = {15},
  eprint = {2209.07471},
  eprinttype = {arXiv},
  eprintclass = {hep-th},
  pages = {779--869},
  doi = {10.4171/qt/223}
}

@article{Freedman:2001eqc,
  title = {Topological Quantum Computation},
  author = {Freedman, Michael H. and Kitaev, Alexei and Larsen, Michael J. and Wang, Zhenghan},
  date = {2003},
  journaltitle = {Bull. Amer. Math. Soc.},
  volume = {40},
  eprint = {quant-ph/0101025},
  eprinttype = {arXiv},
  pages = {31--38},
  doi = {10.1090/S0273-0979-02-00964-3}
}

@article{Fuchs:1996dd,
  title = {A matrix $S$ for all simple current extensions},
  author = {Fuchs, Jürgen and Schellekens, A. N. and Schweigert, Christoph},
  date = {1996},
  journaltitle = {Nucl. Phys. B},
  volume = {473},
  number = {1-2},
  pages = {323--366},
  eprint = {hep-th/9601078},
  eprinttype = {arXiv},
  doi = {10.1016/0550-3213(96)00247-7}
}

@article{Gaiotto:2020iye,
  title = {Orbifold Groupoids},
  author = {Gaiotto, Davide and Kulp, Justin},
  date = {2021},
  journaltitle = {J. High Energ. Phys.},
  volume = {02},
  eprint = {2008.05960},
  eprinttype = {arXiv},
  eprintclass = {hep-th},
  pages = {132},
  doi = {10.1007/JHEP02(2021)132}
}

@article{gaiottoSpinTQFTsFermionic2016,
  title = {Spin {{TQFTs}} and Fermionic Phases of Matter},
  author = {Gaiotto, Davide and Kapustin, Anton},
  date = {2016-10-20},
  journaltitle = {Int. J. Mod. Phys. A},
  volume = {31},
  eprint = {1505.05856},
  eprinttype = {arXiv},
  eprintclass = {cond-mat.str-el},
  pages = {1645044},
  doi = {10.1142/S0217751X16450445},
  issue = {28n29}
}

@article{gannonAlgebraicNumberFields2019,
  title = {Algebraic Number Fields Generated by Dimensions in Fusion Rings},
  author = {Gannon, Terry and Schopieray, Andrew},
  date = {2024},
  journaltitle = {Commun. Number Theory Phys.},
  volume = {18},
  number = {4},
  eprint = {1912.12260},
  eprinttype = {arXiv},
  eprintclass = {math.QA},
  pages = {705--743},
  doi = {10.4310/cntp.241202233757},
  langid = {english}
}

@online{gannonExoticQuantumSubgroups2023,
  title = {Exotic Quantum Subgroups and Extensions of Affine {{Lie}} Algebra {{VOAs}} -- Part {{I}}},
  author = {Gannon, Terry},
  date = {2023-01-18},
  eprint = {2301.07287},
  eprinttype = {arXiv},
  eprintclass = {math.QA},
  doi = {10.48550/arXiv.2301.07287},
  langid = {english},
  pubstate = {prepublished}
}

@article{Halperin:1984fn,
  title = {Statistics of Quasiparticles and the Hierarchy of Fractional Quantized {{Hall}} States},
  author = {Halperin, B. I.},
  date = {1984},
  journaltitle = {Phys. Rev. Lett.},
  volume = {52},
  pages = {1583--1586},
  doi = {10.1103/PhysRevLett.52.1583},
  addendum = {Erratum: Phys. Rev. Lett. 52, 2390 (1984)}
}

@article{haltonEfficiencyCertainQuasirandom1960,
  title = {On the Efficiency of Certain Quasi-Random Sequences of Points in Evaluating Multi-Dimensional Integrals},
  author = {Halton, J. H.},
  date = {1960-12-01},
  journaltitle = {Numer. Math.},
  volume = {2},
  number = {1},
  pages = {84--90},
  doi = {10.1007/BF01386213},
  langid = {english}
}

@article{hongEXOTICMODULARTENSOR2008,
  title = {On Exotic Modular Tensor Categories},
  author = {Hong, Seung-Moon and Rowell, Eric and Wang, Zhenghan},
  date = {2008-11},
  journaltitle = {Commun. Contemp. Math.},
  volume = {10},
  eprint = {0710.5761},
  eprinttype = {arXiv},
  eprintclass = {math.GT},
  pages = {1049--1074},
  doi = {10.1142/S0219199708003162},
  issue = {supp01},
  langid = {english}
}

@article{Iqbal:2023wvm,
  title = {Non-{{Abelian}} Topological Order and Anyons on a Trapped-Ion Processor},
  author = {Iqbal, Mohsin and Tantivasadakarn, Nathanan and Verresen, Ruben and Campbell, Sara L. and Dreiling, Joan M. and Figgatt, Caroline and Gaebler, John P. and Johansen, Jacob and Mills, Michael and Moses, Steven A. and Pino, Juan M. and Ransford, Anthony and Rowe, Mary and Siegfried, Peter and Stutz, Russell P. and Foss-Feig, Michael and Vishwanath, Ashvin and Dreyer, Henrik},
  date = {2024},
  journaltitle = {Nature},
  volume = {626},
  number = {7999},
  eprint = {2305.03766},
  eprinttype = {arXiv},
  eprintclass = {quant-ph},
  pages = {505--511},
  doi = {10.1038/s41586-023-06934-4}
}

@article{izumiStructureSectorsAssociatedLongo2000,
  title = {The {{Structure}} of {{Sectors Associated}} with {{Longo}}–{{Rehren Inclusions I}}. {{General Theory}}},
  author = {Izumi, Masaki},
  date = {2000-09-01},
  journaltitle = {Commun. Math. Phys.},
  volume = {213},
  number = {1},
  pages = {127--179},
  doi = {10.1007/s002200000234},
  langid = {english}
}

@article{johnson-freydMinimalNondegenerateExtensions2023,
  title = {Minimal Nondegenerate Extensions},
  author = {Johnson-Freyd, Theo and Reutter, David},
  date = {2024},
  journaltitle = {J. Amer. Math. Soc.},
  volume = {37},
  number = {1},
  eprint = {2105.15167},
  eprinttype = {arXiv},
  eprintclass = {math.QA},
  pages = {81--150},
  doi = {10.1090/jams/1023}
}

@article{jonesRankfinitenessGcrossedBraided2019,
  title = {Rank-Finiteness for {{G}}-Crossed Braided Fusion Categories},
  author = {Jones, Corey and Morrison, Scott and Nikshych, Dmitri and Rowell, Eric C.},
  date = {2021},
  journaltitle = {Transform. Groups},
  volume = {26},
  number = {3},
  eprint = {1902.06165},
  eprinttype = {arXiv},
  eprintclass = {math.QA},
  pages = {915--927},
  doi = {10.1007/s00031-020-09576-2}
}

@article{Kaidi:2021ent,
  title = {Holomorphic Modular Bootstrap Revisited},
  author = {Kaidi, Justin and Lin, Ying-Hsuan and Parra-Martinez, Julio},
  date = {2021},
  journaltitle = {J. High Energ. Phys.},
  volume = {12},
  eprint = {2107.13557},
  eprinttype = {arXiv},
  eprintclass = {hep-th},
  pages = {151},
  doi = {10.1007/JHEP12(2021)151}
}

@article{Kaidi:2022cpf,
  title = {Symmetry {{TFTs}} for Non-Invertible Defects},
  author = {Kaidi, Justin and Ohmori, Kantaro and Zheng, Yunqin},
  date = {2023},
  journaltitle = {Commun. Math. Phys.},
  volume = {404},
  number = {2},
  eprint = {2209.11062},
  eprinttype = {arXiv},
  eprintclass = {hep-th},
  pages = {1021--1124},
  doi = {10.1007/s00220-023-04859-7}
}

@article{kaidiHigherCentralCharges2022,
  title = {Higher Central Charges and Topological Boundaries in 2+1-Dimensional {{TQFTs}}},
  author = {Kaidi, Justin and Komargodski, Zohar and Ohmori, Kantaro and Seifnashri, Sahand and Shao, Shu-Heng},
  date = {2022-09-26},
  journaltitle = {SciPost Phys.},
  volume = {13},
  number = {3},
  eprint = {2107.13091},
  eprinttype = {arXiv},
  eprintclass = {hep-th},
  pages = {067},
  doi = {10.21468/SciPostPhys.13.3.067}
}

@article{kimingLiftsProjectiveCongruence2011,
  title = {Lifts of Projective Congruence Groups},
  author = {Kiming, Ian and Schuett, Matthias and Verrill, Helena},
  date = {2011-02},
  journaltitle = {Journal of the London Mathematical Society},
  volume = {83},
  number = {1},
  eprint = {0905.4798},
  eprinttype = {arXiv},
  eprintclass = {math.NT},
  pages = {96--120},
  doi = {10.1112/jlms/jdq062}
}

@article{Kitaev:1997wr,
  title = {Fault-Tolerant Quantum Computation by Anyons},
  author = {family=Kitaev, given=A. Yu., given-i=A{{Yu}}},
  date = {2003},
  journaltitle = {Ann. Phys.},
  volume = {303},
  eprint = {quant-ph/9707021},
  eprinttype = {arXiv},
  pages = {2--30},
  doi = {10.1016/S0003-4916(02)00018-0}
}

@article{Kitaev:2011dxc,
  title = {Models for Gapped Boundaries and Domain Walls},
  author = {Kitaev, Alexei and Kong, Liang},
  date = {2012},
  journaltitle = {Commun. Math. Phys.},
  volume = {313},
  number = {2},
  eprint = {1104.5047},
  eprinttype = {arXiv},
  eprintclass = {cond-mat.str-el},
  pages = {351--373},
  doi = {10.1007/s00220-012-1500-5}
}

@article{kitaevAnyonsExactlySolved2006,
  title = {Anyons in an Exactly Solved Model and Beyond},
  author = {Kitaev, Alexei},
  date = {2006-01},
  journaltitle = {Annals of Physics},
  volume = {321},
  number = {1},
  eprint = {cond-mat/0506438},
  eprinttype = {arXiv},
  pages = {2--111},
  doi = {10.1016/j.aop.2005.10.005}
}

@article{Komargodski:2017keh,
  title = {A Symmetry Breaking Scenario for {{QCD}}$_3$},
  author = {Komargodski, Zohar and Seiberg, Nathan},
  date = {2018},
  journaltitle = {J. High Energ. Phys.},
  volume = {01},
  eprint = {1706.08755},
  eprinttype = {arXiv},
  eprintclass = {hep-th},
  pages = {109},
  doi = {10.1007/JHEP01(2018)109}
}

@article{kongAnyonCondensationTensor2021,
  title = {Anyon Condensation and Tensor Categories},
  author = {Kong, Liang},
  date = {2014},
  journaltitle = {Nucl. Phys. B},
  volume = {886},
  eprint = {1307.8244},
  eprinttype = {arXiv},
  eprintclass = {cond-mat.str-el},
  pages = {436--482},
  doi = {10.1016/j.nuclphysb.2014.07.003},
  addendum = {Erratum: Nucl. Phys. B 973 (2021) 115607}
}

@article{Lan:2014uaa,
  title = {Gapped Domain Walls, Gapped Boundaries, and Topological Degeneracy},
  author = {Lan, Tian and Wang, Juven C. and Wen, Xiao-Gang},
  date = {2015},
  journaltitle = {Phys. Rev. Lett.},
  volume = {114},
  number = {7},
  eprint = {1408.6514},
  eprinttype = {arXiv},
  eprintclass = {cond-mat.str-el},
  pages = {076402},
  doi = {10.1103/PhysRevLett.114.076402}
}

@article{Lan:2016seg,
  title = {Classification of (2+1)-Dimensional Topological Order and Symmetry-Protected Topological Order for Bosonic and Fermionic Systems with on-Site Symmetries},
  author = {Lan, Tian and Kong, Liang and Wen, Xiao-Gang},
  date = {2017},
  journaltitle = {Phys. Rev. B},
  volume = {95},
  number = {23},
  eprint = {1602.05946},
  eprinttype = {arXiv},
  eprintclass = {cond-mat.str-el},
  pages = {235140},
  doi = {10.1103/PhysRevB.95.235140}
}

@article{lanTheory2+1DFermionic2016,
  title = {Theory of (2+1)-Dimensional Fermionic Topological Orders and Fermionic/Bosonic Topological Orders with Symmetries},
  author = {Lan, Tian and Kong, Liang and Wen, Xiao-Gang},
  date = {2016-10-10},
  journaltitle = {Phys. Rev. B},
  volume = {94},
  number = {15},
  eprint = {1507.04673},
  eprinttype = {arXiv},
  eprintclass = {cond-mat.str-el},
  pages = {155113},
  doi = {10.1103/PhysRevB.94.155113}
}

@article{Laughlin:1983fy,
  title = {Anomalous Quantum {{Hall}} Effect: {{An}} Incompressible Quantum Fluid with Fractionally Charged Excitations},
  author = {Laughlin, R. B.},
  date = {1983},
  journaltitle = {Phys. Rev. Lett.},
  volume = {50},
  pages = {1395--1398},
  doi = {10.1103/PhysRevLett.50.1395}
}

@article{Leinaas:1977fm,
  title = {On the Theory of Identical Particles},
  author = {Leinaas, J. M. and Myrheim, J.},
  date = {1977},
  journaltitle = {Nuovo Cim. B},
  volume = {37},
  pages = {1--23},
  doi = {10.1007/BF02727953}
}

@article{Lu:2023zow,
  title = {Fractional Quantum Anomalous {{Hall}} Effect in Multilayer Graphene},
  author = {Lu, Zhengguang and Han, Tonghang and Yao, Yuxuan and Reddy, Aidan P. and Yang, Jixiang and Seo, Junseok and Watanabe, Kenji and Taniguchi, Takashi and Fu, Liang and Ju, Long},
  date = {2024},
  journaltitle = {Nature},
  volume = {626},
  eprint = {2309.17436},
  eprinttype = {arXiv},
  eprintclass = {cond-mat.mes-hall},
  pages = {759--764},
  doi = {10.1038/s41586-023-07010-7}
}

@article{Mathur:1988gt,
  title = {Reconstruction of Conformal Field Theories from Modular Geometry on the Torus},
  author = {Mathur, Samir D. and Mukhi, Sunil and Sen, Ashoke},
  date = {1989},
  journaltitle = {Nucl. Phys. B},
  volume = {318},
  number = {2},
  pages = {483--540},
  doi = {10.1016/0550-3213(89)90615-9}
}

@article{Mathur:1988na,
  title = {On the Classification of Rational Conformal Field Theories},
  author = {Mathur, Samir D. and Mukhi, Sunil and Sen, Ashoke},
  date = {1988},
  journaltitle = {Phys. Lett. B},
  volume = {213},
  number = {3},
  pages = {303--308},
  doi = {10.1016/0370-2693(88)91765-0}
}

@online{maurerFsymbolsRsymbolsDrinfeld2026,
  title = {F-Symbols and {{R}}-Symbols for the {{Drinfeld}} Center of the {{Haagerup}} Subfactor},
  author = {Mäurer, Fabian and Thiel, Ulrich and Vercleyen, Gert},
  date = {2026-01-27},
  eprint = {2601.20012},
  eprinttype = {arXiv},
  eprintclass = {math.QA},
  doi = {10.48550/arXiv.2601.20012},
  pubstate = {prepublished}
}

@article{mignardModularCategoriesAre2021,
  title = {Modular Categories Are Not Determined by Their Modular Data},
  author = {Mignard, Michaël and Schauenburg, Peter},
  date = {2021},
  journaltitle = {Lett. Math. Phys.},
  volume = {111},
  number = {3},
  eprint = {1708.02796},
  eprinttype = {arXiv},
  eprintclass = {math.QA},
  pages = {60},
  doi = {10.1007/s11005-021-01395-0}
}

@online{ModularTensorCategories,
  title = {Modular {{Tensor Categories}} from {{Chern-Simons Plus Simple Current Condensation}}},
  author = {Verhaeghe, Alexander and Simon, Steven H.},
  note = {Unpublished, \url{https://www-thphys.physics.ox.ac.uk/people/SteveSimon/mtcfromkac.html}}
}

@article{Moore:1988qv,
  title = {Classical and Quantum Conformal Field Theory},
  author = {Moore, Gregory W. and Seiberg, Nathan},
  date = {1989},
  journaltitle = {Commun. Math. Phys.},
  volume = {123},
  pages = {177--254},
  doi = {10.1007/BF01238857}
}

@article{Moore:1991ks,
  title = {Nonabelions in the Fractional Quantum {{Hall}} Effect},
  author = {Moore, Gregory W. and Read, N.},
  date = {1991},
  journaltitle = {Nucl. Phys. B},
  volume = {360},
  pages = {362--396},
  doi = {10.1016/0550-3213(91)90407-O}
}

@article{mooreTamingConformalZoo1989,
  title = {Taming the {{Conformal Zoo}}},
  author = {Moore, Gregory W. and Seiberg, Nathan},
  date = {1989},
  journaltitle = {Phys. Lett. B},
  volume = {220},
  pages = {422--430},
  doi = {10.1016/0370-2693(89)90897-6}
}

@incollection{muegerModularCategories2012,
  title = {Modular {{Categories}}},
  booktitle = {Quantum {{Physics}} and {{Linguistics}}: {{A Compositional}}, {{Diagrammatic Discourse}}},
  author = {Müger, Michael},
  editor = {Heunen, Chris and Sadrzadeh, Mehrnoosh and Grefenstette, Edward},
  date = {2013},
  eprint = {1201.6593},
  eprinttype = {arXiv},
  eprintclass = {math.CT},
  pages = {146--183},
  publisher = {Oxford University Press},
  location = {Oxford},
  doi = {10.1093/acprof:oso/9780199646296.003.0006}
}

@inproceedings{Mukhi:2019xjy,
  title = {Classification of {{RCFT}} from Holomorphic Modular Bootstrap: {{A}} Status Report},
  booktitle = {Pollica {{Summer Workshop}} 2019: {{Mathematical}} and {{Geometric Tools}} for {{Conformal Field Theories}}},
  author = {Mukhi, Sunil},
  date = {2019},
  eprint = {1910.02973},
  eprinttype = {arXiv},
  eprintclass = {hep-th}
}

@article{Nakamura:2020mok,
  title = {Direct Observation of Anyonic Braiding Statistics},
  author = {Nakamura, James and Liang, Shuang and Gardner, Geoffrey C. and Manfra, Michael J.},
  date = {2020},
  journaltitle = {Nature Phys.},
  volume = {16},
  number = {9},
  eprint = {2006.14115},
  eprinttype = {arXiv},
  eprintclass = {cond-mat.mes-hall},
  pages = {931--936},
  doi = {10.1038/s41567-020-1019-1}
}

@article{Nayak:2008zza,
  title = {Non-{{Abelian}} Anyons and Topological Quantum Computation},
  author = {Nayak, Chetan and Simon, Steven H. and Stern, Ady and Freedman, Michael and Das Sarma, Sankar},
  date = {2008},
  journaltitle = {Rev. Mod. Phys.},
  volume = {80},
  eprint = {0707.1889},
  eprinttype = {arXiv},
  eprintclass = {cond-mat.str-el},
  pages = {1083--1159},
  doi = {10.1103/RevModPhys.80.1083}
}

@online{ngClassificationModularData2025,
  title = {Classification of Modular Data up to Rank 12},
  author = {Ng, Siu-Hung and Rowell, Eric C. and Wen, Xiao-Gang},
  date = {2025-07-03},
  eprint = {2308.09670},
  eprinttype = {arXiv},
  eprintclass = {math.QA},
  doi = {10.48550/arXiv.2308.09670},
  pubstate = {prepublished}
}

@article{ngCongruenceSubgroupsGeneralized2010,
  title = {Congruence {{Subgroups}} and {{Generalized Frobenius-Schur Indicators}}},
  author = {Ng, Siu-Hung and Schauenburg, Peter},
  date = {2010-11},
  journaltitle = {Commun. Math. Phys.},
  volume = {300},
  number = {1},
  eprint = {0806.2493},
  eprinttype = {arXiv},
  eprintclass = {math.QA},
  pages = {1--46},
  doi = {10.1007/s00220-010-1096-6}
}

@article{ngHigherCentralCharges2022,
  title = {Higher Central Charges and {{Witt}} Groups},
  author = {Ng, Siu-Hung and Rowell, Eric C. and Wang, Yilong and Zhang, Qing},
  date = {2022-08},
  journaltitle = {Advances in Mathematics},
  volume = {404},
  eprint = {2002.03570},
  eprinttype = {arXiv},
  eprintclass = {math.QA},
  pages = {108388},
  doi = {10.1016/j.aim.2022.108388}
}

@incollection{ngHigherFrobeniusSchurIndicators2007,
  title = {Higher {Frobenius-Schur} indicators for pivotal categories},
  author = {Ng, Siu-Hung and Schauenburg, Peter},
  booktitle = {Hopf Algebras and Generalizations},
  series = {Contemporary Mathematics},
  volume = {441},
  publisher = {American Mathematical Society},
  year = {2007},
  pages = {63--90},
  ISBN = {9780821881200},
  ISSN = {0271-4132},
  DOI = {10.1090/conm/441/08500},
  url = {https://doi.org/10.1090/conm/441/08500}
}

@article{ngHigherGaussSums2019,
  title = {Higher {{Gauss}} Sums of Modular Categories},
  author = {Ng, Siu-Hung and Schopieray, Andrew and Wang, Yilong},
  date = {2019-10},
  journaltitle = {Sel. Math. New Ser.},
  volume = {25},
  number = {4},
  eprint = {1812.11234},
  eprinttype = {arXiv},
  eprintclass = {math.QA},
  pages = {53},
  doi = {10.1007/s00029-019-0499-2}
}

@article{ngReconstructionModularData2023,
  title = {Reconstruction of Modular Data from $\mathrm{SL}_2(\mathbb{Z})$ Representations},
  author = {Ng, Siu-Hung and Rowell, Eric C. and Wang, Zhenghan and Wen, Xiao-Gang},
  date = {2023-09},
  journaltitle = {Commun. Math. Phys.},
  volume = {402},
  number = {3},
  eprint = {2203.14829},
  eprinttype = {arXiv},
  eprintclass = {math.QA},
  pages = {2465--2545},
  doi = {10.1007/s00220-023-04775-w}
}

@online{ngRecoveringRsymbolsModular2024,
  title = {Recovering {{R}}-Symbols from Modular Data},
  author = {Ng, Siu-Hung and Rowell, Eric C. and Wen, Xiao-Gang},
  date = {2024-08-05},
  eprint = {2408.02748},
  eprinttype = {arXiv},
  eprintclass = {math.QA},
  doi = {10.48550/arXiv.2408.02748},
  pubstate = {prepublished}
}

@article{ngSymmetricRepresentations$textSL_2mathbbZ$2023,
  title = {On Symmetric Representations of $\mathrm{SL}_2(\mathbb{Z})$},
  author = {Ng, Siu-Hung and Wang, Yilong and Wilson, Samuel},
  date = {2023-01-26},
  journaltitle = {Proc. Amer. Math. Soc.},
  volume = {151},
  number = {4},
  eprint = {2203.15701},
  eprinttype = {arXiv},
  eprintclass = {math.QA},
  pages = {1415--1431},
  doi = {10.1090/proc/16205}
}

@article{nobsIrreduziblenDarstellungenGruppen1976,
  title = {Die irreduziblen Darstellungen der Gruppen $\mathrm{SL}_2(\mathbb{Z}_p)$, insbesondere $\mathrm{SL}_2(\mathbb{Z}_2)$. I. Teil},
  author = {Nobs, Alexandre},
  date = {1976-12-01},
  journaltitle = {Comment. Math. Helv.},
  volume = {51},
  number = {1},
  pages = {465--489},
  doi = {10.1007/BF02568170},
  langid = {ngerman}
}

@article{nobsIrreduziblenDarstellungenGruppen1976b,
  title = {Die irreduziblen Darstellungen der Gruppen $\mathrm{SL}_2(\mathbb{Z}_p)$, insbesondere $\mathrm{SL}_2(\mathbb{Z}_2)$. II. Teil},
  author = {Nobs, Alexandre and Wolfart, Jürgen},
  date = {1976-12-01},
  journaltitle = {Comment. Math. Helv.},
  volume = {51},
  number = {1},
  pages = {491--526},
  doi = {10.1007/BF02568171},
  langid = {ngerman}
}

@article{ostrikModuleCategoriesWeak2001,
  title = {Module Categories, Weak {{Hopf}} Algebras and Modular Invariants},
  author = {Ostrik, Viktor},
  date = {2003},
  journaltitle = {Transform. Groups},
  volume = {8},
  number = {2},
  eprint = {math/0111139},
  eprinttype = {arXiv},
  pages = {177--206},
  doi = {10.1007/s00031-003-0515-6}
}

@inproceedings{owenRandomlyPermutedTmsNets1995,
  title = {Randomly {{Permuted}} (t,m,s)-{{Nets}} and (t, s)-{{Sequences}}},
  booktitle = {Monte {{Carlo}} and {{Quasi-Monte Carlo Methods}} in {{Scientific Computing}}},
  author = {Owen, Art B.},
  editor = {Niederreiter, Harald and Shiue, Peter Jau-Shyong},
  date = {1995},
  series = {Lecture {{Notes}} in {{Statistics}}},
  number = {106},
  pages = {299--317},
  publisher = {Springer},
  location = {New York, NY},
  doi = {10.1007/978-1-4612-2552-2_19},
  langid = {english}
}

@article{Park:2023observation,
  title = {Observation of Fractionally Quantized Anomalous {{Hall}} Effect},
  author = {Park, Heonjoon and Cai, Jiaqi and Anderson, Eric and Zhang, Yinong and Zhu, Jiayi and Liu, Xiaoyu and Wang, Chong and Holtzmann, William and Hu, Chaowei and Liu, Zhaoyu and Taniguchi, Takashi and Watanabe, Kenji and Chu, Jiun-Haw and Cao, Ting and Fu, Liang and Yao, Wang and Chang, Cui-Zu and Cobden, David and Xiao, Di and Xu, Xiaodong},
  date = {2023},
  journaltitle = {Nature},
  volume = {622},
  eprint = {2308.02657},
  eprinttype = {arXiv},
  eprintclass = {cond-mat.mes-hall},
  pages = {74--79},
  doi = {10.1038/s41586-023-06536-0}
}

@article{reshetikhinInvariants3manifoldsLink1991,
  title = {Invariants of 3-Manifolds via Link Polynomials and Quantum Groups},
  author = {Reshetikhin, N. and Turaev, V. G.},
  date = {1991-12},
  journaltitle = {Invent Math},
  volume = {103},
  number = {1},
  pages = {547--597},
  doi = {10.1007/BF01239527},
  langid = {english}
}

@article{Rowell:2007dge,
  title = {On Classification of Modular Tensor Categories},
  author = {Rowell, Eric and Stong, Richard and Wang, Zhenghan},
  date = {2009},
  journaltitle = {Commun. Math. Phys.},
  volume = {292},
  number = {2},
  eprint = {0712.1377},
  eprinttype = {arXiv},
  eprintclass = {math.QA},
  pages = {343--389},
  doi = {10.1007/s00220-009-0908-z}
}

@article{rowellMathematicsTopologicalQuantum2018,
  title = {Mathematics of Topological Quantum Computing},
  author = {Rowell, Eric and Wang, Zhenghan},
  date = {2018-01-16},
  journaltitle = {Bull. Amer. Math. Soc.},
  volume = {55},
  number = {2},
  eprint = {1705.06206},
  eprinttype = {arXiv},
  eprintclass = {math.QA},
  pages = {183--238},
  doi = {10.1090/bull/1605},
  langid = {english}
}

@article{schopierayPrimeDecompositionModular2018,
  title = {Prime Decomposition of Modular Tensor Categories of Local Modules of Type {{D}}},
  author = {Schopieray, Andrew},
  date = {2020},
  journaltitle = {Quantum Topol.},
  volume = {11},
  number = {3},
  eprint = {1810.09057},
  eprinttype = {arXiv},
  eprintclass = {math.QA},
  pages = {489--524},
  doi = {10.4171/QT/140},
  langid = {english}
}

@article{seibergDualityWeb2+12016,
  title = {A {{Duality Web}} in 2+1 {{Dimensions}} and {{Condensed Matter Physics}}},
  author = {Seiberg, Nathan and Senthil, T. and Wang, Chong and Witten, Edward},
  date = {2016-11},
  journaltitle = {Annals of Physics},
  volume = {374},
  eprint = {1606.01989},
  eprinttype = {arXiv},
  eprintclass = {hep-th},
  pages = {395--433},
  doi = {10.1016/j.aop.2016.08.007}
}

@article{slagleQuantumFieldTheory2017,
  title = {Quantum {{Field Theory}} of {{X-Cube Fracton Topological Order}} and {{Robust Degeneracy}} from {{Geometry}}},
  author = {Slagle, Kevin and Kim, Yong Baek},
  date = {2017-11-20},
  journaltitle = {Phys. Rev. B},
  volume = {96},
  number = {19},
  eprint = {1708.04619},
  eprinttype = {arXiv},
  eprintclass = {cond-mat.str-el},
  pages = {195139},
  doi = {10.1103/PhysRevB.96.195139}
}

@article{thorngrenAnomaliesBosonization2020,
  title = {Anomalies and {{Bosonization}}},
  author = {Thorngren, Ryan},
  date = {2020-09},
  journaltitle = {Commun. Math. Phys.},
  volume = {378},
  number = {3},
  eprint = {1810.04414},
  eprinttype = {arXiv},
  eprintclass = {cond-mat.str-el},
  pages = {1775--1816},
  doi = {10.1007/s00220-020-03830-0},
  langid = {english}
}

@article{turaevModularCategories3Manifold1992,
  title = {Modular {{Categories}} and 3-{{Manifold Invariants}}},
  author = {Turaev, Vladimir G.},
  date = {1992-06},
  journaltitle = {Int. J. Mod. Phys. B},
  volume = {06},
  pages = {1807--1824},
  doi = {10.1142/S0217979292000876},
  issue = {11n12},
  langid = {english}
}

@article{vafaClassificationConformalTheories1988,
  title = {Toward Classification of Conformal Theories},
  author = {Vafa, Cumrun},
  date = {1988-05-01},
  journaltitle = {Phys. Lett. B},
  volume = {206},
  pages = {421--426},
  publisher = {Elsevier},
  doi = {10.1016/0370-2693(88)91603-6}
}

@article{vercleyenLowRankMultiplicityFreeComplex2026,
  title = {On Low-Rank Multiplicity-Free Complex Fusion Categories},
  author = {Vercleyen, Gert and Slingerland, Joost},
  date = {2026},
  journaltitle = {Commun. Math. Phys.},
  volume = {407},
  number = {11},
  pages = {231},
  eprint = {2405.20075},
  eprinttype = {arXiv},
  eprintclass = {math-ph},
  doi = {10.1007/s00220-026-05705-2}
}

@article{vercleyenLowRankFusion2023,
  title = {On Low Rank Fusion Rings},
  author = {Vercleyen, Gert and Slingerland, Joost K.},
  date = {2023},
  journaltitle = {J. Math. Phys.},
  volume = {64},
  number = {9},
  pages = {091703},
  eprint = {2205.15637},
  eprinttype = {arXiv},
  eprintclass = {math-ph},
  doi = {10.1063/5.0148848}
}

@software{vercleyenAnyonWiki2023,
  title = {{{AnyonWiki}}},
  author = {Vercleyen, Gert and Slingerland, Joost},
  date = {2023},
  version = {1.0.2},
  publisher = {Zenodo},
  doi = {10.5281/zenodo.15863113},
  url = {https://anyonwiki.org}
}

@article{vijayFractonTopologicalOrder2016,
  title = {Fracton {{Topological Order}}, {{Generalized Lattice Gauge Theory}} and {{Duality}}},
  author = {Vijay, Sagar and Haah, Jeongwan and Fu, Liang},
  date = {2016-12-28},
  journaltitle = {Phys. Rev. B},
  volume = {94},
  number = {23},
  eprint = {1603.04442},
  eprinttype = {arXiv},
  eprintclass = {cond-mat.str-el},
  pages = {235157},
  doi = {10.1103/PhysRevB.94.235157}
}

@article{Wen:1989iv,
  title = {Topological Orders in Rigid States},
  author = {Wen, X. G.},
  date = {1990},
  journaltitle = {Int. J. Mod. Phys. B},
  volume = {4},
  pages = {239--271},
  doi = {10.1142/S0217979290000139}
}

@article{Wen:1989zg,
  title = {Vacuum Degeneracy of Chiral Spin States in Compactified Space},
  author = {Wen, X. G.},
  date = {1989},
  journaltitle = {Phys. Rev. B},
  volume = {40},
  pages = {7387--7390},
  doi = {10.1103/PhysRevB.40.7387}
}

@article{Wen:2015qgn,
  title = {A Theory of 2+{{1D}} Bosonic Topological Orders},
  author = {Wen, Xiao-Gang},
  date = {2016},
  journaltitle = {Natl. Sci. Rev.},
  volume = {3},
  number = {1},
  eprint = {1506.05768},
  eprinttype = {arXiv},
  eprintclass = {cond-mat.str-el},
  pages = {68--106},
  doi = {10.1093/nsr/nwv077}
}

@article{Wilczek:1982wy,
  title = {Quantum Mechanics of Fractional-Spin Particles},
  author = {Wilczek, Frank},
  date = {1982},
  journaltitle = {Phys. Rev. Lett.},
  volume = {49},
  pages = {957--959},
  doi = {10.1103/PhysRevLett.49.957}
}

@article{wittenQuantumFieldTheory1989,
  title = {Quantum Field Theory and the {{Jones}} Polynomial},
  author = {Witten, Edward},
  date = {1989-01},
  journaltitle = {Commun. Math. Phys.},
  volume = {121},
  number = {3},
  pages = {351--399},
  publisher = {Springer},
  doi = {10.1007/BF01217730},
  langid = {english}
}

@online{gannonOnlineDatabaseVOA,
  title = {The Online Database of Vertex Operator Algebras and Modular Categories},
  author = {Gannon, Terry and Höhn, Gerald and Yamauchi, Hiroshi and others},
  version = {0.5},
  url = {https://www.math.ksu.edu/~gerald/voas/},
  urldate = {2026-10-17}
}
